\documentclass[11pt, a4paper, oneside]{book}

\usepackage{mathrsfs}
\usepackage{amsmath}
\usepackage{amssymb}
\usepackage{stmaryrd}
\usepackage{amsfonts}
\usepackage{ifthen}     

\usepackage{yfonts}
\usepackage{graphicx, color}

\usepackage{rotating}
\usepackage{makeidx}                          
\usepackage{yhmath} 
\usepackage{theorem}
\usepackage{multicol}
\usepackage{subfigure}

\usepackage[all]{xy}

\theoremstyle{break}
\newtheorem{theorem}{Theorem}[chapter]
\newtheorem{definition}[theorem]{Definition}
\newtheorem{lemma}[theorem]{Lemma}
\newtheorem{proposition}[theorem]{Proposition}
\newtheorem{corollary}[theorem]{Corollary}
\newtheorem{conjecture}[theorem]{Conjecture}

\newtheorem{example}{Example}[chapter]
\newtheorem{problem}{Problem}
\theoremstyle{plain}
\newtheorem{problem-footnote}[problem]{Problem}

\definecolor{gray1}{gray}{0.9}
\definecolor{gray2}{gray}{0.7}
\definecolor{gray3}{gray}{0.5}

\def\real{\mathbb{R}}
\def\rational{\mathbb{Q}}
\def\complex{\mathbb{C}}
\def\integers{\mathbb{Z}}
\def\naturals{\mathbb{N}}

\def\01{\{0,1\}}
\def\oo{\"o}

\def\endo{\mbox{End}}  
\def\sgn{\mbox{sgn}}  

\newcommand{\floor}[1]{\lfloor{#1}\rfloor}          
\newcommand{\ket}[1]{|#1\rangle}                    
\newcommand{\bra}[1]{\langle#1|}                    
\newcommand{\proj}[1]{|#1\rangle\langle#1|}         
\newcommand{\braket}[2]{\langle #1|#2\rangle}       

\newcommand{\be}{\begin{equation}}
\newcommand{\ee}{\end{equation}}
\newcommand{\bea}{\begin{eqnarray}}
\newcommand{\eea}{\end{eqnarray}}
\newcommand{\bestar}{\begin{equation*}}
\newcommand{\eestar}{\end{equation*}}
\newcommand{\beastar}{\begin{eqnarray*}}
\newcommand{\eeastar}{\end{eqnarray*}}

\renewcommand{\ker}{\text{Ker}\;}
\newcommand{\diag}{\text{Diag}\;}
\newcommand{\Span}{\text{Span}\;}
\newcommand{\im}{\text{Im}\;}
\newcommand{\spec}{\text{Spec}\;}
\newcommand{\tr}{\text{Tr}\, }
\def\openone{\leavevmode\hbox{\small1\normalsize\kern-.33em1}}
\def\notimplies{\hspace{0.15cm}\Arrownot\hspace{-0.15cm}\implies}
\newcommand{\id}{\openone}    
\newcommand{\idmap}{\id}    
\newcommand{\half}{\frac{1}{2}}
\newcommand{\rank}{\text{Rank}}

\newcommand{\I}{I_{acc}}        
\newcommand{\gl}{\mathfrak{gl}}  
\newcommand{\su}{\mathfrak{su}}   
\newcommand{\fu}{\mathfrak{u}}
\newcommand{\ft}{\mathfrak{t}}

\newcommand{\fg}{\mathfrak{g}}
\newcommand{\fh}{\mathfrak{h}}

\newcommand{\kron}{{\rm{Kron}}}                    
\newcommand{\KRON}{{\rm{KRON}}}                    
\newcommand{\qmp}{{\rm{CLS}}}                     
\newcommand{\holmap}{\cP(\GL(\complex^d))}

\renewcommand{\min}{{\rm min}}
\newcommand{\bounded}{{\rm{B}}}                     
\newcommand{\positive}{{\rm{B}^+}}

\newcommand{\states}{{\rm{S}}}
\newcommand{\sym}{{\rm{Sym}}}    
\newcommand{\sep}{\states_{\rm{Sep}}}
\newcommand{\ppt}{\states_{\rm{PPT}}}

\newcommand{\ext}{\states_{\rm{Ext}}}

\newcommand{\ens}{\states_{\rm{Ens}}}
\newcommand{\locc}{\rm{LOCC}}
\newcommand{\lopc}{\rm{LOPC}}
\newcommand{\loq}{\rm{LOq}}
\newcommand{\U}{{\rm U}}
\newcommand{\GL}{{\rm{GL}}}
\newcommand{\SL}{{\rm{SL}}}
\newcommand{\SO}{{\rm{SO}}}
\newcommand{\M}{{\rm{M}}}
\newcommand{\unitaries}{{\rm U}}
\newcommand{\B}{{\rm{B}}}
\newcommand{\T}{{\rm{T}}}
\newcommand{\sunitaries}{{\rm SU}}
\newcommand{\SU}{{\rm SU}}

\newcommand{\sq}{E_{sq}}

\newcommand{\hil}{{\cal H}}
\newcommand{\hilH}{{\cal H}}                    
\newcommand{\hilK}{{\cal K}}

\newcommand{\kpsi}{\ket{\psi}}

\newcommand{\ppsiminus}{P_{\psi^-}}
\newcommand{\bpsi}{\bra{\psi}}
\newcommand{\kphi}{\ket{\phi}}
\newcommand{\bphi}{\bra{\phi}}

\newcommand{\cA}{{\cal A}}                           
\newcommand{\cB}{{\cal B}}
\newcommand{\cC}{{\cal C}}
\newcommand{\cD}{{\cal D}}
\newcommand{\cE}{{\cal E}}

\newcommand{\cH}{{\cal H}}
\newcommand{\cI}{{\cal I}}

\newcommand{\cK}{{\cal K}}

\newcommand{\cN}{{\cal N}}
\newcommand{\cO}{{\cal O}}
\newcommand{\cP}{{\cal P}}

\newcommand{\cR}{{\cal R}}
\newcommand{\cS}{{\cal S}}

\newcommand{\cU}{U}
\newcommand{\cV}{V}

\newcommand{\cX}{{\cal X}}
\newcommand{\cY}{{\cal Y}}
\newcommand{\cZ}{{\cal Z}}

\newcommand{\F}{F} 

\newenvironment{proof}
{\noindent {\bf Proof. }} {{\hfill $\Box$} \\
\noindent}

\newcommand{\captionfonts}{\footnotesize}

\makeatletter  
\long\def\@makecaption#1#2{%
  \vskip\abovecaptionskip
  \sbox\@tempboxa{{\captionfonts #1: #2}}%
  \ifdim \wd\@tempboxa >\hsize
    {\captionfonts #1: #2\par}
  \else
    \hbox to\hsize{\hfil\box\@tempboxa\hfil}%
  \fi
  \vskip\belowcaptionskip}
\makeatother   

\newlength{\quoteLength}
\newenvironment{quotewho}
{
\begin{flushright}
\begin{minipage}[r]{\quoteLength}
\begin{flushright}
\it
}
{
\end{flushright}
\end{minipage}
\end{flushright}
\smallskip
}

\newcounter{protoCount}
\newcounter{protoList}
\newsavebox{\tmpbox}
\newlength{\protobox}
\newenvironment{protocol}[2]{
\bigskip
\addtocounter{protoCount}{1} \noindent \begin{lrbox}{\tmpbox}
\setlength{\protobox}{\textwidth} \addtolength{\protobox}{-0.5cm}
\begin{minipage}[c]{\protobox}
\begin{bfseries}Protocol \theprotoCount: #1\end{bfseries}
\ifthenelse{\equal{#2}{\empty}}{}{\\Prerequisite: #2}
\begin{list}{\begin{bfseries}\arabic{protoList}:\end{bfseries}}
{\usecounter{protoList}} }{
\end{list}
\end{minipage}\end{lrbox}
\fbox{\usebox{\tmpbox}}
}

\begin{document}

\frontmatter
\begin{titlepage}
    \begin{center}
    \vspace*{0cm}

    {\LARGE \bf The Structure of Bipartite Quantum States}


    \vspace{0.5cm}

    { \Large \bf Insights from Group Theory and
    Cryptography}

    \vspace{2cm}

\begin{figure}[h]
\begin{center}
\includegraphics[height=12cm]{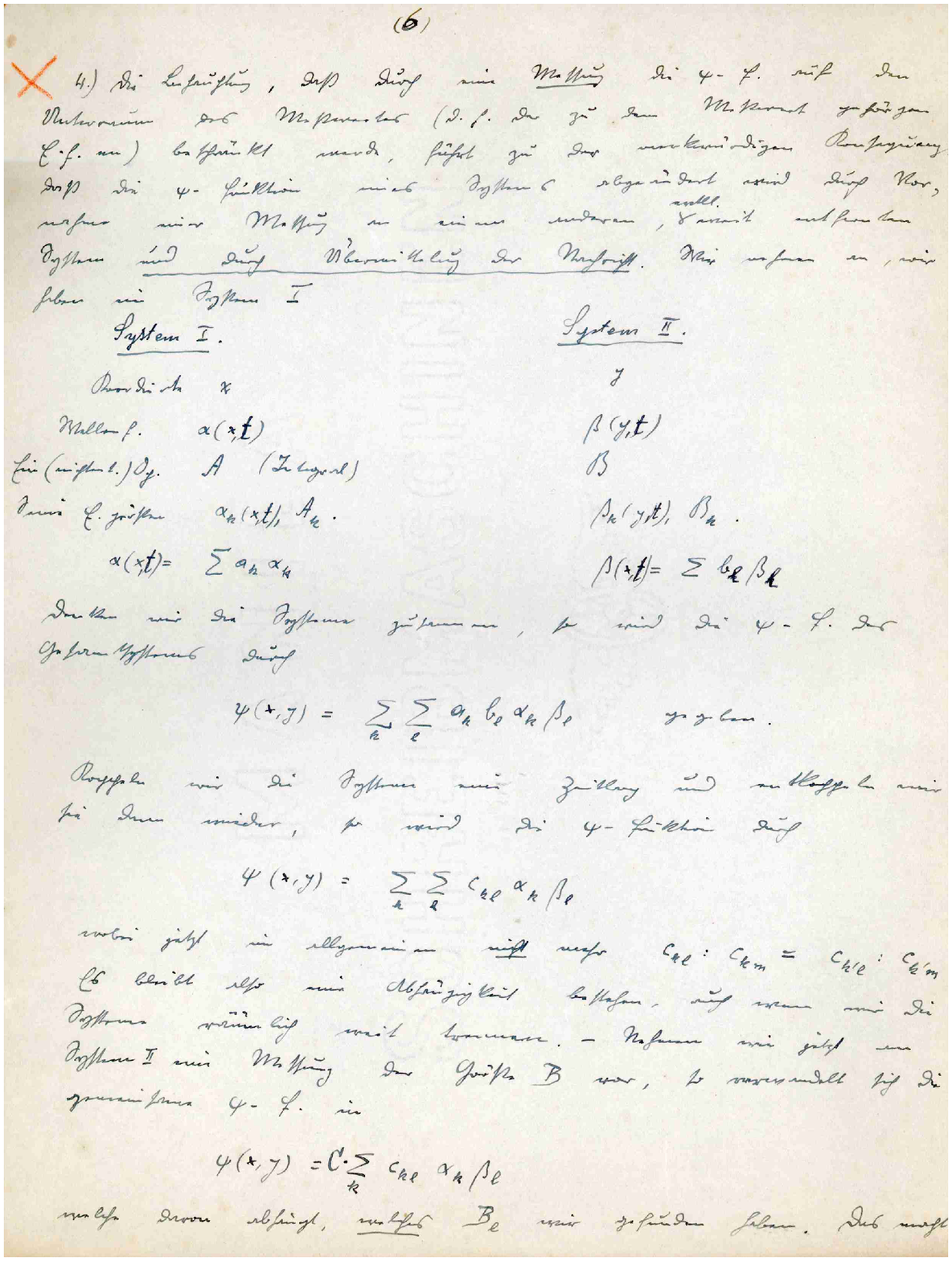}
\end{center}
\end{figure}

\end{center}

\vspace{\stretch{1}}  

\begin{flushright}
This dissertation is submitted for the \\
degree of Doctor of Philosophy \vspace{0.5cm}

Matthias Christandl \\
Selwyn College \\
University of Cambridge
\end{flushright}
%
\end{titlepage}
\chapter*{}
\thispagestyle{empty}

\vspace{3cm}

This dissertation is the result of my own work and includes
nothing which is the outcome of work done in collaboration except
where specifically indicated in the text.

\chapter{Abstract}
Currently, a rethinking of the fundamental properties of quantum
mechanical systems in the light of quantum computation and quantum
cryptography is taking place. In this PhD thesis, I wish to
contribute to this effort with a study of the \emph{bipartite
quantum state}. Bipartite quantum-mechanical systems are made up
of just two subsystems, $A$ and $B$, yet, the quantum states that
describe these systems have a rich structure. The focus is
two-fold: Part~\ref{part-group} studies the relations between the
spectra of the joint and the reduced states, and in
part~\ref{part-crypto}, I will analyse the amount of entanglement,
or quantum correlations, present in a given state.

In part~\ref{part-group}, the mathematical tools from group theory
play an important role, mainly drawing on the representation
theory of finite and Lie groups and the Schur-Weyl duality. This
duality will be used to derive a one-to-one relation between the
spectra of a joint quantum system $AB$ and its parts $A$ and $B$,
and the Kronecker coefficients of the symmetric group. In this way
the two problems are connected for the first time, which makes it
possible to transfer solutions and gain insights that illuminate
both problems.

Part~\ref{part-crypto} of this thesis is guided by the question:
How can we measure the strength of entanglement in bipartite
quantum states? The search for an answer starts with an extensive
review of the literature on entanglement measures. I will then
approach the subject from a cryptographic point of view. The parts
$A$ and $B$ of a bipartite quantum state are given to the
cooperative players Alice and Bob, whereas a purifying system is
given to the eavesdropper Eve, who aims at reducing the
correlation between Alice and Bob. The result is a new measure for
entanglement: \emph{squashed entanglement}. Squashed entanglement
is the only known strongly superadditive, additive and
asymptotically continuous entanglement measure. These properties,
as well as the simplicity of their proofs, position squashed
entanglement as a unique tool to study entanglement.

\chapter{Acknowledgements}
I first want to thank my supervisor Artur Ekert for his support
and encouragement, for our collaborations and for the inspiration
that his thinking has provided for my work. I wish to thank Graeme
Mitchison for many enjoyable weekend discussions which were
essential to the group-theoretic part of this thesis, and to
Andreas Winter with whom I invented \emph{squashed entanglement}.
I also wish to thank Harry Buhrman, Nilanjana Datta, Renato Renner
and Stephanie Wehner for sharing their unique perspectives and
ideas, as well as Claudio Albanese, Patrick Hayden, Dagomir
Kaszlikowski, Alastair Kay, Andrew Landahl, Hoi-Kwong Lo, Falk
Unger and Stefan Wolf for our joint collaborations.

\sloppy My time as a research student has benefited from many
scientific discussions with colleagues. For discussions on the
topics presented in this thesis I would like to thank Jonathan
Barrett, Jens Eisert, Joseph Emerson, Sugi Furuta, Aram Harrow,
Karol, Micha{\l} and Pawe{\l} Horodecki, Adrian Kent, Alexander
Klyachko, Debbie Leung, Norbert L\"ut\-ken\-haus, Toshio Ohshima,
Jonathan Oppenheim, Martin Plenio, Rob Spekkens, Tony Sudbery and
Reinhard Werner for their open exchange of ideas. Special thanks
go to John Preskill for inviting me to Caltech, and to Allen
Knutson whom I met there and who shared with me his insights into
the geometric aspects of this work.

I thank all members of the Centre for Quantum Computation,
especially Dimitris Angelakis and Lawrence Ioannou, and my friends
in Cambridge and at Selwyn College for accepting me and creating
such a pleasant atmosphere. I am also grateful to Alexander Zartl
from the University of Vienna for his assistance with
Schr\"odinger's manuscripts, and Maria Diemling and Josef G\"otten
for helping to transcribe \emph{Kurrentschrift}.

This work has improved tremendously with respect to structure and
language over the past months through the help of Henriette
Steiner. I am indebted to her but also to Jens Eisert, Aram
Harrow, Sean McHugh, Graeme Mitchison, Stephanie Wehner and
Andreas Winter for their careful reading and commenting on drafts
of this thesis. Furthermore, I acknowledge the financial support
of the UK's EPSRC and a Doktorandenstipendium of the German
Academic Exchange Service.

I dedicate this thesis to my family and to Henriette for their
love and support.


\chapter{Foreword}
Quantum theory was born in the early days of the 20th century and
developed into a mature physical theory in the 1920s and 1930s. It
was to influence the 20th century, not only with respect to world
politics, but also on a philosophical level: the elementary
constituents that make up our world obey quantum mechanical laws
which run counter what a human being is naturally exposed to.
Despite immense practical success, the understanding of quantum
mechanics is far from complete and remains an important goal of
physics research. With this thesis I wish to make a contribution
to the understanding of a quantum system made up of two subsystems
that challenges our imagination: the bipartite quantum system. An
early account of such a composite quantum system and its puzzling
behaviour is the manuscript by Erwin Schr\"odinger from 1932,
which is reproduced on the cover of this thesis. The possibility
of paradoxical behaviour of this system was pointed out by Albert
Einstein, Boris Podolsky and Nathan Rosen in 1935~\cite{EiPoRo35},
named \emph{entanglement} by Schr\"odinger in the same
year~\cite{Schroedinger35a} and qualitatively analysed by John
Bell in 1964~\cite{Bell64}. First experiments to test the nonlocal
nature of quantum mechanics were proposed and performed by John F.
Clauser and coworkers~\cite{CHSH69, ClaShi78} and Alain Aspect and
coworkers~\cite{AsDaRo82}. Later in the 1980s and 1990s
entanglement was taken up again and now forms an essential part of
quantum information theory, the field to which this research
belongs. Quantum computation, quantum information theory and
quantum cryptography seek to unify quantum mechanics with their
respective classical fields of research in order to understand the
physics of information and to develop viable technological
applications.

This PhD thesis analyses bipartite quantum states from a
group-theoretic and a cryptographic perspective. This double focus
is reflected in the structure of this thesis:

\begin{itemize}
\item Part~\ref{part-group} -- Insights from Group Theory: a study of the relation between representation theory
and quantum states.
\item Part~\ref{part-crypto} -- Insights from Cryptography: the proposal of \emph{squashed entanglement}, an cryptographically motivated additive
entanglement measure.
\end{itemize}

\noindent Part~\ref{part-group} presents a way to transform
relations between spectra of quantum states into relations of
representations of the symmetric and unitary groups. This
transformation not only sheds light on quantum states from the
unexpected angle of group theory, but also leads to consequences
for group representations. I have divided part~\ref{part-group}
into two chapters. Chapter~\ref{chapter-structure} provides the
group-theoretical background needed in
chapter~\ref{chapter-relation}, focusing on representation theory
of the symmetric and unitary group and their relation known as
Schur-Weyl duality. In chapter~\ref{chapter-relation}, I will
present my main research result: the asymptotic equivalence of the
problem of determining the spectral structure of bipartite quantum
states and the problem of deciding whether or not an irreducible
representation of the symmetric group is contained in the tensor
product of two irreducible representations of the same group. The
chapter also contains results on issues of convexity and finite
generation of the mentioned problems as well as on the relation
between Horn's problem and the Littlewood-Richardson coefficients.

The importance of spectra for the characterisation of pure state
entanglement was noted by Nielsen in 1999. In the case of mixed
quantum states, spectra alone do not determine the structure of
the states and therefore other approaches are needed to reach an
understanding of entanglement. I propose to embed a quantum state
consisting of part $A$ and part $B$ in a cryptographic scenario:
parts $A$ and $B$ are given to cooperative players, whereas a
third and malicious player holds an extension of the system. The
malicious player is allowed to squash out the quantum correlations
between the cooperating players and thereby defines a new measure
for entanglement: squashed entanglement. Part~\ref{part-crypto} of
this thesis is divided into two chapters. The first chapter,
chapter~\ref{chapter-entanglement}, is an extensive review of the
theory of entanglement measures, and provides the background for
chapter~\ref{chapter-squashed}, in which I will define squashed
entanglement. The remainder of part~\ref{part-crypto} is devoted
to the study of properties of squashed entanglement and its
consequences for quantum information theory and cryptography.

%
\vspace{0.5cm} \noindent This thesis contains work which I have
done during the three years as a PhD student at Selwyn College and
the Centre for Quantum Computation, University of Cambridge.
During this period I have had the pleasure to collaborate with
many researchers in the field. Results of some of these
collaborations are included in this dissertation and are marked
out in the text itself. In part~\ref{part-group}, I discuss the
relation of group theory and quantum states. Part of this
collaborative work has appeared in,
\begin{itemize}
\item M.~Christandl and G.~Mitchison.
\newblock The spectra of density operators and the {K}ronecker coefficients of
  the symmetric group.
\newblock {\em Communications in Mathematical Physics}, 2005.
\newblock to appear, quant-ph/0409016.
\end{itemize}
Theorem~\ref{theorem-converse} has been obtained in collaboration
with Aram Harrow and Graeme Mitchison. In part~\ref{part-crypto}
of this dissertation, I discuss the impact of cryptography on
entanglement. The work with Andreas Winter on squashed
entanglement has been the topic of two publications.
\begin{itemize}
\item M.~Christandl and A.~Winter.
\newblock Squashed entanglement -- an additive entanglement measure.
\newblock {\em Journal of Mathematical Physics}, 45(3):829--840, 2004.
\newblock quant-ph/0308088.
\item M.~Christandl and A.~Winter.
\newblock Uncertainty, monogamy and locking of quantum correlations.
\newblock {\em IEEE Transactions on Information Theory}, 51(9):3159-3165, 2005,
\newblock and in Proceedings of ISIT 2005, pp. 879-883.
\newblock quant-ph/0501090.
\end{itemize}
Patches of work from other collaborations have also been included
into this thesis: the historical note in
subsection~\ref{section-history} appeared in a joint paper with
Daniel Oi~\cite{OiChr03} and the work on string commitment,
subsection~\ref{subsec-cheat-sensitive}, was done in collaboration
with Harry Buhrman, Patrick Hayden, Hoi-Kwong Lo and Stephanie
Wehner~\cite{BCHLW05}.

Cover and back of this thesis feature a two-page manuscript by
Erwin Schr\"odinger, which Lawrence Ioannou and I discovered in
the Schr\"odinger archive Vienna in March 2003. The manuscript
describes the phenomenon entanglement very accurately and is
contained in a folder of the year
1932\index{entanglement!Schr\"odinger manuscript}. To my knowledge
it is the first known written document of this phenomenon.
Reproduced with kind permission of the {\"O}sterreichische
Zentralbibliothek f\"ur Physik.
\tableofcontents
%
%
%
%

\chapter{Preliminaries} \label{chapter-intro}

The Preliminaries consist of two sections. In the first one, the
basic formalism of non-relativistic quantum mechanics is
introduced and fundamental results concerning quantum states and
entropy are reviewed. Good introductions to quantum mechanics are
the book by Asher Peres~\cite{Peres93} and the lecture notes by
Klaus Hepp (in German)~\cite{Hepp00}. As a reference for quantum
computation and quantum information theory I recommend the book by
Michael A. Nielsen and Isaac L. Chuang~\cite{NieChu00Book} as well
as John Preskill's lecture notes~\cite{preskill98}. The second
section introduces background material from the theory of finite
groups that can be found in many standard textbooks. I used the
books by Barry Simon~\cite{Simon96Book}, William Fulton and Joe
Harris~\cite{FultonHarris91}, and Roe Goodman and Nolan R.
Wallach~\cite{GooWal98}, which also provided the representation
theory needed in part~\ref{part-group}.

\section*{Quantum Mechanics\index{quantum mechanics}} \label{sec-intro-QM}

\subsection*{The Wavefunction and Schr\oo dinger's Equation}

In quantum mechanics, the state of a physical system at time $t$
is given by a \emph{wavefunction\index{wavefunction}} or
\emph{pure quantum state\index{quantum state!pure}}  $\psi(t)$, a
norm-one vector in a separable complex Hilbert space\index{Hilbert
space} $\hil$ with scalar product $(\cdot, \cdot)$. The time
evolution of the wavefunction is given by the \emph{Schr\oo dinger
equation} ($\hbar=1$)
\be \label{eq-intro-Schroedinger} H(t) \psi(t) = \partial_t \psi(t), \ee
where the \emph{Hamiltonian\index{Hamiltonian}} $H(t)$ is a
Hermitian operator\index{operator!Hermitian}, which is piecewise
continuous in $t$. An operator $H$ is Hermitian if and only if
$(v,Hv)=(Hv,v)$ for all $v \in \cH$.


\sloppy Throughout this thesis I will only consider one type of
Hilbert space: finite-dimensional complex vector spaces with the
inner product given by the dot-product, i.e. $\cH\cong \complex^d$
for some $d < \infty$. In \emph{Dirac notation\index{Dirac
notation}}, a vector $\psi \in \hilH$ is written as $\kpsi$
whereas the adjoint vector to $\kpsi$ is denoted by $\bpsi$. The
scalar product $(v,w)$ then reads $\braket{v}{w}$ and the
projector onto $\psi$, $P_{\kpsi}$, is often written as
$\proj{\psi}$.

\fussy The evolution of a state at time $t_0$ to a state at later
time $t_1$, $-\infty<t_0\leq t_1 <\infty$, assumes the form \bestar
\label{eq-into-unitary} U(t_1, t_0) \ket{\psi(t_0)}
=\ket{\psi(t_1)},\eestar where \bestar  U(t_1, t_0)=\lim_{\Delta t
\rightarrow 0} e^{-i H(t_1-\Delta t) \Delta t} \cdots e^{-i
H(t_0+\Delta t)\Delta t} e^{-i H(t_0)\Delta t} \eestar is unitary.
In quantum computation and quantum information theory it often
suffices to restrict the attention to discrete time steps and,
rather than referring to a Hamiltonian, regard the evolution as a
sequence of unitary operations.

\subsection*{The Measurement Postulate\index{measurement!postulate}}

The previous section defined states and evolution of
quantum-mechanical systems. As human beings, however, we appear to
be classical objects. This raises the question of how we interact
with quantum mechanical systems or how we gather information from
a quantum mechanical system. The standard approach to this problem
postulates the measurement. A
\emph{measurement\index{measurement}} is defined by a Hermitian
operator $O$ acting on $\cH$. Upon measuring the system in state
$\kpsi$ with $O$, we obtain an outcome $o_i$, which is one of the
eigenvalues of $O$. If $P_i$ denotes the projector onto the
eigenspace of $O$ with eigenvalue $o_i$, the probability with
which outcome $o_i$ will appear is given by $p_i=\bpsi P_i \kpsi$.
The quantum state after the measurement conditioned on obtaining
outcome $o_i$ is known as the \emph{post measurement state} and
given by $\ket{\psi_i}=P_i \kpsi / \sqrt{o_i}.$

\subsection*{Tensor Products\index{tensor product} and Composite Systems\index{composite system}}

The ancient Greeks postulated that the material world consists of
microscopic indivisible parts, which they called atoms. Since
then, the study of science has been reduced to the study of its
smallest parts and theoretical rules for combining and dividing
the basic elements have been put forward. In the following, the
quantum mechanical rules for combining and dividing will be
introduced: the tensor product and the partial trace operation.

A system consisting of several parts is called a
\emph{composite\index{quantum state!composite}} or
\emph{multipartite system\index{quantum state!multipartite}}. The
state space of a composite system is the tensor product of the
Hilbert spaces of the individual subsystems. A formal definition
of the tensor product is as follows:

\begin{definition}
Let $\cH, \cK, \cX$ be vector spaces and suppose $f: \cH \times
\cK \rightarrow \cX$ is a bilinear mapping. The pair $(\cX, f)$ is
called a \emph{tensor product} of $\cH$ and $\cK$ if
\begin{itemize}
\item $\im f=\cX$
\item If $h: \cH \times \cK \rightarrow \cY$ is a bilinear mapping
into a vector space $\cY$, then there is a linear map $g: \cX
\rightarrow \cY$ such that $h=g \circ f$.
\end{itemize}
Cast in the form of a diagram, the second condition demands that
\begin{displaymath}
\xymatrix{ \cH \times \cK  \ar[d]^f \ar[r]^h &  \cY\\
\cX \ar[ur]_g}
\end{displaymath}
commutes. Since all such tensor products are equivalent up to
isomorphism it is customary to drop the map $f$ and simply write
$\cH \otimes \cK$ for the tensor product of $\cH$ and $\cK$.
\end{definition}

\noindent There are a number of equivalent definitions of the
tensor product; a nice introduction can be found in the book by
Werner H. Greub on multilinear algebra~\cite{Greub67}. The
advantage of the above definition is its independence of a choice
of basis for $\cH$ and $\cK$, and the disadvantage is the indirect
definition, which makes the actual construction difficult to
envision. Reversing drawbacks and advantages, I will now give a
more concrete construction for finite-dimensional spaces. Let
$\ket{v}, \ket{v_1}, \ket{v_2} \in \cH$, $\ket{w}, \ket{w_1},
\ket{w_2} \in \cK$ and $\alpha \in \complex$. Then, $f$ is given
by
$$ f: \cH \times \cK \rightarrow \cH \otimes \cK$$
$$ f(\ket{v}, \ket{w})=\ket{v} \otimes \ket{w},$$
\noindent where the symbols $\ket{v} \otimes \ket{w}$ obey the
following linearity properties
$$ (\ket{v_1}+\ket{v_2}) \otimes \ket{w}= \ket{v_1} \otimes \ket{w}+\ket{v_2} \otimes \ket{w}$$
$$ \ket{v} \otimes (\ket{w_1}+\ket{w_2}) = \ket{v} \otimes \ket{w_1} +\ket{v} \otimes \ket{w_2}$$
$$ \alpha (\ket{v} \otimes \ket{w})= (\alpha \ket{v}) \otimes \ket{w}=\ket{v} \otimes (\alpha
\ket{w})
$$
Given orthonormal bases $\{ \ket{e_i}\}$ of $V$ and
$\{\ket{f_j}\}$ of $W$, a basis for $V\otimes W$ is given by
$\ket{e_i} \otimes \ket{f_j}$ often abbreviated by
$\ket{e_i}\ket{f_j}$. Every $\kpsi \in \hilH\otimes \hilK$ then
assumes the form
\be \sum_i \alpha_{ij} \ket{e_i}\ket{f_j},\ee
for some $\alpha_{ij}$. Quantum states of the form
\be \label{eq-product-state} \kpsi^{\cH
\cK}=\kpsi^\cH\otimes \kpsi^\cK \ee are called \emph{pure product
states\index{quantum state!pure product}}. Here, the state
describing subsystem $\cH$ is given by $\kpsi^\cH$. Note, however,
that it is not immediately clear how to describe the state of
$\cH$ if the total system is not in a product state. As we will
see below, the appropriate description is given by a \emph{mixed
quantum state}, generalising the notion of a pure quantum state.

\subsection*{Mixed Quantum States\index{quantum state!mixed}}
\label{mixed-state} Since classical information about the state
$\kpsi$ can only be obtained from measurements and since
measurements only depend on $P_{\kpsi}$, the projector onto
$\kpsi$, the latter is often referred to as the state of the
system. $P_{\kpsi}$ is an operator on $\cH$ and as such the
concept of pure quantum states will generalise to mixed quantum
states. Since $\hilH$ is finite dimensional, the set of operators
on $\hilH$ coincides with the set \emph{bounded
operators\index{operator!bounded}} and will be denoted by
$\bounded(\hilH)$. The \emph{trace} of an operator $O$ is most
conveniently defined in terms of an orthonormal basis
$\{\ket{e_i}\}$ of $\hilH$,
\bestar \tr \ O=\sum_i \bra{e_i} O \ket{e_i},\eestar
and the \emph{Hilbert-Schmidt inner product\index{Hilbert-Schmidt
inner product}} of two operators $O$ and $P$ is
\bestar (O, P)_{HS}= \tr \ O^\dagger P.\eestar
A bounded operator $O$ is \emph{positive} if and only if
$\bra{v}O\ket{v}\geq 0$ for all $v \in \hilH$ and the set of
positive operators\index{operator!positive} is denoted by
$\positive(\hilH)$.

Let us come back to the question of how to define the state on
$\cH$ when there is a non-product state $\kpsi$ describing the
total system $\cH\otimes \cK$. The answer is induced by the
\emph{partial trace\index{partial trace}} which is given in terms
of a basis for $\bounded(\hilH)$ by
\be \label{eq-intro-partialtrace}
\tr_\cK \ket{e_i}\bra{e_j} \otimes \ket{f_k}\bra{f_l}=
\ket{e_i}\bra{e_j} \tr \ket{f_k}\bra{f_l} \ee and extends to the
set of all operators by linearity. Since $\proj{\psi}$ is a
positive operator, the \emph{reduced state\index{quantum
state!reduced}} or simply \emph{state} on system $\cH$,
$$ \rho^\cH=\tr_\cK \proj{\psi},$$
is positive, too. The partial trace operation is unique in the
sense that it is the only operation that results in the correct
measurement statistics on the reduced states~\cite[page
107]{NieChu00Book}. $\rho^{\cH}$ has trace one, since $\tr_{\cH}
\tr_{\cK} \proj{\psi}= \tr \proj{\psi}=1$. In fact, the converse
is also true: for every trace-one positive operator $\rho$ on
$\cH$, there is a Hilbert space $\hilK$ and a pure quantum state
$\kpsi \in \hilH\otimes \hilK$ such that $\rho=\tr_\cK P_{\kpsi}$.
To see this write $\rho$ in its eigenbasis $\rho=\sum_i \lambda_i
\proj{e_i}$ and define $\kpsi=\sum_i \sqrt{\lambda} \ket{e_i}
\otimes \ket{f_i} \in \hilH\otimes \hilK$ for $\hilK=\complex^n$,
$\hilH=\complex^m$ and $m \leq n$. $\kpsi$ is called a
\emph{purification\index{purification}} for $\rho$. It is an
important detail that all purifications are equivalent, i.e. given
two purifications $\ket{\psi_1} \in \cH\otimes \cK_1$ and
$\ket{\psi_2} \in \cH\otimes \cK_2$ of $\rho$, there is an
isometry $U: \cK_1 \rightarrow \cK_2$ such that
$$ \id \otimes U \ket{\psi_1} =\ket{\psi_2}.$$
In summary, a \emph{mixed quantum state}, hereby referred to as a
\emph{quantum state}, is a positive operator in $\bounded(\hilH)$
with trace equal to one. The set of mixed quantum states is
denoted by $\states(\hilH)$.

\subsection*{Classical Quantum States}

On several occasions in part~\ref{part-crypto} classical and
quantum scenarios will be encountered side-by-side. The way to
deal with this situation is to regard probability distributions as
quantum states. The probability distribution $P_X$ of a random
variable $X$ with range $\cX, |\cX| =d<\infty$, can naturally be
written as a \emph{classical state\index{classical state}} in the
form
$$ \rho^X=\sum_{x \in \cX} P_X(x) \proj{x},$$
where $\{\ket{x}\}$ is an orthonormal basis of $\complex^d$.
Conversely, every quantum state $\rho$ can be regarded as a
classical state if $\{\ket{x}\}$ is an eigenbasis of $\rho$. When
multiple parties are involved and the basis choice is restricted
to local orthogonal bases -- a natural restriction if only local
operations or local operations assisted by classical communication
are allowed -- not every quantum state can be regarded as a
classical state. It is therefore convenient to make the following
hybrid definition: $\rho^{C_1\cdots C_m Q_1 \cdots Q_n}$ is a
$\underbrace{c\cdots c}_m\underbrace{q\cdots
q}_n$-state\index{quantum state!cq-state}\label{def-ccq-state} if
there exist local bases on the first $m$ systems such that
    $$
        \rho^{C_1\cdots C_m Q_1 \cdots Q_n}=\sum_{c_1, \cdots, c_m}
        P_{C_1\cdots C_m}(c_1\cdots c_m) \proj{c_1\cdots c_m}\otimes
        \rho^{Q_1\cdots Q_n}_{c_1\cdots c_m},
    $$
for random variables $C_1, \ldots, C_m$ and quantum states
$\rho^{Q_1\cdots Q_n}_{c_1\cdots c_m}$.

\subsection*{Quantum Operations}

Physical operations transforming quantum states are modeled by
\emph{completely positive trace preserving (CPTP)}\index{CPTP}
maps. A positive map $\Lambda$ is a linear map
$$ \Lambda: \cB^+(\cH) \rightarrow \cB^+(\cH').$$
$\Lambda$ is called \emph{completely positive
(CP)}\index{completely positive (CP)} if $\Lambda \otimes
\idmap_\cK$ is positive for all $\cK$. A completely positive trace
preserving map is therefore a map
$$ \Lambda: \states(\cH) \rightarrow \states(\cH'),$$
where $\states(\cH)$ is the set of quantum states on $\cH$. CPTP
maps are precisely the maps that can be composed out of the
following three steps:
\begin{enumerate}
\item appending an uncorrelated pure state: $\Lambda_{append}: \states(\cH)
\rightarrow \states(\cH\otimes \cK)$ with $\rho \mapsto
\Lambda_{append}(\rho)=\rho\otimes \proj{\psi}$.
\item applying a unitary transformation $\Lambda_{unitary}: \states(\cH\otimes \cK)
\rightarrow \states(\cH\otimes \cK)$ with $\rho \mapsto
\Lambda_{unitary}(\rho)=U\rho U^\dagger$ for some $U \in
\unitaries(\cH\otimes \cK)$.
\item tracing out over a subsystem: $\Lambda_{trace}: \states(\cH\otimes \cK)
\rightarrow \states(\cH)$ with $\rho \mapsto
\Lambda_{trace}(\rho)=\rho^{\cH}=\tr_\cK \rho $.
\end{enumerate}
A theorem by Kraus gives a different characterisation: a map is
completely positive (CP) if and only if it can be written in the
form
\be \label{eq-intro-Kraus}\Lambda(\rho)=\sum_i M_i \rho M_i^\dagger,\ee
$M_i$ are called the \emph{Kraus operators\index{operator!Kraus}}
of $\Lambda$ and \emph{trace preserving (TP)}\index{trace
preserving} if in addition $$\sum_i M_i^\dagger M_i =\id.$$

If the measurement postulate is taken into account\footnote{Note
that the postulated measurement consists of a set of projectors
$\{P_i\}$. The outcomes are given by values $o_i$ and are only
referred to by their index $i$. This type of measurement is known
as \emph{projective\index{measurement!projective}} or \emph{von
Neumann measurement\index{measurement!von Neumann}}.}, maps are
obtained that transform a state $\rho$ into an ensemble of states
$\rho_i$ that appear with probability $p_i$. Such maps are known
as \emph{quantum instruments\index{quantum instrument}} and are
described by a set of positive maps $\{\Lambda_i\}$, such that
$\sum_i \Lambda_i$ is a CPTP map.

If one is only interested in the measurement outcome and the
corresponding probability
$$ p_i = \tr \Lambda_i (\rho)= \tr \sum_j M_{i,j} \rho M_{i,j}^\dagger= \tr \big( \sum_j M_{i,j}^\dagger M_{i,j} \big) \rho,$$
it suffices to consider a set of positive operators $E_i:=\sum_j
M_{i,j}^\dagger M_{i,j}$ with $\sum_i E_i =\id$ and consequently
speak of a \emph{positive operator valued measure
(POVM)}\index{POVM}\index{measurement!POVM} $\{E_i\}$.

Finally, let us draw a connection between measurement and density
operators and see how one can prepare a quantum state with density
matrix $\rho \in \states(\cH)$. There are two conceptually
different approaches to this. The first is the so-called
\emph{improper mixture\index{mixture!improper}}. Here one prepares
a purification $\kpsi \in \cH \otimes \cK$ of $\rho$ by means of a
unitary operator and traces out over $\hilK$. Since all
purifications are equivalent, every choice of the purification
will lead to the same result. The second procedure is the
\emph{proper mixture\index{mixture!proper}}. Here one chooses a
$\rho$-\emph{ensemble\index{ensemble}\label{ensemble}} of pure
states\footnote{It will also be useful to introduce
$\rho$-ensembles of mixed states, i.e. sets $\{p_i, \rho_i\}$
where the $\rho_i$'s are mixed states with $\sum_i p_i
\rho_i=\rho$.} i.e. an ensemble $\{p_i, \proj{\psi_i}\}_{i=1}^m$
such that $\rho=\sum_i p_i \proj{\psi_i}$, flips an $m$-valued
coin with distribution $p_i$ and prepares the state $\ket{\psi_i}$
if the coin shows $i$. Note that there is freedom in choosing the
ensemble. Proper and improper preparation procedures are
mathematically equivalent in the sense that a person receiving the
prepared state $\rho$ will not be able to decide which procedure
has been applied to generate $\rho$.

\subsection*{Distance Measures}

Intuitively, two quantum states are close to each other if we can
hardly tell the difference in an experiment. This intuition can be
made precise in a variety of different ways. Relevant for this
thesis are three measures: the trace distance, the fidelity and
the relative entropy.

The quantum analog of the variational distance for random
variables is the \emph{trace distance\index{trace distance}} of
two operators $\rho$ and $\sigma$
\be \label{eq-intro-tracedistance} \delta(\rho, \sigma) =\half ||\rho-\sigma||_1=\half \tr |\rho-\sigma|,\ee
where $|A|=\sqrt{A^\dagger A}$. In operational terms, the trace
distance equals the variational distance of the probability
distribution that results from a POVM when maximised over all
possible POVMs
    \be \label{eq-intro-trace}
        \delta(\rho, \sigma)=\max_{M} \delta(P, Q),
    \ee
where the maximisation is taken over all POVMs $M$ applied to
$\rho$ and $\sigma$. The resulting probability distributions are
$P$ and $Q$ respectively. As an application, consider the problem
of deciding whether a given quantum state equals $\rho$ or
$\sigma$ when no prior knowledge is given. Most generally, this
can be done with a two outcome POVM $\{E_0, E_1\}$, where outcome
$0 \, (1)$ corresponds to the guess $\rho \, (\sigma)$. By
equation~(\ref{eq-intro-trace}), the solution to the problem, i.e.
the maximal probability of guessing correctly is
\be \label{eq-guess-correctly} \half + \frac{\delta(\rho, \sigma)}{2}.\ee
Another consequence of equation~(\ref{eq-intro-trace}) is the
monotonicity of the trace distance under CPTP maps
\be \delta(\Lambda(\rho), \Lambda(\sigma)) \leq \delta(\rho,
\sigma)\ee and, by a similar argument, the \emph{strong convexity}
of the trace distance,
    \bestar
        \delta(\sum_i p_i \rho_i, \sum_i q_i \sigma_i) \leq \delta(P, Q) +\sum_i p_i
        \delta(\rho_i, \sigma_i).
    \eestar
Both properties make the trace distance a convenient tool in
quantum information theory.

\begin{figure}\label{fig-intro-channel}
\unitlength 0.9mm
\begin{picture}(100.00,50.00)(0,35)

\linethickness{0.15mm}
\put(20.00,70.00){\line(1,0){10.00}}
\put(20.00,70.00){\line(0,1){10.00}}
\put(30.00,70.00){\line(0,1){10.00}}
\put(20.00,80.00){\line(1,0){10.00}}

\put(23,73){\makebox{$\cK$}}

\linethickness{0.15mm}
\put(20.00,50.00){\line(1,0){10.00}}
\put(20.00,50.00){\line(0,1){10.00}}
\put(30.00,50.00){\line(0,1){10.00}}
\put(20.00,60.00){\line(1,0){10.00}}

\put(23,53){\makebox{$\cH$}}

\linethickness{0.15mm}
\multiput(25.00,60.00)(0,1.82){6}{\line(0,1){0.91}}

\linethickness{0.15mm}
\put(50.00,40.00){\line(1,0){20.00}}
\put(50.00,40.00){\line(0,1){20.00}}
\put(70.00,40.00){\line(0,1){20.00}}
\put(50.00,60.00){\line(1,0){20.00}}

\put(58,48){\makebox{$U$}}

\linethickness{0.15mm}
\put(70.00,55.00){\line(1,0){20.00}}

\put(93,53){\makebox{$\cH'$}}

\linethickness{0.15mm} \put(30.00,45.00){\line(1,0){20.00}}

\linethickness{0.15mm}
\put(70.00,45.00){\line(1,0){20.00}}

\linethickness{0.15mm}
\put(30.00,55.00){\line(1,0){20.00}}
%

\linethickness{0.15mm}
\put(90.00,50.00){\line(1,0){10.00}}
\put(90.00,50.00){\line(0,1){10.00}}
\put(100.00,50.00){\line(0,1){10.00}}
\put(90.00,60.00){\line(1,0){10.00}}

\end{picture}
\caption{A state $\rho$ on system $\cH$ is sent through a channel
$\Lambda$ implemented by the unitary $U$. Ancilla systems are the
lose ends of lines. The state on $\cH\otimes \cK$ is $\kpsi$ and
$\Lambda(\rho)$, the state on $\cH'$, is the channel output.}
\label{figure-channel}
\end{figure}
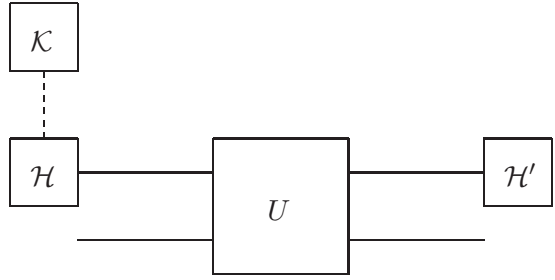

A different measure for distinguishing quantum states is the
\emph{fidelity\index{fidelity}}, which quantifies the
\emph{overlap} of two quantum states. In case of pure states:
    \bestar
        F(P_{\psi}, P_{\phi})=|\braket{\psi}{\phi}|^2.
    \eestar
Operationally, the fidelity equals the probability for $\kpsi$ to
pass the test whether or not it equals $\kphi$. In the case where
one quantum state is mixed, this formula immediately generalises
to
    \bestar
        F(P_{\kpsi}, \sigma)= \tr P_{\kpsi} \sigma,
    \eestar
whereas a less obvious generalisation to mixed states is given by
    \bestar
        F(\rho, \sigma)=(\tr
        \sqrt{\sqrt{\rho}\sigma\sqrt{\rho}})^2.
    \eestar
Note that $F(\rho, \sigma)$ is in fact symmetric with respect to
its arguments. Monotonicity under CPTP maps,
    \bestar
        F(\Lambda(\rho), \Lambda(\sigma) \geq F(\rho, \sigma),
    \eestar
follows from the following variational formula due to
Uhlmann\index{Uhlmann's theorem}~\cite{Uhlmann76} (see also
\cite{NieChu00Book} and~\cite{Jozsa94}):
    \be \label{eq-uhlmann}
        F(\rho, \sigma) =\max_{\phi, \psi}
        |\braket{\phi}{\psi}|^2,
    \ee
where $\kphi$ and $\kpsi$ are purifications of $\rho$ and $\sigma$
respectively. The fidelity also serves as a measure of how well
quantum states are preserved when they pass through a quantum
channel. Since not only the quality of the transmission of quantum
or classical signals but also the ability to establish quantum
correlations between sender and receiver is an important property
of quantum channels, an appropriate measure is needed. The
\emph{entanglement fidelity\index{fidelity!entanglement}} is such
a measure and is defined as the fidelity of a purification $\kpsi
\in \cH\otimes \cK$ of $\rho$ on $\cH$ and the channel state $\id
\otimes \Lambda (\proj{\psi})$, i.e. the combined system of
reference and output (see figure~\ref{figure-channel}):
    \bestar
        F_e(\rho, \Lambda) =\bpsi (\id \otimes \Lambda(
        \proj{\psi}))\kpsi
    \eestar
or in Kraus operator form
    \bestar
        F_e(\rho, \Lambda) =\sum_i |\tr M_i \rho|^2.
    \eestar

\noindent The entanglement fidelity with respect to the maximally
mixed state $\rho=\frac{\id}{d}$, where $d$ is the input dimension
of the channel, has a nice expression in terms of the average
fidelity of pure quantum states that are sent through the channel:
    \bestar
        \bar{F}(\Lambda)=\int_\phi \bphi \Lambda(\proj{\phi})\kphi
        d\phi
    \eestar
The average is taken with respect to the Haar measure\index{Haar
measure} of $\unitaries(d)$, i.e. the invariant measure on
$\unitaries(d)$, with normalisation $\int_{\phi} d\phi=1$. A handy
formula that connects the average fidelity with the entanglement
fidelity was found by Micha{\l}, Pawe{\l} and Ryszard
Horodecki~\cite{HoHoHo99}:
\be \bar{F}(\Lambda)=\frac{d F_e(\frac{\id}{d}, \Lambda)+1}{d+1}.\ee
Finally, there are a few inequalities connecting the trace
distance and the fidelity.
    \be \label{eq-fidelity-55}
        1-\sqrt{F(\rho, \sigma)}\leq \delta(\rho, \sigma) \leq
        \sqrt{1-F(\rho, \sigma)}
    \ee
and in the special case, where $\rho=\proj{\psi}$ is pure,
    \be \label{fidelity-trace-inequ} 1-F(P_{\kpsi}, \sigma)\leq \delta(P_{\kpsi}, \sigma)\ee
with equality if also $\sigma$ is a pure state.

\subsection*{Entropy}

\sloppy Entropy is a fundamental properties in physics. Its
quantum version for finite-dimensional systems is the \emph{von
Neumann entropy\index{von Neumann entropy}}\index{entropy!von
Neumann}. For a density operator $\rho$ it is given by
$$S(\rho)=-\tr \rho \log \rho,$$
with logarithm taken to base $2$ here and throughout this thesis.
It is an immediate consequence that the von Neumann entropy takes
the form
$$ S(\rho)=-\sum_i \lambda_i \log \lambda_i,$$
and therefore equals the \emph{Shannon entropy\index{Shannon
entropy}}\index{entropy!Shannon} of the eigenvalues $\lambda_i$ of
$\rho$. If systems are composed of several parts, say $ABC\ldots$,
it is natural to compare the entropies of the reduced states and
of the overall states. For a subset $X \in \{A, B, C, \ldots\}$, I
write $S(X)_\rho$ for $S(\rho^X)$ or even more conveniently,
$S(X)$ if the underlying state is clear from the context.
Fundamental limitations on the possible set of such entropies are
known as \emph{entropy inequalities}; examples are
\emph{subadditivity}\index{von Neumann entropy!subadditivity}
$$  S(A)+S(B) \geq S(AB)$$
and \emph{strong subadditivity}\index{von Neumann entropy!strong
subadditivity}\label{strong-subadditivity}
    \be \label{eq-strong-subadd}
        S(AB)+S(BC)\geq S(B)+S(ABC).
    \ee
The latter is of fundamental importance to the physics of
condensed matter systems and especially quantum information
theory. Initially proved by Elliott H. Lieb and Mary Beth Ruskai
using operator convexity results~\cite{LieRus73PRL, LieRus73JMP},
a number of different proofs have appeared in the literature most
recently using techniques from quantum information
theory~\cite{HoOpWi05, GrPoWi05}. In part~\ref{part-group}, I will
discuss parallels between re\-pre\-sen\-ta\-tion-theoretic
inequalities and entropy inequalities and give a new proof for
subadditivity (corollary \ref{corollary-subadditive}). Strong
subadditivity plays a vital role in chapter~\ref{chapter-squashed}
in the definition of the newly proposed entanglement measure:
squashed entanglement. General properties of entropy inequalities
have been discussed in~\cite{Pippenger03} and most recently, a new
(restricted) entropy inequality has been discovered by
\cite{LinWin04}.

\fussy A related quantity is the \emph{relative
entropy\index{relative entropy}}\label{relative-entropy}
$$ S(\rho||\sigma):= \tr \rho (\log \rho -\log \sigma),$$
which measures the entropy of $\rho$ relative to $\sigma$. The
relative entropy is less of an entropy but more of a distance
measure and satisfies $S(\rho|| \sigma) =0$ if and only if
$\rho=\sigma$. In particular, it is not symmetric under
interchange of its arguments, hence not a distance in the
mathematical sense. The defining property of the relative entropy
is its monotonicity under a CPTP map $\Lambda$, i.e.
$$ S(\Lambda(\rho)||\Lambda(\sigma))\leq S(\rho||\sigma),$$
a fact which implies strong subadditivity of von Neumann entropy.
Observe that
    \be \label{eq-cond-rel}
        S(AB)-S(B)=\log \dim A - S(\rho^{AB}||\tau^A \otimes \rho^B),
    \ee
where $\tau^A=\frac{\id}{\dim A}$, in order to rewrite
inequality~(\ref{eq-strong-subadd}) in the form
$$ S(\rho^{AB}|| \tau^A \otimes \rho^B) \leq S(\rho^{ABC}|| \tau^{A}\otimes \rho^{BC}). $$
This inequality is true by the monotonicity of the relative
entropy under the CPTP map $\Lambda: \rho^{ABC} \rightarrow
\rho^{AC} \tr \rho^B$. Another useful property of
$S(\rho||\sigma)$ is the joint convexity (see
e.g.~\cite[(11.135)]{NieChu00Book}):
\begin{lemma}[Joint convexity of relative
entropy\index{relative entropy!joint
convexity}]\label{joint-convexity-rel-ent} Let $\{p_i, \rho_i\}$
and $\{q_i, \sigma_i\}$ be two ensembles of mixed states, then
$$ S(\rho||\sigma) \leq D(P||Q) +\sum_i p_i S(\rho_i||\sigma_i),$$
where $D(P||Q)=\sum_i p_i (\log p_i -\log q_i)$, the
\emph{Kullback-Leibler distance\index{Kullback-Leibler
distance}\label{Kullback-Leibler distance}} of two probability
distributions. Since $D(P||Q)=0$ for $P=Q$, this implies convexity
of $S(\rho||\sigma)$.
\end{lemma}
Note that convexity of the relative entropy implies via
equation~(\ref{eq-cond-rel}) the concavity of the conditional von
Neumann entropy\index{von Neumann entropy!conditional}
$S(A|B)=S(AB)-S(B)$. A consequence, but also an easier result is
the following:
\begin{lemma}[Concavity of von Neumann entropy\index{von Neumann entropy!concavity}] Let $\{p_i, \rho_i\}$ be an ensemble of $\rho$ and
$P_X(i)=p_i$ be the distribution of a random variable $X$. Then
$$ \sum_i p_i S(\rho_i)\leq S(\rho) \leq H(X)+\sum_i p_i S(\rho_i)$$
with equality if and only if the $\rho_i$ have pairwise orthogonal
support. Note also that
$$ S(\sum_i \rho_i \otimes \proj{i}) = H(X)+\sum_i p_i S(\rho_i).$$
\end{lemma}
Concavity of von Neumann entropy relates to the following result,
known as \emph{Holevo's theorem} or \emph{Holevo's bound}:
\begin{theorem}[Holevo's bound\index{Holevo's
bound}~\cite{Holevo73a}]\label{theorem-holevo} Let $\cE=\{p_x,
\rho_x\}$ be a $\rho$-ensemble of quantum states. For every
measurement on $\cE$ with outcome saved in the random variable $Y$
it is true that
$$ I(X;Y)\leq \chi(\cE),$$
where $\chi(\cE)=S(\rho)-\sum_x p_x S(\rho_x)$ is the \emph{Holevo
$\chi$ information} or \emph{Holevo $\chi$ quantity}\index{Holevo
$\chi$ information}.
\end{theorem}

\noindent The maximum over all possible measurements is also known
$M$ with outcome saved in $Y$,
$$\I(\cE)=\max_M I(X;Y),$$
is known as \emph{accessible information\index{accessible
information}}. Hence, by Holevo's bound
$$ \I(\cE) \leq \chi(\cE).$$
Finally, let us review a couple of inequalities that relate the
trace distance and the relative entropy as well as the von Neumann
entropy and the trace distance.
\begin{lemma}[{\cite[theorem
1.15]{OhyPet04Book}}]\label{lemma-pinsker}
\be \delta(\rho, \sigma)^2 \leq \frac{\ln 2}{2} S(\rho||\sigma),\ee
where $\ln 2$ denotes the natural logarithm of $2$. If $\rho$ and
$\sigma$ are probability distributions, this inequality is known
as \emph{Pinsker's inequality\index{Pinsker's inequality}}
(cf.~\cite{FeHaTo03}).
\end{lemma}

\begin{lemma}[Fannes' inequality\index{Fannes' inequality}~\cite{Fannes73}]\label{lemma-fannes}
Let $\rho$ and $\sigma$ be supported on a $d$-dimensional Hilbert
space and $\delta(\rho, \sigma)\leq \epsilon$. Then
$$ |S(\rho)-S(\sigma)|\leq 2 \epsilon \log d + \mu(\epsilon) $$
where $\mu(x):=\min \{-x\log x, \frac{1}{e}\}$.
\end{lemma}
In chapter~\ref{chapter-squashed}, I will discuss a recent
extension of this inequality for conditional entropies
(lemma~\ref{lemma-cond-fannes}).

This concludes the introduction to the quantum mechanical tools
that will be used repeatedly throughout.

\section*{Group Theory}
\label{sec-group-theory} This section summarises some basics of
group theory of finite groups and establishes the notation used in
part~\ref{part-group}.

\subsection*{Groups and Representations}

Let $(G, \circ)$ be a group. If clear from the context $g \circ h$
will be abbreviated by $gh$. A
\emph{homomorphism\index{homomorphism!group}} between groups $G$
and $H$ is a map $f: G \rightarrow H$ such that $f(gh)=f(g)f(h)$
for all $g, h \in G$. A
\emph{representation\index{representation}} of a finite group $G$
on a finite-dimensional complex vector space $V$ is a homomorphism
$f: G \rightarrow \GL(V)$ of $G$ to the \emph{general linear
group} of $V$, i.e. the invertible elements in $\endo(V)$. Given a
space $V$, the underlying representation of a particular group is
often clear from the context. $V$ will then be referred to as the
representation of $G$, or a space with an action of $G$. For a
vector $\ket{v} \in V$, it is also understood that $g\ket{v}\equiv
f(g)\ket{v}$. Two representations $V$ and $W$ are
\emph{equivalent\index{representation!equivalent}} if there is a
map $\phi$ such that the diagram
\begin{displaymath}
\xymatrix{ V \ar[r]^\phi \ar[d]_g & W \ar[d]^g \\
V \ar[r]^{\phi} &W}
\end{displaymath}
commutes. A representation $V$ of $G$ is
\emph{irreducible\index{representation!irreducible}} if it has no
non-trivial invariant subspace under the action of the group,
i.e.~the only subspaces $W \subseteq V$ for which $g\ket{w} \in W$
for all $\ket{w} \in W$ and $g \in G$ are $W =\emptyset$ or $W=V$.
The importance of irreducible representations comes from the fact
that in the case of finite groups every representation is
equivalent to a direct sum of
irreducible representations. 

\begin{theorem} \label{theorem-irrep-decomp}
Let $W$ be a representation of a finite group $G$. Then $W$ is
isomorphic to a direct sum of irreducible representations of $G$,
i.e. $W \cong \bigoplus_i V_i$ for irreducible representations
$V_i$ of $G$.
\end{theorem}

\begin{proof}
Let $(w_1,w_2 )$ be a scalar product on $W$, then
$$[ w_1, w_2 ]= \frac{1}{|G|} \sum_{g \in G} (gw_1, gw_2)$$
is a $G$-invariant scalar product on $W$. If $V$ is an invariant
subspace of $W$, then $V^\perp$, the orthogonal complement of $V$
in $W$, is also an invariant subspace: for $v \in V$ and $v^\perp
\in V^\perp$, $[ g v^\perp, v]=[ v^\perp, g^{-1} v ]=0$, since
$g^{-1}v \in V$ and $V$ is $G$-invariant. In this way one can keep
on breaking up the space of $W$ into invariant subspaces. This
procedure will terminate, because $W$ is finite-dimensional.
\end{proof}

\noindent Probably the most frequently used result in
representation theory is the famous lemma by Isaac Schur.

\begin{lemma}[Schur's lemma\index{Schur's
lemma}]\label{lemma-Schur} Let $V$ and $W$ be irreducible
representations of $G$. If the homomorphism $\phi: V \rightarrow
W$ commutes with the action of $G$, then
\begin{itemize}
\item either $\phi$ is an isomorphism, or $\phi=0$.
\item if $V=W$, then $\phi=\lambda \id$ for some $\lambda \in
\complex$.
\end{itemize}
\end{lemma}

\begin{proof}
$\ker \phi$ ($\im \phi$) is an invariant subspace of $V$ ($W$) and
since $V$ ($W$) is irreducible it can only be equal to $\emptyset$
or $V$ ($W$). Hence $\phi$ is either an isomorphism or it
vanishes. Since $\complex$ is algebraically closed, $\phi$ must
have an eigenvalue $\lambda \in \complex$. Then $\ker
(\phi-\lambda \id) \neq \emptyset$ and $\phi-\lambda \id$ is not
an isomorphism, which implies by the first part of the lemma that
$\phi-\lambda \id=0$.
\end{proof}

\noindent Schur's lemma implies that the decomposition in theorem
\ref{theorem-irrep-decomp} is unique up to isomorphism. The
classification of representations of a finite group $G$ is
therefore reduced to the classification of all irreducible
representations.

\label{remark-carry-over} Most theorems in this section carry over
almost unchanged to compact groups. Most importantly this is true
for theorem~\ref{theorem-irrep-decomp} and Schur's lemma, lemma
\ref{lemma-Schur}.

\subsection*{Group Algebra} 
Rather than working with a group itself, it is sometimes more
convenient to work with the \emph{group algebra}
$\cA(G)$\index{group algebra}, the complex vector space spanned by
the group elements. Formally, let us define the vector space
consisting of the elements $a=\sum_{g \in G} a(g) g$, where $a(g)
\in \complex$ and the $g$'s are linearly independent basis
vectors. The dimension of this space equals the order of the
group. In addition to the vector space structure, there is a
product structure given by the group operation $(g, h) \mapsto gh$
that turns the space into an algebra: for two elements $a$ and $b$
in the just defined space, one has
    \bestar
        ab=\left( \sum_g a(g) g \right) \left(\sum_h b(h) h \right)=\sum_{g, h} a(g)
        b(h) (gh)=\sum_x \left( \sum_y a(xy^{-1})b(y) \right)x.
    \eestar
The elements of the group algebra can also be regarded as
complex-valued functions $a: G \rightarrow \complex$ taking a
complex value for each element of the group. For two such
functions $a(x)$ and $b(x)$, their product is
    \be (a \star b) (x)=\sum_y a(xy^{-1}) b(y)\ee
        and the adjoint of $a(x)$ is given by \be a^\star
        (g)=\overline{a(g^{-1})}.
    \ee
The concept of a representation
extends from a group to its group algebra in the following way:
\begin{theorem} \label{theorem-algebra-representation}
Let $V$ be a unitary representation of $G$. The definition of a
representation of elements of $G$ extends by linearity to elements
$a \in \cA(G)$:
\be V(a)=\sum_g a(g) V(g),\ee
which obey
\beastar
V (a+b)&=& V (a)+V(b)\\
V (a \star b)&=& V(a) V(b)\\
V (a^\star)&=& V(a)^\star\\
V (e)&=&\id.
\eeastar
Conversely, if $V$ obeys these conditions, then its restriction to
$G$ is a unitary representation of $G$.
\end{theorem}
Let
$$ (f, g)=\frac{1}{|G|}\sum_{x \in G} \overline{f(x)} g(x), \qquad f, g \in \cA(G)$$
be the inner product of $\cA(G)$. The set of equivalence classes
of irreducible representations of $G$ is denoted by $\hat{G}$. For
each $\alpha \in \hat{G}$ pick a unitary representative
irreducible representation $V_\alpha$ and denote its matrix
entries by $V_{\alpha, ij}$ where $i,j \in \{1, \ldots,
d_\alpha:=\dim V_\alpha\}$.

\begin{theorem}[Orthogonality Relations] \label{theorem-D-ortho}
The functions $\sqrt{d_\alpha} V_{\alpha, ij}(g)$ are an
orthonormal basis of $\cA(G)$, i.e.
    \be \label{eq-group-algebra-1}
        \frac{1}{|G|} \sum_{g \in G} \overline{V_{\alpha, ij}(g)}
        V_{\beta, kl}(g)=\frac{1}{d_\alpha} \delta_{\alpha \beta}
        \delta_{ik} \delta_{jl}.
    \ee
and the elements corresponding to a fixed $\alpha$ are closed
under multiplication:
    \be
         \left(\frac{d_\alpha}{|G|}V_{\alpha, ij}\star
        \frac{d_\beta}{|G|}V_{\beta, kl}\right)(x)=\delta_{\alpha,
        \beta} \delta_{jk} \frac{d_\alpha}{|G|}V_{\alpha, il}(x)
    \ee
\end{theorem}
In the following I will introduce the character of a
representation, a useful tool for studying the equivalence
properties of representations.

\subsection*{Characters}
 $f$ is a \emph{class functions} if it is an
element in the group algebra $\cA(G)$ that is constant on
conjugacy classes, i.e.
\be f(y)=f(xyx^{-1}) \quad \forall \, x ,y \in G.\ee
Let $\cZ(G)$ be the set of \emph{class functions}. It is then not
difficult to see that $\cZ(G)$ is in fact the centre of $\cA(G)$,
i.e. it consists of all elements $f \in \cA(G)$ with $f \star g=g
\star f$ for all $g \in \cA(G)$. The
\emph{character\index{character}} of a representation $V$ of $G$
is defined as
\be \chi(g)=\tr V(g).\ee
The characters of two representations are identical if the
representations are equivalent. In fact it follows directly from
theorem~\ref{theorem-D-ortho}:
\begin{corollary} \label{cor-char-orthonormal}
The characters $\chi_\alpha$ form an orthonormal basis for
$\cZ(G)$, i.e.
$$ \frac{1}{|G|} \chi_\alpha(g) \overline{\chi_\beta(g)}
=\delta_{\alpha \beta}.$$
\end{corollary}
Since $\cZ(G)$ has dimension equal to the number of conjugacy
classes $ |\hat{G}|=\dim \cZ(G)$ which equals the number of
conjugacy classes of $G$.

The decomposition of a representation can be analysed in terms of
its characters:
\begin{corollary}
$V$ decomposes as $V\cong \bigoplus_\alpha V_\alpha^{\oplus
m_\alpha}$ if and only if $ \chi_V=\sum_i m_\alpha \chi_\alpha.$
\end{corollary}
An important representation is the \emph{regular
representation\index{representation!regular}} which we will denote
by $R$. Here, $G$ that acts by conjugation on the group algebra.
$R$ contains each irreducible representation $V_\alpha$ of $G$
exactly $d_\alpha$ times, i.e.
$$ R\cong\bigoplus_\alpha V_\alpha^{\oplus d_\alpha}.$$
This can be seen as follows: the character of $R$ is given by
$\chi_R(e)=|G|$ and $\chi_R(g)=0$ for $g \neq e$, the identity
element of $G$. The multiplicity of $V_\alpha$ in $R$ is given by
$m_\alpha=(\chi_R, \chi_\alpha)=\chi_\alpha(e)=d_\alpha.$ The
dimension formula for finite groups $|G|=\sum_\alpha d_\alpha^2$
is an immediate corollary.

\subsection*{Tensor Product Representations}
\label{sec-intro-clebsch} Let $V$ and $W$ be represntations of
$G$. The \emph{tensor product
representation\index{representation!tensor product}} $V \otimes W$
of $V$ and $W$ is defined as
\bestar (V \otimes W)(g) =V(g) \otimes W(g). \eestar
Note that $G$ acts simultaneously on $V$ and $W$, therefore $V
\otimes W$ is a representation of $G$ and not of $G\times G$. This
is also the reason why the representation $V \otimes W$, even for
irreducible $V$ and $W$, is in general a reducible representation
and can be decomposed:
\be \label{eq-CG}V_\alpha \otimes V_\beta \cong \bigoplus_{\gamma \in \hat{G}}
n^\gamma_{\alpha \beta} V_\gamma.
\ee
The coefficients $n^\gamma_{\alpha \beta}$ are known as the
\emph{Clebsch-Gordan integers\index{Clebsch-Gordan!integers}} of
the group $G$. Clebsch-Gordan integers count how many copies of an
irreducible representation $V_\gamma$ are contained in $V_\alpha
\otimes V_\beta$ and should not be confused with the
\emph{Clebsch-Gordan
coefficients\index{Clebsch-Gordan!coefficients}}. The latter are
entries in the unitary matrix that transforms from the product
basis of two spins, i.e. irreducible representations of $\SU(2)$,
into a basis of the total spin of the system (see
subsection~\ref{subsec-spin-states}).

\subsection*{Representations of Direct Product Groups}
\label{section-boxtimes} 

Let $G$ and $H$ be groups and $G\times H$ the \emph{group direct
product}
\bestar (g_1,h_1) \times (g_2,h_2) =(g_1 g_2, h_1 h_2)\eestar
induced from the group operations on $G$ and $H$.
\label{pageref-tensor-product-reps} If $V$ and $W$ are
representations of $G$ and $H$, one can define the \emph{external
product representation\index{representation!external product}} $V
\boxtimes W (g, h)$ by $V(g) \otimes W(h)$. If $V$ and $W$ are
irreducible, then $V \boxtimes W$ is also irreducible and
conversely, all irreducible representations of $G \times H$ arise
as external product representations of irreducible representations
of $G$ and $H$. Note that the space of the representations is the
tensor product of the spaces of the individual representations
$V\otimes W$. Therefore care needs to be taken, so as not to
confuse the representations of the direct product group $G\times
H$ with tensor product representation of $G$ in the case $G=H$.

This concludes the basic material from the group theory of finite
groups. More definitions and theorems as well as extensions to
compact groups will be discussed in
chapter~\ref{chapter-structure}.

\mainmatter

\part{Insights from Group Theory}\label{part-group}
\chapter*{Prologue}

Bipartite quantum systems form an important resource in quantum
cryptography and quantum teleportation and also play a prominent
role in condensed matter systems that are governed by
nearest-neighbour interactions. Just as the spectrum of a single
quantum state contains information about its entropic, energetic
and information-theoretical properties, the spectral properties of
a bipartite quantum state can elucidate correlations, energies and
entanglement amongst its two parts.

\sloppy This part of my PhD thesis investigates the spectral
structure of bipartite quantum states. More precisely, I examine
the constraints on a triple of spectra $(r^A, r^B, r^{AB})$ that
are necessary to ensure the existence of a bipartite quantum state
compatible with this triple, that is, the existence of $\rho^{AB}$
on $\cH_A \otimes \cH_B$ with $\spec \rho^{AB}=r^{AB}$ as well as
$\spec \rho^A=r^A$ and $\spec \rho^B=r^B$ for the reduced states.

\fussy My main result is the discovery of the equivalence -- in a
precise asymptotic sense -- between the problem of determining the
spectral structure of bipartite quantum states and a well-known
representation-theoretic problem of the symmetric group: Given an
irreducible representation, is this representation contained in
the tensor product of two other selected irreducible
representations? The coefficients governing the decomposition of
the tensor product are known as the Kronecker coefficients and no
closed combinatorical algorithm for their computation is known.
The result presented here offers a way to investigate the spectral
structure of bipartite quantum states using tools from group
theory. Conversely, spectral properties of quantum states can be
used to illuminate the calculation of Kronecker coefficients.

Part~\ref{part-group} is split into two chapters.
Chapter~\ref{chapter-structure} introduces concepts from the
theory of the symmetric and Lie groups, with focus on Weyl's
tensorial construction and the Schur-Weyl duality.
Chapter~\ref{chapter-relation} formally presents the two problems
listed above, proves their equivalence and discusses issues of
convexity and finite generation. In parallel, a novel proof is
given for the well-known asymptotic equivalence of Horn's problem
and the problem of calculating the Littlewood-Richardson
coefficients.

\chapter{The Symmetric and the Unitary Groups}\label{chapter-structure}

This chapter provides the background material for
chapter~\ref{chapter-relation} which can be found in introductory
texts on the subject~\cite{Simon96Book, GooWal98, FultonHarris91}.
Section~\ref{sec-Schur-Weyl-I} introduces the famous Schur-Weyl
duality, the pairing of irreducible representations of the
symmetric and the unitary group, and presents part of its proof.
The remainder of this chapter is then devoted to completing the
proof and furnishing the result with details. In
section~\ref{section-symm-group} the irreducible representations
of the symmetric group are constructed and in
section~\ref{section-unitary-group} we touch upon some Lie group
theory to introduce the notion of a highest weight. This completes
the proof of Schur-Weyl duality in
section~\ref{sec-schur-weyl-duality-2}. The chapter concludes with
a construction of an orthogonal basis for irreducible
representations of the symmetric and unitary groups in
section~\ref{sec-on-basis-subgroup-chains}.

\section{Schur-Weyl Duality I}
\label{sec-Schur-Weyl-I}

Subsection~\ref{sec-minimal-proj} reviews results on the
construction of irreducible representations of finite groups with
the help of projections into the group algebra. In
subsection~\ref{sec-duality-theorem} a duality theorem for
representations of a finite group and its commutant are derived.
The duality theorem is then applied in
subsection~\ref{sec-schur-weyl-duality-1} to the symmetric and
unitary group. The resulting Schur-Weyl duality is stated as
theorem~\ref{theorem-Schur-duality} and the first part of the
proof is given.

\subsection{Minimal Projections and Irreducible Representations}
\label{sec-minimal-proj}

Throughout this section, $G$ will be a finite group. The necessary
background material of the representation theory of finite groups
can be found in Preliminaries, page~\pageref{sec-group-theory}.

\begin{definition}
A \emph{projection} $p$ is an element in $\cA(G)$ with $p^2=p$. A
projection $p \neq 0$ is called
\emph{minimal}\index{projection!minimal} if it cannot be
decomposed into projections $q \neq 0$ and $r \neq 0$ as $p=q+r$.
Two projections $p, q$ are called equivalent, if there exists an
invertible $u, v$ such that $u p v=q$ and disjoint if and only if
$p u q=0$ for all $u$.
\end{definition}

\begin{definition}
A \emph{central projection}\index{projection!central} $p$ is an
element in $\cZ(G)$ with $p^2=p$. A central projection $p \neq 0$
is called \emph{minimal} if it cannot be decomposed into central
projections $q \neq 0$ and $r \neq 0$ as $p=q+r$.
\end{definition}

\begin{theorem} \label{theorem-equiv-minproj-irreps}
There is a one-to-one correspondence between equivalence classes
of minimal projections and irreducible representations.
Furthermore, there is a one-to-one correspondence between minimal
central projections and irreducible representations. The minimal
central projections are given by
\be \frac{d_\alpha}{|G|}\chi^{\alpha},\ee
where $\chi_\alpha$ is the character corresponding to an
irreducible representation from the equivalence class $\alpha \in
\hat{G}$.
\end{theorem}

\begin{proof}
By theorem \ref{theorem-D-ortho}, the group algebra is isomorphic
to a direct sum of matrix algebras
$$ \cA(G) \cong \bigoplus_\alpha \endo(\complex^{d_\alpha}),$$
where $\endo(\complex^{d_\alpha})$, the endomorphisms of
$\complex^{d_\alpha}$ is the algebra generated by $V_{\alpha
,ij}$. Thus any $p \in \cA(G)$ is a sum of components
$(p_\alpha)_{\alpha \in \hat{G}}$. Since multiplication is
componentwise, a projection must satisfy $p_\alpha^2=p_\alpha$.
Minimality is achieved if only one $p_\alpha \neq 0$. Regarded as
an element of $\endo(\complex^{d_\alpha})$, a minimal projection
$p_\alpha$ is a rank one projector and vice versa, every rank one
projector in $\endo(\complex^{d_\alpha})$ is a minimal projection.
Rank one projectors $p_\alpha$ and $q_\beta$ are then equivalent
if and only if $\alpha=\beta$. This establishes the one-to-one
correspondence between equivalence classes of minimal projections
and irreducible representations.

As an element of the group algebra, a central projection $c$ is of
the form $(c_\alpha)_{\alpha \in \hat{G}}$, where each component
$c_\alpha$ is a central projection. Minimality requires that only
one $c_\alpha$ may be nonzero. As it is a projection into the
center of $\endo(\complex^{d_\alpha})$, which is one-dimensional
and spanned by $\chi_\alpha$, one still requires a proportionality
constant. After a short calculation one obtains
$c_\alpha=\frac{d_\alpha}{|G|} \chi^\alpha$.
\end{proof}

\noindent Directly from theorem \ref{theorem-equiv-minproj-irreps}
and the properties of representations of the group algebra
(theorem \ref{theorem-algebra-representation}) one has:

\begin{corollary} \label{cor-irrep-minproj}
Let $V$ be a representation of $G$ and $p_\alpha$ be a minimal
central projection in the group algebra of $G$. If $V\cong
\bigoplus_\alpha V_\alpha^{\oplus m_\alpha}$ is the decomposition
of $V$ into irreducible representations $V_\alpha$ with
multiplicity $m_\alpha$, then $V(p_\alpha)$ is the projector onto
$V_\alpha^{\oplus m_\alpha}.$
\end{corollary}

\noindent The construction of the irreducible representations of
the symmetric and the unitary groups will make direct use of the
above results. But first, I will review a duality theorem, which
is the basis of the Schur-Weyl duality.

\subsection{A Duality Theorem} \label{sec-duality-theorem}

The \emph{commutant}\index{commutant} $\cA'$ of a subset $\cA$ of
the algebra $\cC$ is the set of elements in $\cC$ that commute
with all elements in $\cA$,
$$\cA'=\{ b \in \cC | \, ab=ba \mbox{ for all } a \in \cA\}$$

\begin{lemma} \label{lemma-commutant-hom}
Let $V$ and $W$ be finite dimensional complex vector spaces. The
commutant of $\cA=\endo(V) \otimes \id$ in $\endo(V \otimes W)$ is
$\cB=\id \otimes \endo(W)$
\end{lemma}

\begin{proof}
Clearly $\cB \subset \cA'$. To show $\cA' \subset \cB$ consider
general elements $A \otimes \id_W \in \cA$ and $B \in \cA'$ and
write them in block diagonal form with $\dim W$ blocks each of
size $\dim V$:
\begin{displaymath}
A\otimes \id_W= \left(\begin{array}{c|c|c|c}
A & 0  & \cdots  & \\
\hline
0  & \ddots &   & \\
  \hline
\vdots    & &  \ddots  & \\
    \hline
  &   &  & A  \\
\end{array} \right)
\quad B=\left(\begin{array}{c|c|c|c}
B_{11} & B_{12} &   & B_{1n}   \\
\hline
B_{21}  & \ddots &    \\
\hline
  &   & \ddots &  \\
\hline B_{n1} &  &  & B_{nn}
\end{array} \right)
\end{displaymath}
The commutant requirement $(A \otimes \id_W) \; B=B \; (A \otimes
\id_W)$ reads in matrix form

\begin{displaymath}
\left(\begin{array}{c|c|c|c}
A B_{11} & A B_{12} &   & A B_{1n}   \\
\hline
A B_{21}  & \ddots &    \\
\hline
  &   & \ddots &  \\
\hline A B_{n1} &  &  & A B_{nn}
\end{array} \right)
=
\left(\begin{array}{c|c|c|c}
B_{11} A & B_{12} A &   & B_{1n} A   \\
\hline
B_{21} A  & \ddots &    \\
\hline
  &   & \ddots &  \\
\hline B_{n1} A &  &  & B_{nn} A
\end{array} \right)
\end{displaymath}

\noindent In summary, $[A, B_{ij}]=0$ for all $ij$. Consider the
group $\GL(V)$ as a representation of $\GL(V)$ on $V$. This
representation is clearly irreducible and spans all of $V$. Since
$B_{ij}$ commutes with all elements, by Schur's lemma this
representation must be proportional to the identity, i.e.~there
are values $b_{ij}$ such that $B_{ij}=b_{ij} \id_V$. $B$ therefore
assumes the form $B=\id_V \otimes b$ with a matrix $b \in
\endo(W)$ with entries $(b)_{ij}=b_{ij}$.
\end{proof}
This brings us to the main theorem in this subsection, a duality
result for $\cA$ and its commutant $\cA'$.

\begin{theorem} \label{theorem-duality}
Let $V$ be a representation of a finite group with decomposition
$V=\bigoplus_\alpha V_\alpha \otimes \complex^{n_\alpha}$. Let
$\cA$ be the algebra generated by $V$ and $\cB=\cA'$ its
commutant. Then
\be \label{lemma-algebra-1} \cA\cong\bigoplus_\alpha \endo(V_\alpha) \otimes
\id_{\complex^{n_\alpha}} \ee
\be \label{lemma-algebra-2} \cB\cong \bigoplus_\alpha \id_{V_\alpha} \otimes \endo(\complex^{n_\alpha})\ee
Furthermore we have $\cB'=\cA$, where $\cB'$ is the commutant of
$\cB$ (double commutant theorem).
\end{theorem}

\begin{proof}
The operator
$$ d_\alpha \sum_{g \in G} \overline{V_{\alpha, ij}(g)} V(g)$$
is an element in $\cA$. By the orthonormality of the functions
$V_{\alpha, ij}$ and the decomposition of $V$ into irreducible
components it equals $E_{\alpha, ij}\otimes \id$, where
$E_{\alpha, ij}$ is the matrix with a one at position $(i,j)$ and
zeros otherwise. This shows that $\cA \supset$ RHS of
(\ref{lemma-algebra-1}). But every element in $\cA$ is an element
of the RHS of (\ref{lemma-algebra-1}), so (\ref{lemma-algebra-1})
follows.

Clearly the RHS of (\ref{lemma-algebra-2}) is contained in
$\cA'=\cB$. To see that every element in $\cB$ is of this form,
consider a projection $P_\alpha$ onto $V_\alpha \otimes
\complex^{n_\alpha}$. The projectors $P_\alpha$ form a resolution
of the identity and as an element of $\cA$, $P_\alpha$ commutes
with any $B \in \cB$. This leads to
$$ B=(\sum_\alpha P_\alpha) B= \sum_\alpha  P_\alpha B
P_\alpha=\sum_\alpha B_\alpha.$$ Lemma \ref{lemma-commutant-hom}
implies that $B_\alpha=\id_{V_\alpha} \otimes b_\alpha$.
\end{proof}

\subsection{Schur-Weyl Duality} \label{sec-schur-weyl-duality-1}
In this section, it is shown that $S_k$ and $\unitaries(d)$ are
double commutants, and their representation space therefore has a
nice decomposition according to theorem~\ref{theorem-duality}. But
before we get too far ahead of ourselves, let us define the action
of the groups on the tensor product space. If $\complex^d$ denotes
a $d$-dimensional complex vector space, $S_k$ operates on
$(\complex^d)^{\otimes k}$ by
\be \label{equation-symmetry-action}
\pi : \ket{e_{i_1}} \otimes \ket{e_{i_2}} \otimes \ldots \otimes
\ket{e_{i_k}} \mapsto \ket{e_{i_{\pi^{-1}(1)}}} \otimes
\ket{e_{i_{\pi^{-1}(2)}}} \otimes \ldots \otimes
\ket{e_{i_{\pi^{-1}(k)}}},
\ee for $\pi \in S_k$, where the $\ket{e_1}, \ldots \ket{e_d}$ are elements of some basis
of $\complex^d$. The group $\unitaries(d)$ acts by
\be \label{equation-SU-action} U: \ket{e_{i_1}} \otimes
\ket{e_{i_2}} \otimes \ldots \otimes \ket{e_{i_k}} \mapsto
U\ket{e_{i_1}} \otimes U \ket{e_{i_2}} \otimes \ldots \otimes
U\ket{e_{i_k}},\ee for $U \in \unitaries(d)$.
 These actions of $S_k$ and
$\unitaries(d)$ on $(\complex^d)^{\otimes k}$ define
representations of each group, but both representations are
reducible. I will refer to this representation of $\unitaries (d)$
as the \emph{tensor product representation}. The following lemma
plays a significant role in establishing the double commutant
theorem and the converse theorems presented in
chapter~\ref{chapter-relation}.

\begin{lemma} \label{lemma-symmetric} Let $V$ be a vector space and let
$\sym^k(V)$ be the $k$'th symmetric power of $V$, i.e.~the vector
space generated by the projection $\frac{1}{n!}\sum_\pi \pi$
applied to $V^{\otimes k}$. Then
$$ \sym^k(V)=\Span \{ \ket{v}^{\otimes k} | \ket{v} \in V\}.$$
\end{lemma}

\begin{proof}
By definition, $\sym^k(V)$ is spanned by the vectors $\ket{v_{i_1
\ldots i_k}} =\sum_\pi \ket{e_{i_{\pi^{-1}(1)}}} \otimes \cdots
\otimes \ket{e_{i_{\pi^{-1}(k)}}}$, where the indices $i_j$ run
through $\{1, \ldots, d\}$.

Clearly $\Span \{ \ket{v}^{\otimes k} | \ket{v} \in V\} \subset
\sym^k(V)$, it therefore suffices to show that every $\ket{v_{i_1
\ldots i_k}}$ can be written in terms of tensor products
$\ket{v}^{\otimes k}$. This is done as follows: Consider the
derivative
    $$
        \ket{w_{i_1 \ldots i_k}}:=\frac{\partial}{\partial \lambda_2 \ldots \lambda_k} (\ket{e_1}+
        \sum_{i=1}^k \lambda_k \ket{e_k})^{\otimes k}
        \big|_{\lambda_2=\cdots = \lambda_k=0},
    $$
which can be realised by subsequently applying
    $$
        \frac{\partial}{\partial \lambda_j} (\ket{v}+\lambda_j
        \ket{e_j})^{\otimes k}\big|_{\lambda_j=0}= \lim_{\lambda_j \rightarrow
        0}  \frac{(\ket{v}+\lambda_j \ket{e_j})^{\otimes k}-\ket{v}^{\otimes
        k}}{\lambda_j},
    $$ iteratively going from $j=k$ all the way to
$j=2$. Then $\ket{w_{i_1 \ldots i_k}}$ takes the form of a limit
of sums of tensor powers. Since $\Span \{ \ket{v}^{\otimes k} |
\ket{v} \in V\}$ is a finite dimensional vector space this limit
is contained in $\Span\{ \ket{v}^{\otimes k} | \ket{v} \in V\}$.
On the other hand, a direct calculation shows that $\ket{w_{i_1
\ldots i_k}}$ equals $\ket{v_{i_1 \ldots i_k}}$ and hence all
vectors $\ket{v_{i_1 \ldots i_k}}$ are contained in $\Span \{
\ket{v}^{\otimes k} | \ket{v} \in V\}$.
\end{proof}

\noindent \sloppy Note that $\endo(\complex^d)$ is a Hilbert space
with Hilbert-Schmidt inner product $(A, B)=\tr A^\dagger B$ where
the action of $S_k$ is induced by its action on
$(\complex^d)^{\otimes k}$ and given by
\be \label{eq-inv} V(\pi) C V^{-1}(\pi),\ee where $C \in
\endo(V^{\otimes k})$ and $V(\pi)$ is the tensor product
representation on $(\complex^d)^{\otimes k}$.
Lemma~\ref{lemma-symmetric} then says that the vector space of all
$C$ that are invariant under the action given in (\ref{eq-inv}) is
spanned by $\{ X^{\otimes k}| X \in
\endo(\complex^d)\}$. This brings us to the next step on the
way to Schur-Weyl duality:

\fussy \begin{theorem}[$\SU(d)$ and $S_k$ are commutants]
\label{theorem-unitary-symmetric-duality} Let $\cA$ denote the
algebra generated by $V(\pi)$ for all $\pi \in S_k$, where $V$ is
the representation given above. Let further $\cB$ be the algebra
generated by $U(x)$, the elements in the representation of
$\SU(d)$ (or $\unitaries(d)$ or $\GL(d)$). Then $\cA'=\cB$ and
$\cB'=\cA$.
\end{theorem}
\begin{proof}
The proof will be given for $\SU(d)$; the cases of $\unitaries(d)$
and $\GL(d)$ are then imminent from the proof. Clearly $\cB
\subset \cA'$. It remains to show that every element in $\cA'$ is
also an element of $\cB$. But let us first take a look at $\cB$
itself. Clearly $\{A^{\otimes k} | A \in \SU(d)\} \subset \cB$ --
and -- adding a phase: $\{A^{\otimes k} | A \in U(d)\} \subset
\cA'$. Let us now consider the element $dU(X)=\frac{d}{dt} \left(
(e^{tX})^{\otimes k}\right)|_{t=0} \in \cB$ for $X \in \fu(d)$,
where $\fu(d)$ is the Lie algebra of $\U(d)$, which can be
identified with the set of skew Hermitian operators (see
subsection~\ref{subsec-weights}):
    $$
        d U: X \mapsto X \otimes \id \otimes \cdots \otimes \id+
        \cdots + \id \otimes \id \otimes \cdots \otimes X.
    $$
Clearly $dU(X)+idU(Y)=dU(X+iY)$ is an element of $\cB$. Applying
the exponential map results in $\{ A^{\otimes k} | A \in \GL(d)\}
\subset \cB$ and since $GL(d)$ is dense in $\endo(\complex^d)$ and
$\cB$ is closed, $H:=\{  A^{\otimes k} | A \in
\endo(\complex^d)\} \subset \cB$.

Now consider an element $B \in \cA'$. By definition $B$ commutes
with all $V(\pi)$. Therefore $B$ is an element of the symmetric
subspace of $\endo(\complex^d)^{\otimes k}$, which by
lemma~\ref{lemma-symmetric} equals $\{  A^{\otimes k} | A \in
\endo(\complex^d)\}=H$. In summary: $ \cB \subset \cA'$ as well as
$ \cA'=H \subset \cB$ hold. This shows $\cA'=\cB$. $\cB'=\cA$ then
follows from corollary \ref{theorem-duality}.
\end{proof}

\noindent The following theorem reduces the problem of determining
irreducibility from $\U(d)$ to $\GL(d)$, the complexification of
$\U(d)$. For a proof see~\cite[chapter
12]{CarterSegalMacDonald95}. Related are the statements of
theorems~\ref{theorem-Lie-group-facts}
and~\ref{theorem-lie-algebra-complexification}.

\begin{theorem} \label{theorem-unitary-gl}
 A representation of
$\unitaries(d)$ is irreducible if and only if the corresponding
representation of $\GL(d)$ is irreducible.
\end{theorem}

\noindent This brings us to the main result in this section.

\begin{theorem}[Schur-Weyl duality\index{Schur-Weyl duality}] \label{theorem-Schur-duality}
Let $\cH\cong (\complex^d)^{\otimes k}$ and let $V(\pi)$ be the
natural representation of the symmetric group $S_k$ on $\cH$ and
$U(x)$ the tensor representation of $\U(d)$. Then i) \bea
\label{eq-schur-weyl-1}\cH&\cong &\bigoplus_\lambda U_\lambda \otimes V_\lambda\\
\label{eq-schur-weyl-2}U(x)&=&\bigoplus_\lambda U_\lambda(x) \otimes \id_{V_\lambda} \\
\label{eq-schur-weyl-3}V(\pi)&=&\bigoplus_\lambda \id_{U_\lambda}
\otimes V_\lambda(\pi),
\eea
where $U_\lambda$ and $V_\lambda$ are irreducible representations
of $\U(d)$ and $S_k$ respectively. ii) The sum is taken over Young
frames $\lambda \vdash (k, d)$.
\end{theorem}
The second half will be proven in section
\ref{sec-schur-weyl-duality-2}.
\begin{proof}[Part i)]
The application of theorem~\ref{theorem-duality} to $G=S_k$ (and
to its dual partner $\unitaries(d)$,
theorem~\ref{theorem-unitary-symmetric-duality}) proves
equations~(\ref{eq-schur-weyl-1})-(\ref{eq-schur-weyl-3}), where
$V_\lambda$ are irreducible representations of $S_k$. The
representation of $\unitaries(d)$ that is paired with $V_\lambda$
is denoted by $U_\lambda$. It remains to show that the
$U_\lambda$s are in fact irreducible. A brief but elegant argument
follows~\cite[p.112]{CarterSegalMacDonald95}: by
theorem~\ref{theorem-unitary-gl} $U_\lambda$ is irreducible if and
only if its extension to $\GL(d)$ is irreducible. So, it suffices
to show that $U_\lambda$ is indecomposable under $\GL(d)$. By
Schur's lemma this is equivalent to showing that the maps in
$\endo(U_\lambda)$ that commute with the action of $\GL(d)$ are
proportional to the identity; in other words, that
$\endo_{\GL(d)}(U_\lambda)=\complex$. Now $\complex$ is the center
of the matrix algebra $\endo(U_\lambda)$. From Schur's lemma we
have
$$ \endo_{S_k}(V^{\otimes k})\cong \bigoplus_\lambda \endo(U_\lambda)$$
$$ \endo_{\GL(d) \times S_k}(V^{\otimes k})\cong \bigoplus_\lambda \endo_{\GL(d)}(U_\lambda)$$
If $\endo_{\GL(d) \times S_k}(V^{\otimes k})$ is in the center of
$\endo_{S_k}(V^{\otimes k})$, then so is
$\endo_{\GL(d)}(U_\lambda)$ in the center of $\endo(U_\lambda)$.
$\endo(U_\lambda)$ in turn equals $\complex$ which was what we set
out to prove. But since $\GL(d)$ and $S_k$ are double commutants
(theorem \ref{theorem-unitary-symmetric-duality}),
$\endo_{S_k}(V^{\otimes k})=\Span \{ A^{\otimes k}| A \in
\GL(d)\}$ and thus clearly $\endo_{\GL(d) \times S_k}(V^{\otimes
k})$ is contained in the center of $\endo_{S_k}(V^{\otimes k})$.
\end{proof}
This concludes the first part of the Schur-Weyl duality. In the
next two sections, the constructions of the irreducible
representations of $\SU(d)$ and $S_k$ will be in the center of
attention and explain the labeling by Young frames.

\section[The Irreducible Representations of the Symmetric Group]{The Irreducible Representations of the Symmetric \protect\newline Group}
\label{section-symm-group}

\fussy In subsection~\ref{sec-minimal-proj} it was shown that the
irreducible representations of a finite group stand in one-to-one
relation to minimal projections in the group algebra. Here, I will
construct the minimal projections for the symmetric group. Let
$e_T\equiv e_{T_\lambda}$ be a minimal projection corresponding to
$V_\lambda$ and $e_\lambda$ the minimal central projection. It
then follows that $U_T:=V(e_T)\cH$ is a subspace of $V(e_\lambda)
\cH$. By theorem~\ref{theorem-Schur-duality} there are two cases:

\begin{enumerate}
\item[i)] $V(e_\lambda) \cH \cong U_\lambda \otimes V_\lambda$
\item[ii)] $V(e_\lambda)\cH=0$
\end{enumerate}
Since the application of $e_T$ commutes with the action of
$\unitaries(d)$, $U_T$ is a representation of $\unitaries(d)$.
Since $U_T \subset U_\lambda \otimes V_\lambda$ and $U_\lambda$ is
irreducible, $U_T$ is equivalent to a number of copies of
$U_\lambda$. Since $e_T$ has rank one, the number of copies can be
maximally one. This shows that $V(e_T)\cong U_\lambda$ if and only
if $U_\lambda$ ($V_\lambda$) is a subrepresentation of
$\unitaries(d)$ ($S_k$).

The irreducible representations of $S_k$ are obtained by first
constructing one nonzero vector in $U_\lambda$ and then applying
the permutations to this vector. Alternatively, one can construct
the minimal central projections, and so $U_\lambda \otimes
V_\lambda$, and subsequently project with a rank one operator to
obtain a representation isomorphic to $U_\lambda$.

There are two tasks to fulfill in this section: The first task is
to find a convenient labeling for the elements of $\hat{G}$ and to
construct a minimal projection for each $\lambda \in \hat{G}$.
This will lead to the concept of Young frames and Young tableaux
(subsection~\ref{sec-young-symmetriser}). The second task is to
study the size of the representations $U_\lambda$ and $V_\lambda$,
a subject which relates to the combinatorics of Young tableaux
(subsection~\ref{subsection-symm-group-young}).

\subsection{The Young Symmetriser}
\label{sec-young-symmetriser} A set $\lambda=(\lambda_1,
\lambda_2, \ldots, \lambda_d)$ of nonincreasing integers is called
a \emph{Young frame\index{Young frame}} or \emph{Young
diagram\index{Young diagram}} and is usually illustrated by a
diagram consisting of empty boxes arranged in rows, which are left
adjusted. The $i$-th row, counted from the top, consists of
$\lambda_i$ boxes. The purpose of the empty boxes is to be filled
in with numbers. A Young frame containing integers is then also
known as \emph{Young tableau} $T$ with Young frame $\lambda \equiv
\F(T)$. The \emph{size} of a Young frame or tableau is defined as
the number of boxes $k$ in its diagram, i.e.~$k=\sum_i
\lambda_i=|\lambda|$. $\lambda \vdash k$ (or $T \vdash k$)
indicates that $\lambda$ (or $\F(T)$) is a partition of $k$. The
\emph{depth} of a diagram or tableau is the number of rows $d$.
Often there are restrictions on the maximal number of rows of a
diagram; for brevity $\lambda \vdash (k, d)$ designates that
$\lambda$ has no more than $d$ rows. Two types of Young tableaux
will be relevant:

\begin{itemize}
\item A \emph{standard Young tableau}\index{Young tableau!standard} $T$ is a Young tableau with
the numbers $1$ to $k$ each in one box such that the numbers are
increasing to the right and downwards.
\item A \emph{semistandard Young tableau}\index{Young tableau!semistandard} $T$ is a Young tableau
containing numbers, possibly repeatedly, that weakly increase to
the right and strictly increase downwards.
\end{itemize}

\begin{figure}
\begin{center}
\begin{picture}(60,40)
\setlength{\unitlength}{0.30mm}
\put(0,20){\framebox(20,20){$1$}}\put(20,20){\framebox(20,20){$2$}}\put(40,20){\framebox(20,20){$3$}}
\put(0,0){\framebox(20,20){$4$}}\put(20,0){\framebox(20,20){$5$}}
\end{picture}
\begin{picture}(60,40) \setlength{\unitlength}{0.30mm}
\put(0,20){\framebox(20,20){$1$}}\put(20,20){\framebox(20,20){$2$}}\put(40,20){\framebox(20,20){$4$}}
\put(0,0){\framebox(20,20){$3$}}\put(20,0){\framebox(20,20){$5$}}
\end{picture}
\begin{picture}(60,40)
\setlength{\unitlength}{0.30mm}
\put(0,20){\framebox(20,20){$1$}}\put(20,20){\framebox(20,20){$3$}}\put(40,20){\framebox(20,20){$4$}}
\put(0,0){\framebox(20,20){$2$}}\put(20,0){\framebox(20,20){$5$}}
\end{picture}
\begin{picture}(60,40)
\setlength{\unitlength}{0.30mm}
\put(0,20){\framebox(20,20){$1$}}\put(20,20){\framebox(20,20){$2$}}\put(40,20){\framebox(20,20){$5$}}
\put(0,0){\framebox(20,20){$3$}}\put(20,0){\framebox(20,20){$4$}}
\end{picture}
\begin{picture}(60,40)
\setlength{\unitlength}{0.30mm}
\put(0,20){\framebox(20,20){$1$}}\put(20,20){\framebox(20,20){$3$}}\put(40,20){\framebox(20,20){$5$}}
\put(0,0){\framebox(20,20){$2$}}\put(20,0){\framebox(20,20){$4$}}
\end{picture}
\end{center}
\begin{center}
\begin{picture}(60,40) \setlength{\unitlength}{0.30mm}
\put(0,20){\framebox(20,20){$1$}}\put(20,20){\framebox(20,20){$1$}}\put(40,20){\framebox(20,20){$1$}}
\put(0,0){\framebox(20,20){$2$}}\put(20,0){\framebox(20,20){$2$}}
\end{picture}
\begin{picture}(60,40)
\setlength{\unitlength}{0.30mm}
\put(0,20){\framebox(20,20){$1$}}\put(20,20){\framebox(20,20){$1$}}\put(40,20){\framebox(20,20){$2$}}
\put(0,0){\framebox(20,20){$2$}}\put(20,0){\framebox(20,20){$2$}}
\end{picture}
\end{center}
\caption{Young frame $(3,2)$, top row: standard Young tableaux,
bottom row: semistandard Young tableaux with numbering $\{1,
2\}$.} \label{figure-tableaux}
\end{figure}
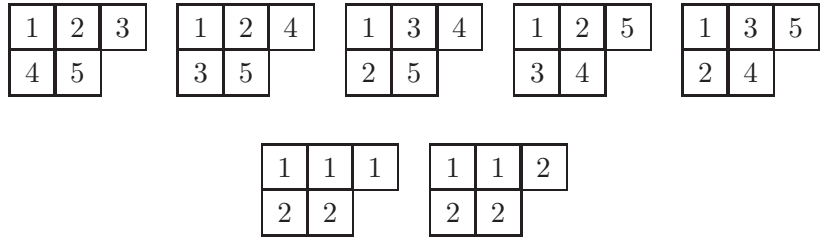

\noindent Standard Young tableaux will be important in the
construction of the Young symmetriser and, ultimately, the set of
standard Young tableaux with frame $\lambda$ will provide a
numbering for an orthonormal basis of the irreducible
representation $V_\lambda$ of $S_k$. Similarly, semistandard Young
tableaux, when filled with numbers $\{1, \ldots, d\}$, can be used
to enumerate an orthonormal basis of the irreducible
representation $U_\lambda$ of $\unitaries(d)$. Examples of
standard and semistandard Young tableaux can be found in
figure~\ref{figure-tableaux}.

To each standard Young tableau $T$, associate two elements of the
group algebra,
$$ c_T=\sum_{\pi \in \cC(T)} \text{sgn } (\pi) \pi \qquad r_T=\sum_{\pi \in \cR(T)} \pi,$$
where $\cR(T)$ and ${\cal C}(T)$ are sets of permutations of
$S_k$, $\cR(T)$ being those that are obtained by permuting the
integers within each row of $T$, and $\cC(T)$ those obtained by
permuting integers within each column of $T$. The \emph{Young
symmetriser}\index{Young symmetriser} $e_T$ is given by
\be \label{symmetry-operator} e_T=r_T c_T.
\ee
The goal of the section is to show that $e_T$ is proportional to a
minimal projection in $\cA(S_k)$, more precisely,
$$ (e_T)^2=\frac{k!}{f_\lambda} e_T, \qquad \mbox{where }
f_\lambda:=\dim V_\lambda.$$ The construction of irreducible
representations with Young symmetrisers is also known as Weyl's
tensorial construction. Throughout the remaining part of this
section, $T$ and $T'$ are standard Young tableaux.

\begin{lemma} \label{lemma-same-row-col}
If $T'=gT$ and if there are no two integers that are in the same
row of $T$ and also in the same column of $T'$, then $g=rc$ for
some transpositions $r \in \cR(T)$ and $c \in \cC(T)$. In
particular, if $g \neq rc$ then there is an element $t \in \cR(T)$
with $g^{-1}tg \in \cC(T)$.
\end{lemma}
\begin{proof}
Let $r_1 \in \cR(T)$ and $c'_1 \in \cC(T')$ such that $r_1 T$ and
$c'_1T'$ have the same first row. It is possible to find such
$r_1$ and $c'_1$ since by assumption there are no elements that
are in the same row of $T$ and also in the same column of $T'$.
Now repeat this process for row 2 until row $d$. Then $r=r_d
\cdots r_1$ and $c'=c'_d \cdots c'_1$ are elements with
$rT=c'T'=c'gT$. Hence $r=c'g$ and thus $g=rc$ with
$c=g^{-1}(c')^{-1}g \in \cC(T)$.
\end{proof}

\begin{lemma} \label{lemma-young-shuffling}
Let $r \in \cR(T)$ and $c \in \cC(T)$, then
\bea \label{eq-young-shuff-1}r r_T&=&r_T r=r_T\\
\label{eq-young-shuff-2}\sgn(c) c c_T&=&\sgn(c) c_T c=c_T\\
\label{eq-young-shuff-3} r e_T \sgn(c) c&=&e_T \eea and $e_T$ is
the only such element (up to scalar multiplication) for which
equation~(\ref{eq-young-shuff-3}) holds for all $r \in \cR(T)$ and
$c \in \cC(T)$.
\end{lemma}

\begin{proof}
The first two assertions as well the correctness of the last
equation are straightforward. It remains to show the uniqueness of
$e_T$ in the last equation. If $e=\sum_g n_g g$ satisfies
\be \label{eq-young-1} r e \ \sgn(c) c=e,\ee
for all $r \in \cR(T)$ and $c \in \cC(T)$, then $n_{rgc}=\sgn(c)
n_g$ for all $g$. It will suffice to show that $n_g=0$ for all $g
\neq rc$ for $r \in \cR(T)$ and $c \in \cC(T)$, since then
$e=n_{\id} \sum_{r \in \cR, c \in \cC} \sgn(c) rc=n_{\id} e_T$,
which concludes the proof. But lemma~\ref{lemma-same-row-col}
implies that for every $g\neq rc$, there is a transposition $t \in
\cR(T)$ such that $t'=g^{-1}tg \in \cC$ and therefore $g=tgt'$,
which by equation~(\ref{eq-young-1}) implies $n_g=-n_g=0$.
\end{proof}

\noindent In \emph{lexicographical
order\label{lexicographical}\index{lexicographical order}}, we say
that $\lambda >\mu$ if the first nonzero difference
$\lambda_i-\mu_i$ is larger than zero.

\begin{corollary} \label{cor-min-proj}
If $\lambda >\mu$, then $e_T e_{T'}=0$ for all $T, T'$ with
$\lambda=F(T)$ and $\mu=F(T')$. Furthermore $ (e_T)^2=n_T e_T$ for
some $n_T \in \complex$.
\end{corollary}

\begin{proof}
First part: If $\lambda>\mu$ then there are two integers in the
same row of $T$ and the same column of $T'$. For the transposition
$t$ of those two integers, $t \in \cR(T) \cap \cC(T')$ and thus by
lemma~\ref{lemma-young-shuffling} $r_Tc_{T'}=r_T t^2c_{T'}=(r_T t)
(t c_{T'})=-r_Tc_{T'}$. Second part: From
equation~(\ref{eq-young-shuff-1}) and
equation~(\ref{eq-young-shuff-2}) follows
$r(e_T)^2\sgn(c)=(e_T)^2$ and with
equation~(\ref{eq-young-shuff-3}): $(e_T)^2=n_T e_T$.
\end{proof}

\begin{theorem}
Let $T$ be a standard Young tableau to the frame $\lambda$. Then
$\frac{f_\lambda}{k!} e_T$ is a minimal projection associated with
the irreducible representation $V_\lambda$ of $S_k$, where $\dim
V_\lambda=f_\lambda$. The $V_\lambda$'s for $\lambda \vdash k$
form a complete set of irreducible representations of $S_k$.
\end{theorem}

\begin{proof}
According to corollary~\ref{cor-min-proj}, $e_T$ is proportional
to a projection, say $e$. Now assume that $e=e_1+e_2$ for
projections $e_1$ and $e_2$. Since $0=e^2-e=e_1e_2+e_2e_1$, one
has $0=e_1(e_1e_2+e_2e_1)e_1=2e_1e_2e_1$. This shows that
$e_1e_2=e_1^2e_2=e_1(e_1e_2+e_2e_1)-e_1e_2e_1=0$ holds. Since
$e_1e_2+e_2e_1=0$ this also implies $e_2e_1=0$, and proves
$$e e_1e=e_1.$$
According to lemma~\ref{lemma-young-shuffling}, the LHS of this
equation absorbs the multiplication by $r \in \cR(T)$ from the
left and by $c \in \cC(T)$ from the right, therefore
$re_1c=\sgn(c)e_1$. Furthermore, this lemma implies that $e_1$ is
proportional to $e_T$, which is therefore proportional to a
minimal projection.

The proportionality constant can be worked out as follows: for the
left action of $g \in G$ on $a \in \cA(G)$, $R_g: a \mapsto ga$,
one has $\tr R_{\id}=k!$, $\tr R_g=0$ for $g\neq \id$ and
therefore $\tr R_{e_T}=k!$. But also $e_T f=n_T f$ for $f \in e_T
\cA(G)$ and 0 otherwise. Hence $k!=n_T \dim e_T \cA(G)$, where
$e_T\cA(G)=\Span\{a e_T| a \in \cA(g)\}$. By theorem
\ref{theorem-equiv-minproj-irreps}, $\dim e_T \cA(G)=\dim
V_\lambda$.

$e_T$ and $e_{T'}$ are equivalent if $T$ and $T'$ have the same
frame, since then $T'=gT$ for some $g\in G$ and thus $e_{T'}=ge_T
g^{-1}$. They are inequivalent when their Young frames are
different, a fact that follows from corollary \ref{cor-min-proj}.

The number of conjugacy classes of a finite group equals the
number of inequivalent irreducible representations (see corollary
\ref{cor-char-orthonormal} below). Above, to each Young frame we
have constructed an inequivalent irreducible representations. To
conclude the proof of completeness it will therefore suffice to
give a one to one mapping of conjugacy classes of $S_k$ and Young
frames. Every permutation can be written as a unique product of
disjoint cycles, i.e.~$\pi =(i_{11} \ldots i_{1j_1})(i_{21} \ldots
i_{2j_2}) \ldots (i_{m1} \ldots i_{mj_m})$. Two permutations are
conjugate if and only if they have the same number of cycles of
length $j$ for all $j$. The correspondence of cycles to Young
frames is then apparent.
\end{proof}

\subsection{Combinatorics of Young Tableaux}
\label{subsection-symm-group-young}

In this subsection I state a few facts about the combinatorics of
Young tableaux which are elegantly exhibited in
Fulton~\cite{Fulton97}. $f^\lambda$ had been defined as the
dimension of the irreducible representation $V_\lambda$ of $S_k$.
Likewise let $t_\lambda(d)$ be the dimension of the corresponding
irreducible representation of $\unitaries (d)$. Both numbers can
be expressed as sums over Young tableaux.

\begin{theorem} \label{theorem-combinatorics}
The following formulas hold:
\bea \label{eq-comb-1} f^\lambda&=& | \{ \; T \; | \; T \mbox{ standard Young tableau with
} F(T)=\lambda\; \}|\\
 \label{eq-comb-2} t_\lambda(d)&=&  | \{ \; T \; | \; T \mbox{ semistandard Young
tableau }   \\
&& \nonumber \quad \mbox{with } F(T)=\lambda \mbox{ and numbers }
\{1,
\ldots, d\}\}| \\
 \label{eq-comb-3} d^k&=&\sum_{\lambda \vdash (k, d)} t_\lambda(d) f^\lambda.\eea
\end{theorem}

\noindent The minimal central projection is proportional to a sum
over the corresponding minimal projections. This shows that the
RHS of equation~(\ref{eq-comb-1}) is an upper bound on the LHS.
That equality holds stems from the independence of the minimal
projections. This is a consequence of
lemma~\ref{lemma-same-row-col}, which is explained
in~\cite[proposition VI.3.12]{Simon96Book}.
Formula~(\ref{eq-comb-2}) will be discussed in
section~\ref{sec-on-basis-subgroup-chains}; a full proof is not
given, but the problem is reduced to the well-known branching rule
for the unitary group. Equation~(\ref{eq-comb-3}) follows from
equations~(\ref{eq-comb-1}) and~(\ref{eq-comb-2}) as well as
Schur-Weyl duality, theorem~\ref{theorem-Schur-duality}.

\begin{figure}[h]
\begin{center}
\begin{picture}(40,60) \setlength{\unitlength}{0.30mm}
\put(0,40){\framebox(20,20){$$}} \put(20,40){\framebox(20,20){$$}} \put(40,40){\framebox(20,20){$$}}\put(60,40){\framebox(20,20){$$}}
\put(0,20){\framebox(20,20){$$}} \put(20,20){\framebox(20,20){$$}} \put(40,20){\framebox(20,20){$$}}
\put(0,0){\framebox(20,20){$$}}
\put(30,50){\vector(1,0){40}}
\put(30,50){\vector(0,-1){20}}
\end{picture}
\vspace{-0.5cm}
\end{center}
\caption{Hook of box $(1,2)$ in Young frame $(4,3,1)$}
\label{figure-hook-length}
\end{figure}
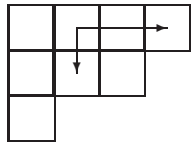

For both $f^\lambda$ and $t_\lambda(d)$, the combinatorical sum
can be evaluated and results in so-called \emph{hook length
formulae}. The \emph{hook} of box $(i,j)$ in a diagram is given by
the box itself, the boxes to its right and below. The \emph{hook
length\index{hook length}} is the number of boxes in a hook. An
illustration of the hook length is shown in figure
\ref{figure-hook-length}

\begin{theorem}[Hook length formulae\index{hook length!formulae}]
    \bea \label{eq-hook}
        f^\lambda &=& \frac{k!}{\prod_{(i,j) \in \lambda}h(i,j)}, \\
        t_\lambda(d) &=& \prod_{(i,j) \in \lambda}
        \frac{d+j-i}{h(i,j)}=\frac{f^\lambda}{k!} \prod_{(i,j) \in
        \lambda} (d+j-i)
    \eea
This formula is also a reformulation of Weyl's dimension formula,
which is best known in the form
\be t_\lambda(d)= \prod_{i<j} \frac{\lambda_i-\lambda_j+j-i}{j-i}\ee
\end{theorem}
For the purpose of this work, rather than precise formulae, the
asymptotic growth of $f_\lambda$ and $t_\lambda(d)$ in the limit
of large $k$ but with fixed ratios of the row lengths will be
important.

For the construction of a standard Young tableau there are
restrictions both on the rows and on the columns. Only considering
the restrictions on the rows clearly gives an upper bound on the
number of tableaux, namely,
    \be \label{eq-tableaux-bound}
        f^\lambda \leq \binom{k}{\lambda_1 \cdots
        \lambda_d}.
    \ee
For $t_\lambda(d)$ only a bound in terms of $k$ and $d$ will be
needed. Consider the diagram with all $k$ boxes in the first row.
Each semistandard Young tableau is characterised by the positions
where strict increase happens. These positions are the $k+1$
places \emph{between the boxes}. For an alphabet of size $d$, the
tableau is specified by $d-1$ such (possibly repeating) places,
giving $t_k(d) \leq (k+1)^{d-1}$.

For a general diagram $\lambda$, the above bound is applied to
each row individually, but note that in row $j$ only the letters
$j, j+1, \ldots, d$ can appear. Thus,
    \be \label{eq-semi-tableaux-bound}
        t_\lambda(d) \leq (\lambda_1+1)^{d-1}(\lambda_2+1)^{d-2}  \cdots  (\lambda_d+1)^{0}\leq (k+1)^{d(d-1)/2}.
    \ee
Let $\lambda \vdash (k, d)$ and define
$\bar{\lambda}:=(\frac{\lambda_i}{k}, \ldots,
\frac{\lambda_d}{k})$ with all the $\bar{\lambda}_i$ distinct.
Then Weyl's dimension formula gives
    \be
        t_\lambda(d) = O(k^{d(d-1)/2}).
    \ee
The estimate~(\ref{eq-semi-tableaux-bound}) is therefore optimal
in its dependence on $k$ for fixed $d$.

\section{The Irreducible Representations of the Unitary Group}
\label{section-unitary-group} This section starts by reviewing
definitions and fundamental results of Lie group and Lie algebra
theory (subsection~\ref{subse-lie-group-algebra}). Subsequently,
the focus will be narrowed to the unitary group, where the
labeling of an irreducible representation with a Young frame will
receive a natural interpretation as the \emph{highest weight}
(subsection~\ref{subsec-weights}). This also connects the Young
symmetriser to majorisation, a vital ingredient in
chapter~\ref{chapter-relation}
(subsection~\ref{subsec-highest-weight}). The results are stated
without proof; the reader is referred to standard literature
(e.g.~\cite{FultonHarris91, Simon96Book, GooWal98}).

\subsection{Lie Groups and Lie Algebras}
\label{subse-lie-group-algebra}
\begin{definition}
A \emph{Lie group\index{Lie group}} $G$ is a $d$-dimensional real
$C^\infty$-manifold as well as a group such that product and
inverse are $C^\infty$-maps. Additionally there should exist
countably many open sets in $G$ that cover $G$.
\end{definition}

\noindent The Lie algebra $\fg$ of a Lie group $G$ is the tangent
space of $G$ at the unit element $e$, when equipped with a Lie
bracket $[\; , \; ]: \fg \times \fg \rightarrow \fg$. For all $A,
B \in \fg$:
    $$
        [A, B]:=\lim_{t \rightarrow 0} \frac{(a(t)b(t)a^{-1}(t)
        b^{-1}(t)-e)}{t^2},
    $$
where $t \mapsto a(t), b(t)$ are $C^1$-curves on $G$ with
$a(0)=b(0)=e$, $a'(0)=A$ and $b'(0)=B$. Any Lie algebra obtained
in this way will satisfy the following definition.

\begin{definition}
A \emph{Lie algebra\index{Lie algebra}} over $\real$ (or
$\complex$) is a vector space $\fg$ together with a \emph{Lie
bracket}\index{Lie bracket} $ [\; , \; ]: \fg \times \fg
\rightarrow \fg$ such that for all $A, B, C \in \fg$ and $\alpha,
\beta \in \real (\complex)$ the following three hold:
\beastar \mbox{Linearity} &&[\alpha A+\beta B, C]= \alpha[A,C]+\beta[B,C] \\
\mbox{Anticommutativity}&&  [A,B]=-[B,A]\\
 \mbox{Jacobi identity\index{Jacobi identity}} &&   [A, [B, C]]+[B, [C, A]]+[C, [A,
B]]=0.
\eeastar
\end{definition}

\noindent Lie algebras capture the local properties of a Lie group
and provide an elegant way for their analysis. Since Lie groups
are also topological spaces, it is natural to discuss certain
topological properties.

\begin{itemize}
\item Compactness\index{compact}: $X$ is compact if for every set of open sets that cover $X$ there is a finite subset that also covers $X$.
\item Connectedness\index{connected}: $X$ is connected if it cannot be divided in
two disjoint nonempty closed sets.
\item Simply connectedness\index{simply connected}: $X$ is simply connected if it is path
connected (i.e.~for every $x, x'$ there is a continuous function
$f:[0,1] \rightarrow X$ with $f(0)=x$ and $f(1)=x'$) and every
continuous map $f: S^1 \rightarrow X$ can be continuously
contracted to a point.
\end{itemize}
An extensive table of Lie groups, their Lie algebras and
topological properties can be found in~\cite{WikipediaLieGroups}.
The Lie groups that are relevant for this thesis are summarised in
table~\ref{table-lie-groups}.
\begin{table}
\begin{center}
\begin{small}
\begin{tabular}{llll}
Lie group            & definition                        & topology \slash \\
 \, \slash Lie algebra   &                                   & \,  real vs. complex\\
\hline\hline\\
$\GL(d, \complex)$   & $g \in \M(d, \complex), \det g\neq 0$       & not compact, connected,\\
                     &                                            & not simply connected\\
$\gl(d, \complex)$   & $A \in \M(d, \complex)$                      & $\complex$\\
\hline
$\SL(d, \complex)$     & $g \in \M(d, \complex), \det g=1$           & not compact ($d\geq2$), \\
                     &                                            &  simply connected\\
$\mathfrak{sl}(d, \complex)$ & $A  \in \M(d, \complex), \tr A=0$    & $\complex$\\
\hline
$ \unitaries(d)$     & $g \in \M(d, \complex), gg^\dagger=\id $    & compact, connected, \\
                       &                                          & not simply connected  \\
$ \mathfrak{u}(d)$   & $A \in \M(d, \complex), A=-A^\dagger$     & $\real$\\
\hline
$ \sunitaries(d)$    & $g \in \M(d, \complex), gg^\dagger=\id, \det g=1$ &  compact,  \\
                     &                                             & simply connected\\
$ \mathfrak{su}(d)$   & $A \in \M(d, \complex), A=-A^\dagger, \tr A=0$ & $\real$\\
\hline
$ \SO(d)$             & $g \in \M(d, \real), gg^T=\id, \det g=1 $     & compact, connected, \\
                       &                                            & not simply connected ($d \geq 2$)\\
$\mathfrak{so}(d)$   & $A \in \M(d, \real), A=-A^T$               & $\real$ \\
\end{tabular}
\end{small}
\end{center}
\caption{Table of Lie groups and their Lie algebras that appear in
this text.} \label{table-lie-groups}
\end{table}

The topological property that will be used frequently is
compactness since it implies the existence of an invariant measure
on the group: for every compact Lie group $G$ there exists a
left-invariant measure $d\mu(g)$, the \emph{Haar
measure\index{Haar measure}}, which is unique up to normalisation.
Here, the normalisation is chosen such that $\int_{g \in G}
d\mu(g) =1$. The Haar measure is one reason why the representation
theory of compact Lie groups is in many ways analogous to the
representation theory of finite groups.

A \emph{representation of a Lie group $G$\index{representation!Lie
group}} on a vector space $V$ is a homomorphism $\phi: G
\rightarrow \GL(V)$ such that the group operation is preserved,
i.e.
$$ \phi(gh)=\phi(g)\phi(h) \qquad \mbox{for all } g, h, k \in G,$$
The dimension of the representation is given by the dimension of
$V$. Informally, a representation of a Lie algebra is a
homomorphism of a Lie algebra into a matrix Lie algebra that
preserves the Lie bracket. Formally, a \emph{representation of a
Lie algebra} $\fg$ on a vector space $V$ is a homomorphism $\phi:
\fg \rightarrow
\endo(V)$ such that
$$ \phi([A,B])=[\phi(A),\phi(B)] \qquad \mbox{for all } A, B \in \fg,$$
where $[A, B]:=AB-BA$ is the Lie bracket and $[\phi(A),
\phi(B)]:=\phi(A)\phi(B)-\phi(B)\phi(A)$ is the commutator derived
from the matrix product in $\endo(V)$. The dimension of the
representation is given by the dimension of $V$.

With regard to local properties, the representation theories of
Lie algebras and Lie groups parallel each other. Here is a list of
some pertinent facts.

\begin{theorem} \label{theorem-Lie-group-facts}
Let $\phi: G \rightarrow \GL(V)$ be a representation of a Lie
group $G$.
\begin{itemize}
\item $ L \phi: \mathfrak{g} \rightarrow \gl (V)$ is a representation of the corresponding
Lie algebras, where $L \phi$ is the linearisation at the identity
element.
\item Conversely, for every representation $\psi: \mathfrak{g} \rightarrow \gl (V)$ of the Lie algebra $\mathfrak{g}$, there is a local
representation~\footnote{A \emph{local representation} of a Lie
group $G$ on a vector space $V$ is a group homomorphism $\phi:
\cO(e) \subset G \rightarrow \cO'(\id) \subset \GL(V),$ where
$\cO(e)$ and $\cO'(\id)$ are opens sets containing the respective
identity elements.} $\phi: G \rightarrow \GL(V)$ such that $L
\phi=\psi$.
\item If $G$ is connected, then the representation $\phi$ is
irreducible whenever $L\phi$ is irreducible.
\item If $G$ is simply connected, then for every
$ \psi: \mathfrak{g} \rightarrow \gl (V)$ there is exactly one
representation $ \phi: G \rightarrow \GL(V)$ with $L \phi=\psi$.
\end{itemize}
\end{theorem}
All groups that will be relevant for us are subgroups of $\GL(d,
\complex)$. Likewise the Lie algebras will be subalgebras of the
Lie algebra $\gl(d, \complex)$, which equals $\M(d, \complex)$,
the complex $d\times d$ matrices. The group operation is matrix
multiplication and the Lie bracket is given by the commutator $[A,
B]:=AB-BA,$ where $AB$ is the matrix multiplication in $\M(d,
\complex)$. Table~\ref{table-lie-groups} gives a short list of
relevant Lie groups and their Lie algebras. The following theorem
allows to reduce the study of irreducible representation of a Lie
algebra to the study of its complexified Lie algebra. For a Lie
subalgebra $\mathfrak{g}$ of the real Lie algebra $\gl(n, \real)$,
define
$$\mathfrak{g}_\complex:=\Span \{zA| z \in \complex, A \in \mathfrak{g}\},$$
the \emph{complexified Lie algebra}\index{Lie
algebra!complexification} of $\mathfrak{g}$.
\begin{theorem}\label{theorem-lie-algebra-complexification}
\label{theorem-lie-complexify} Let $\psi: \mathfrak{g} \rightarrow
\gl(V, \real) $ be a representation of a real Lie algebra
$\mathfrak{g}$ on the complex space $V$. Then
$$\psi_\complex: \fg_\complex \rightarrow \gl(V,
\complex),$$ given by $ \psi_\complex(zA):=z \psi(A)$ for $z \in
\complex$ is a representation of $\fg_\complex$. Further, $\psi$
is irreducible if and only if $\psi_\complex$ is irreducible.
\end{theorem}

\subsection{The Lie Algebra of $\U(d)$, Weights and Weight Vectors}
\label{subsec-weights} In this subsection, weights and weight
vectors will be introduced and the irreducible representations of
the unitary group will be characterised by their lexicographically
highest weight.

Let $V$ be a representation of $\unitaries(d)$ and restrict the
action to the diagonal elements of $\unitaries(d)$. The
commutative group of diagonal elements is known as
\emph{torus}\index{torus} $T$ and is isomorphic to
$\underbrace{\unitaries(1)\times \unitaries(1)\times \cdots \times
\unitaries(1)}_d$. The holomorphic irreducible representations of
$\unitaries(1)$, i.e.~the irreducible representations whose matrix
entries are holomorphic functions in $\unitaries(1)$ are given by
$u^k$ for $k \in \integers, u \in \unitaries(1)$. $V$, as a
representation of $T$, then decomposes into a direct sum of
one{\-}-di\-men\-sio\-nal irreducible representations. For
$U=\diag(u_1, \ldots, u_d)$,
    $$
        U\ket{v}=u_1^{f_1} \cdots u_d^{f_d}\ket{v}=u^f\ket{v} \quad \mbox{ for } f=(f_1, \ldots,
        f_d) \in \integers^d,
    $$
where $\ket{v}$ is a basis vector for one of the irreducible
representations of $T$. $\ket{v}$ is called a \emph{weight
vector\index{weight!vector}} with \emph{weight\index{weight}} $f$.

As for Young diagrams the set of weights is ordered
lexicographically, i.e.~$f$ is of higher weight than $f'$ if
$f>f'$ in lexicographical order\index{lexicographical order}. It
will turn out that every irreducible representation has a unique
\emph{highest weight vector\index{weight!highest}}. But let us
first see how one operates on the set of weights.

The Lie algebra $\mathfrak{u}(d)$ of $\unitaries(n)$ consists of
skew Hermitian matrices. Any element in $\mathfrak{u}(d)$ can be
written as a real linear combination of $i$ times a Hermitian
matrix. Since any complex $d\times d$ matrix is of the form
$A+iB$, where $A$ and $B$ are Hermitian matrices, the
complexification\index{Lie group!complexification} of
$\mathfrak{u}(d)$ is just $\M(d, \complex)$, the Lie algebra
$\gl(d)$ of $\GL(d)$. An action of $\unitaries(d)$ thus extends to
an action of $\GL(d)$. Let $E_{ij}$ be the matrix with a one at
position $(i,j)$ and zero otherwise. The $E_{ij}$ form a basis for
$\gl(d)$. The action of an element $g \in \GL(d)$ that is
infinitesimally close to $\id$ can be linearly approximated by
$\id + \sum_{ij} \epsilon_{ij} E_{ij}$, which leads to the action
of $E_{ij}$ on $V$. Let $\ket{v}$ be a weight vector with weight
$f$, then
$$ UE_{ij}\ket{v}=UE_{ij}U^{-1}U\ket{v}=u_iu^{-1}_jE_{ij}U\ket{v}=u^{f+\epsilon_{ij}} E_{ij}\ket{v},$$
where $U=\diag(u_1, \ldots, u_d)$ as above and $E_{ij}\ket{v}$ is
either zero or a weight vector with weight $f+\epsilon_{ij}$,
where $\epsilon_{ij}$ is a vector with a $1$ at position $i$, a
$-1$ at position $j$ and zero otherwise.

\subsection{The Relation between $\sunitaries(d)$, $\unitaries(d)$
and $\GL(d)$}\label{sec-relation-SU-U}

\begin{quote} ``The importance of the full
linear group $\GL(n)$ lies in the fact that any group $\Gamma$ of
linear transformations is a subgroup of $\GL(n)$ and hence
decomposition of the tensor space with respect to $\GL(n)$ must
precede decomposition relative to $\Gamma$. One should, however
not overemphasize this relationship; for after all each group
stands in its own right and does not deserve to be looked upon
merely as a subgroup of something else, be it even Her
All-embracing Majesty $\GL(n)$.''
\end{quote}
\begin{quotewho}
H. Weyl, 1939 in ``The Classical Groups''~\cite{Weyl50Book}
\end{quotewho}

\noindent The purpose of this subsection is to clarify the
relation between $\GL(d)$, $\U(d)$ and $\SU(d)$ and how it
manifests itself in its unitary representations and Young frames.

In the previous section it was noted that the action of $\U(d)$
extends to an action of its complexification $\GL(d)$. Since both
groups are connected, theorem~\ref{theorem-Lie-group-facts}
and~\ref{theorem-lie-algebra-complexification} imply that a
representation of $\U(d)$ is irreducible if and only if the
corresponding representation of $\GL(d)$ is irreducible.

This brings us to the relation between $\U(d)$ and $\SU(d)$. Since
$\U(d)=\U(1) \times \SU(d)$, the irreducible representations of
$\U(d)$ can be realised as products of irreducible representations
of $\U(1)$ and $\SU(d)$. Since $\U(1)$ is an Abelian group, its
irreducible representations are one{\-}-di\-men\-sio\-nal and, as
discussed previously, depend on an integer $m$, i.e.~$u \mapsto
u^m$, where $u \in \U(1)$ are all inequivalent irreducible
representations of $\U(1)$. Weyl's tensorial construction only
results in the representations with nonnegative $m$, i.e.~only
constructed the irreducible \emph{polynomial} representations of
$\U(1)$ (and so for $\U(d)$). The case of a negative integer $m$
can be included into Weyl's construction by defining
$$ U_{\lambda'}(g) := (\det g)^m U_\lambda(g).$$
$\lambda'=(m+\lambda_1, \ldots, m+\lambda_d)$ is an element of
$\integers^d$ and it can be shown that the representations of this
form are the set of all irreducible \emph{holomorphic}
representations of $\U(d)$. Since every element $g \in
\sunitaries(d)$ has determinant equal to one, all representations
$U_{\lambda+(m, \ldots, m)}$ are equivalent and a complete set can
be indexed diagrams $\lambda$ with only $d-1$ rows.

The study of $\sunitaries(2)$ representations is then merely the
study of representations with one row diagrams $\lambda=k$, the
\emph{spin representations}\index{representation!spin}. The
dimension of $U_k$ is just $k+1$ and by physicists usually
identified with a spin $j$ particle, where $j=\frac{k}{2}$. The
following section offers an alternative way of looking at the
representations of $\sunitaries(d)$, probably more familiar to
physicists: the irreducible representations are here constructed
from the commutation relations of the Lie algebra $\su(2) \cong
\mathfrak{so}(3)$.

\subsection{Spin States and the Clebsch-Gordan Transformation}
\label{subsec-spin-states}
\begin{quote}
``The analogy between the spinor spanner and the neutron suggests
that the state of the latter depends not only on its position and
momentum but on which of two topologically distinct ways it is
tied to its surroundings. A full turn about an axis leaves its
position and momentum unchanged but reverses \emph{its topological
relation to the rest of the universe}.''
\end{quote}

\begin{quotewho}
Ethan D. Bolker in ``The Spinner Spanner''~\cite{Bolker73}
\end{quotewho}

\noindent In quantum information theory one usually speaks of $d$
level systems, whereas in physics a quantum system with a finite
number of degrees of freedom is denoted by a nonnegative half
integer $j$, which lives in $d=2j+1$ complex dimensions. As we
have discussed at the end of the previous section, the natural Lie
algebraic notation is $k=2j=d-1$.

Let us start by reviewing the usual construction for the
irreducible representation of $\sunitaries(2)$ by constructing the
irreducible representations of its Lie algebra $\su(2)$. Since
$\su(2)$ is isomorphic to $\mathfrak{so}(3)$ the irreducible
representations of $SO(3)$ will be obtained as well. Whereas in
the case of the simple Lie group $SU(2)$ each irreducible
representation of $\su(2)$ leads to an irreducible representation
of $SU(2)$ this cannot be expected for $SO(3)$. In fact only the
ones with integral $j$ do. This is related to the double covering
of $SO(3)$ by $SU(2)$. The quote at the start of this section
refers to a famous illustration of the double covering due to
Dirac, known as \emph{Dirac's spanner}\index{Dirac's spanner} or
the \emph{Spinor spanner}: one attaches three or more ropes to a
spanner and fixes the other ends of the ropes at positions in
space. If the spanner is turned around 360 degrees, the ropes
become tangled up and it is not possible to bring them into the
original configuration without rotating the spanner or cutting the
ropes. If, however, one continues to rotate the spanner in the
same direction a further 360 degrees, it magically becomes
possible to disentangle the mess of ropes.\footnote{One needs a
bit of patience and I am particularly grateful to Henriette
Steiner who helped me conduct the experiment twice!}

\begin{figure} \label{figure-clebsch-gordon}
\hspace{1cm}
\begin{picture}(50, 50)
\setlength{\unitlength}{0.2mm}
\put(-20,20){\framebox(20,20){$$}}\put(0,20){\framebox(20,20){$$}}\put(20,20){\framebox(20,20){$$}}
\put(40,20){\framebox(20,20){$$}}
\put(70,25){$\otimes$}\put(90,20){\framebox(20,20){$$}}\put(110,20){\framebox(20,20){$$}}
\put(140,25){$=$}\put(160,20){\framebox(20,20){$$}}\put(180,20){\framebox(20,20){$$}}\put(200,20){\framebox(20,20){$$}}\put(220,20){\framebox(20,20){$$}}\put(240,20){\framebox(20,20){$$}}\put(260,20){\framebox(20,20){$$}}

\put(290,25){$\oplus$}\put(310,20){\framebox(20,20){$$}}\put(330,20){\framebox(20,20){$$}}\put(350,20){\framebox(20,20){$$}}\put(370,20){\framebox(20,20){$$}}\put(390,20){\framebox(20,20){$$}}
\put(310,0){\framebox(20,20){$$}}
\put(140,-15){$\oplus$}\put(160,-20){\framebox(20,20){$$}}\put(180,-20){\framebox(20,20){$$}}\put(200,-20){\framebox(20,20){$$}}\put(220,-20){\framebox(20,20){$$}}
\put(160,-40){\framebox(20,20){$$}}\put(180,-40){\framebox(20,20){$$}}

\put(140,-80){$=$}\put(160,-85){\framebox(20,20){$$}}\put(180,-85){\framebox(20,20){$$}}\put(200,-85){\framebox(20,20){$$}}\put(220,-85){\framebox(20,20){$$}}\put(240,-85){\framebox(20,20){$$}}\put(260,-85){\framebox(20,20){$$}}
\put(290,-80){$\oplus$}\put(310,-85){\framebox(20,20){$$}}\put(330,-85){\framebox(20,20){$$}}\put(350,-85){\framebox(20,20){$$}}\put(370,-85){\framebox(20,20){$$}}
\put(400,-80){$\oplus$}\put(420,-85){\framebox(20,20){$$}}\put(440,-85){\framebox(20,20){$$}}
\end{picture}
\vspace{2cm} \caption{Clebsch-Gordan decomposition: $\mbox{spin }
2 \,\otimes \mbox{ spin } 1= \mbox{ spin } 3\, \oplus \mbox{ spin
} 2\, \oplus \mbox{ spin } 1$}
\end{figure}
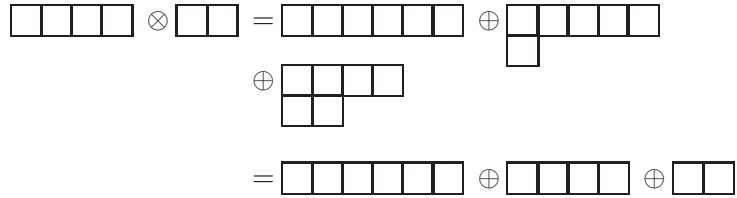

A basis for $\su(2)$ is then given by $\{i \sigma_x, i \sigma_y,
i\sigma_z\}$, where
\[
\sigma_x=\left( \begin{array}{cc} 0&1\\ 1&0 \end{array} \right)
\quad \sigma_y=\left( \begin{array}{cc} 0&-i\\ i&0 \end{array}
\right) \quad \sigma_z=\left( \begin{array}{cc} 1&0\\ 0&-1
\end{array} \right),
\]
are the \emph{Pauli operators\index{Pauli operator}}. Physicists
usually absorb the imaginary unit into the commutation relations,
which then read
$$ [\sigma_x, \sigma_y]=i 2 \sigma_z$$
and similarly for cyclic permutations of $(x, y, z)$. A
representation of the Lie algebra is then a mapping of Pauli
operators to Hermitian operators $J_x, J_y$ and $J_z$ that obeys
    \be \label{commutation-relations}
        [J_x, J_y]=i 2J_z.
    \ee
Since the operators $J^2=J_x^2+J_y^2+J_z^2$ and $J_z$ commute they
have a common eigenbasis $\ket{j, m}$, where $m=-j, -j+1, \ldots,
j-1, j$ and
    \beastar
        J_z \ket{j, m}= m \ket{j, m}\\
        J^2 \ket{j, m}= j(j+1)\ket{j, m}.
    \eeastar
In fact for every half integer $j$, i.e.~$j \in \{0, \half, 1,
\frac{3}{2}, \ldots \}$ there is an irreducible representation of
this form, usually denoted by $\cD^{(j)}$. The dimension of
$\cD^{(j)}$ is $2j+1$ and the eigenstates transform by means of
lowering and raising operators $J_\pm=J_x \pm iJ_y$ according to
$$ J_\pm \ket{j, m}=\sqrt{j(j+1)-m(m\pm1)} \ket{j, m\pm1}.$$
The relation to the tensorial construction using Young
symmetrisers reads
$$ \ket{j,m}=\frac{1}{\sqrt{k! (j+m)! (k-(j+m))!}} \sum_{\pi \in S_k}
\pi \ket{\uparrow\uparrow\ldots \uparrow \underbrace{\downarrow
\downarrow \cdots \downarrow }_{j+m}},$$ where I have used the
common notation for the basis states of a spin $\half$ particle,
`spin up': $\ket{\half,\half}=\ket{1}=\ket{\uparrow\nolinebreak}$
and `spin down':
$\ket{\half,-\half}=\ket{0}=\ket{\nolinebreak\downarrow\nolinebreak}$.
For $j=2$ this gives
\beastar \ket{2,2}&=&\ket{\uparrow  \uparrow  \uparrow  \uparrow }\\
\ket{2,1}&=&\frac{1}{2}\left( \ket{ \downarrow \uparrow  \uparrow
\uparrow }+\ket{ \uparrow \downarrow  \uparrow
 \uparrow }+\ket{ \uparrow   \uparrow   \downarrow \uparrow }+\ket{ \uparrow   \uparrow  \uparrow  \downarrow }\right) \\
\ket{2,0}&=&\frac{1}{\sqrt{6}}\left( \ket{\uparrow \uparrow \downarrow \downarrow}+  \ket{\uparrow \downarrow \downarrow \uparrow }+\ket{\uparrow \downarrow  \uparrow\downarrow} +\ket{\downarrow\uparrow \downarrow  \uparrow}+\ket{\downarrow \uparrow \uparrow \downarrow }+\ket{ \downarrow \downarrow \uparrow\uparrow }\right)\\
 \ket{2,-1}&=&\frac{1}{2}\left( \ket{ \downarrow
\uparrow \uparrow \uparrow }+\ket{ \uparrow \downarrow  \uparrow
 \uparrow }+\ket{ \uparrow   \uparrow   \downarrow \uparrow }+\ket{ \uparrow   \uparrow  \uparrow  \downarrow }\right) \\
\ket{2,-2}&=&\ket{\downarrow\downarrow\downarrow\downarrow }.
\eeastar
Often, in physical systems the Hamiltonian does not so strongly
depend on the individual spins it is made up of, $J_1$ and $J_2$,
but rather on their total spin $J=J_1+J_2$. It is therefore
customary to change from the tensor product basis $\ket{j_1 m_1}
\ket{j_2 m_2}$ to the eigenstates of $J^2, J_1^2, J_2^2$ and
$J_z=J_{z1}+J_{z2}$, given by $\ket{j, j_1, j_2, m}$. The unitary
matrix making this change of basis is known as the
\emph{Clebsch-Gordan transformation} and given by the
\emph{Clebsch-Gordan
coefficients\index{Clebsch-Gordan!coefficients}} $\braket{j m j_1
j_2 m}{j_1 m_1 j_2 m_2}$. Focusing on the decomposition of the
tensor product in irreducible components alone gives
    \be \label{eq-clebsch-gordon} \cD^{(j_1)} \otimes
        \cD^{(j_2)}\cong \bigoplus_{j=|j_1-j_2|}^{j_1+j_2} \cD^{(j)}.
    \ee
The correspondence to representations constructed with Young
symmetrisers, $\cD^{(j)} \cong U_{2j}$, leads to the illustration
of formula~(\ref{eq-clebsch-gordon}) in terms of Young diagrams
(see figure~\ref{figure-clebsch-gordon}). The integer indicating
the multiplicity of the irreducible representation (here 0 or 1)
is known as \emph{Clebsch-Gordan integer},
\[ c_{2j_1,2j_2}^{2j}=\left\{ \begin{array}{l} 1, \qquad  j \in \{|j_1-j_2|, \ldots, j_1+j_2\} \\ 0, \qquad \mbox{ otherwise} \end{array} \right\}, \]
and paves the notational way for generalisation: for a general
irreducible representation of $\U(d)$ the multiplicity in this
decomposition will often be higher than 0 or 1. The Clebsch-Gordan
integer for this case is known as the
\emph{Little\-wood-Richard\-son coefficient} and will -- together
with its symmetric group analogue, the \emph{Kronecker
coefficient} -- play a central role in
chapter~\ref{chapter-relation}.

\section{Schur-Weyl Duality II}
\label{sec-schur-weyl-duality-2} Here, the pieces will be put
together. In subsection~\ref{subsec-highest-weight} the highest
weights are brought into relation to the Young symmetriser, and in
subsection~\ref{subsec-summary} a brief conclusion is drawn and
then the results are highlighted that are most relevant for
chapter~\ref{chapter-relation}.

\subsection{Highest Weights, Young Symmetrisers and Majorisation}
\label{subsec-highest-weight} In this subsection the highest
weight of the representations $U_\lambda$, constructed with help
of $e_\lambda$, will be shown to equal $\lambda$. Further, it will
be shown that $U_\lambda$ has a unique highest weight vector. It
is the main purpose of this subsection, however, to explain the
connection between majorisation and irreducible representations --
a connection that points to the relation between irreducible
representations and typical subspaces.

Consider a basis vector $\ket{v}=\ket{e_{i_1}} \otimes \cdots
\otimes \ket{e_{i_k}}$ and a Young tableau $T$. To the pair $(T,
\ket{v})$ associate a tableau $T_v$ by replacing $j$ in $T$ with
$i_j$; also define $\ket{v_T}$ as the basis vector with $i$'s at
the positions in row $i$ of $T$.

\begin{lemma} \label{lemma-major}
Let $\ket{v}$ be a basis vector with frequency distribution $f$.
Then $e_T\ket{v}=0$ for all standard Young tableaux $T$ with frame
$\lambda$ and $f \not\prec \lambda$.
\end{lemma}

\begin{proof}
Let $f=(f_1, \ldots, f_d)$, where $f_i$ is the frequency with
which symbol $i$ appears in $\ket{v}$. Without loss of generality,
let $f_i\geq f_{i+1}$ for all $i$ and the row length $\lambda_i$
be the number of columns of size $i$ or larger. Note that
$e_T\ket{v}=0$ if a column in $T_v$ contains two identical
elements, since $c_T=c_T(\id-(kl))$, where $k$ and $l$ are the
positions of the identical elements.

The total number of columns is given by $\lambda_1$, hence, there
cannot be more than $\lambda_1$ elements of the same kind,
i.e.~$f_1 \leq \lambda_1$, if $e_T\ket{v}\neq 0$. Similarly, if
one inserts the two most abundant elements, one can insert pairs
of different elements in all the columns of length 2 or larger. In
addition, one can insert some elements in the columns of length
one. Thus, $f_1+f_2\leq 2
\lambda_2+(\lambda_1-\lambda_2)=\lambda_1+\lambda_2$. Continuing
in this way, it follows that for all basis vectors $\ket{v}$ with
$e_T\ket{v}\neq 0$, it must hold that $f \prec \lambda$.
\end{proof}

\begin{lemma} \label{lemma-highest-weight}
Let $\ket{v_T}$ be the basis vector with $i$'s at the positions in
row $i$ of $T$. Then $e_T\ket{v_T}$ is the unique vector of
highest weight in the representation associated to $T$.
\end{lemma}

\begin{proof}
Note that the application of $e_T$ does not change the weight of a
vector (even though the vector might send to zero).
Lemma~\ref{lemma-major} states that any basis vector with $f
\not\prec \lambda$ is sent to zero by $e_T$. Since $\ket{v_T}$ has
$f=\lambda$, it will be either sent to zero by $e_T$ or
$e_T\ket{v_T}$ is the desired highest weight vector (note that
vectors $\tilde{f}$ with $\tilde{f}_{i}=f_{\pi(i)}$ for $\pi \in
S_d$ and $\pi \neq \id$ are of lower lexicographical order when
compared to $f$). Since $c \ket{v_T}$ are orthogonal for all $c
\in \cC(T)$ and $r \ket{v_T}=\ket{v_T}$ for all $r \in \cR(T)$,
\beastar
\braket{e_Tv_T}{e_Tv_T} &=&\braket{v_T}{e_T^2v_T} \\
                    &=&n_T\bra{v_T}e_T\ket{v_T}\\
                    &=&n_T (\bra{v_T}r_T)(c_T\ket{v_T})\\
                    &=&n_T \braket{v_T}{v_T}=n_T|\cR(T)| \neq
                    0.
\eeastar
So, $e_T\ket{v_T}$ is a highest weight vector in $U_\lambda$ as
constructed from $e_T$.

To see that $e_T\ket{v_T}$ is unique let $T'=gT$ be a different
tableau. If $g \neq rc$, then by lemma~\ref{lemma-same-row-col}
there are two numbers in the same row of $T$ and the same column
of $T'$, thus
$e_T\ket{v_{T'}}=ge_{T'}g^{-1}\ket{v_{T'}}=ge_{T'}\ket{v_T}=0,$
since the two identified values in the same row of $T$ will be
antisymmetrised by $e_T$. Finally, if $g=rc$, then
$e_T\ket{v_{T'}}=e_T g \ket{v_T}=e_T rc \ket{v_T}=e_T c
\ket{v_T}=0$ if $c \neq \id$ and otherwise
$e_T\ket{v_{T'}}=e_T\ket{v_T}$.\end{proof}

\begin{corollary} \label{cor-last-bit-in-puzzle}
The irreducible representation $U_\lambda$ of $\U(d)$ is contained
in the tensor product representation on $\cH \cong
(\complex^d)^{\otimes k}$ if and only if $\lambda$ is a diagram
with no more than $d$ rows containing a total of $k$ boxes,
i.e.~$\lambda \vdash (k, d)$
\end{corollary}

\begin{proof}
From the representation theory of symmetric groups we know that
only Young frames with $k$ boxes denote irreducible representation
of $S_k$, hence $\lambda \vdash k$.

Trivially each basis vector in $(\complex^d)^{\otimes k}$ has a
frequency vector $f$ with no more than $d$ rows. A diagram
$\lambda$ with more than $d$ rows can therefore never majorise
$f$, i.e.~$f \not\prec \lambda$. By lemma \ref{lemma-major}, the
Young projectors $e_T$ corresponding to $\lambda$ give $e_T\cH=0$
and therefore $U_\lambda$ is not contained in the tensor
representation of $\U(d)$ in $\cH$. Conversely the vectors
$\ket{v_T}$ with no more than $d$ rows are contained in $\cH$ and
by lemma \ref{lemma-highest-weight} they lead to nonzero
projections $e_T \ket{v_T} \neq 0$.
\end{proof}

\subsection{Summary}
\label{subsec-summary} Let us summarise the achievements so far.
With the proof of corollary~\ref{cor-last-bit-in-puzzle}, it has
been determined which irreducible representations appear in the
tensor product decomposition. This completes the proof of
Schur-Weyl duality (theorem~\ref{theorem-Schur-duality}).
Furthermore, an explicit construction of irreducible
representation of $S_k$ and $\U(d)$ by means of Young symmetrisers
has been obtained. That is, for $T$ a tableau with frame
$\lambda$, where $\lambda \vdash (k, d)$,
\begin{itemize}
\item $e_T\cH \cong U_\lambda$ the irreducible representation of
$\U(d)$ with highest weight $\lambda$ (where $\cH \cong
(\complex^d)^{\otimes k}$.)
\item $\Span\{\pi e_T v_T | \pi \in S_k \} \cong V_\lambda$ the irreducible
representation of $S_k$ corresponding to the conjugacy class given
by the set of cycles $\lambda$.
\end{itemize}
In chapter~\ref{chapter-relation}, asymptotic properties of the
irreducible representations will play a significant role. The
asymptotic is considered for Young frames with fixed or converging
ratios of row lengths. Bounds for the dimension of the irreducible
representations of $\U(d)$ and $S_k$ are
    \bestar
        \dim U_\lambda \leq (k+1)^{d(d-1)/2} \qquad \dim V_\lambda \leq e^{k
        H(\bar{\lambda})},
    \eestar
where $\bar{\lambda}=(\frac{\lambda_i}{k}, \ldots,
\frac{\lambda_d}{k})$ with $k=|\lambda|$ and will be used to
relate the irreducible representations, with help of the
majorisation property -- if $\ket{v}$ is a basis vector with
frequency $f$, then $e_T\ket{v}=0$ if $f \not\prec \lambda$ -- to
typical subspaces of density operators.

\section{Orthogonal Bases and Subgroup Chains}
\label{sec-on-basis-subgroup-chains} Unfortunately, the Young
symmetriser construction for the basis vectors of the symmetric
and unitary group representations does not in general lead to
orthogonal vectors. In many applications this is a disadvantage.
This section is devoted to enhancing the previous construction (or
for that matter any other construction) in order to recursively
obtain an orthogonal basis. This is accomplished by employing the
properties of a \emph{subgroup chain}, a chain of proper subgroups
each contained in the previous one,
$$G =G_0 \supset G_1 \supset \ldots \supset G_n.$$
The general idea is the following: start with an irreducible
representation of a group $G$ and consider the reduction into
irreducible representation of $G_1$. The space of this (in general
reducible) representation can be written as a direct sum of
subspaces. Repeating this process for subgroups $G_2$ until $G_n$
results in a decomposition of the space into irreducible
representations of $G_n$. If each step in the chain is
multiplicity free, the representations of $G_n$ will have a unique
label given by a path of irreducible representations. If $G_n$ is
Abelian, then all its irreducible representation are
one-dimensional and the vectors spanning the representations form
the desired orthonormal basis up to an arbitrary choice of phase
for each basis vector.

In case of the symmetric group the main player in this procedure
is the \emph{branching rule} that governs the reduction from $S_k$
to $S_{k-1}$ (for a proof see~\cite[page 108]{Simon96Book}).

\begin{theorem}[Branching rule for $S_k$\index{branching rule!symmetric group}]
\be \cV_\lambda \downarrow^{S_k}_{S_{k-1}} \cong \bigoplus_{\lambda'
\vartriangleleft \lambda} \cV_{\lambda'}, \ee where $\lambda'
\vartriangleleft \lambda$ holds if $\lambda'$ can be obtained from
$\lambda$ by removing one box.
\end{theorem}

\noindent By repeating this process along the subgroup chain
$$ S_k \supset S_{k-1} \supset \cdots \supset S_1$$
we obtain a unique orthogonal basis, given by the spaces of the
one{\-}-di\-men\-sio\-nal irreducible representation of $S_1$. The
resulting basis is known as \emph{Young's orthogonal
basis}\index{Young's orthogonal basis} or the
\emph{Young-Yamanouchi basis}~\cite{JamKer81Book}.

Let us now consider the case of the unitary group. Note that the
basis vectors with different weight will remain vectors with
different weight and therefore remain orthogonal. For $\U(2)$,
where to each weight there is only a single vector, this means
that the construction actually produces an orthogonal basis (see
subsection~\ref{subsec-spin-states}). For $\U(d)$ we can move down
the subgroup chain
    $$
        \unitaries(d) \supset \unitaries(d-1) \supset \cdots \supset \unitaries(1)
    $$
with the help of the following branching rule:

\begin{figure}\label{figure-gelfand}
\begin{center}
\begin{picture}(140,70)(0,-25)
\linethickness{0.15mm}
\setlength{\unitlength}{0.30mm}
\color{gray1}
\put(0,20){\rule{20.00\unitlength}{20.00\unitlength}}\put(20,20){\rule{20.00\unitlength}{20.00\unitlength}}\put(40,20){\rule{20.00\unitlength}{20.00\unitlength}}
\color{gray2}
\put(70,10){\rule{20.00\unitlength}{20.00\unitlength}}\put(90,10){\rule{20.00\unitlength}{20.00\unitlength}}
\put(10,-10){\rule{20.00\unitlength}{20.00\unitlength}}\put(30,-10){\rule{20.00\unitlength}{20.00\unitlength}}
\color{gray3}
\put(60,-20){\rule{20.00\unitlength}{20.00\unitlength}}
\put(20,-40){\rule{20.00\unitlength}{20.00\unitlength}}\put(40,-40){\rule{20.00\unitlength}{20.00\unitlength}}
\color{black}
\put(0,20){\framebox(20,20){$1$}}\put(20,20){\framebox(20,20){$1$}}\put(40,20){\framebox(20,20){$1$}}
\put(70,10){\framebox(20,20){$2$}}\put(90,10){\framebox(20,20){$2$}}
\put(10,-10){\framebox(20,20){$2$}}\put(30,-10){\framebox(20,20){$2$}}\put(60,-20){\framebox(20,20){$3$}}
\put(20,-40){\framebox(20,20){$3$}}\put(40,-40){\framebox(20,20){$3$}}
\end{picture}
\begin{picture}(80,70)(0,-15)
\setlength{\unitlength}{0.30mm}
\color{gray1}
\put(0,20){\rule{20.00\unitlength}{20.00\unitlength}}\put(20,20){\rule{20.00\unitlength}{20.00\unitlength}}\put(40,20){\rule{20.00\unitlength}{20.00\unitlength}}
\color{gray2}
\put(60,20){\rule{20.00\unitlength}{20.00\unitlength}}\put(80,20){\rule{20.00\unitlength}{20.00\unitlength}}
\put(0,00){\rule{20.00\unitlength}{20.00\unitlength}}\put(20,00){\rule{20.00\unitlength}{20.00\unitlength}}
\color{gray3}
\put(40,00){\rule{20.00\unitlength}{20.00\unitlength}}
\put(0,-20){\rule{20.00\unitlength}{20.00\unitlength}}\put(20,-20){\rule{20.00\unitlength}{20.00\unitlength}}
\color{black}
\put(0,20){\framebox(20,20){$1$}}\put(20,20){\framebox(20,20){$1$}}\put(40,20){\framebox(20,20){$1$}}\put(60,20){\framebox(20,20){$2$}}\put(80,20){\framebox(20,20){$2$}}
\put(0,00){\framebox(20,20){$2$}}\put(20,00){\framebox(20,20){$2$}}\put(40,00){\framebox(20,20){$3$}}
\put(0,-20){\framebox(20,20){$3$}}\put(20,-20){\framebox(20,20){$3$}}
\end{picture}
\caption{Building a semistandard Young tableau from three skew
diagrams}
\end{center}
\end{figure}
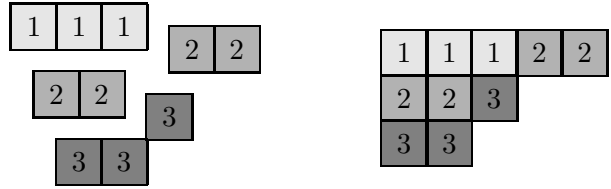

\begin{theorem}[Branching rule
for $\unitaries(d)$\index{branching rule!unitary
group}]\label{theorem-unitary-branching}
$$U_\lambda \downarrow^{U(d)}_{U(d-1)} \cong \bigoplus_{\mu} U_\mu,$$
where the sum is taken over all $\mu$ satisfying the
\emph{betweenness condition\index{betweenness condition}}
$$ \lambda_{i+1} \leq \mu_i\leq \lambda_i  \quad \forall i \in \{1, \ldots, d-1\}$$
\end{theorem}

\noindent I am not aware of any elementary proof and guide the
reader to the arguments in \cite{Molev02} and in \cite[chapter 8
and 12]{GooWal98}. A path down the chain is known as a
\emph{Gelfand-Zetlin pattern\index{Gelfand-Zetlin pattern}}; for
the diagram $\lambda=(\lambda_1, \ldots \lambda_d)$, this is an
array of the form
\begin{center}
\begin{picture}(180,125)(0,20)
\setlength{\unitlength}{0.003\textwidth}
\put(0,100){$\lambda^{(d)}_{d}$} \put(40,100){$\lambda^{(d)}_{d-1}$} \put(80,100){$\hdots$} \put(120,100){$\lambda^{(d)}_{2}$} \put(160,100){$\lambda^{(d)}_{1}$}
\put(20,80){$\lambda^{(d-1)}_{d-1}$} \put(60,80){$$} \put(140,80){$\lambda^{(d-1)}_{1}$}
\put(40,60){$\ddots$} \put(60,60){$$} \put(120,60){$\adots$}
\put(60,40){$\lambda^{(2)}_{2}$} \put(60,40){$$} \put(100,40){$\lambda^{(2)}_{1} $}
\put(80,20){$ \lambda^{(1)}_{1}$}
\end{picture}
\end{center}
\noindent where the top row equals the initial pattern $\lambda$,
i.e.~$\lambda^{(d)}_i=\lambda_i$ and where any row satisfies the
betweenness condition relative to the row above it, i.e.
    $$
        \lambda^{(j)}_{i+1} \leq \lambda^{(j-1)}_{i} \leq \lambda^{(j)}_{i}
        \quad \forall j \in \{d, \ldots, 2\} \mbox{ and } i \in \{1,
        \ldots, j-1\}
    $$
There is a straightforward bijection between Gelfand-Zetlin
patterns and semistandard Young tableaux: define the \emph{skew
diagram} $\theta^{(j)}$ as the set of boxes of $\lambda^{(j)}$
that are not contained in the diagram $\lambda^{(j-1)}$. The union
of these skew diagrams is disjoint and equals again the diagram of
$\lambda$. Now fill each box in $\lambda$ which belongs to
$\theta^{(j)}$ with a $j$, then the resulting diagram is a
semistandard Young tableau (see figure \ref{figure-gelfand}). This
fact as well as the converse, i.e.~the fact that every
semistandard Young tableau will lead to a Gelfand-Zetlin pattern,
can be seen by examining the betweenness condition. Taking the
branching rule for granted, the discussion above amounts to a
proof of the second equation in
theorem~\ref{theorem-combinatorics}: a combinatorical algorithm
for computing the dimension of irreducible representations of the
unitary group.



\chapter{Spectra of Quantum States and Representation Theory}
\label{chapter-relation}

\section{Introduction}
\label{sec-spectra-intro} The spectra of Hermitian operators play
a central role in quantum mechanics, not only in the measurement
postulate, but also as a quantifier of order and disorder in
quantum states. The most well-known example of a measure of
disorder is entropy. Entropy first emerged in the context of
thermodynamics and statistical mechanics, and is one of the
fundamental quantities in physics. Its quantum version for
discrete systems is the von Neumann entropy, which is defined as
$$ S(\rho)=-\tr \rho \log \rho$$
and has an information-theoretic interpretation as the Shannon
entropy of the eigenvalues of $\rho$. As such, the von Neumann
entropy plays the role in quantum information theory that the
Shannon entropy plays in classical information theory.

In this chapter I will investigate spectral properties of
Hermitian operators, in particular of quantum states. The prime
motivation for this research has been the study of von Neumann
entropy and its inequalities, such as strong subadditivity (see
also subsection~\ref{sec-group-intro}). There are, however, a
number of other contexts in which spectra of quantum states play
an important role. In 1999, Nielsen discovered that a bipartite
pure state $\kpsi$ can be transformed into a bipartite pure state
$\kphi$ by local operations and classical communication (LOCC) if
and only if the spectrum of $\rho^A=\tr_B \proj{\psi}$ is
majorised by the spectrum of $\sigma^A=\tr_B
\proj{\phi}$~\cite{Nielsen99}. This gives a simple operational
meaning to the partial ordering of spectra induced by
majorisation, and refines the entropic principle that the local
entropy in a closed system cannot be increased $S(\rho^A)\geq
S(\sigma^A)$. The second example is the separability criterion by
Nielsen and Kempe~\cite{NieKem01}, which asserts that a separable
state $\rho^{AB}$ is more disordered globally than locally. More
precisely, the spectrum of a separable state $\rho^{AB}$ is
majorised by the spectrum of $\rho^A$. This again is a
generalisation of an entropic result, namely the separability
criterion that says that any separable state has $S(\rho^A)\leq
S(\rho^{AB})$.

Both examples show that the spectra of quantum states can reveal
structure of quantum states. The main question that will be
investigated in this section asks for the compatibility of quantum
states: given two states $\rho^A$ and $\rho^B$, is there a state
$\rho^{AB}$ on $\cH_A \otimes \cH_B$ such that $\rho^A=\tr_B
\rho^{AB}$ and $\rho^B=\tr_A \rho^{AB}$? Since an affirmative
answer immediately extends to the whole orbit $(U^A \otimes U^B)
\rho^{AB} (U^{A \dagger} \otimes U^{B \dagger})$ for $U^A \in
\sunitaries (\cH_A), U^B \in \sunitaries(\cH_B)$, any condition on
the compatibility can only depend on the spectra of $\rho^A$ and
$\rho^B$. The question addressed in this chapter in a slightly
more general form reads as follows:
\begin{problem-footnote}[Compatibility of Local Spectra]\footnote{On the one hand Klyachko
has coined the term `Quantum Marginal Problem'\index{quantum
marginal problem}~\cite{Klyachko04} by analogy with the marginal
problem in classical probability theory. On the other hand this
problem parallels Horn's problem and, in the same way as Horn's
problem, comes in a pair with a closely related group-theoretic
problem. In their paper~\cite{KnuTao01} Allen Knutson and Terence
Tao have classified Horn's problem as a `classical' problem and
its group-theoretic variant a `quantum' problem. Both
terminologies make sense in their own right, but to avoid
confusion and to highlight the spectral nature of the problem, I
have avoided the words quantum and classical
altogether.}\label{prob-comp} \mbox{ }
\newline Given three spectra $r^A$, $r^B$ and $r^{AB}$, is there a
quantum state $\rho^{AB}$ with
\beastar \spec \rho^{AB}&=&r^{AB}\\
         \spec \rho^{A}&=&r^{A}\\
         \spec \rho^{B}&=&r^{B} \quad ?
\eeastar
\end{problem-footnote}
A triple of spectra $(r^A, r^B, r^{AB})$ is said to be an
\emph{admissible spectral triple} if the question can be answered
positively. This question, or the more general task of describing
the set of quantum states with given marginal states, is
fundamental to quantum information theory. It appears in
minimisation problems of correlation measures and channel
capacities~\cite{ChrWin04, THLD02, EiAuPl03}, as well as in state
transformations in quantum communication
protocols~\cite{DafHay04}. The foundation for the study of
problem~\ref{prob-comp}, however, can be traced further back to
research done in the 1960s in quantum chemistry and the theory of
condensed matter. Here, a closely related problem appears under
the name of the \emph{$N$-representability
problem\index{$N$-representability problem}}, which asks for the
compatibility of $p$-particle reduced density matrices with a
wavefunction of $N$ identical particles. The importance of this
problem is best illustrated by the case $p=2$. An exact expression
for the energy of a system of $N$ identical particles governed by
a nearest-neighbour Hamiltonian
$$ H=\sum_{i=1}^N H_i+\sum_{i<j} H_{ij},$$
where $H_i=H_1$ and $H_{ij}=H_{12}$, can be given in an expression
only involving the 1- and 2-particle reduced states:
\be \label{eq-energy-reduced} E=N \tr \rho^{(1)} H_1+ \frac{N(N-1)}{2} \tr \rho^{(12)} H_{12}. \ee
Whereas the brute force calculation of the ground state energy of
$H$ involves a minimisation over all possible $N$-particle
wavefunctions, equation~(\ref{eq-energy-reduced}) shows that the
effort can be reduced to a minimisation over all density matrices
$\rho^{(12)}$ that are compatible with an overall wavefunction of
$N$ identical particles. This would require significantly less
computational effort to calculate the ground state energy, but
only if it was not such a difficult problem to find the
compatibility constraints for density operator $\rho^{(12)}$. More
than forty years after Coulson recognised the importance of
finding these constraints, Coleman and Yukalov write, rephrasing
Coulson,
\begin{quote}
``If it were possible to obtain a reasonably accurate expression
for the 2-[particle reduced density] matrix of an $N$-particle
state, without recourse to the wavefunction, nearly all the
properties of matter which are of interest to chemists and
physicists would become accessible to us.''
\end{quote}
\begin{quotewho}
A.~John~Coleman and Vyacheslav.~I.~Yukalov, in ``Reduced Density
Matrices -- Coulson's Challenge''~\cite{ColYuk00}.
\end{quotewho}
Unfortunately, the research presented here will not directly lead
to insights into the solution of the $N$-representability problem
of the 2-particle reduced density matrix, but is more closely
related to the study of the 1-particle reduced density matrix.
Nevertheless, I hope I have convinced the reader of the
fundamental relevance of problem~\ref{prob-comp} to quantum
information theory and to physics as a whole.

The main result of this chapter is the proof of a correspondence
between problem~\ref{prob-comp} and a specific group-theoretic
problem, which will now be described. Consider the tensor product
of two irreducible representations $V_\mu$ and $V_\nu$ of the
symmetric group $S_k$, and its decomposition into irreducible
representations:
$$ V_\mu \otimes V_\nu \cong \bigoplus_\lambda g_{\mu \nu \lambda}
V_\lambda.$$ The Clebsch-Gordan integer of this decomposition is
denoted by $g_{\mu \nu \lambda}$ and known as the \emph{Kronecker
coefficient of the symmetric group}. To date there is no
combinatorical formula known for $g_{\mu \nu \lambda}$ and finding
one is considered difficult. The specific group-theoretic problem
addressed here is to decide when $g_{\mu \nu \lambda}$ is nonzero.
\begin{problem}[Nonvanishing of Kronecker
Coefficient]\label{prob-Kronecker} \sloppy Given three irreducible
representations $V_\mu, V_\nu$ and $V_\lambda$ of $S_k$, is it
true that $V_\lambda \subset V_\mu \otimes V_\nu$? \fussy
\end{problem}
The main result alluded to is the asymptotic
equivalence\label{pageref-asymptotic-equivalence} of problem
\ref{prob-comp} and problem~\ref{prob-Kronecker} (see theorems
\ref{ourtheorem} \&~\ref{theorem-converse} for a precise
statement).

This unexpected relation between previously unrelated problems
opens new avenues to their solutions. A more immediate benefit is
the transfer of results of one problem to the other and vice
versa, an example of which is given in
subsection~\ref{subsec-spec-two-qubit}. The method that is used to
proof this equivalence uses the tools of
chapter~\ref{chapter-structure} and a theorem concerning the
estimation of the spectrum of a density operator, theorem
\ref{theorem-Keyl-Werner}. The method developed here is versatile
and can also be applied in other contexts. An example is the new
and very elementary proof of the asymptotic equivalence of Horn's
problem (problem~\ref{prob-Horn}) and the problem of deciding when
a Littlewood-Richardson coefficient is nonzero (problem
\ref{prob-LR}).

This chapter is composed of four sections. After this introduction
I address the tensor representations of the symmetric and the
unitary group in section~\ref{section-main-clebsch}, and discuss
their Clebsch-Gordan integers: the Kronecker and the
Littlewood-Richardson coefficients. A proof is given showing that
a semigroup is formed by both the nonzero Kronecker and
Littlewood-Richardson coefficients. Whereas the semigroup property
of the Littlewood-Richardson coefficients is well-known, the
result concerning Kronecker coefficients provides a positive
answer to a recent conjecture by Alexander
Klyachko~\cite[conjecture 7.1.4]{Klyachko04}. The next section,
section~\ref{sec-spectra-states}, is less group-theoretical and
works mostly with tools from quantum information theory. It starts
with an account of the spectral estimation theorem and a short
proof thereof. Subsequently, the spectral estimation theorem plays
an important role in deriving the asymptotic equivalence of
problem~\ref{prob-comp} and~\ref{prob-Kronecker}, as well as the
equivalence of problem~\ref{prob-Horn} and~\ref{prob-LR}. Some
applications follow: first, a proof of the convexity of
problem~\ref{prob-comp}, for which I will also provide an
independent proof, second, a new proof of all spectral two-qubits
inequalities, and third, -- connecting back to the study of
entropies --  a proof of subadditivity of von Neumann entropy.

\section{Tensor Product Representations}
\label{section-main-clebsch} This section builds on
chapter~\ref{chapter-structure}, where the irreducible
representations of the symmetric and unitary groups were
introduced. In subsection~\ref{subsec-dual-inv} I will explain how
representations can be viewed as invariants. Subsection
\ref{subsec-kronecker} gives several equivalent definitions of the
Kronecker coefficient and reviews briefly its basic properties and
history. The same is done in subsection~\ref{subsec-LR} for the
Littlewood-Richardson coefficient. Subsection
\ref{subsec-semigroup} then derives the semigroup property for
both coefficients, the main research result of this section.

\subsection{Invariants and the Dual Representation}
\label{subsec-dual-inv} The subsection starts with a simple
isomorphism between homomorphism and vectors, which is an exercise
in most linear algebra courses and has as consequence the
Jamilkowski isomorphism, a well-known result in the quantum
information community. Afterwards I define the dual of a
representation and introduce invariants. In later sections, the
connection between invariants and representations,
theorem~\ref{theorem-invariant}, plays an important role.

\begin{lemma} \label{lemma-space-isomorphism}
Let $V$ and $W$ be two finite-dimensional complex vector spaces,
then $$\hom(V,W)\cong V^\star \otimes W.$$
\end{lemma}
\begin{proof}
Let $\ket{e_i}$ and $\ket{f_j}$ be o.n. bases for $V$ and $W$
respectively and let $\ket{\psi}=\sum_{i=1}^{\dim V} \bra{e_i}
\otimes \ket{e_i} \in V^\star \otimes V$. The map from $\hom(V,
W)$ to $V^\star \otimes W$ is given by
$$ A \mapsto [\id \otimes A] \psi= \sum_i \bra{e_i} \otimes (A \ket{e_i}).$$
The matrix elements $a_{ij}= \bra{f_i} A \ket{e_j} $ of $A$ equal
the coefficients in the expansion of $\id \otimes A \psi$ in the
basis $\bra{e_j} \otimes \ket{f_i}$, because
$$  \ket{e_j} \otimes \bra{f_i} \sum_k \ket{e_k} \otimes A
\ket{e_k}  = \delta_{jk}  \bra{f_i}  A \ket{e_k}  = \bra{f_i} A
\ket{e_j} =a_{ij}.
$$ Conversely, the coefficients of a vector $\ket{\phi} \in V^\star
\otimes W$ expanded in this basis define a map $A$.
\end{proof}

\begin{definition}\label{def-dual}
Let $V$ be a representation of a group $G$. The \emph{dual
representation\index{representation!dual}} $V^\star: G \rightarrow
\GL(V^\star)$ is the representation satisfying
$$  (V^\star(g)\bra{w}) (V(g)\ket{v}) \equiv \bra{w} V^\star(g)^T V(g)\ket{v} =  \braket{w}{v}$$
for all $g \in G$, $\ket{v} \in V$, and $\bra{w} \in V^\star$.
This unique $V^\star(g)$ satisfying the definition is given by
$$ V^\star(g)=V(g^{-1})^T: V^\star \rightarrow V^\star.$$
\end{definition}

\begin{definition}
Let $V$ and $W$ be representations of a group $G$.
$$V^G=\{\ket{v} \in V | V(g)\ket{v}=\ket{v} \quad \forall g \in G\}$$
is the $G$-invariant subspace\index{invariant subspace} of $V$
whereas
    \beastar
        \hom_G(V,W)&:=&\hom(V,W)^G\\
                   &=&\{A \in \hom(V,W)| \; W(g) A V(g^{-1})=A \quad \forall g \in G\},
    \eeastar
are the $G$-invariant homomorphisms from $V$ to $W$.
\end{definition}

\begin{corollary} \label{corollary-invariant}
If $V$ is a representation of $G$ and $\ket{e_i}$ is a basis for
$V$, then $\kpsi=\sum_i \bra{e_i} \otimes \ket{e_i}$ is
$G$-invariant, i.e.
$$ [V(g^{-1})^T \otimes V(g)] \kpsi=\kpsi,$$
or succinctly $g\kpsi=\kpsi$. If $V$ is an irreducible
representation, then $\kpsi$ is the unique $G$-invariant vector up
to scalar multiplication.
\end{corollary}

\begin{proof}
Let $V(g) \ket{e_i} =\sum_j V_{ij}(g) \ket{e_j}$. Then
\beastar [V(g^{-1})^T \otimes V(g)] \sum_i \bra{e_i} \otimes
\ket{e_i}
    &=& \sum_i \big(\sum_k V(g^{-1})_{ki} \bra{e_k} \big) \otimes \big( \sum_j V(g)_{ij} \ket{e_j} \big)\\
    &=& \sum_{jk} \big( \underbrace{\sum_{i} (V(g)^{-1})_{ki} V(g)_{ij}}_{=\delta_{jk}}\big) \bra{e_j} \otimes \ket{e_k}\\
    &=& \sum_j \bra{e_j} \otimes \ket{e_j}.
\eeastar
\end{proof}
Let $V$ and $W$ be two representations of $G$. Then $\hom(V, W)$
is a representation of $G$ defined by
$$ g: A \mapsto W(g)A V(g^{-1})$$
for all $A \in \hom(V, W)$.

\begin{corollary} \label{corollary-invariant-hom}
Let $V$ and $W$ be representations of $G$. Then $\hom_G(V,W)\cong
(V^\star \otimes W)^G$ holds.
\end{corollary}

\begin{proof}
The map
$$ A \mapsto [\id \otimes A] \kpsi$$
from lemma~\ref{lemma-space-isomorphism} defines the isomorphism
$\hom(V,W)\cong V^\star \otimes W$. The following line shows that
any $A \in \hom_G(V, W)$ is mapped to $[\id \otimes A] \kpsi \in
(V^\star \otimes W)^G$:
\beastar [V(g^{-1})^T \otimes W(g)] \; [\id \otimes A] \;\kpsi& =& [\id \otimes W(g)A V(g^{-1})] \;[V(g^{-1})^T \otimes V(g)] \kpsi \\
    &=&[\id \otimes A] \; [V(g^{-1})^T \otimes V(g)] \kpsi \\
    &=&[\id \otimes A] \; \kpsi,
\eeastar
where corollary~\ref{corollary-invariant} was used in the last
equation.
\end{proof}

\noindent The multiplicities of an irreducible representation
appearing in a reducible representation can be expressed as the
dimension of the space of invariants under the action of the
group.

\begin{theorem} \label{theorem-invariant}
Let $U$ be a representation of $G$ and $U=\bigoplus_\alpha
m_\alpha U_\alpha$ be its decomposition into irreducible
representations $\alpha$ with multiplicities $m_\alpha$. Then
$$ m_\alpha=\dim  \left(  U^\star \otimes U_\alpha\right)^G=\dim \hom_G(U, U_\alpha),$$ where $U^\star$
is the dual representation of $U$.
\end{theorem}

\begin{proof}
The second equality follows from
corollary~\ref{corollary-invariant-hom}. By Schur's lemma
$$\hom_G(U_\beta, U_\alpha)\cong \delta_{\alpha, \beta} \complex,$$
and therefore
$$ \hom_G(U, U_\alpha)\cong \hom_G(\bigoplus_\beta U_\beta
\otimes \complex^{m_\beta}, U_\alpha) \cong \complex^{m_\alpha},$$
which shows that the left hand side equals the right hand side.
\end{proof}

\noindent This concludes the general remarks on duals and
invariants. The next subsection applies theorem
\ref{theorem-invariant} as well as Schur-Weyl duality, theorem
\ref{theorem-Schur-duality}, to obtain a number of equivalent
definitions of the Kronecker coefficient.

\subsection{The Kronecker Coefficients of the Symmetric Group}
\label{subsec-kronecker} The first definition of the Kronecker
coefficient is given in terms of unitary groups. Let $m$ and $n$
be natural numbers and $d=mn$. Consider an irreducible
representation $U^{mn}_\lambda$ of the unitary group $\SU(mn)$
embedded into $(\complex^{mn})^{\otimes k}$, i.e.~$k=|\lambda|$.
When restricted to the subgroup $\SU(m) \times \SU(n)$, this
representation becomes reducible can be written as a direct sum of
irreducible representations of $\SU(m) \times \SU(n)$, which are
equivalent to tensor products of irreducible representations of
$\SU(m)$ and $\SU(n)$ as explained in the Preliminaries
(pages~\pageref{remark-carry-over} and
\pageref{section-boxtimes}):
\be \label{eq-kron-unitary} U^{mn}_\lambda \downarrow^{\SU(mn)}_{\SU(m)\times \SU(n)}\cong \bigoplus_{\substack{\mu \vdash (k, m)\\ \nu \vdash (k, n)}} g_{\mu \nu \lambda}
U^m_\mu \boxtimes U^n_\nu, \ee The coefficient $g_{\mu \nu
\lambda}$ denotes the multiplicity of $U^m_\mu \boxtimes U^n_\nu$
in the representation $U^{mn}_\lambda$ when restricted to $\SU(m)
\times \SU(n)$ and is commonly known as \emph{Kronecker
coefficient of the symmetric group}, or simply the \emph{Kronecker
coefficient\index{Kronecker coefficient}}. The connection to the
symmetric group is easily established via Schur-Weyl duality
(theorem~\ref{theorem-Schur-duality}). In terms of the spaces of
the representations, equation~(\ref{eq-kron-unitary}) becomes
\be \label{eq-kron-unitary-spaces} U^{mn}_\lambda \cong \bigoplus_{\mu, \nu} U^m_\mu \otimes U^n_\nu \otimes \complex^{g_{\mu \nu
\lambda}}.
\ee
 As a next step we
consider two different decompositions of the space
$(\complex^{mn})^{\otimes k}$. The first decomposition is obtained
according to the Schur-Weyl duality of $\SU(mn)$ and $S_k$, and
subsequently reducing from $\SU(mn)$ to $\SU(m)\times \SU(n)$ with
equation~(\ref{eq-kron-unitary-spaces}),
\bea
(\complex^{mn})^{\otimes k}&\cong &\bigoplus_{\lambda \vdash (k,
mn)} U^{mn}_\lambda \otimes V_\lambda \\
&\cong & \label{eq-Schur-dual-AB} \bigoplus_{\lambda
\vdash (k, mn)} \big( \bigoplus_{\substack{\mu \vdash (k, m)\\
\nu \vdash (k, n)}} U^m_\mu \otimes U^n_\nu \otimes
\complex^{g_{\mu \nu \lambda}}\big) \otimes V_\lambda.
\eea
The second decomposition takes $(\complex^{mn})^{\otimes k}$,
interprets it as $(\complex^{m})^{\otimes k}\otimes
(\complex^{n})^{\otimes k}$ and applies Schur-Weyl duality of
$\SU(m)$ and $S_k$ to the first and of $\SU(n)$ and $S_k$ to the
second factor. The result is
\be \label{eq-Schur-dual-A-B}(\complex^{mn})^{\otimes k}\cong \left(\bigoplus_{\mu' \vdash (k, m)}  U^m_{\mu'} \otimes
V_{\mu'} \right) \otimes \left(\bigoplus_{\nu' \vdash (k, n)}
U^n_{\nu'} \otimes V_{\nu'} \right).
\ee

\noindent Note that the spaces of all representations equivalent
to $U^m_\mu\boxtimes U^n_\nu$ which are contained in
equation~(\ref{eq-Schur-dual-AB}) are subspaces of $ U^m_{\mu}
\otimes U^n_{\nu} \otimes V_{\mu} \otimes V_{\nu}$. Comparing
equation~(\ref{eq-Schur-dual-AB}) and
equation~(\ref{eq-Schur-dual-A-B}) and projection onto the space
$U^m_{\mu} \otimes U^n_{\nu} \otimes V_{\mu} \otimes V_{\nu}$
results in
    \be \label{eq-proj}
        U^m_{\mu} \otimes U^n_{\nu} \otimes V_{\mu}
        \otimes V_{\nu} \cong \bigoplus_{\lambda
        \vdash (k, mn)} U^m_\mu \otimes U^n_\nu \otimes \complex^{g_{\mu
        \nu \lambda}} \otimes V_\lambda.
    \ee
This shows that
    \be \label{eq-kron-symm}
        V_\mu \otimes V_\nu \cong \bigoplus_{\lambda \vdash (k, mn)} g_{\mu \nu \lambda} V_\lambda.
    \ee
The Kronecker coefficient $g_{\mu \nu \lambda}$ is therefore not
only the multiplicity of $U^n_\mu \boxtimes U^m_\nu$ in
$U^{mn}_\lambda$, when reduced to $\SU(m)\times \SU(n)$, but also
the multiplicity of $V_\lambda$ in $V_\mu \otimes V_\nu$. In the
language of invariants,
    \be \label{eq-kron-invariant}
        g_{\mu \nu \lambda} = \dim
        (U^\star_\lambda \otimes U_\mu \otimes U_\nu)^{\SU(m) \times
        \SU(n)}=\dim (V_\lambda \otimes V_\mu \otimes V_\lambda)^{S_k}
    \ee
\fussy At first, it might seem that the star, which marks the dual
of $V_\lambda$ and which should appear on the RHS of
equation~(\ref{eq-kron-invariant}), has been forgotten; since the
matrices that represent the elements of the permutation group are
unitary and real at the same time, $V(g^{-1})^T=V(g)$ and
$V_\lambda \cong V_\lambda^\star$ hold. This last remark is of
interest to us as it shows that $g_{\mu \nu \lambda}$ is symmetric
under interchange of all three indices $\mu, \nu$ and
$\lambda$\label{pageref-kron-symm}.\footnote{The symmetry of
$g_{\mu \nu \lambda}$ under interchange $\mu, \nu$ and $\lambda$
is the reason for having all three indices in subscript. The
Littlewood-Richardson coefficient $c_{\mu \nu}^\lambda$, in
contrast, is only symmetric under interchange of $\mu$ and $\nu$.}

At present, no satisfying combinatorial expression has been found
for $g_{\mu \nu \lambda}$ and its calculation remains difficult. A
direct way to obtain $g_{\mu \nu \lambda}$ is by computation of
the character $\chi_\lambda$ of the symmetric group and
application of the formula
$$ g_{\mu \nu \lambda} = \langle \chi_\lambda, \chi_\mu
\chi_\nu\rangle=\frac{1}{n!} \sum_{\pi \in S_k}
\chi_\lambda(\pi)\chi_\mu(\pi) \chi_\nu(\pi).$$ A different
algorithm is given in~\cite{HagMac65a, HagMac65b} and a number of
special cases have been examined in~\cite{Dvir93, Rosas01,
BesKle99}. In subsection~\ref{subsection-symm-group-young} I
discuss combinatorial algorithms in terms of Young tableaux for
the dimensions of irreducible representations of the unitary and
symmetric group. Combinatorical expressions have also been
established for a number of other group-theoretic quantities; the
most famous of which is the Littlewood-Richardson rule, which is
an algorithm to calculate the Littlewood-Richardson coefficient.
The finding of such an algorithm for the Kronecker coefficient has
remained elusive, and even the weaker task of efficiently
determining whether or not $g_{\mu \nu \lambda}$ is nonzero
remains an open problem. This problem, which was presented in the
introduction to this chapter as problem~\ref{prob-Kronecker}, is
the main focus later on. The next subsection essentially carries
out the same calculation as above, but instead for the
Littlewood-Richardson coefficient.

\subsection{The Littlewood-Richardson Coefficients}
\label{subsec-LR} Let $U_\mu$ and $U_\nu$ be two irreducible
representations of $\unitaries (d)$. The multiplicities in the
decomposition of the tensor product representation of $U_\mu$ and
$U_\nu$ into irreducible representations
\be \label{eq-littlewood} U_\mu \otimes U_\nu \cong \bigoplus_\lambda c_{\mu \nu}^\lambda
U_\lambda \ee are called \emph{Littlewood-Richardson
coefficients\index{Littlewood-Richardson coefficient}}. Just as
there are two expansions for the $g_{\mu \nu \lambda}$
(eqs.~(\ref{eq-kron-unitary}) and~(\ref{eq-kron-symm})), $c_{\mu
\nu}^\lambda$ can also be defined by the expansion
\be \label{eq-littlewood-symmetric}
V_\lambda\downarrow^{S_k}_{S_{k_1}\times S_{k_2}}\cong
\bigoplus_{\substack{\mu \vdash (k_1, d) \\ \nu \vdash (k_2, d)}}
c_{\mu \nu}^\lambda V_\mu \otimes V_\nu,
\ee
where $k_1+k_2=k$. This follows from a comparison of the following
two:
\begin{itemize}
\item[i)] Schur-Weyl duality applied to $(\complex^d)^{\otimes k}$
followed by restricting $S_k$ to $S_{k_1}\times S_{k_2}$:
    \be \label{eq-littlewood-1}
    \begin{split}
        (\complex^d)^{\otimes k}
        \cong &\bigoplus_{\lambda \vdash (k, d)} U_\lambda \otimes V_\lambda \\
        \cong & \bigoplus_{\lambda \vdash (k, d)}  U_\lambda \otimes \left( V_\lambda\downarrow^{S_k}_{S_{k_1}\times S_{k_2}}\right).
    \end{split}
    \ee
\item[ii)] \sloppy Schur-Weyl duality applied to $(\complex^d)^{\otimes k_1}$ and
$(\complex^d)^{\otimes k_2}$ followed inserting
equation~(\ref{eq-littlewood}):
    \be\label{eq-littlewood-2}
    \begin{split}
        (\complex^d)^{\otimes k_1}\otimes (\complex^d)^{\otimes k_2}
        \cong &\big( \bigoplus_{\mu \vdash (k_1, d)} U_\mu \otimes V_\mu \big)
        \otimes \big( \bigoplus_{\nu \vdash (k_2, d)} U_\nu \otimes V_\nu \big)\\
         \cong & \bigoplus_{\substack{\mu \vdash (k_1, d) \\ \nu \vdash (k_2, d)}}
        \big(\bigoplus_{\lambda \vdash (k, d) } U_\lambda \otimes \complex^{c_{\mu
        \nu}^\lambda}\big) \otimes V_\mu \otimes V_\nu.
    \end{split}
    \ee
\end{itemize}
\fussy Note that all representations similar to $U_\lambda$ in
equation~(\ref{eq-littlewood-2}) appear in the term $U_\lambda
\otimes \left( V_\lambda\downarrow^{S_k}_{S_{k_1}\times
S_{k_2}}\right)$ of equation~(\ref{eq-littlewood-1}).
Equating~(\ref{eq-littlewood-1}) and~(\ref{eq-littlewood-2})
therefore results in
\bea
U_\lambda \otimes \left( V_\lambda\downarrow^{S_k}_{S_{k_1}\times
S_{k_2}}\right)\cong \bigoplus_{\substack{\mu \vdash (k_1, m) \\
\nu \vdash (k_2, n)}} \left( U_\lambda \otimes \complex^{c_{\mu
\nu}^\lambda} \right) \otimes V_\mu \otimes V_\nu.
\eea
This demonstrates equation~(\ref{eq-littlewood-symmetric}). The
invariant-theoretic formulation is
\be \label{eq-LR-invariant}
c_{\mu \nu}^\lambda = \dim ( U_\lambda^\star \otimes U_\mu \otimes
U_\nu)^{U(d)}= \dim (V_\lambda \otimes V_\mu \otimes
V_\nu)^{S_{k_1}\times S_{k_2}},\ee where -- in contrast to the
Kronecker coefficient -- the Littlewood-Richardson coefficient is
only symmetric under exchange of $\mu$ and $\nu$.

The algorithm to compute the $c_{\mu \nu}^\lambda$ is the famous
Littlewood-Richardson rule; for more information
see~\cite{Fulton97} and the recent review by Marc~A.~A. van
Leeuwen~\cite{vanLeeuwen01}. In
subsection~\ref{subsec-spectra-horn}, I discuss the relation of
the Littlewood-Richardson coefficient to the following spectral
problem.
\begin{problem}[Horn's Problem]\label{prob-Horn}
Given three spectra $r^A, r^B$ and $r^{AB}$, do Hermitian
operators $A$ and $B$ exist such that
\beastar \spec A&=&r^{A}\\
         \spec B&=&r^{B}\\
         \spec A+B&=&r^{AB} \quad ?
\eeastar
\end{problem}
More precisely, I will provide a new and particularly simple proof
for the well-known asymptotic equivalence of the Horn's problem
and
\begin{problem}[Nonvanishing of Littlewood-Richardson
Coefficient]\label{prob-LR} Given three irreducible
representations $U_\mu, U_\nu$ and $U_\lambda$ of $\unitaries
(d)$, is it true that $U_\lambda \subset U_\mu \otimes U_\nu$?
\end{problem}

\noindent Before this subject is touched upon, two group-theoretic
properties of $c_{\mu \nu}^\lambda$ and $g_{\mu \nu \lambda}$ are
proven.

\subsection[The Semigroup Property]{The Semigroup Property\protect\footnote{This section contains collaborative work with
Graeme Mitchison. I am grateful to Allen Knutson for sharing his
expertise that has been essential for obtaining the results in
this subsection.}} \label{sec-semigroup} \label{subsec-semigroup}
Instead of looking at individual representations, I will now turn
the attention to the direct sum of all representations of a group
in order to derive a global property for the Littlewood-Richardson
coefficients $c_{\mu \nu}^\lambda$ and the Kronecker coefficients
$g_{\mu \nu \lambda}$. This property is known as \emph{semigroup
or monoid property}\index{semigroup property}.

\begin{theorem}[Semigroup Property] \label{theorem-stability}
The set of triples $(\mu, \nu, \lambda)$ with nonzero $c_{\mu
\nu}^\lambda$ (or $g_{\mu \nu \lambda}$) form a semigroup with
respect to row-wise addition, i.e.~$c_{\mu \nu}^{\lambda} \neq 0$
and $c_{\mu' \nu'}^{\lambda'} \neq 0$ implies $c_{\mu+\mu',
\nu+\nu'}^{\lambda+\lambda'} \neq 0$ (and similarly for $g_{\mu
\nu \lambda}$). \fussy
\end{theorem}
The semigroup property of the Littlewood-Richardson coefficient is
well known~\cite{Elashvili92, Zelevinsky97}. Note that the claim
regarding the Kronecker coefficients is precisely the statement
conjectured in Klyachko's paper~\cite[conjecture
7.1.4]{Klyachko04}, and generalises a recent theorem by Anatol
N.~Kirillov~\cite[theorem 2.11]{Kirillov04} which was announced
without a proof.

A simple corollary to theorem~\ref{theorem-stability} is that
non-vanishing Kronecker coefficients obey entropic relations.

\begin{corollary}
Let $\lambda, \mu, \nu \vdash k$. If $g_{\lambda\mu\nu} \ne 0$,
then $H(\bar \lambda) \le H(\bar \mu) + H(\bar \nu)$, where
$H(\bar \lambda)=-\sum_i \bar \lambda_i \log(\bar \lambda_i)$ is
the Shannon entropy of $\bar \lambda=\lambda/k$.
\end{corollary}

\begin{proof} By theorem~\ref{theorem-stability}, $g_{\lambda\mu\nu} \ne 0$ implies $g_{N\lambda \, N\mu \, N\nu}
\ne 0$, for all $N \in \naturals$, where $N\lambda$ means the
partition with lengths $N\lambda_i$. By definition of $g_{\mu \nu
\lambda}$ (eq.~(\ref{eq-kron-symm})), \be \dim V_{N\lambda} \le \dim
V_{N\mu} \dim V_{N\nu} \ee and by
inequality~(\ref{eq-tableaux-bound}) and Stirling's approximation
$\frac{1}{N} \log(\dim V_{N\lambda})$ tends to
$kH(\overline{\lambda})$ for large $N$ the claim follows.
\end{proof}

\noindent I now introduce the background material needed in order
to prove theorem~\ref{theorem-stability}. In particular, I
introduce the ring of representations and prove that it has no
zero divisors with help of the Borel-Weil theorem.

Recall that
$$ U_\mu \otimes U_\nu \cong \bigoplus_{\lambda} c^\lambda_{\mu \nu}
U_\lambda$$ defines $c_{\mu \nu}^\lambda$ and that in particular
$c_{\mu \nu}^{\mu+\nu} =1$. This holds because the tensor product
of the highest weight vectors $\ket{v_\mu}$ and $\ket{v_\nu}$ of
$U_\mu$ and $U_\nu$ is the vector of highest weight in the tensor
product representation with weight $\mu+\nu.$ By the properties of
the lexicographical ordering this vector is unique. It is
straightforward to check this statement using the construction
with Young symmetrisers.

Consider now the direct sum
\be Q^d:= \bigoplus_\lambda U_\lambda\ee
of irreducible representations  $U_\lambda$ of $\GL(\complex^d)$
with highest weight $\lambda$ ($\lambda_d \geq 0$). $Q^d$ is a
graded ring when equipped with the product $U_\mu \otimes U_\nu
\rightarrow U_{\mu+\nu}$, which is sometimes called the
\emph{Cartan product\index{Cartan product}}~\cite{FultonHarris91}.
In terms of the elements of the ring (the vectors), this product
corresponds to the tensor product of $\ket{v} \in U_\mu$ and
$\ket{w} \in U_\nu$ followed by a projection onto $U_{\mu+\nu}$.
The resulting vector is denoted by $\ket{v} \circ \ket{w} \in
U_{\mu+\nu}$.

The goal of the next few paragraphs is to show that $Q^d$ is a
ring with no \emph{zero divisors}\index{zero divisor}, i.e.~that
there are no nonzero elements $\ket{v}, \ket{w} \in Q^d$ with
$\ket{v} \circ \ket{w}=0$. Before we start let us quickly simplify
the claim. Write $\ket{v}=\sum_\mu \ket{v_\mu}$ and
$\ket{w}=\sum_\nu \ket{w_\nu}$, where $\ket{v_\mu} \in U_\mu$ and
$\ket{w_\nu} \in U_\nu$. Let further $\tilde \mu$ and $\tilde \nu$
be the lexicographically highest Young diagrams for which
$\ket{v_{\tilde \mu}}, \ket{w_{\tilde \nu}} \neq 0$. Then,
    $$
        \ket{v} \circ \ket{w}=\ket{v_{\tilde\mu}} \circ \ket{w_{\tilde\nu}} + \sum_{\lambda < \tilde{\mu} +\tilde{\nu}}
        \ket{x_\lambda},
    $$
where $<$ denotes the lexicographical ordering
(page~\pageref{lexicographical}). Hence, if $\ket{v}\circ \ket{w}$
vanishes, so does $\ket{v_{\tilde\mu}} \circ \ket{w_{\tilde\nu}}$.
In order to prove that $Q^d$ has no zero divisors it therefore
suffices to prove the statement for elements $\ket{v}$ and
$\ket{w}$ in $U_\mu$ and $U_\nu$ respectively, i.e.~for elements
that lie within graded pieces.

Let us start with some notation: $\B(\complex^d)$ is the subgroup
of $\GL(\complex^d)$ consisting of upper triangular matrices,
known as the Borel subgroup, and $\T(\complex^d)$ is the torus of
$\GL(\complex^d)$ consisting of the diagonal matrices. Fix an
irreducible representation $U_\lambda$ with highest weight
$\lambda$ (with $\lambda_d \geq 0$). An element $b \in
B(\complex^d)$ acts on the highest weight vector $\ket{v}$ by
\be \label{eq-Borel-action} U_\lambda(b) \ket{v}=\lambda(b)\ket{v} ,\ee where
$\lambda(b)=b_{11}^{\lambda_1} \cdots b_{dd}^{\lambda_d}$. The key
to seeing that $Q^d$ has no zero divisors is the Borel-Weil
theorem which relates irreducible representation to polynomials in
the matrix entries of $\GL(\complex^d)$ with complex coefficients.
Let $\holmap$ be the polynomial functions on $\GL(\complex^d)$,
i.e.~the functions $f: \GL(\complex^d) \rightarrow \complex$ that
are polynomials in the matrix entries of $\GL(\complex^d)$ and the
inverse of the determinant with complex coefficients.
\begin{theorem}[Borel-Weil theorem\index{Borel-Weil theorem}]\label{theorem-borel-weil}
Let $U_\lambda$ be an irreducible representation of
$\GL(\complex^d)$ with highest weight $\lambda$. Then
$$ U_\lambda^\star \cong \holmap_{\lambda}$$
where $\holmap_{\lambda}$ is the space of polynomial functions in
$\holmap$ which satisfy
\be \label{eq-Borel-Weil} f(gb)=\lambda(b) f(g), \ee
for all $b \in B(\complex^d)$. The action of $\GL(\complex^d)$ on
$\holmap_{\lambda}$ is given by
\bestar (gf)(h)=f(g^{-1}h)\eestar
and the isomorphism is the map
\be \label{eq-Borel-Weil-map} \bra{\alpha} \mapsto f_\alpha(g):= \bra{\alpha}U(g)\ket{v},\ee
where $\ket{v}$ is the highest weight vector of $U_\lambda$.
\end{theorem}
A succinct proof of the theorem is given
in~\cite[p.~115]{CarterSegalMacDonald95}. Since
$$ g f_\alpha(h)=f_\alpha(g^{-1}h)=\bra{\alpha} U(g^{-1})U(h) \ket{v}=\big( U^\star(g)
\bra{\alpha} \big) U(h)\ket{v}=f_{g \alpha},$$ the map defined
in~(\ref{eq-Borel-Weil-map}) is in accordance with the action of
$\GL(\complex^d)$ on the dual space $U^\star_\lambda$ and
satisfies also equation~(\ref{eq-Borel-Weil}) by
equation~(\ref{eq-Borel-action}).

Let $(Q^d)^\star=\bigoplus_\lambda U^\star_\lambda$ be ring with
with Cartan product
$$ U_\mu^\star \otimes U_\nu^\star \mapsto U_{\mu+\nu}^\star.$$
Working with $(Q^d)^\star$ instead of $Q^d$, the previously
artificial-looking product turning $(Q^d)^\star$ into a ring is
now simply the product of two polynomial functions: for
$\bra{\alpha} \in U^\star_\mu$ and $\bra{\beta} \in U^\star_\nu$,
the Cartan product is given by
$$ f_\alpha(g)\times f_\beta(g) \mapsto  f_\alpha(g)f_\beta(g).$$
The product is clearly a function in $\holmap_{\mu+\nu}$.

Now comes the key argument in proving that $R$ has no zero
divisors.
Define the two subsets of $\GL(\complex^d)$: $A:=\{g \in
\GL(\complex^d)| f_\alpha(g)=0\}$ and $B:=\{g \in \GL(\complex^d)|
f_\beta(g)=0\}$. Their union $A \cup B=\{g \in
\GL(\complex^d)|f_\alpha(g) f_\beta(g)=0\}$ equals
$\GL(\complex^d)$ since by assumption $f_\alpha(g)f_\beta(g)$
vanishes on all of $\GL(\complex^d)$. In \emph{Zariski
topology\index{Zariski topology}}, a closed set is the set of
common zeros of a set of polynomials, hence $A$ and $B$ are
closed. Therefore we have shown that $\GL(\complex^d)$ is the
union of two closed proper subsets in Zariski topology. By
definition, this means that $\GL(\complex^d)$ is not irreducible.
However, connected algebraic groups, such as $\GL(\complex^d)$,
are known to be irreducible~\cite[p.~147]{CarterSegalMacDonald95}.
Our assumptions must therefore have been incorrect and
$f_\alpha(g)f_\beta(g)$ is a nonzero function of $g$ whenever
$f_\alpha(g)$ and $f_\beta(g)$ are nonzero. This shows that the
product $\bra{v} \circ \bra{w}$ does not vanish for $\bra{v},
\bra{w} \neq 0$. Thus $Q^d$ has no zero divisors.

Given any ring $R$ with an action of $G$ on it, $R^G$ denotes the
ring of $G$-invariants in $R$\index{ring of invariants}. As shown
below, $R^G$ is a ring without zero divisors if $R$ has no zero
devisors.

\begin{lemma} \label{lemma-graded-knutson}
Let $R$ be a graded ring without zero divisors, i.e.~for all $a, b
\neq 0$ the product $a b \neq 0$, and denote the graded pieces of
$R$ by $R_\tau$. Let $G$ also act on $R$, preserving the grading.
For any two $G$-invariant graded pieces: if $(R^G)_\tau \neq 0$
and $(R^G)_{\tau'} \neq 0$, then $(R^G)_{\tau+\tau'} \neq 0$.
\end{lemma}
\begin{proof}
Since $R^G \subset R$ and $R$ has no zero divisors, also $R^G$ has
no zero divisors. Then $a \in (R^G)_\tau \backslash \{0\}$ and $b
\in (R^G)_{\tau'} \backslash \{0\}$ implies $a b \in
(R^G)_{\tau+\tau'} \backslash \{0\}$.
\end{proof}

\noindent We now have the tools at hand to prove the main claim.

\begin{proof}[Proof of theorem~\ref{theorem-stability}]
Let us first consider the case of the Littlewood-Richardson
coefficients. Here, consider the ring $R=Q^d \otimes Q^d \otimes
(Q^d)^\star$. Its ring of invariants is given by
$$R^{\U(d)}=\left((\bigoplus_\mu U_\mu)\otimes (\bigoplus_\mu U_\nu)\otimes (\bigoplus_\mu
U_\lambda)\right)^{\U(d)}=\bigoplus_{\mu \nu \lambda} (U_\mu
\otimes U_\nu \otimes U_\lambda)^{\U(d)},$$ where $\U(d)$ acts
simultaneously (or diagonally) on the three factors of the ring.
The product operation in the ring takes triples $(\mu, \nu,
\lambda)$ and $(\mu', \nu', \lambda')$ to the triple $(\mu+\mu',
\nu+\nu', \lambda+\lambda')$. Any nonzero piece $(U_\mu \otimes
U_\nu \otimes U_\lambda)^{\U(d)}$ corresponds to a nonzero
coefficient $c_{\mu \nu}^\lambda$. Hence by
lemma~\ref{lemma-graded-knutson} $c_{\mu \nu}^\lambda \neq 0$ and
$c_{\mu' \nu'}^{\lambda'} \neq 0$ imply
$c_{\mu+\mu',\nu+\nu'}^{\lambda+\lambda'}\neq 0$.

For the second part of the theorem, which concerns the Kronecker
coefficients, we consider the ring
$$ R=Q^m \otimes Q^n \otimes (Q^{mn})^\star$$ and the invariant
ring under the action of the group $\SU(m)\times \SU(n)$.
$\SU(m)\times \SU(n)$ acts by inclusion in $\SU(mn)$ on
$U^\star_\lambda$
$$ R^G=\bigoplus_{\mu \nu \lambda} (U_\mu \otimes U_\nu \otimes
U^\star_\lambda)^{\SU(m)\times \SU(n)}.$$ The product operation in
the ring takes triples $(\mu, \nu, \lambda)$ and $(\mu', \nu',
\lambda')$ to the triple $(\mu+\mu', \nu+\nu', \lambda+\lambda')$.
Any nonzero piece $(U_\mu \otimes U_\nu \otimes
U^\star_\lambda)^{\SU(m)\times \SU(n)}$ corresponds to a nonzero
coefficient $g_{\mu \nu \lambda}$. Hence by lemma
\ref{lemma-graded-knutson} $g_{\mu \nu \lambda} \neq 0$ and
$g_{\mu' \nu' \lambda'} \neq 0$ imply
$g_{\mu+\mu',\nu+\nu',\lambda+\lambda'} \neq 0$.
\end{proof}

\sloppy \noindent It is well-known that the set of nonzero
Littlewood-Richardson coefficients is \emph{finitely
generated}\index{finitely generated}~\cite{Elashvili92}, i.e.~that
there is a finite set of triples $(\mu_i, \nu_i, \lambda_i)$ with
$c_{\mu_i \nu_i}^{\lambda_i} \neq 0$ such that every triples
$(\mu, \nu, \lambda)$ with $c_{\mu \nu}^\lambda \neq 0$ is an
integral linear combination of the $(\mu_i, \nu_i, \lambda_i)$.
Below, I will show that the semigroup of nonzero Kronecker
coefficients is finitely generated, too. This result was expected
by Klyachko~\cite[statement below conjecture 7.1.4]{Klyachko04}.

\fussy

Consider the ring $$R=Q^m \otimes Q^n \otimes (Q^{mn})^{\star}$$
as well as the ring of invariants $R^{\SU(m)\times \SU(n)}$. Since
$R^{\SU(m)\times \SU(n)}=R^{SL(m)\times SL(n)}$, the problem of
finite generation can be dealt with from the perspective of linear
algebraic groups. A subgroup $G$ of $GL(\complex^d)$ is a
\emph{linear algebraic group}\index{linear algebraic group} if it
is the set of common zeros of a set of polynomials $A$ in the
matrix entries of $\M(\complex^d)$,
    $$
        G=\{ g \in \GL(\complex^d) | \; f(g)=0 \mbox{ for all } f \in
        A\}.
    $$
A linear algebraic group is \emph{reductive}\index{reductive} if
every finite-dimensional regular representation is completely
reducible. Let $V$ be a $q$-dimensional complex vector space with
coordinates $x_i$ on $V$. The algebra of polynomial functions in
the $x_i$ is denoted by $\cS(V)=\complex[x_1, \ldots, x_{q}]$. I
state without proof the following theorem, which is a consequence
of Hilbert's basis theorem\index{Hilbert's basis theorem} (see
e.g.~\cite[theorem 2.4.9]{Springer77}), and which will imply the
finite generation of $R^G$.

\begin{theorem} \label{theorem-hilbert}
Suppose $G$ is a reductive linear algebraic group acting
polynomially on $V$, hence on $\cS(V)$, and preserving an ideal
$\cI$ of $\cS(V)$. Then $(\cS(V)/\cI)^G$ is finitely generated,
i.e.~there are $\phi_1, \ldots, \phi_p \in (\cS(V)/\cI)^G$ such
that every $\phi \in (\cS(V)/\cI)^G$ is a polynomial in the
$\phi_1, \ldots, \phi_p$ with complex coefficients.
\end{theorem}

\noindent In order to apply theorem~\ref{theorem-hilbert} to the
ring $R$ and $SL(m)\times SL(n)$, I now explain how one can
identify $Q^d$ -- and then $R$ -- with a quotient as required in
the theorem.

 Let $U^d_{\omega^i}$ with $\omega_i=(\underbrace{1,
\ldots, 1}_{i}, 0, \ldots, 0)$ be the \emph{fundamental
representations\index{representation!fundamental}} of
$\GL(\complex^d)$ for $i \leq d$ and consider the representation
$V^d:=\bigoplus_{i\leq d} U^d_{\omega_i}$ of $\GL(\complex^d)$.
The symmetric algebra of $V^d$,
$$ S^{d \bullet}:=\sym (V^d)= \sym (U^d_{\omega_1} \oplus \cdots \oplus U^d_{\omega^d}),$$
is the direct sum of the representations
$$ S^{d, \bf a} := \sym^{a_1} U^d_{\omega_1} \otimes \cdots \otimes \sym^{a_d} U^d_{\omega_d}.$$
$S^{d, \bf a}$ is a representation with highest weight $\lambda$,
where $a_i=\lambda_i-\lambda_{i+1}$, and contains the irreducible
representation $U^d_\lambda$ exactly once. Let $I^{d, \bf a}$ be
the direct sum of all irreducible representations contained in
$S^{d, \bf a}$ apart from $U_\lambda$. It can be shown that $I^{d
\bullet}:=\bigoplus_{\bf a} I^{d, \bf a}$ is an ideal in $S^{d
\bullet}$ with respect to the Cartan product~\cite[page
428]{FultonHarris91}. Factoring out this ideal from $S^{d
\bullet}$ leaves us with the ring $Q^d$:
$$ Q^d \cong S^{d \bullet}/I^{d \bullet} \cong \bigoplus_\lambda
U^d_\lambda.
$$
Let us now regard the representations as spaces of polynomials.
Let $\ket{e_i}$ be an orthonormal basis for $\complex^d$. An
orthonormal basis for the $U_{\omega_k}\equiv \wedge^k \complex^d$
is given by $\ket{e_{i_1}} \wedge \cdots \wedge \ket{e_{i_k}},$
where $i_j \in \{1, \ldots, d\}$. The coordinates of
$U_{\omega_k}$ relative to the chosen basis are the maximal
minors, i.e.~minors of size $k \times k$, of a $d \times k$ matrix
filled with indeterminates. The minors are also known as
\emph{Pl\"ucker coordinates}\index{Pl\"ucker coordinates} and the
action on these coordinates is determined by $\GL(\complex^d)$
acting on the $d \times k$-matrix from the left. Denote by
$x_{kl}$ for $l=1, \ldots, \dim U^d_{\omega_k}$ the minors
corresponding to $U^d_{\omega_k}$, i.e.~$U^d_{\omega_k}$'s
coordinates.

The symmetric algebra $S^{d \bullet}$ can now be identified with
the ring of polynomials in the variables $x_{kl}$, $k \in \{1,
\ldots d\}$ and $l \in \{1, \ldots, \dim U^d_{\omega_k}\}$:
$$ \cS(V^d) := \complex[\{x_{kl}\}].$$
Likewise, one can identify the representation $I^{d, \bf a}$ with
the set of polynomials $\cI^{d, \bf a}$ spanning $I^{d, \bf a}$.
Hence, $I^{d, \bullet}$ is identified with the polynomial ideal
$\cI^d:=\{ f \in \cI^{d, \bf a} \mbox{ for some } \bf{a}\}$ in
$\complex[\{x_{kl}\}]$. This shows that
$$ Q^d \cong \cS(V^d)/\cI^d.$$
Consider now the tensor product ring
    $$
        R:=Q^m \otimes Q^n \otimes (Q^{mn})^{\star}.
    $$
It follows from the discussion above that $R$ is isomorphic to
    $$
        \cS(V) /\cI,
    $$
where $V:=V^m \otimes V^n \otimes V^{mn}$ and
$\cS(V)=\complex[\{x_{kl}y_{pq}z_{rs}\}]$ for coordinates $y_{pq}$
and $z_{rs}$ of $\complex^n$ and $\complex^{mn}$, respectively.
The ideal $\cI$ in $\cS(V)$ is generated by all polynomials in
$\cI^m$, $\cI^n$ and $\cI^{mn}$.

I will now apply theorem~\ref{theorem-hilbert} to $\cS(V)/\cI$ and
the group $G= SL(m) \times SL(n)$. $G$ acts by the defining
representation on the coordinates $\{ x_{kl}y_{mn}\}$ of
$V^m\otimes V^n$. Note that $\GL(\complex^{mn})$ acts with the
dual representation on $z_{rs}$ and that $G$ acts on $z_{rs}$ via
inclusion into $\GL(\complex^{mn})$. It remains to show that $G$
is reductive or -- since any semisimple algebraic group is
reductive -- that $G$ is semisimple\index{Lie group!semisimple}. A
connected Lie group (such as $G$) is \emph{semisimple} if its Lie
algebra is. By definition, a Lie algebra is semisimple\index{Lie
algebra!semisimple} if it is a direct sum of simple Lie algebras
and a \emph{simple} Lie algebra is one that has no nontrivial
ideals and is not Abelian. This is evident in our case because the
Lie algebra of $SL(m)\times SL(n)$ is the direct sum of the Lie
algebras of $SL(m)$ and $SL(n)$, which are both
simple~\cite{WikiReductive}. By theorem~\ref{theorem-hilbert}
therefore, the ring of $G$-invariants $( \cS(V) /\cI)^G$ is
finitely generated. This results in the following corollary.

\begin{corollary} \label{corollary-finitely-generated}
The triples $(\mu, \nu, \lambda)$ with $g_{\mu \nu \lambda}\neq 0$
form a finitely generated semigroup under row-wise addition.
\end{corollary}

\section{Spectra of Quantum States}
\label{sec-spectra-states} This section is the core of
chapter~\ref{chapter-relation}. The topic of the first subsection
(\ref{subsec-spec-estimate}) is the estimation of a spectrum \`a
la Keyl and Werner. After a brief history of this result I give a
short proof of it with emphasis on the connection to
representation theory. This connection provides the link to the
asymptotic equivalence of problem~\ref{prob-comp} and
problem~\ref{prob-Kronecker}, which is proven in
subsection~\ref{subsec-spec-comp}. A short excursus to the
classical analogue of this result is presented in
subsection~\ref{subsection-classical-analogue}. This is followed
up by a brief intermezzo, which guides us to the second result of
this section: a new proof of the asymptotic equivalence of
problem~\ref{prob-Horn} and problem~\ref{prob-LR}
(subsection~\ref{subsec-spectra-horn}).
Subsection~\ref{subsec-spec-convexity} draws on the previously
established equivalences and uses the semigroup property from
subsection~\ref{sec-semigroup} to infer convexity of the spectral
problem. An independent proof of convexity based on a theorem by
Frances Kirwan brings this discussion to a close. The final
subsection (\ref{subsec-spec-comp}) uses the established
equivalences to derive all spectral inequalities from a
group-theoretic result due to Klemm, Dvir, and Clausen and Maier.

\subsection[Spectrum Estimation]{Spectrum Estimation\protect\footnote{The results presented in this subsection have appeared in~\cite{ChrMit05}.}} \label{subsec-spec-estimate}
Given $k$ identical copies of a quantum state $\rho$, i.e.~given
$\rho^{\otimes k}$, what is the optimal way to estimate $\rho$?
This question appears frequently in quantum information theory and
is essential to quantum cryptography.\footnote{In a general
quantum cryptographic setting the tensor product structure might
not be given. In almost all situations, however, it forms the
important special case and sometimes even, the problem at hand
reduces to this form.} Michael Keyl and Reinhard Werner split the
question in two: i) the estimation of the spectrum of $\rho$, and
ii) the estimation of the corresponding eigenvectors. Combining
the answers leads to an estimate of $\rho$. In 2001, Keyl and
Werner published a solution to i)~\cite{KeyWer01PRA}. Subsequently
Keyl showed the optimality of this solution using Stein's lemma
and addressed ii)~\cite{Keyl04}. The solution to i) brought up a
remarkable connection between Young frames and density operators.
Using large deviations theory, the authors showed that, for large
$k$, the quantum state $\rho^{\otimes k}$ will project with high
probability into the Young subspaces $\lambda \vdash k$ such that
$\bar{\lambda}$ approximates the spectrum of $\rho$
(theorem~\ref{theorem-Keyl-Werner}). In Spring 2002, Graeme
Mitchison and I discovered a short proof based on the majorisation
property of lemma~\ref{lemma-major}. Shortly after, we realised
that this proof appears in an appendix to Hayashi and Matsumoto's
quantum source coding paper~\cite{HayMat02}. This proof will be
given here correcting an algebraic slip in Hayashi and Matsumoto's
work. Recently, Koenraad Audenaert has pointed out to me that, as
early as 1988, Robert Alicki, S\l awomir Rudnicki and S\l awomir
Sadowski had already discovered that the probability distribution
over Young frames peaks around the spectrum of
$\rho$~\cite{AlRuSa87}. Interestingly, the authors describe their
work as part of the theoretical study of collective phenomena in
quantum optics. I believe that the history of
theorem~\ref{theorem-Keyl-Werner}, rather than diminishing the
individual contributions, highlights its fundamental importance.

\begin{theorem}[Spectrum Estimation\index{spectrum estimation}] \label{theorem-Keyl-Werner}
Let $\rho$ be a density operator with spectrum $r=\spec \rho$, and
let $P_\lambda$ be the projection onto $U_\lambda \otimes
V_\lambda$. Then \be \label{eq-KeylWerner-1}\tr P_\lambda
\rho^{\otimes k} \leq (k+1)^{d(d-1)/2} \exp \left(-k D
(\bar{\lambda}||r)\right)\ee with $D(\cdot ||\cdot)$ the
Kullback-Leibler distance of two probability distributions, which
has been defined in the Preliminaries (page
\pageref{Kullback-Leibler distance}).
\end{theorem}

\begin{proof}
Let $\{r_i, \ket{i}\}$ be a set of eigenvalues and corresponding
eigenvectors for $\rho$, ordered according to size: $r_i \geq
r_{i+1}$. A basis for $(\complex^d)^{\otimes k}$ and an eigenbasis
for $\rho^{\otimes k}$ is given by the tensor products of the
previously chosen eigenbasis for $\rho$. According to the
Schur-Weyl duality, theorem~\ref{theorem-Schur-duality},
$(\complex^d)^{\otimes k}$ decomposes as a direct sum of pairs of
irreducible representations of $\U(d)$ and $S_k$,
$$(\complex^d)^{\otimes k}\cong \bigoplus_{\lambda \vdash (k, d)} U_\lambda \otimes V_\lambda,$$
where a basis for $U_\lambda\otimes V_\lambda$ is constructed by
applying the Young symmetrisers $e_T$ to the chosen basis of
$(\complex^d)^{\otimes k}$, where $T$ runs over all tableaux to
the frame $\lambda$. According to lemma~\ref{lemma-major}, all
vectors whose frequency $f$ is not majorised by the $\lambda$ will
be sent to zero. The frequency of the surviving eigenvectors
therefore obeys $f \prec \lambda$ and the corresponding
eigenvalues $\prod_i r_i^{f_i}$ are smaller than or equal to
$\prod_i r_i^{\lambda_i}$. Using the
bounds~(\ref{eq-tableaux-bound})
and~(\ref{eq-semi-tableaux-bound}), it follows that
\beastar \tr P_\lambda \rho^{\otimes k} &\leq& \dim \cU_\lambda \dim
\cV_\lambda \prod_i r_i^{\lambda_i} \\
&\leq& (k+1)^{d(d-1)/2} \binom{k}{\lambda_1 \cdots \lambda_d} \prod_i r_i^{\lambda_i} \\
&\leq& (k+1)^{d(d-1)/2} \exp \left(-k D (\bar{\lambda}||r)\right).
\eeastar This completes the proof.
\end{proof}

\noindent To show one direction of the equivalence of
problem~\ref{prob-comp} and~\ref{prob-Kronecker} (and likewise of
problem~\ref{prob-Horn} and~\ref{prob-LR}) the exponential decay
is crucial. In the opposite direction, however, a weaker result
suffices; namely the result that the projection onto Young frames
is actually an estimation scheme. The following simple corollaries
capture this fact.

\begin{corollary}
\label{eq-KeylWerner2} If $\rho$ is a density operator with
spectrum $r=\spec \rho $,
\be \tr P_X \rho^{\otimes k} \leq
(k+1)^{d(d+1)/2} \exp (-k \ \min_{\lambda \vdash k: \bar{\lambda}
\in {\cal S}}D (\bar{\lambda}||r)),
\ee
where $P_X:= \sum_{\lambda \vdash k: \bar{\lambda} \in {\cal
S}}P_\lambda$ for a set of spectra $\cS$ .
\end{corollary}

\begin{proof}
This follows from theorem~\ref{theorem-Keyl-Werner} by picking the
Young frame with the slowest convergence and multiplying it by the
total number of possible Young frames with $k$ boxes in $d$ rows.
This number is smaller than $(k+1)^d$.
\end{proof}

\noindent Let ${\cal B}_\epsilon(r):=\{r': \sum |r'_i-r_i|<
\epsilon\}$ be the $\epsilon$-ball around the spectrum $r$. If we
take ${\cS}$ to be the complement of ${\cB}_\epsilon(r)$, it
becomes clear that for large $k$, $\rho^{\otimes k}$ will project
onto a Young subspace $\lambda$ with $\bar{\lambda}$ close to $r$
with high probability. More precisely:

\begin{corollary}
\label{corollary-Keyl-2} Let $\rho$ be a state with spectrum
$r=\spec \rho$ and $\epsilon, \delta>0$ small numbers and let
$P_X:=\sum_{\lambda \vdash k: \bar{\lambda} \in \cB_{\epsilon}(r)}
P_\lambda$. Then there exists $k_0 \equiv k_0(\epsilon, \delta)>0$
such that for all $k\geq k_0$,
\be \tr P_X \rho^{\otimes k}>1-\delta. \ee
\end{corollary}

\subsection[Problem~\ref{prob-comp} vs. Problem~\ref{prob-Kronecker}]
{Problem~\ref{prob-comp} vs.
Problem~\ref{prob-Kronecker}~\protect\footnote{Here, I present
work which originated in a collaboration with Graeme Mitchison,
part of which has been published in~\cite{ChrMit05}.
Theorem~\ref{theorem-converse} has jointly been obtained with Aram
Harrow and Graeme Mitchison.}} \label{subsec-spec-comp}

This subsection contains the proof of a close connection between
problem~\ref{prob-comp} and~\ref{prob-Kronecker}, which was
described as asymptotic equivalence in the introduction
(page~\pageref{pageref-asymptotic-equivalence}). More precisely,
it is shown in theorem~\ref{ourtheorem} that for every density
operator $\rho^{AB}$, there is a sequence of nonvanishing $g_{\mu
\nu \lambda}$ such that the triple of normalised Young diagrams
$(\bar{\mu}, \bar{\nu}, \bar{\lambda})$ converges to $(\spec
\rho^{A}, \spec \rho^B, \spec \rho^{AB})$. The converse,
theorem~\ref{theorem-converse}, constructs for every $g_{\mu \nu
\lambda} \neq 0$ a density operator $\rho^{AB}$ with $(\spec
\rho^A, \spec \rho^B, \spec \rho^{AB})$ equal to $(\bar{\mu},
\bar{\nu}, \bar{\lambda})$. Hence, only one direction involves an
asymptotic statement, whereas the other is direct.

\begin{theorem}\label{ourtheorem}
For every density operator $\rho^{AB}$, there is a sequence
$(\lambda_j, \mu_j, \nu_j)$ of partitions, labeled by natural
numbers $j$, with $|\lambda_j|=|\mu_j|=|\nu_j|$ such that \be
g_{\lambda_j \mu_j \nu_j} \neq 0 \quad \mbox{ for all }\ j\ee and
    \bea
        \lim_{j \to \infty} \bar{\lambda}_j&=&\spec \rho^{AB}\\
        \lim_{j \to \infty} \bar{\mu}_j&=&\spec \rho^{A}\\
        \lim_{j \to \infty} \bar{\nu}_j&=&\spec \rho^{B}
    \eea
\end{theorem}

\begin{proof}
Let $r^{AB}=\spec \rho^{AB}, r^{A}=\spec \rho^{A}, r^{B}=\spec
\rho^{B}$. $P^{AB}_\lambda$ denotes the projector onto the Young
subspace $U_\lambda \otimes V_\lambda$ in system $AB$, and
$P^A_\mu$, $P^B_\nu$ are the corresponding projectors onto Young
subspaces in $A$ and $B$, respectively. By corollary
\ref{corollary-Keyl-2}, for given $\epsilon>0$, one can find a
$k_0$ such that the following inequalities hold simultaneously for
all $k \geq k_0$,
\bea \label{equation-sum1} \tr P_X (\rho^A)^{\otimes k} &\geq& 1-\epsilon, \quad P_X:=\sum_{\mu: \bar{\mu} \in
\cB_\epsilon(r^A)} P^{A}_\mu\\
 \label{equation-sum2}\tr P_Y (\rho^B)^{\otimes k} &\geq& 1-\epsilon, \quad P_Y:=\sum_{\nu: \bar{\nu} \in \cB_\epsilon(r^B)}
P^{B}_\nu\\
\label{equation-sum3} \tr P_Z (\rho^{AB})^{\otimes k} &\geq&
1-\epsilon, \quad P_Z:=\sum_{\lambda: \bar{\lambda} \in
\cB_\epsilon(r^{AB})} P^{AB}_\lambda.
\eea
The estimates~(\ref{equation-sum1}) and~(\ref{equation-sum2}) can
be combined to yield
\be \label{equation-sum4} \tr (P_X \otimes P_Y) (\rho^{AB})^{\otimes k} \geq
1-2\epsilon.
\ee
This follows from
\bestar \label{trace-inequality} \tr (P \otimes Q) \xi^{AB}
\ge \tr P \xi^A + \tr Q \xi^B-1,
\eestar
which holds for all projectors $P$ and $Q$ and density operators
$\xi^{AB}$ since $\tr [(\openone -P) \otimes (\openone-Q) \xi^{AB}
]\geq 0$. Because $(\rho^{AB})^{\otimes k}$ maps each Young frame
onto itself, writing $\sigma=(\rho^{AB})^{\otimes k}$, we have
\be \label{eq-project}\sum_{\lambda \vdash k} P_\lambda
\sigma P_\lambda=\sigma.
\ee
Defining $P_{\bar Z}:=\openone-P_Z$, the
estimates~(\ref{equation-sum4}) and~(\ref{eq-project}) imply
\bestar \tr [(P_X \otimes P_Y) (P_Z\sigma P_Z + P_{\bar Z}\sigma P_{\bar
Z})] \geq 1-2 \epsilon.
\eestar
Inserting $\tr [(P_X \otimes P_Y) P_{\bar Z}\sigma P_{\bar Z}]
\leq \epsilon$ (from eq.~(\ref{equation-sum3})) gives
\bestar \tr [(P_X \otimes P_Y) P_Z\sigma P_Z] \geq 1-3 \epsilon.
\eestar
Clearly, there must be at least one triple $(\mu, \nu, \lambda)$
with $\bar{\mu} \in \cB_\epsilon(r^A), \bar{\nu} \in
\cB_\epsilon(r^B)$ and $\bar{\lambda} \in \cB_\epsilon(r^{AB})$
with $\tr[(P^A_\mu \otimes P^B_\nu) P^{AB}_\lambda \sigma
P^{AB}_\lambda] \neq 0$. Thus
\be \label{eq-nonzero} (P^A_\mu \otimes P^B_\nu) P^{AB}_\lambda \neq 0.\ee
The LHS, reminding ourselves of equation~(\ref{eq-proj}), is the
projector onto the space
\bestar \bigoplus_\lambda \bigoplus_{i=1}^{g_{\mu \nu \lambda}} U^{\lambda,i}_{\mu \nu}  \otimes V_\lambda
\cong
 U_{\mu} \otimes V_{\mu} \otimes U_{\nu} \otimes V_{\nu},
\eestar
which then, invoking~(\ref{eq-nonzero}), gives $g_{\mu \nu
\lambda} \neq 0$.
\end{proof}

\noindent \sloppy Shortly after posting \cite{ChrMit05}, which
contains theorem~\ref{ourtheorem}, on a preprint server, Klyachko,
unaware of~\cite{ChrMit05}, announced his work on the quantum
marginal problem~\cite{Klyachko04}. He solves
problem~\ref{prob-comp} in the framework of geometric invariant
theory by showing how to calculate inequalities that define the
polytope of admissible spectral triples. The fact that the
solution is indeed a polytope will be proven in
subsection~\ref{subsec-spec-convexity}. Klyachko also discovers
the connection between problem~\ref{prob-comp}
and~\ref{prob-Kronecker}, and includes a theorem~\cite[theorem
5.3.1]{Klyachko04} which is similar to theorem~\ref{ourtheorem}.

\fussy \begin{theorem}\label{theorem-klyachko-our} Let $\rho^{AB}$
be a density operator with rational spectral triple $(r^A, r^B,
r^{AB}),$ then there is an integer $m>0$ such that $g_{m r^A, m
r^B, m r^{AB}}\neq 0$.
\end{theorem}

\noindent Even though very similar, neither of these statements
follows directly from the other: theorem~\ref{ourtheorem} remains
in an approximate form even if the triple of spectra is rational,
and theorem~\ref{theorem-klyachko-our} only constructs a single
nonzero Kronecker coefficient and not a whole sequence.

But in fact both theorems are equivalent. The missing link is the
fact that the nonvanishing Kronecker coefficients form a finitely
generated semigroup (theorem~\ref{theorem-stability} and
corollary~\ref{corollary-finitely-generated}). I now show how
theorem~\ref{theorem-klyachko-our} follows from
theorem~\ref{ourtheorem}.

\begin{proof}[Proof of theorem~\ref{theorem-klyachko-our}]
$\qmp$ denotes the set of all admissible triples of spectra $(r^A,
r^B, r^{AB})$ and $\kron$ the set of all $(\bar{\mu}, \bar{\nu},
\bar{\lambda})$ for which $g_{\mu \nu \lambda} \neq 0$. As an
immediate corollary of theorem~\ref{ourtheorem}
$$ \qmp \subset \overline{\kron},$$ where $\overline{\kron}$ denotes the
closure of $\kron$. The next step is to show that taking the
closure of $\kron$ only adds irrational spectra, i.e.~to show that
$$\overline{\kron} \cap \rational^{m+n+mn}=\kron.$$
The inclusion $\kron \subset \overline{\kron} \, \cap \,
\rational^{m+n+mn}$ is obvious (and also not needed here). To see
that the converse is true, recall that $\KRON$, the set of
nonvanishing Kronecker coefficients, is a finitely generated
semigroup. Let $(\mu_i, \nu_i, \lambda_i)$ be a finite set of
generators. $\overline{\kron}$ is a convex polytope consisting of
all convex combinations of $(\bar{\mu}_i, \bar{\nu}_i,
\bar{\lambda}_i)$ and let its dimension be $t$. Every point $(r^A,
r^B, r^{AB}) \in \overline{\kron}$ can be written as
\be \label{eq-vertex}
(r^A, r^B, r^{AB})= \sum_i x_i (\bar{\mu}_i, \bar{\nu}_i,
\bar{\lambda}_i),
\ee
\sloppy for a set of nonnegative numbers $x_i$ which sum to one.
Since the union of the $t+1$-vertex simplices equals the whole
polytope, every point in $\overline{\kron}$ can be taken to be the
sum of just $t+1$ normalised generators $(\bar{\mu}_i,
\bar{\nu}_i, \bar{\lambda}_i)$ (cf. Carath\'eodory's theorem).
From the set of $m+n+mn$ equations in the variables $x_i$ in
equation~(\ref{eq-vertex}), choose a set of $t$ linearly
independent ones, add the $t+1$'th constraint $\sum_i x_i=1$ and
write the set of equations as $ M\vec{x} = \vec{r},$
i.e.~$\vec{r}=(r_1, \ldots, r_t, 1)$ for $r_j \in \{ r^A_1,
\ldots, r^A_m, r^B_1, \ldots r^B_n, r^{AB}_1, \ldots,
r^{AB}_{mn}\}$ and $x=(x_1, \ldots, x_{t+1})$.

\fussy If $(r^A, r^B, r^{AB}) \in \overline{\kron}$ is rational,
the $x_i$ will be rational as well, since $M$ is rational. This
shows that $(r^A, r^B, r^{AB})=\sum_i \frac{n_i}{n}(\bar{\mu}_i,
\bar{\nu}_i, \bar{\lambda}_i)$, where we set $x_i=\frac{n_i}{n}$
for $n_i, n \in \naturals$. Multiplication with $|\mu|$ and $n$
results in
    \bestar \label{eq-rational}
        |\mu|n(r^A, r^B, r^{AB})=\sum_i n_i( \mu_i, \nu_i,
        \lambda_i)
    \eestar
Since the RHS of the previous equation is certainly an element of
$\KRON$ this shows that for rational $(r^A, r^B, r^{AB}) \in \qmp$
(since $\qmp \subset \overline{\kron}$) there is a number
$m:=|N||\mu|$ such that $g_{m r^A, m r^B, m r^{AB}}\neq 0$.
\end{proof}

\noindent It remains to prove the converse, namely that
theorem~\ref{ourtheorem} is a corollary to
theorem~\ref{theorem-klyachko-our}. According to
theorem~\ref{theorem-klyachko-our}, for every rational spectral
triple $(r^A, r^B, r^{AB})$ there is an integer $m>0$ such that
$g_{mr^A, mr^B, mr^{AB}}\neq 0$. The semigroup property,
theorem~\ref{theorem-stability}, extends this single nonvanishing
coefficient to a whole sequence, $\{g_{nmr^A, nmr^B,
nmr^{AB}}\}_{n\geq 1}$ of nonzero coefficients, just as required.
Irrational triples can be dealt with through an approximation by
rational triples.

\begin{figure} \label{figure-red-dots-black-dots}
\hspace{0cm}
\includegraphics[width=0.5\textwidth]{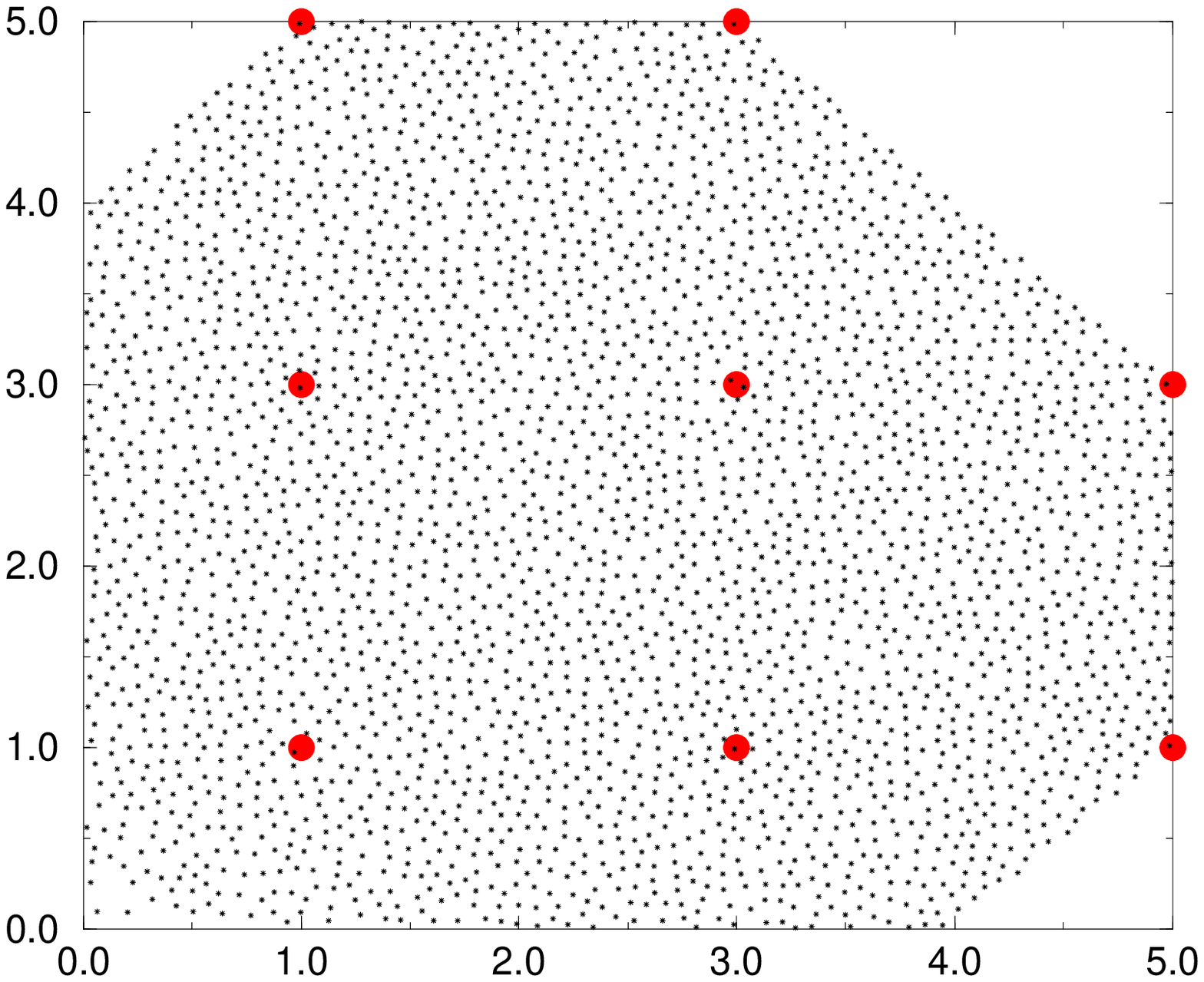}
\includegraphics[width=0.5\textwidth]{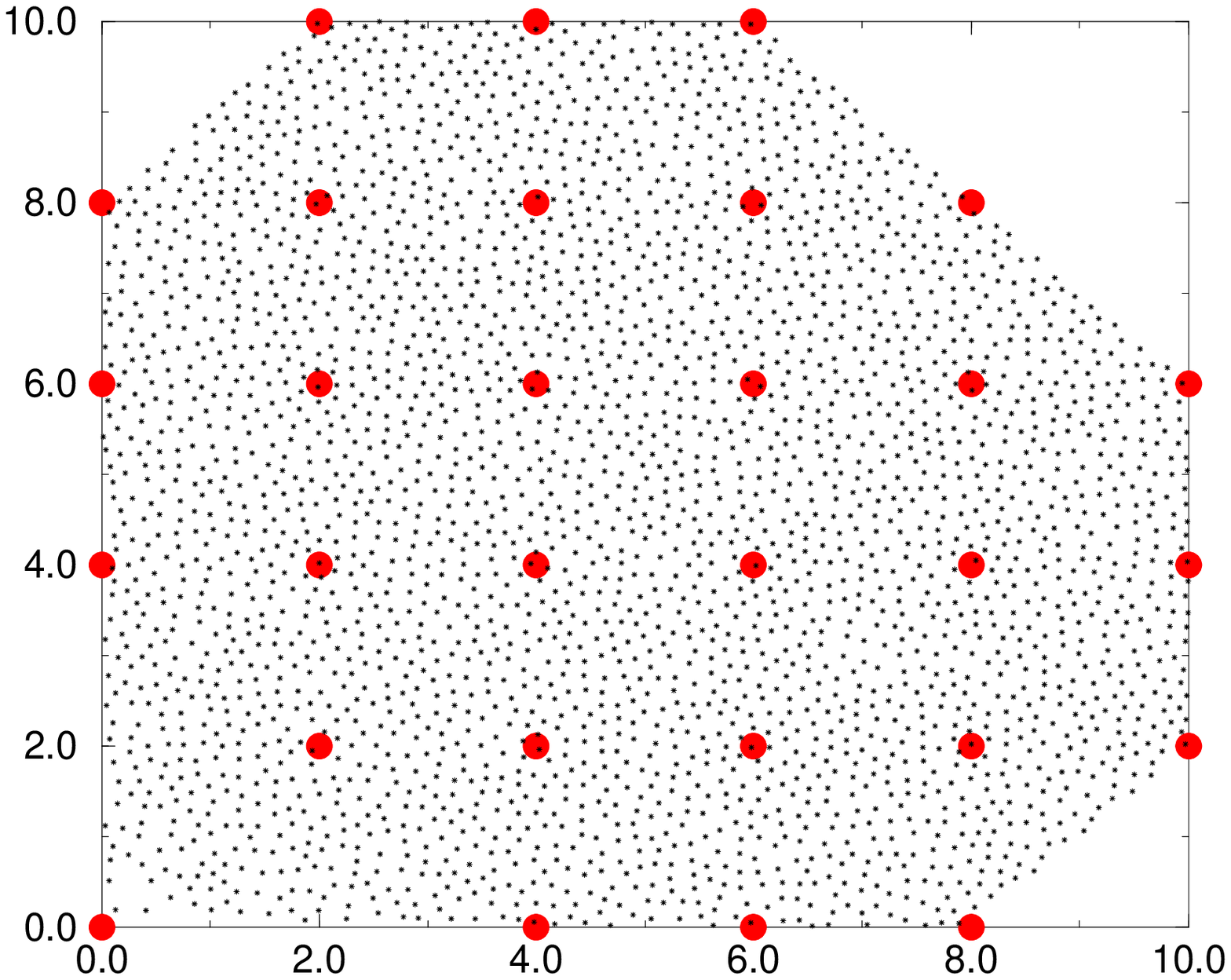}
\vspace{-1cm} \caption{Red dots correspond to $g_{\mu \nu
\lambda}\neq 0$, black dots correspond to admissible spectral
triple $(r^A, r^B, r^{AB})$. Two plots of the case $\SU(2)\times
\SU(2) \subset \SU(4)$ case, or two-qubit case, are on display.
The $x$-axis label shows $\mu_1-\mu_2$ and $k(r^{A}_1-r^A_2)$, the
$y$-axis shows $\nu_1-\nu_2$ and $k(r^B_1-r^B_2)$. The plot on the
left contains data for $\lambda=(4,2,1)=7 \,r^{AB}$, which is to
be contrasted with the plot on the right, $\lambda=(8,4,2)=14
\,r^{AB}$.}
\end{figure}

The plots for the $\SU(2)\times \SU(2) \subset \SU(4)$ -- or the
two-qubit -- case suggested that not only is $\qmp \subset
\overline{\kron}$, but also conversely $\kron \subset \qmp$ (see
figure~\ref{figure-red-dots-black-dots}). It turned out to be more
difficult than expected to prove this conjecture and a proof was
only given after Klyachko announced his paper~\cite[theorem
5.3.1]{Klyachko04}. This work was done in collaboration with Aram
Harrow and Graeme Mitchison and a joint publication is in
preparation.

\begin{theorem} \label{theorem-converse}
Let $\mu, \nu$ and $\lambda$ be diagrams with $k$ boxes and at
most $m$, $n$ and $mn$ rows, respectively. If $g_{\mu \nu \lambda}
\neq 0$, then there exists a density operator $\rho^{AB}$ on
$\cH_A \otimes \cH_B=\complex^{m} \otimes \complex^n$ with spectra
\bea \spec \rho^{A}&=&\bar{\mu}\\
\spec \rho^{B}&=&\bar{\nu}\\
\spec \rho^{AB}&=&\bar{\lambda}
\eea
\end{theorem}

\begin{proof}
Let $\cH_C\cong \cH_A \otimes \cH_B=\complex^{mn}$. It will
suffice to construct a pure state $\ket{\varphi} \in \cH_A\otimes
\cH_B \otimes \cH_C$ with margins $\rho^A, \rho^B$ and $\rho^C$
satisfying $\spec \rho^A=\bar{\mu}, \spec \rho^{B}=\bar{\nu}$ and
$\spec \rho^{C}=\bar{\lambda}$, since $\rho^{AB}$'s spectrum
automatically equals the spectrum of $\rho^C$, which is
$\bar{\lambda}$.

According to theorem~\ref{theorem-stability}, $g_{\mu \nu \lambda}
\neq 0$ implies that $g_{N \mu, N \nu, N \lambda}\neq 0$ for all
integers $N \geq 1$. Let us fix $N$, and observe that $g_{N \mu, N
\nu, N \lambda}\neq 0$ implies that there exists a vector
$\ket{\psi_N} \in V_{N\mu} \otimes V_{N\nu} \otimes
V_{N\lambda}\subset (\complex^m \otimes \complex^n \otimes
\complex^{mn})^{\otimes Nk}$ which is invariant under permutations
of its $Nk$ subsystems, hence $\ket{\psi_N}$ spans the space of an
irreducible representation of $S_{Nk}$ with Young frame
$\kappa=(Nk, 0, \ldots, 0)$, the trivial representation. According
to Schur-Weyl duality there is an associated representation
$U_{\kappa}$ of $\U(l)$, where $l:=(mn)^2$. The space of
$U_{\kappa}$ is spanned by $\{ U^{\otimes Nk} \ket{\psi_N}: U \in
U(l)\}$ (lemma~\ref{lemma-symmetric}). Since $V_\kappa$ is the
trivial representation of $S_k$, it is one-dimensional and it
holds $U_\kappa \cong U_\kappa \otimes V_\kappa$. The latter,
however, is spanned by the product vectors $\ket{v}^{\otimes Nk}$
(see lemma~\ref{lemma-symmetric}). Fix one of these, call it
$\ket{\phi}^{\otimes Nk}$ and consider the operator
$$P_{\kappa}=\dim U_\kappa \int_{U \in \U(l)} dU \left( U \proj{\phi} U^\dagger
\right)^{\otimes Nk},$$ where the measure $dU$ is a Haar measure
on $\U(l)$ with normalisation $\int_{U \in U(l)} dU=1$. $P_\kappa$
is in fact the projector onto $U_\kappa$: the $\U(l)$ invariance
of $dU$ implies the $\U(l)$ invariance of $P_\kappa$. Since
$P_\kappa$ is supported on the space of the irreducible
representation $U_\kappa$, Schur's lemma implies that $P_\kappa$
is proportional to the identity on that space; and the
normalisation ensures that $P_\kappa$ is the projector onto
$U_\kappa$.

If now for all $U \in \U(l)$, $|\bra{\psi_N}\big(
(U\ket{\phi})^{\otimes Nk}\big)|^2 < \frac{1}{\dim U_\kappa}$,
then
\beastar 1&=&\tr P_\kappa \proj{\psi_N} \\
    &=& \dim U_\kappa \int_{U \in \U(d)} dU
\; \tr \proj{\psi_N} (U \proj{\phi}U^\dagger)^{\otimes Nk} \\
    &<& \dim U_\kappa \int_{U \in \U(d)} \frac{1}{\dim
U_\kappa}\\
    &=&1.
\eeastar
This is a contradiction, which leads to the conclusion that there
is a vector $\ket{\phi_N}\equiv U \ket{\phi}$ with
$|\bra{\psi_N}\left( \ket{\phi_N}^{\otimes Nk}\right)|^2 \geq
\frac{1}{\dim U_\kappa}.$ The dimension of representations of the
unitary group obeys the bound~(\ref{eq-semi-tableaux-bound})
\bestar \dim U_\kappa \leq (Nk+1)^{l(l-1)/2}=:p(Nk), \eestar
which is a polynomial in $Nk$. Combining these facts, this shows
that $\ket{\psi_N}$ and $\ket{\phi_N}^{\otimes Nk}$ have a
polynomial overlap. Translating this into the spectral estimation
of $Nk$ copies of $\ket{\phi_N}$, one sees that the triple $(N\mu,
N\nu, N\lambda)$ appears with at most polynomially decaying
probability,
\beastar \tr [P_{N \mu} \otimes P_{N \nu} \otimes P_{N \lambda}] \proj{\phi_N^{\otimes
Nk}} &\geq& \tr P_\kappa \proj{\phi_N^{\otimes Nk}} \\
&\geq & \tr \proj{\psi_N} \proj{\phi_N^{\otimes Nk}} \\
&\geq & \frac{1}{\dim U_\kappa} \\
&\geq & \frac{1}{p(Nk)}.\eeastar Similar statements are
straightforward for the reduced density operators, since $P_{N\mu}
\otimes P_{N\nu} \otimes P_{N\lambda} \subset P_{N\mu} \otimes \id
\otimes \id$ and likewise for systems $B$ and $C$:
\bea \label{eq-in-th-5-lower}
\tr P_{N\mu} (\rho_N^A)^{\otimes Nk}&\geq & \frac{1}{p(Nk)} \qquad \rho_N^A=\tr_{BC} \proj{\phi_N}\\
    \tr P_{N\nu} (\rho_N^B)^{\otimes Nk}&\geq &\frac{1}{p(Nk)} \qquad \rho_N^B=\tr_{AC} \proj{\phi_N}\\
   \label{eq-in-th-5-lower-2} \tr P_{N\lambda} (\rho_N^C)^{\otimes Nk}&\geq &\frac{1}{p(Nk)} \qquad \rho_N^C=\tr_{AB} \proj{\phi_N}.\eea

\noindent The crux of the argument is the comparison of the
estimates~(\ref{eq-in-th-5-lower})-(\ref{eq-in-th-5-lower-2}) with
the implications of the estimation theorem, theorem
\ref{theorem-Keyl-Werner}, when applied to $A$, $B$ and $C$,
respectively,
\beastar  \label{eq-in-th-5-upper}\tr P_{N\mu} (\rho_N^A)^{\otimes Nk} &\leq& p(Nk) \exp ( -
Nk D(\bar{\mu}||\spec \rho_N^A ))\\
\tr P_{N\nu} (\rho_N^B)^{\otimes Nk} &\leq &p(Nk) \exp ( -
Nk D(\bar{\nu}|| \spec \rho_N^B ))\\
 \label{eq-in-th-5-upper-2}\tr P_{N\lambda} (\rho_N^C)^{\otimes Nk} &\leq & p(Nk) \exp ( - Nk
D(\bar{\lambda}||\spec \rho_N^C )).\eeastar

\noindent The lower and upper bounds clash, since for all $N \geq
N_0$ there is an $\epsilon_N$ such that for all $\epsilon \geq
\epsilon_N$
$$ \exp (-Nk \epsilon) \geq \frac{1}{p(Nk)^2}$$
is violated. Thus for all $N \geq N_0$ the state $\ket{\phi_N}$
has
\beastar D(\bar{\mu}||\spec \rho_N^A) &\leq &\epsilon_N\\
     D( \bar{\nu}||\spec \rho_N^B) &\leq& \epsilon_N\\
     D(\bar{\lambda}||\spec \rho_N^C ) &\leq& \epsilon_N.
\eeastar
As $N$ increases, $\epsilon_N$ approaches $0$, hence the sequence
of states $\{\ket{\phi_{N}}\}$ obeys
\bea \label{eq-in-th-5-final}\lim_{N\rightarrow \infty} \spec \rho^A_N &=& \bar{\mu}\\
    \lim_{N\rightarrow \infty} \spec \rho^B_N &=& \bar{\nu}\\
   \label{eq-in-th-5-final-2} \lim_{N\rightarrow \infty} \spec \rho^C_N &=& \bar{\lambda}.\eea

\noindent The set of states in $\complex^m \otimes \complex^n
\otimes \complex^{mn}$ is compact. Therefore, the sequence
$\{\ket{\phi_N}\}_{N\geq N_0}$ contains a convergent subsequence
$\{\ket{\phi_{N_i}}\}_{i}$ with limiting state
$\ket{\varphi}:=\lim_{i \rightarrow \infty} \ket{\phi_{N_i}}$. The
limits~(\ref{eq-in-th-5-final})-(\ref{eq-in-th-5-final-2}) assert
that $\ket{\varphi}$ has spectra $(\bar{\mu}, \bar{\nu},
\bar{\lambda})$, which was set out to prove.
\end{proof}

\noindent Notice that in the above proof the use of the
exponential convergence of theorem~\ref{theorem-Keyl-Werner} is of
utmost importance. In contrast, for the proof of
theorem~\ref{ourtheorem} the statement of the much weaker
corollary~\ref{corollary-Keyl-2} sufficed.

Surprisingly, apart from partial results, some of which will be
discussed in subsection~\ref{subsec-spec-two-qubit}, little is
known about the general characteristics of the Kronecker
coefficient. Let us go back to the asymptotic character of the
presented equivalence and try to reach a better understanding. By
multiplying $\rho^{AB}$ in theorem~\ref{theorem-converse} by
$|\lambda|$, one can replace the rational spectra by integral
spectra and quantum states by positive operators. This leaves us
with a key question: given $\rho^{AB}$ with integral spectrum
$(\mu, \nu, \lambda)$, is $g_{\mu \nu \lambda} \neq 0$? A positive
answer would remove the asymptote of theorem~\ref{ourtheorem}
and~\ref{theorem-klyachko-our} entirely and establish a one-to-one
correspondence between triples of integral spectra and Kronecker
coefficients. The answer, however, is negative. To verify this,
reformulate the question with help of
theorem~\ref{theorem-klyachko-our}:
    $$ \mbox{Does } g_{N\mu, N\nu N\lambda} \neq 0 \mbox{ for some } N \in \naturals \mbox{ imply } g_{\mu
        \nu \lambda}\neq 0?
    $$
Then, take a glance at figure~\ref{figure-red-dots-black-dots}.
The lattice of red dots in the plot on the right hand side has two
holes, namely the point $(\mu, \nu, \lambda) \equiv \left((8,6),
(7,7), (8,4,2)\right)$ and, by symmetry, $\left((7,7),(8,6),
(8,4,2)\right)$; points which do not belong to $\KRON$.
Asymptotically such holes will be filled in by the semigroup
property. In this example, doubling the length will do and the
calculation
\beastar
(2\mu, 2\nu, 2\lambda)  &=&\left((16,12), (14,14), (16,8,4)\right)\\
                        &=&\left((7,7), (7,7), (8,4,2)\right)+\left((9,5), (7,7),
    (8,4,2)\right)
\eeastar
shows that $(2\mu, 2\nu, 2\lambda)$ must be in $\KRON$, since both
$$\left((7,7), (7,7), (8,4,2)\right) \mbox{ and } \left((9,5),
(7,7), (8,4,2)\right)$$ are (see
figure~\ref{figure-red-dots-black-dots}). In spite of the fact
that $g_{\mu \nu \lambda}=0$, $g_{2\mu, 2\nu, 2\lambda} \neq 0$
and the answer to the above question must be `no'. This contrasts
the case of the Littlewood-Richardson coefficients, where an
analogue question has been settled in the affirmative by Allen
Knutson and Terence Tao (see end of
subsection~\ref{subsec-spectra-horn},
page~\pageref{saturation-conjecture}). The precise understanding
of the non-asymptotic relation between $\KRON$ and the set of
positive operators remains a challenge which originates in this
work. Before concluding the section, two short and neat
corollaries can be drawn.

\begin{corollary}[Subadditivity\index{von Neumann entropy!subadditivity}] \label{corollary-subadditive} Von Neumann entropy is subadditive; i.e.~for all
$\rho^{AB}$, $S(\rho^{AB}) \le S(\rho^A) + S(\rho^B)$.\\
\end{corollary}
\begin{proof}
Theorem~\ref{ourtheorem} says that for every operator $\rho^{AB}$
there is a sequence of non-vanishing $g_{\lambda_j \mu_j \nu_j}$
with $\bar \lambda_j$, $\bar \mu_j$, $\bar \nu_j$ converging to
the spectra of $\rho^{AB}$, $\rho^A$ and $\rho^B$. By definition
of the Kronecker coefficient,
$$ V_{\lambda_j} \subset V_{\mu_j}\otimes V_{\nu_j},$$
and it is therefore clear that
\be \label{space2}
\dim V_{\lambda_j} \le \dim V_{\mu_j} \dim V_{\nu_j}.
\ee
For large $j$, Stirling's approximation and inequality
(\ref{eq-tableaux-bound}) imply that $\frac{1}{k}\log(\dim
V_{\lambda_j})$ tends to $S(\rho^{AB})$, where $k=|\lambda_j|$,
and similarly for systems $A$ and $B$. Inspection of
inequality~(\ref{space2}) concludes the proof. \end{proof}

\begin{corollary}[Triangle Inequality\index{von Neumann entropy!triangle inequality}]
\label{corollary-triangle} Von Neumann entropy obeys the triangle
inequality, i.e.~for all $\rho^{AB}$, $S(\rho^{AB}) \ge |S(\rho^A)
- S(\rho^B)|$.
\end{corollary}
\begin{proof} On page~\pageref{pageref-kron-symm} it was argued that $g_{\mu \nu \lambda}$ is symmetric under exchange of the
indices. In addition to inequality~(\ref{space2}) therefore the
cyclical permutations of $\mu, \nu$ and $\lambda$ produce two
extra inequalities,
\beastar \dim V_{\mu_j} &\le& \dim V_{\lambda_j} \dim
V_{\nu_j}\\
\dim V_{\nu_j} &\le& \dim V_{\lambda_j} \dim V_{\mu_j}.
\eeastar The triangle inequality then follows by applying
the reasoning in the proof of the preceding corollary.
\end{proof}

\noindent Note that the proof of the triangle inequality is very
different in spirit from the conventional one that applies
subadditivity to the purification of the state. Unfortunately,
this method does not directly extend to prove the more difficult
strong subadditivity of von Neumann entropy. Finding a
group-theoretic proof for strong subadditivity remains one of the
main challenges raised in this thesis. Success in this direction
could lead the way to a new understanding of entropy inequalities
and the discovery of new ones.

In the next subsection, an excursus to the classical realm of
random variables will be made. I will explain how random variables
are connected with sizes of cosets of finite groups and what the
classical analogues of the results presented in this section are.

\subsection{The Classical Analogue}
\label{subsection-classical-analogue}

\label{sec-group-intro} Recently, Terence H. Chan and Raymond W.
Yeung have discovered a remarkable connection between group theory
and random variables~\cite{ChaYeu02}. They show that there is a
one-to-one correspondence between inequalities of entropies of a
set of random variables and inequalities of orders of a set of
subgroups of a finite group. This result was one of the
motivations for the research presented in the previous
subsections. Here, I will review Chan and Yeung's result and
derive a classical analogue to the equivalence of
problem~\ref{prob-comp} and problem~\ref{prob-Kronecker}.

Let $X_1, \ldots X_n$ be a set of random variables and define the
joint distributions $X_\alpha=\bigcup_{i \in \alpha} X_i$ for a
subset $\alpha \subset \cN=\{1, \ldots, n\}$. Further let $\Omega$
be the set of subsets of $\cN$. An \emph{information inequality}
is an inequality of the form
    \be \label{eq-random-1}
        \sum_{\alpha \in \Omega }
        \beta_\alpha H(X_\alpha) \geq 0
    \ee
where $\beta_\alpha$ are real coefficients and $H(X_\alpha)$
denotes the Shannon entropy of the joint distribution $X_\alpha$.
An example of such an inequality is strong subadditivity of
Shannon entropy
    \be \label{eq-random-ex}
        H(X_1X_3)+H(X_2X_3)-H(X_3)-H(X_1X_2X_3) \geq 0.
    \ee
Let $G$ be a group with subgroups $G_i$ and define for $\alpha \in
\Omega$ the intersection $G_\alpha=\bigcap_{i \in \alpha} G_i$. A
\emph{group inequality} is an inequality that relates the orders
of the subgroups, here denoted by $|G_\alpha|$,
\be \label{equation-group-ineq} \prod_{\alpha \in \Omega}
|G_\alpha|^{\beta_\alpha} \leq 1\ee for real numbers
$\beta_\alpha$. An example is the following inequality
\be \label{eq-group-ex}
\frac{|G_{13}||G_{23}|}{|G_{123}||G_3|}\leq 1. \ee The surprising
fact is inequalities~(\ref{eq-random-ex}) and~(\ref{eq-group-ex}),
and indeed both sets of inequalities (\ref{eq-random-1}
and~\ref{equation-group-ineq}), are equivalent. The formal analogy
can already be observed when the logarithm is taken on both sides
of the group inequality. Conceptually clearer, however, is the
introduction of the cosets of $G_I$ in $G$,
$$C_I:=\{ g G_I| g \in
G\},$$ which are sets of size $|G|/|G_I|$.
Inequality~(\ref{equation-group-ineq}) is then equivalent to
$$  \sum_{\alpha \in \Omega} \beta_\alpha \log |C_\alpha| \geq 0.$$
Chan and Yeung's equivalence is stated in terms of entropy
functions: $h=(h_\alpha)_{\alpha \in \Omega}$ is an \emph{entropy
function\index{entropy!function}} if there exist random variables
$X_1 \ldots X_n$ such that $h_\alpha = H(X_\alpha)$ for all
$\alpha$. Further they say that $g=(g_\alpha)_{\alpha \in \Omega}$
is \emph{group characterisable\index{group characterisable}} if
there exists a group $G$ and subgroups $G_i$ such that $g_\alpha
=\log C_\alpha$ for cosets $C_\alpha$ of $G_\alpha$ in $G$.

\begin{theorem}[{Chan and Yeung~\cite{ChaYeu02}}]
\label{theorem-chan} If $h$ is group characterisable then it is an
entropy function. Conversely, for any entropy function $h$ there
exists a sequence of group characterisable functions $f_j$ with
$$\lim_{j \rightarrow \infty} \frac{f_j}{j}=h.$$
\end{theorem}

\noindent This theorem establishes a firm connection between group
inequalities and entropy inequalities. I will now explain the
classical analogue to theorems~\ref{ourtheorem}
and~\ref{theorem-converse}.

The analogue of the $k$-fold product of $\complex^d$ will be
played by strings of length $k$ with alphabet $\cA=\{1, \ldots,
d\}$, $\cA^{\times k}$. On $\cA^{\times k}$ the symmetric group
$S_k$ permutes the symbols and the symmetric group $S_d$ permutes
the letters in the alphabet. Let $x \in \cA^{\times k}$ be a
string with frequency $f=(f_1, \ldots, f_d)$, i.e.~symbol $i$
occurs $f_i$ times in $x$, and decompose
$$ \cA^{\times k} = \bigsqcup_f \cS_f$$
where $\cS_f$ denotes the set of strings with frequency $f$ and
$\sqcup$ the disjoint union. Every frequency $f$ can be described
as a pair of a partition $\lambda$ (i.e.~$\lambda \vdash k$) with
$\lambda_{\pi(i)}=f_i$ and the permutation $\pi \in S_d$. This
way, the set of strings with fixed $\lambda$ assumes a product
structure
\be \label{eq-sequence-decomp} \cA^{\times k}=\bigsqcup_{f}
\cS_f=\bigsqcup_{\lambda \vdash (k, d)} \cR_\lambda \times
\cS_\lambda,
\ee
where $\cR_\lambda$ is the set of frequencies $f$ with partition
$\lambda$ ($f=\pi(\lambda)$). One can regard this as the analogue
to Schur-Weyl duality, where the set $\cR_\lambda$ is the analogue
of an irreducible representation $U_\lambda$ of $\U(d)$ and
$\cS_\lambda$ is the analogue of a representation $V_\lambda$ of
the $S_k$. It is easy to compute the size of these sets:
\beastar d^k=\sum_{\lambda \vdash (k, d)} |\cR_\lambda| \times |\cS_\lambda|=
\sum_{\lambda \vdash (k, d)} \binom{k}{\lambda_1 \cdots \lambda_d}
\binom{d}{l_1 \cdots l_m}
\eeastar
where $m$ is the number of different lengths $\lambda_i$ and $l_j$
is the number of times the length $\lambda_j$ appeared in
$\lambda$.\footnote{At this point one could view the $\lambda$ and
$l \equiv l(\lambda)=(l_1, \ldots, l_m)$ as a pair of Young frames
and also introduce pairs of Young tableaux corresponding to
strings in $\cA^{\times k}$. A similar but more involved
correspondence between a string (word) and two Young tableaux is
known as the Robinson-Schensted
correspondence\index{Robinson-Schensted correspondence} and
closely related to the combinatorics of Young diagrams and to
Schur-Weyl duality~\cite{Fulton97}.}

The analogue of the tensor product of two representations of the
symmetric group is given by the following product defined for two
sets $\cS_f$ and $\cS_g$ by
\be
\cS_f \ast \cS_g=\{ (x_1y_1, \ldots, x_ky_k)| x \in \cS_g, y \in
\cS_g\}.
\ee
This product can also be looked upon as the direct product $\cS_f
\times \cS_g$ followed by restriction of the action of the
symmetric group to permute the symbols in both strings
simultaneously.\footnote{The direct product $\times$ plays the
role of the product $\boxtimes$ of two representations, whereas
$\ast$ is the analogue of the tensor product representations
indicated by $\otimes$ (see Preliminaries,
page~\pageref{sec-intro-clebsch}).} As a next step, the strings in
the set $\cS_f \ast \cS_g$ are sorted according to their frequency
distribution.

To see how it works, pick one string from $\cS_\mu$ and one from
$\cS_\nu$ then pair the $i$'th letters. The symmetric group acts
by permuting the pairs. This way one can identify the different
frequency distributions. It is of course also possible that there
are several different (i.e.~not connected through a permutation)
ways of obtaining the same frequency distribution over the
alphabet $\cA\times \cA$ of size $d^2$. This is illustrated in the
following example:

\begin{example}
Let $\mu=(3,3)$ and $\nu=(4,2)$. The set of strings $\cS_\mu \ast
\cS_\nu$ falls into sets of strings corresponding to two
partitions: $\lambda_1=(3,2,1)$ and $\lambda_2=(2,2,1,1)$. Pick
the strings $(a, a, a, b, b, b) \in \cS_\mu$ and $(c, c, c, c, d,
d) \in \cS_\nu$ and write them in different ways underneath each
other. Two different frequencies give rise to the first
possibility, $\lambda_1$;
\beastar
a \ a \ a \ b \ b \ b  \qquad a \ a \ a \ b \ b \ b \\
c \ d \ d \ c \ c \ c  \qquad c \ c \ c \ c \ d \ d
\eeastar
and one possibility to the second, $\lambda_2$:
\beastar
a \ a \ a \ b \ b \ b \\
c \ c \ d \ c \ c \ d.
\eeastar
In summary
\bea \cS_\mu \ast \cS_\nu&=& \{
\pi_1, \pi_2\} \times S_{\lambda_1} \sqcup  \{\pi_3\}\times S_{\lambda_2} \\
&=& 2 S_{\lambda_1}\sqcup S_{\lambda_2},\eea where $\pi_1, \pi_2$
and $\pi_3$ permute the partition into the correct alphabet, and
the second line only indicates the number of frequencies leading
to the same partition (multiplicity of a partition). Below we will
formally introduce the coefficient determining these
multiplicities, which in this case is $h_{\mu \nu}^{ \lambda_1}=2$
and $h_{\mu \nu}^{\lambda_2}=1$. Finally, a check of set sizes
gives
$$ \frac{6!}{3!3!}\cdot \frac{6!}{4!2!} =2 \frac{6!}{3! 2! 1!} +1
\frac{6!}{2!2!1!1!}$$ which is true.
\end{example}

In general, the classical analogue of the Clebsch-Gordan
decomposition of the symmetric group is the decomposition of the
product of strings:
$$\cS_\mu \ast \cS_\nu =\sum_\lambda h_{\mu \nu}^\lambda \
\cS_\lambda$$ for some non-negative integers $h_{\mu
\nu}^\lambda$, where the sum is over all $\lambda$ with
$|\lambda|=k$. The coefficient denotes the number of different
product alphabets that lead to the same diagram $\lambda$.

By looking at the set $\cS_\lambda$ as the set of cosets of the
group $S_{\lambda_1}\times S_{\lambda_2} \times \cdots \times
\cS_{\lambda_d}$ in $S_k$, one finds the following corollary:

\begin{corollary} \label{dim-cor}
$$ \frac{|G|}{|G_\mu|}\cdot \frac{|G|}{|G_\nu|}= \sum_\lambda
h_{\mu \nu}^{ \lambda} \frac{|G|}{|G_\lambda|}, $$ where $G\equiv
S_k$ and $G_\lambda \equiv S_{\lambda_1}\times S_{\lambda_2}
\times \cdots \times \cS_{\lambda_d}$.
\end{corollary}

Before we move on, let us pause and discuss some straightforward
properties of $h_{\mu \nu}^{\lambda}$. $h_{\mu \nu}^\lambda$ is
symmetric with respect to interchange of $\mu$ and $\nu$, however
it is not symmetric with interchange of $\mu$ and $\lambda$. This
contrasts with its analogue, the Kronecker coefficient of the
symmetric group, which is symmetric under interchange of all three
indices. Another interesting property is that the nonzero
coefficients form a semigroup, both classically and
quantum-mechanically. In the classical case this is
straightforward to verify, whereas the quantum case requires
somewhat more effort (see section~\ref{sec-semigroup}).

\begin{theorem}[Semigroup Property]
\label{theorem-classical-semigroup} The triples $(\mu, \nu,
\lambda)$ with nonzero $h_{\mu \nu}^{ \lambda}$ form a semigroup
with respect to row-wise addition, i.e.~for $h_{\mu \nu}^{
\lambda}\neq 0$ and $h_{\mu' \nu'}^{ \lambda'}\neq 0$ we have
$h_{\mu+\mu', \nu+\nu'}^{\lambda+\lambda'}\neq 0$.
\end{theorem}

\begin{proof}
To $(\mu, \nu, \lambda)$ as well as $(\mu', \nu', \lambda')$ write
down corresponding triple of strings as done in the examples and
pair the $i$'th triples. The resulting triple of string is
representative for $(\mu+\mu', \nu+\nu', \lambda+\lambda')$.
\end{proof}

\noindent The following theorem is the analogue to
theorems~\ref{ourtheorem} and~\ref{theorem-converse}.

\begin{theorem} \label{theorem-corre}
If there is a pair of random variables $X_1X_2$ with (rational)
distributions $(\mu, \nu, \lambda)$, then there is a natural
number $m$ such that $h_{m \mu, m \nu}^{ m \lambda}\neq 0$.
Conversely, if $h_{\mu \nu}^\lambda\neq0$, there is a pair of r.v.
$X_1X_2$ with distributions $(\bar{\mu}, \bar{\nu},
\bar{\lambda})$. (The statement extends to irrational
distributions by taking the appropriate limits.)
\end{theorem}

\begin{proof}
Since the distributions are rational we can construct a string of
length $m$ (for some $m$) such that the frequency distribution of
the string equals the probability distribution of $X_1X_2$. This
string defines $\cS_{m \lambda}$ and its marginals define $\cS_{m
\mu}$ and $\cS_{m \nu}$. Since the product $\cS_{m\mu}\ast \cS_{m
\nu}$ contains all possible product distributions, also
$\cS_{m\lambda}$ must be contained in $\cS_{m\mu}\ast \cS_{m
\nu}$.

Conversely, if $h_{\mu \nu}^\lambda\neq 0$, simply build the
string corresponding to $\lambda, \mu, \nu$ and take the frequency
distribution of the product alphabet as the definition of the
distribution of $X_1X_2$. Note that this way, all possible
frequency distributions can be constructed.
\end{proof}

\noindent A direct consequence of theorem~\ref{theorem-corre},
corollary~\ref{dim-cor} and the fact that $|G|/|G_\mu| \approx
2^{k H(\bar{\mu})}$ for large $k$ is subadditivity of Shannon
entropy. More notable, however, is that this analysis was carried
out on a level of probability distributions, rather than
entropies.

\subsection{Intermezzo} \label{section-intermezzo} The
starting point for the research presented in this section was Chan
and Yeung's paper on the connection between Shannon entropy
inequalities and group inequalities~\cite{ChaYeu02} (see
subsection~\ref{subsection-classical-analogue}). The initial goal
was to find a quantum analogue of their fundamental classical
result. It soon became clear that the sizes of group
representations should play the role of the sizes of the cosets in
Chan and Yeung's construction, but a direct translation seemed
difficult. A first link between quantum states and representations
was provided by the estimation theorem,
theorem~\ref{theorem-Keyl-Werner}~\cite{AlRuSa87, KeyWer01PRA}.
This link, however, worked on the level of spectra of quantum
states rather than on the level of their entropies; this shifted
the topic of this work from entropies to the spectra themselves.
With focus on the bipartite case, Graeme Mitchison and I began a
search for literature on the local symmetry groups $\SU(m)\times
\SU(n)$ embedded in the global symmetry group $\SU(mn)$ which led
us into the literature of particle physics. Here, in the mid-1960s
Murray Gell-Mann's and Yuval Ne'eman's eightfold
way~\cite{Gell-Mann61,Gell-Mann62,Neeman61} had motivated a study
of these groups by C.~Richard Hagen and Alan MacFarlane as well as
Claude Itzykson and Michael Nauenberg~\cite{HagMac65a, ItzNau66},
which circled around the decomposition
    $$
        U_\lambda \downarrow^{\SU(mn)}_{\SU(m)\times
        \SU(n)} \cong \bigoplus_{\mu \nu} g_{\mu \nu \lambda} U_\mu \boxtimes
        U_\nu.
    $$
The plots presented in figure~\ref{figure-red-dots-black-dots}
contain the first data that supported our conjecture of a
connection between admissible spectral triple and nonvanishing
Kronecker coefficients.

The results in this section have been obtained in the framework of
quantum information theory and have relied almost exclusively on
standard textbook material from the group theory of the unitary
and symmetric groups. The connection that has been established
between problem~\ref{prob-comp} and problem~\ref{prob-Kronecker}
might be viewed as surprising. More astonishing, however, is the
basic nature of the proof of this connection. Klyachko's results,
which parallel theorems~\ref{ourtheorem} \&~\ref{theorem-converse}
closely, have entirely different proofs. Furthermore, these proofs
are cast in the framework of geometric invariant theory and
require advanced knowledge of the subject. Geometric invariant
theory is a very powerful tool and besides the stated results, it
allowed Klyachko to come up with an algorithm to calculate the
inequalities which describe the polytope of solutions to
problem~\ref{prob-comp}.

In 1998, Klyachko used the same method to give a set of
inequalities describing the polytope of solutions to Horn's
problem, problem~\ref{prob-Horn}~\cite{Klyachko98}. In this paper
he also gave a detailed account of the connection between spectra
of sums of Hermitian operators and the Littlewood-Richardson
coefficients. This connection was first stated by
B.~V.~Lidskii~\cite{Lidskii82} (see also
G.~J.~Heckman~\cite{Heckman82}). The interested reader should
consult Knutson's excellent account Horn's problem, in which he
places the problem in the wider context of symplectic geometry and
geometric invariant theory~\cite{Knutson00}. The paper can
therefore serve as a \emph{Leitfaden} to the problem
pair~\ref{prob-comp}~and~\ref{prob-Kronecker} as well, whereby one
`replaces' the groups that are involved (see
table~\ref{table-prob-comparison}); i.e.~one considers the
inclusion of $\SU(m)\times \SU(n)$ in $\SU(mn)$ rather than the
diagonal action of $\U(d)$ in $\U(d)\times \U(d)$. This will be
done explicitly in subsection~\ref{subsec-spec-convexity} in order
to give an alternative proof of the fact that the solution to
problem~\ref{prob-comp} is a convex polytope. With this
understanding of the context in mind, let us return to quantum
information theory and show how straightforward a proof of the
asymptotic equivalence of
problems~\ref{prob-Horn}~and~\ref{prob-LR} can be.
\begin{table}\label{table-prob-comparison}
\begin{tabular}{cll}
Problems                                  &  Groups                          & Action\\
\ref{prob-comp} \&~\ref{prob-Kronecker}   & $\U(d) \subset \U(d)\times \U(d)$  & diagonal: $U \mapsto U\otimes U$\\
\ref{prob-Horn} \&~\ref{prob-LR}          & $\SU(m)\times \SU(n)
\subset \SU(mn)$& inclusion: $(U^A, U^B) \mapsto U^A \otimes U^B$
\end{tabular}
\caption{The problems, the groups and their actions.}
\end{table}

\subsection{Problem~\ref{prob-Horn} vs. Problem~\ref{prob-LR}}
\label{subsec-spectra-horn}

At the end of the previous section, a few words were said about
the history of the connection between Horn's problem and the
Littlewood-Richardson coefficients (problem~\ref{prob-Horn}
and~\ref{prob-LR}). This section provides a novel and compact
proof of this result and provides an analogy to
subsection~\ref{subsec-spec-comp} on Kronecker coefficients and
the compatibility of local spectra. As in
subsection~\ref{subsec-spec-comp}, the proofs here are also based
on the estimation theorem for spectra of quantum states
(theorem~\ref{theorem-Keyl-Werner}). The presentation will start
with theorem~\ref{theorem-horn-lidskii-Klyachko} and its proof:
the construction of a sequence of nonvanishing
Littlewood-Richardson coefficients whose normalised index triple
converges to the spectra of two operators and their weighted sum.
A proof of the converse, theorem~\ref{theorem-Horn-converse},
follows. Starting from a nonzero Littlewood-Richardson coefficient
three density operators $A$, $B$ and $C=pA+(1-p)B$ will be
constructed such that the normalised index triple equals the
spectral triple.

\begin{theorem}\label{theorem-horn-lidskii-Klyachko}
For all density operators $A$, $B$ and $C=pA+(1-p)B$ on
$\complex^d$ with spectra $\mu, \nu$ and $\lambda$ and $p \in
[0,1]$, there is a sequence $\mu_j, \nu_j$ and $\lambda_j$, such
that
\bestar c_{\mu_j  \nu_j}^{\lambda_j} \neq 0 \eestar
and
\beastar
\lim_{j \to \infty} \bar{\mu}_j&=&\spec A\\
\lim_{j \to \infty} \bar{\nu}_j&=&\spec B\\
\lim_{j \to \infty} \bar{\lambda}_j&=&\spec C.
\eeastar
\end{theorem}

\begin{proof} Let $r^X=\spec X$ for $X \in \{A,
B, C\}$. For all $\epsilon>0$, according to corollary
\ref{corollary-Keyl-2}, there is an $n_0\equiv n_0(\epsilon)$ such
that for all $n\geq n_0$ and all $k$ with $k/n \in
\cB^\epsilon(p)$
\bea
\label{eq-horn-1} \tr P^k_X A^{\otimes k} &\geq& 1-\epsilon,
\qquad P^k_X:=\sum_{\mu: \bar{\mu} \in \cB^\epsilon(r^A)} \tr P^{k}_\mu\\
\label{eq-horn-2} \tr P^{n-k}_Y B^{\otimes
(n-k)}  &\geq& 1-\epsilon, \qquad P^n_Y:=\sum_{\nu: \bar{\nu} \in \cB^\epsilon(r^B)} \tr P^{n-k}_\nu\\
\label{eq-horn-3} \tr P^n_Z C^{\otimes n}  &\geq& 1-\epsilon,
\qquad P^n_Z:=\sum_{\lambda: \bar{\lambda} \in \cB^\epsilon(r^C)}
\tr P^{n}_\lambda
\eea
hold simultaneously. $P^n_\lambda$ (and similarly for $k, \mu$ and
$n-k, \nu$) is the projector onto $U_\lambda \otimes V_\lambda$ of
the Schur-Weyl duality induced decomposition of
$(\mathbb{C}^d)^{\otimes n}$ (see
theorem~\ref{theorem-Schur-duality}). The number of factors is
indicated in the superscript to make it easier keep to track of
them. It is also convenient to choose $n$ large enough, such that
\be \label{eq-horn-4} (n+1) e^{-n \ \min_{k/n \notin \cB^{\epsilon}(p)} D(k/n||p)} \leq
\epsilon \ee also holds. This is possible, because Pinsker's
inequality $D(k/n||p) \geq \frac{2}{\ln 2} \delta(k/n, p)^2$
(lemma~\ref{lemma-pinsker}) implies
$$
(n+1) e^{-n \min_{k/n \notin \cB^{\epsilon}(p)} D(k/n||p)} \leq
(n+1) e^{-n \frac{2}{\ln 2} \epsilon^2}.
$$
Denote the LHS of~(\ref{eq-horn-3}) by $R$ and express it as
\beastar
 R=\tr (pA+(1-p)B)^{\otimes n}P^n_Z.
\eeastar
The binomial expansion gives
\be \label{eq-horn-5}
R=\sum_k \binom{n}{k} \tr p^k (1-p)^{n-k} [A^{\otimes k} \otimes
B^{\otimes (n-k)}]P^n_Z.
\ee
Here, the $A$'s can be sorted to the left and the $B$'s to the
right using the fact that $P^n_\lambda$ is invariant under
permutation since $U_\lambda \otimes V_\lambda$ is a
representation of $S_n$. Then use the relative entropy to bound
the multinomial distribution
\bea
R &\leq & \sum_k  e^{-n D(k/n||p)} \tr [A^{\otimes k} \otimes
B^{\otimes (n-k)}] P^n_Z,
\eea
separate the untypical factors and apply
inequality~(\ref{eq-horn-4}):
\bea
\nonumber R&\leq & (n+1) e^{-n \ \min_{k: k/n  \notin
\cB^{\epsilon}(p)}
D(k/n||p)}\\
\nonumber &&+ \sum_{k: k/n \in \cB^{\epsilon}(p)}\tr [A^{\otimes
k} \otimes B^{\otimes (n-k)}]
P^n_Z\\
\label{eq-horn-8} R &\leq & \epsilon+ \sum_{k: k/n \in
\cB^{\epsilon}(p)}\tr [A^{\otimes k} \otimes B^{\otimes (n-k)}]
P^n_Z.
\eea
Let $\sum_{\mu \vdash (k, d)} P_\mu=\id_{(\complex^d)^{\otimes
k}}$ be the decomposition of the first $k$ factors according to
Schur-Weyl duality and observe that $\sum_{\mu \vdash (k, d)}
P_\mu A^{\otimes k} P_\mu=A^{\otimes k}$ (and likewise for
$B^{\otimes (n-k)}$):
\be \quad R \leq  \epsilon +\sum_{k: k/n \in \cB^{\epsilon}(p)}\tr \sum_{\substack{\mu \vdash (k, d),\\ \nu \vdash (n-k, d)}}
[P^k_\mu \otimes P^{n-k}_\nu] [A^{\otimes k} \otimes B^{\otimes
(n-k)}][ P^k_\mu \otimes P^{n-k}_\nu] P^n_Z.
\ee
Divide the $\mu$-summation into the $\epsilon$-ball around $\spec
A$ and its complement and estimate the complement
with~(\ref{eq-horn-1}) (and likewise for the remaining $n-k$
factors around $\spec B$ with~(\ref{eq-horn-2})),
\be \label{eq-horn-10}
\leq 3 \epsilon+\sum_{k: k/n \in \cB^\epsilon(p)}  \tr [P^k_X
\otimes P^{n-k}_Y][A^{\otimes k} \otimes B^{\otimes (n-k)}][ P^k_X
\otimes P^{n-k}_Y ]P^n_Z.
\ee
Keeping the estimate~(\ref{eq-horn-3}) in mind, the right term
of~(\ref{eq-horn-10}) is bounded away from zero for all $\epsilon<
\frac{1}{4}$. Hence for such $\epsilon$, there is a triple $(\mu,
\nu, \lambda)$ with $k=|\mu|$ such that
\bestar k/n \in
\cB^{\epsilon}(p), \bar{\lambda} \in \cB^\epsilon(r^{C}),
\bar{\mu} \in \cB^\epsilon(r^A) \mbox{ and } \bar{\nu} \in
\cB^\epsilon(r^B)
\eestar
and
\be \label{eq-horn-11} [P^k_\mu \otimes P^{n-k}_\nu] P^n_{\lambda} \neq 0.\ee
The last step is to invoke decomposition~(\ref{eq-littlewood})
$$U_\mu \otimes U_\nu \cong \bigoplus_{\lambda'} c_{\mu
\nu}^{\lambda '} U_{\lambda'},$$ where we have dropped the
superscript. This decomposition transforms~(\ref{eq-horn-11}) into
    \beastar
            [P_{U_\mu} \otimes P_{U_\nu} \otimes P_{V_\mu} \otimes
            P_{V_\nu}]&\cdot & [ P_{U_\lambda} \otimes P_{V_\lambda}] \\
        &&= [\big( \sum_{\lambda'} \sum_{i=1}^{c_{\mu \nu}^{\lambda '}}
            P_{U^i_{\lambda'}} \big)\otimes P_{V_\mu} \otimes P_{V_\nu}][
            P_{U_\lambda} \otimes P_{V_\lambda} ]\\
        &&= [\sum_{\lambda'} \big( P_{U_{\lambda'}}\otimes
            \sum_{i=1}^{c_{\mu \nu}^{\lambda '}} (P_{V^i_\mu} \otimes
            P_{V^i_\nu}) \big)][
            P_{U_\lambda} \otimes P_{V_\lambda} ]\\
        &&= P_{U_{\lambda}} \otimes \big( \sum_{i=1}^{c_{\mu
            \nu}^{\lambda}} P_{V^i_\mu} \otimes P_{V^i_\nu}\big) .
\eeastar
Since this projector cannot vanish, $c_{\mu \nu}^\lambda \neq 0$
must hold. This concludes the proof of the theorem, because
$\epsilon>0$ is arbitrary.
\end{proof}

\noindent The next theorem completes the asymptotic equivalence
and proves the analogue to theorem~\ref{theorem-converse}.

\begin{theorem}\label{theorem-Horn-converse}
If $c_{\mu \nu}^\lambda \neq 0$, there exist quantum states $A$
and $B$ such that
\beastar
\spec  A &=&\bar{\mu}\\
\spec  B &=& \bar{\nu}\\
\spec C &=& \bar{\lambda},
\eeastar
where $p=\frac{|\mu|}{|\lambda|}$ and $C=pA+(1-p)B$.
\end{theorem}

\begin{proof}
Let $d$ be the number of rows of $\lambda$. Theorem
\ref{theorem-stability} asserts that $c_{\mu \nu}^\lambda \neq 0$
implies $c_{N\mu N\nu}^{N\lambda} \neq 0$ for all $N$. For every
$N$, a triple of density matrices $A_N$, $B_N$ and $C_N$ will be
constructed and it will be shown that their limits, as $N$
approaches infinity, defines operators that satisfy the claim of
the theorem.

Fix $N$ and set $n=N|\lambda|$ as well as $k=pn=N|\mu|$. By the
invariant-theoretic characterisation of the Littlewood-Richardson
coefficient, equation~(\ref{eq-LR-invariant}),
$$c_{\mu \nu}^\lambda =\dim (V_{N\mu} \otimes V_{N\nu} \otimes
V_{N\lambda})^{S_{k}\times S_{n-k}},$$
where $S_k$ acts on
$V_{N\mu}$, $S_{n-k}$ on $V_{N\nu}$ and $S_k \times S_{n-k}\subset
S_n$ on $V_{N\lambda}$. Now pick a nonzero $\ket{\Psi_N} \in
(V_{N\mu} \otimes V_{N\nu} \otimes V_{N\lambda})^{S_{k}\times
S_{n-k}}$. Consider
\be \label{eq-horn-conv-5}\begin{split} &\cH^{(1)}\otimes \cdots \otimes \cH^{(k)} \otimes
\cH^{(k+1)}\otimes \cdots \otimes \cH^{(n)} \\
&\otimes \cK^{(1)}\otimes \cdots \otimes \cK^{(k)} \otimes
\cK^{(k+1)}\otimes \cdots \otimes \cK^{(n)}, \end{split} \ee
 where $\cH^{(i)}$ and $\cK^{(j)}$ are isomorphic to
 $\complex^d$. Embed the representation $V_{N\mu}$ in $\cH^{(1)}\otimes \cdots \otimes
 \cH^{(k)}$, $V_{N\nu}$ in $\cH^{(k+1)}\otimes \cdots \otimes
 \cH^{(n)}$ and $V_{N\lambda}$ in $\cK^{(1)}\otimes \cdots \otimes
 \cK^{(n)}$. The symmetric group $S_n$ permutes the pairs $\cH^{(i)}\otimes \cK^{(i)}\cong
 \complex^{d^2}$ and its subgroup $S_k\times S_{n-k}$ permutes the first $k$ and the last $n-k$ pairs separately.

As mentioned in the Preliminaries
(page~\pageref{pageref-tensor-product-reps}), an irreducible
representation of the group $S_k\times S_{n-k}$ is isomorphic to a
tensor product of irreducible representations of $S_k$ and
$S_{n-k}$. $\ket{\Psi_N}$ is a trivial representation of
$S_k\times S_{n-k}$ and can therefore only be isomorphic to the
tensor product $V_k\otimes V_{n-k}$ of the trivial representations
$V_k \equiv V_{(k, 0, \ldots, 0)}$ of $S_k$ and $V_k \equiv
V_{(n-k, 0, \ldots, 0)}$ of $S_{n-k}$. On the first $k$ pairs the
$k$-fold tensor product of $U \in \U(d^2)$ commutes with the
action of $S_k$, and on the remaining pairs it is the $n-k$-fold
tensor product of $U\in \U(d^2)$ which commutes with $S_{n-k}$.
Schur-Weyl duality decomposes the space in~(\ref{eq-horn-conv-5})
into
$$ \bigoplus_{\substack{\mu \vdash (k, d^2)\\ \nu \vdash (n-k, d^2)}} U^{d^2}_\mu \otimes V_\mu \otimes U^{d^2}_\nu \otimes V_\nu,$$ so
that
$$ \ket{\Psi_N} \in U^{d^2}_k \otimes V_k \otimes U^{d^2}_{n-k} \otimes V_{n-k} \cong  U^{d^2}_k \otimes U^{d^2}_{n-k}.$$
The isomorphism stems from the triviality of $V_k$ and $V_{n-k}$.
By lemma~\ref{lemma-symmetric}, a basis for $U^{d^2}_k\otimes
U^{d^2}_{n-k}$ is given by $\ket{v}^{\otimes k}\otimes
\ket{w}^{\otimes (n-k)}$. Now comes the key step, namely a proof
of existence for vectors $\ket{\phi_N}$ and $\ket{\psi_N}$, both
in $\complex^{d^2}$, which satisfy
\be \label{eq-horn-converse-1} |\bra{\Psi_N} \left(\ket{\phi_N}^{\otimes
k}\otimes \ket{\psi_N}^{\otimes (n-k)}\right)|^2 \geq
\frac{1}{\dim U^{d^2}_k \dim U^{d^2}_{n-k}}.\ee

\noindent This will be done by contradiction. Let $P^{d^2}_k
\otimes P^{d^2}_{n-k}$ be the projector onto $U^{d^2}_k\otimes
U^{d^2}_{n-k}$ and invoke Schur's lemma to write this projector in
integral form
    \bestar
    \begin{split}
    P^{d^2}_k \otimes P^{d^2}_{n-k} = &\dim U^{d^2}_k \dim U^{d^2}_{n-k} \\ & \times \int_{U, V \in \U(d^2)} dUdV[ U \proj{v}
    U^\dagger]^{\otimes k} \otimes [ V \proj{w} U^\dagger]^{\otimes
    (n-k)},
    \end{split}
    \eestar
where $dU$ and $dV$ are both Haar measures on $\U(d^2)$ with
normalisation $\int_U dU =1$ and $\int_V dV =1$. Assume by
contradiction that for all states $\ket{v}, \ket{w} \in
\complex^{d^2}$, inequality~(\ref{eq-horn-converse-1}) is
violated, i.e.
    $$
        |\bra{\Psi_N}\left(\ket{v}^{\otimes k}\otimes
        \ket{w}^{\otimes (n-k)}\right)|^2 < \frac{1}{\dim U^{d^2}_k \dim U^{d^2}_{n-k}}.
    $$
and estimate
\beastar 1&=&\tr [P^{d^2}_k \otimes P^{d^2}_{n-k} ] \proj{\Psi_N} \\
    &=& \dim U^{d^2}_k \dim U^{d^2}_{n-k} \\
    &&\times \int_{U, V \in \U(d)} dUdV \tr \proj{\Psi_N} [U \proj{v}
U^\dagger]^{\otimes k} \otimes [V \proj{w} V^\dagger]^{\otimes (n-k)} \\
    &<& \dim U^{d^2}_k \dim U^{d^2}_{n-k} \int \frac{1}{ \dim U^{d^2}_k \dim U^{d^2}_{n-k}}\\
    &=&1.
\eeastar
Since this estimation led to a contradiction there must exist
vectors $\ket{\phi_N}$ and $\ket{\psi_N}$ satisfying
inequality~(\ref{eq-horn-converse-1}). From these vectors, the
operators $A_N, B_N$ and $C_N$ will be constructed.

From inequality~(\ref{eq-semi-tableaux-bound}) we have the
following bound, which is polynomial in $n$:
\bestar \dim U^{d^2}_k \dim U^{d^2}_{n-k} \leq (n+1)^{d^2(d^2-1)} =:p(n).
\eestar
This shows that
\begin{align*}
     \tr [P_{N \mu} \otimes P_{N \nu} \otimes P_{N \lambda} ]& [\proj{\phi_N}^{\otimes
pn}  \otimes   \proj{\psi_N}^{\otimes (1-p)n} ] \\
        &\geq  \tr \proj{\Psi_N}[\proj{\phi_N}^{\otimes
pn}  \otimes  \proj{\psi_N}^{\otimes (1-p)n}]  \\
        &\geq  \frac{1}{\dim U_k \dim U_{n-k}} \\
        &\geq  \frac{1}{p(n)}, \end{align*}
since $\proj{\psi_N} \subset P_{N\mu} \otimes P_{N\nu} \otimes
P_{N\lambda}$. Recall that $\ket{\phi_N}$ is a vector on one of
the first $k$ pairs $\cH^{(i)}\otimes \cK^{(i)}$ and
$\ket{\psi_N}$ is a vector on one of the last $n-k$ pairs. Tracing
out over one part of a pair defines a density operator on the
other and leads to
\bea \label{eq-horn-converse-8}\tr P_{N\mu} A_N^{\otimes pn}\geq \frac{1}{p(n)}, &&\quad A_N=\tr_{\cK^{(1)}} \proj{\phi_N} \\
    \label{eq-horn-converse-9} \tr P_{N\nu} B_N^{\otimes (1-p)n}\geq \frac{1}{p(n)}, &&\quad B_N=\tr_{\cK^{(k+1)}} \proj{\psi_N}\\
    \label{eq-horn-converse-10} \tr P_{N\lambda} \; C_{AN}^{\otimes pn} \otimes C_{BN}^{\otimes (1-p)n}\geq \frac{1}{p(n)}, &&\quad C_{AN}=\tr_{\cH^{(1)}} \proj{\phi_N}\\
\nonumber  && \quad C_{BN}= \tr_{\cH^{(k+1)}} \proj{\psi_N}.
\eea
This concludes the construction of $A_N$ and $B_N$ and it remains
to discuss $C_N=pA_N+(1-p)B_N$. Binomial expansion, together with
the fact that $P_{N\lambda}$ is invariant under permutation,
allows for the estimate
\be \label{eq-horn-converse-11}
\begin{split} \tr P_{N\lambda} C_N^{\otimes n}
            &= \sum_i p^i (1-p)^{n-i} \binom{n}{i} \tr P_{N\lambda} [A_N^{\otimes i} \otimes B_N^{\otimes
            (n-i)}]\\
            &\geq \frac{1}{n+1} \tr P_{N\lambda} [A_N^{\otimes k} \otimes B_N^{\otimes
            (n-k)}]\geq \frac{1}{(n+1) p(n)}. \end{split}
\ee
The first inequality arises because the binomial distribution
takes a maximum at $k=pn$ and since there are only $n+1$ different
possibilities for $k$. The second inequality is a simple insertion
of the estimate~(\ref{eq-horn-converse-10}). The remainder of the
proof is now identical to the proof of
theorem~\ref{theorem-converse}: assume that the spectra of $A_N,
B_N$ and $C_N$ do not converge to $\mu, \nu$ and $\lambda$. Then
the probabilities
$$\tr P_{N\mu} A_N^{\otimes k},  \quad \tr P_{N\nu} B_N^{\otimes (n-k)}
\;\mbox{ and }\; \tr P_{N\lambda} C_N^{\otimes n}$$ decrease
exponentially according to the estimation theorem,
theorem~\ref{theorem-Keyl-Werner}. This, however, contradicts the
bounds~(\ref{eq-horn-converse-8}), (\ref{eq-horn-converse-9}) and
(\ref{eq-horn-converse-11}), which say that the decay can be at
most polynomial. It has therefore been shown that the spectra of
$A_N$, $B_N$ and $C_N$ converge to $\bar{\mu}$, $\bar{\nu}$ and
$\bar{\lambda}$ as desired. The limiting operators therefore
satisfy the claim of the theorem.
\end{proof}

\noindent A different proof for
theorem~\ref{theorem-horn-lidskii-Klyachko}
and~\ref{theorem-Horn-converse} can be obtained via the following
characterisation of the Littlewood-Richardson coefficient in terms
of representations of $\SU(m)\times SU(n)$ embedded in $\SU(m+n)$:
    \be \label{eq-seesaw-LR} U^{m+n}_\lambda \downarrow^{\SU(m+n)}_{\SU(m)\times \SU(n)}
         \cong \bigoplus_{\substack{k_1+k_2=|\lambda|\\ \mu \vdash (k_1, m) \\ \nu \vdash (k_2, n)}} c_{\mu \nu}^\lambda U^m_\mu \boxtimes
        U^n_\nu.
    \ee
In contrast to the proof given, which regards the density matrices
$A$, $B$ and $C$ as operators on $\complex^d$, an argument using
equation~(\ref{eq-seesaw-LR}) makes more efficient use of the
dimensions. Here, $A$ is embedded into $\complex^m$ and $B$ into
$\complex^n$ for $m$ and $n$ the rank of $A$ and $B$,
respectively.

The fact that the coefficients in equation~(\ref{eq-seesaw-LR})
are Littlewood-Richardson coefficients is a consequence of the
groups $G\times G'$ and $K \times K'$, where
\begin{center}
\begin{tabular}{ll}
$ G=\GL(d, \complex)\times \GL(d, \complex)$&$G'=\GL(k,
\complex)\times \GL(m, \complex)$\\
$ K=\GL(d, \complex)$&$ K'=\GL(m+n, \complex),$
\end{tabular}
\end{center}
being a so-called seesaw pair. A detailed definition and clear
exposition of seesaw pairs can be found in~\cite[chapter
9.2]{GooWal98}.

The characterisation of $c_{\mu \nu}^\lambda$ in terms of
equation~(\ref{eq-seesaw-LR}) also relates back to the physics
literature and the work by Hagen and MacFarlane, who studied not
only the subgroup reduction $\SU(m)\times \SU(n) \subset
\SU(mn)$~\cite{HagMac65a}, but also in a second paper the
reduction $\SU(m)\times \SU(n) \subset \SU(m+n)$,
i.e.~equation~(\ref{eq-seesaw-LR}), however, without mentioning
the connection to Littlewood-Richardson
coefficients~\cite{HagMac65b}. Proof of
theorems~\ref{theorem-horn-lidskii-Klyachko}
and~\ref{theorem-Horn-converse} in terms of $\SU(m)\times \SU(n)
\subset \SU(m+n)$ will be presented elsewhere.

Readers familiar with the way the connection between spectra and
Littlewood-Richardson coefficients is usually
presented~\cite{Klyachko98, Knutson00} will have observed two
divergences in this presentation. The first of these, which is
similar to the one discussed in the subsection on Kronecker
coefficients and the compatibility of local spectra, can be found
in theorem~\ref{theorem-horn-lidskii-Klyachko}. Previously, the
mismatch was resolved by invoking the semigroup property.
Likewise, it is possible to prove that
theorem~\ref{theorem-horn-lidskii-Klyachko} is equivalent to the
statement that a triple of states $A, B$ and $C=pA+(1-p)B$ with
rational spectra $(r^A, r^B, r^{C})$ and rational $p$ leads to a
nonzero $c_{Npr^A, N(1-p)r^B}^{Nr^C}$ for some integral $N$. The
second divergence occurs because this presentation focuses on
quantum states and not on the more general case of Hermitian
operators $A, B$ and $C=A+B$ on $\complex^d$. There is an easy
two-step process to extend the result:
\begin{itemize}
\item[i)] the spectra are shifted to be
positive, i.e.
\be
\label{eq-horn-transfer-1}
\begin{split}   A &\mapsto A':=A-r^A_{\min} \id_{\complex^d} \\
                B & \mapsto B':=B-r^B_{\min} \id_{\complex^d},
\end{split}
\ee
for $r^A_{\min}$ and $r^B_{\min}$ the smallest eigenvalues of $A$
and $B$.
\item[ii)] positive operators are rescaled to quantum
states, i.e.
\beastar A' \mapsto A'':= \frac{A'}{\tr A'}\\
B' \mapsto B'':= \frac{B'}{\tr B'}\\
C' \mapsto C'':= \frac{C'}{\tr C'}
\eeastar and a weight $p:=\frac{\tr A'}{\tr C'}$ is defined such that
\bestar C''=pA''+(1-p)B''.\eestar
\end{itemize}
How do these steps translate into the language of
Littlewood-Richardson coefficients?
\begin{itemize}
\item[i)]
the spectral shift corresponds to a shift of the Young
diagrams explained in section~\ref{sec-relation-SU-U}. It carries
through to the Littlewood-Richardson coefficients as
    \be \label{eq-horn-transfer-2} c_{\mu \nu}^\lambda \;  \mapsto
        c_{\mu'\nu'}^{\lambda'}=c_{\mu \nu}^\lambda
    \ee
for
    \be
        (\mu', \nu', \lambda'):=(\mu+R\tau, \nu+S\tau, \lambda+(S+R)\tau),
    \ee
where $\tau =\underbrace{(1, \ldots, 1)}_{d}$ and $S, R \in
\mathbb{Z}$. Note that no positivity constraint of the type
$\mu_d, \nu_d, \lambda_d \geq 0$, is imposed on the Young frames.
This makes sense as the polynomial representations constructed
with Young symmetrisers lead to holomorphic representations when
multiplied by $(\det U)^{R}$ for $R \in \mathbb{Z}$ (see
section~\ref{sec-semigroup} as well as~\cite[chapter
14]{CarterSegalMacDonald95}).
\item[ii)] the rescaling absorbs the
relative difference of the trace of $A'$ and $B'$ in $p$ and will
only correspond to a relabeling on the level of
Littlewood-Richardson coefficients. The overall scaling factor
$\tr A'$ will turn out to be irrelevant, since we are only
concerned with the asymptotic nature of the representations.
\end{itemize}
The two steps and their analogues are sufficient to formulate
theorem~\ref{theorem-Horn-converse} in the usual way:

\begin{corollary}\label{corollary-Horn-converse}
If $c_{\mu \nu}^\lambda \neq 0$, there exist Hermitian operators
$A$ and $B$ such that
\beastar
\spec  A &=&\mu\\
\spec  B &=&\nu\\
\spec A+B &=&\lambda.
\eeastar
\end{corollary}

\begin{proof}
Define $(\mu', \nu', \lambda')$ as above with $R=-\mu'$ and
$S=-\nu_d$. According to theorem~\ref{theorem-Horn-converse},
there are density operators $A'', B''$  with
\beastar
\spec  A'' &=&\bar{\mu}'\\
\spec  B'' &=&\bar{\nu}'\\
\spec pA''+(1-p)B'' &=&\bar{\lambda}',\eeastar where
$p=\frac{|\mu'|}{|\lambda'|}$. With the definitions $A'=|\mu'|A''$
and $B'=|\nu'|B''$ the previous equation can be rewritten in the
form
\beastar
\spec  A' &=&\mu'\\
\spec  B' &=&\nu'\\
\spec A'+B' &=&\lambda',\eeastar and, finally, shifted back into
the -- possibly -- negative with the
transformations~(\ref{eq-horn-transfer-1}) and
(\ref{eq-horn-transfer-2}):
\beastar
\spec  A &=&\mu\\
\spec  B &=&\nu\\
\spec A+B &=&\lambda. \eeastar This completes the proof of the
corollary.
\end{proof}
This concludes the discussion on the relation of Horn's problem
and the \linebreak Littlewood-Richardson coefficients. In contrast
to the relation between the compatibility of local spectra and
Kronecker coefficients, however, this story does not end here. In
1999, Knutson and Tao proved the \emph{saturation
conjecture\index{saturation
conjecture}\label{saturation-conjecture}} for $\GL(n, \complex)$,
i.e.~they proved that
    $$ c_{N\mu, N\nu}^{N\lambda} \neq 0 \mbox{ for some } N \in \naturals \mbox{ implies } c_{\mu
        \nu}^\lambda\neq 0.
    $$
The proof appeared in~\cite{KnuTao99} and introduces the
\emph{honeycomb} model. A more compact version of this proof based
on the \emph{hive} model was given by~\cite{Buch00}, and a more
accessible discussion can be found in~\cite{KnuTao01}.

\subsection[Convexity of Spectral Problems]{Convexity of Spectral Problems\protect\footnote{Apart from the alternative proof of theorem~\ref{theorem-convex-polytope} the results in this section have been obtained in collaboration with Graeme Mitchison.}}
\label{subsec-spec-convexity} In
subsection~\ref{subsec-semigroup}, the set of nonzero Kronecker
coefficients, $\KRON$, was proved to be a finitely generated
semigroup. This is a statement about the shape of $\KRON$, which
will be taken up in this subsection to show that $\qmp$, the set
of admissible spectral triple, is a convex polytope
(theorem~\ref{theorem-convex-polytope}). This theorem is the major
finding of this subsection and, as will be shown later, falls into
a general framework of convexity results in Lie algebra theory,
the simplest instance of which is the Schur-Horn theorem.

\begin{theorem}[Schur-Horn]
Let $A$ be a Hermitian operator on $\complex^d$ with spectrum
$\lambda$. The set of diagonals of the matrices $UAU^\dagger$ with
$U \in \U(d)$ is a convex polytope whose extreme points are the
permutations of $\lambda$.
\end{theorem}

\noindent In Lie algebra theory a general understanding of
convexity has been reached and the result of interest here is a
theorem by Kirwan (theorem~\ref{theorem-kirwan}). Knutson's
exposition~\cite{Knutson00} shows how to apply Kirwan's theorem to
Horn's problem: given two spectra $r^A$ and $r^B$, the set of
possible spectra $r^C$, such that there are Hermitian operators
$A$ and $B$ with
\beastar \spec A&=&r^{A}\\
         \spec B&=&r^{B}\\
         \spec A+B&=&r^{C},
\eeastar
is a convex polytope.

The last part of this section follows Knutson's paper closely,
where the groups involved in Horn's problem will be carefully
replaced with the ones occurring in problem~\ref{prob-comp} (see
table~\ref{table-prob-comparison}). In this way it will become
clear how to apply Kirwan's theorem in order to obtain an
alternative proof for theorem~\ref{theorem-convex-polytope}. It
should also be noted that Sumit Daftuar and Patrick Hayden have
observed that it is possible to apply Kirwan's theorem to the
problem of the compatibility of a bipartite spectrum with the
spectrum of one margin~\cite{DafHay04}.

\vspace{0.5cm}

\noindent Let us now start with the implications of
theorem~\ref{ourtheorem} and theorem~\ref{theorem-converse}:
$\qmp$ denotes the set of all admissible spectral triple $(r^A,
r^B, r^{AB})$, and $\kron$ is the set of all $(\bar{\mu},
\bar{\nu}, \bar{\lambda})$ for which $g_{\mu \nu \lambda} \neq 0$.
As an immediate corollary of theorem~\ref{ourtheorem} and
\ref{theorem-converse} one finds that
    $$ \kron \subset \qmp \subset \overline{\kron},$$
where $\overline{\kron}$ denotes the closure of $\kron$. Since
$\kron$ only consists of rational triples, and since by an easy
example it can be shown that there exists an admissible irrational
spectral triple, it follows that $\kron$ cannot equal
$\overline{\kron}$. Moreover, to every sequence in $\kron$, by
theorem~\ref{theorem-converse}, we can choose a corresponding
sequence of density operators. Since the set of density operators
is a compact set it is possible to select a convergent
subsequence. The spectrum of the limiting density operator equals
the limit of the previously chosen sequence in $\kron$. The
following theorem summarises this discussion.

\begin{theorem}
$$\kron \subsetneq \qmp=\overline{\qmp}=\overline{\kron},$$
where $\overline{\qmp}$ $(\overline{\kron})$ denotes the closure
of $\qmp$ $(\kron)$.
\end{theorem}
But one can say more about the shape of the set $\qmp$ by turning
to representation theory once more. As it had been shown in
theorem~\ref{theorem-stability}, $\KRON$ is a semigroup under
addition, i.e.~if $(\mu, \nu, \lambda), (\mu', \nu', \lambda') \in
\KRON$ and $P, Q \in \naturals$, then $(P\mu+Q\mu', P\nu+Q\nu',
P\lambda+Q\lambda') \in \KRON$. Hence the convex combination of
$(\bar{\mu}, \bar{\nu}, \bar{\lambda}), (\bar{\mu}', \bar{\nu}',
\bar{\lambda}') \in \kron$ with rational weight $p$,
$$(p\bar{\mu}+(1-p)\bar{\mu}', p\bar{\nu}+(1-p)\bar{\nu}',
p\bar{\lambda}+(1-p)\bar{\lambda}')$$ is in $\kron$. This implies
that $\overline{\kron}$ is convex. Furthermore it is true that
$\KRON$ is finitely generated (corollary
\ref{corollary-finitely-generated}); this means that there is a
finite number of triples $(\mu_i, \nu_i, \lambda_i) \in \KRON$
with the property that every other triple is of the form
$$ (\mu, \nu, \lambda)=\sum_i P_i (\mu_i, \nu_i, \lambda_i)=(\sum_i P_i \mu_i, \sum_i P_i \nu_i, \sum_i P_i \lambda_i),$$
where $P_i \in \naturals$. $\kron$ therefore equals the set of
rational convex combinations of the finite set of points
$\{(\bar{\mu}_i, \bar{\nu}_i, \bar{\lambda}_i)\}$. Hence
$\overline{\kron}$ (and thus $\qmp$) is a convex polytope.

\begin{theorem} \label{theorem-convex-polytope}
$\qmp$, the set of admissible spectral triple, is a convex
polytope.
\end{theorem}

\noindent In the remaining part of this section I give an
alternative proof for theorem~\ref{theorem-convex-polytope}
following Knutson's exposition for the convexity of Horn's problem
in~\cite{Knutson00}.

Some concepts and notation of Lie theory and symplectic geometry
need to be introduced (see~\cite{daSilva01}). Let $G$ be a
connected Lie group, $\fg$ its Lie algebra and $\fg^\star$ the
dual of the Lie algebra, acting on a symplectic manifold $M$ with
symplectic form $\omega$. For $A \in \fg$ let $A^{\#}$ be the
vector field on $M$ generated by the one-parameter subgroup $\{
\exp t A : t \in \real \}\subset G$. For a function $f$ on $M$ let
$X_f\equiv X_f (m)$ be the \emph{symplectic gradient} of $f$
defined by $D_{\vec{v}(m)} f=\omega(\vec{v}(m), X_f(m))$, where
$D_{\vec{v}(m)} f$ is the directional derivative of $f$ at $m$ in
direction $\vec{v}(m)$. Then $\Theta: M \rightarrow \fg^\star$ is
a \emph{moment map\index{moment map}} for the action of $G$ on
$M$, if the following two hold:
\begin{itemize}
\item $\Theta$ is $G$-equivariant, i.e.~for all $g \in G$ and $m \in M$,
$\Theta(g(m))=g\Theta(m)g^{-1}$.
\item For all $A \in \fg$, the symplectic
gradient $X_{f_A}$ of $f_A(m):=\langle A, \Theta(m) \rangle$ is
equal to the vector field $A^{\#}$, where $\langle \ , \ \rangle$
is the natural pairing of $\fg$ and $\fg^\star$.
\end{itemize}

The \emph{coadjoint representation} given by $m \mapsto
g(m)=gmg^{-1}$ is the natural action of $g \in G$ on $m \in
\fg^\star$. Let $M$ denote an orbit of this action, a
\emph{coadjoint orbit}. It is then a general result that every
coadjoint orbit has a unique symplectic structure such that the
inclusion map
$$ \Phi: M \hookrightarrow \fg^\star$$
is a moment map for the action of $G$ on $M$~\cite{Bryant91}. Next
we will compose the coadjoint action with a Lie group
homomorphism, $\Psi: H \rightarrow G$, so that now $H$ also acts
on $M$. Let $\psi: \mathfrak{h} \rightarrow \mathfrak{g}$ be the
corresponding map of Lie algebras and $\psi^\star:
\mathfrak{g}^\star \rightarrow \mathfrak{h}^\star$ be the dual
map. Another general result asserts that $\psi^\star \circ \Phi: M
\rightarrow \mathfrak{h}^\star$ is a moment map for the action of
$H$ on $M$. This suffices as precursor for Kirwan's convexity
theorem.
    \begin{theorem}[Kirwan\index{Kirwan's
theorem}]\label{theorem-kirwan} Let $M$ be a symplectic manifold
with Lie group $H$ acting on it. Let $\ft_+$ be the positive Weyl
chamber of $H$ and $\Theta$ a moment map for the action of $H$ on
$M$. The image of the composition of $\Theta$ with the
$H$-invariant map $\fh^\star \rightarrow \ft^\star_+$ that maps an
element of $\fh^\star$ to a unique point in its $H$-orbit in
$\ft_+$, is a convex polytope.
    \end{theorem}
To give an alternative proof of
theorem~\ref{theorem-convex-polytope}, it therefore suffices to
formulate problem~\ref{prob-comp} so that Kirwan's theorem applies
to it.

\vspace{0.3cm}

\begin{proof}[Alternative proof of theorem~\ref{theorem-convex-polytope}\label{alternative-proof}]
Let $H=\U(m)\times \U(n)$ and $G=\U(mn)$. The corresponding Lie
algebras are $\fh=\fu(m)\oplus \fu(n)$ and $\fg=\fu(mn)$.
$\fu(mn)$ is the set of skew-Hermitian matrices (see
table~\ref{table-lie-groups}). One can identify a Hermitian matrix
$\rho$ on $\complex^{mn}$ by $\rho \rightarrow \tr i \rho \ \cdot
$ with an element in $\fu^\star(mn)$, and conversely every linear
form on $\fu(mn)$ can be brought into this form, thereby
specifying a unique Hermitian matrix $\rho$. A coadjoint orbit of
the action of $\U(mn)$ on $\fu^\star(mn)$ seen in this light is
nothing but a set of Hermitian matrices of a given spectrum, say
$\lambda$, and denoted by $\cO_\lambda$.

The map for this correspondence is the moment map of the unitary
group. It is denoted by $\Phi: \cO_\lambda \rightarrow
\fu(d)^\star$. Consider $\Psi$, the natural inclusion
\beastar
    \Psi: \U(m)\times \U(n) &\rightarrow &\U(mn)\\
          (U, V)& \mapsto & U\otimes V
\eeastar
and its derivative, the map of the corresponding Lie algebras
\bea
\psi: \fu(m) \oplus \fu(n) &\rightarrow & \fu(mn)\\
(A, B) &\mapsto & A \otimes \id_n + \id_m \otimes B.
\eea
With respect to the Hilbert-Schmidt inner product, choose an
orthonormal basis $\{\sigma_j\}_{j=1}^{m^2}$ of $\fu(m)$ such that
$\sigma_1=i \id_m$ and $\sigma_j$ traceless for $j>1$. Likewise,
choose an orthonormal basis $\{ \tau_k\}_{k=1}^{n^2}$ for $\fu(n)$
with $\tau_1=i \id_n$ and $\tau_k$ traceless for $k>1$. A basis
for $\fu(mn)$ is given by $\{-i \sigma_j \otimes \tau_k\}_{j=1,
k=1}^{m^2,n^2}$. In terms of the bases
$$
\{ \sigma_1, \ldots, \sigma_{m^2}, \tau_1, \ldots, \tau_{n^2} \} \mbox{ for } \fu(m) \oplus
\fu(n)
$$
and
$$
\{ -i \sigma_1 \otimes \tau_1, \ldots, -i \sigma_{m^2} \otimes
\tau_1, \ldots \ldots , -i \sigma_1 \otimes \tau_{n^2}, \ldots, -i
\sigma_{m^2} \otimes \tau_{n^2}\}
$$
for $\fu(mn)$, $\psi$ can be expressed as the matrix (note that
e.g. $-i \sigma_j \otimes \tau_1=\sigma_j \otimes \id_n$):
\begin{displaymath}
\psi=\begin{scriptsize}\left(
\begin{tabular}{cccccc}
      &   &                      &\; 1 & 0 & $\hdots$ \\
      &  {\large \id} &                   &\; 0 & 0 & $\hdots$ \\
       &   &     &\; $\vdots$ & $\vdots$ & $\ddots$\\
\\
&       &   &       \; 0 & 1 & $\hdots$ \\
& {\large 0}   &   & \; 0 & 0 & $\hdots$ \\
&       &   & \; $\vdots$ & $\vdots$ & $\ddots$\\
\\
&       &   & \; & &\\
& {\large 0}   &   & \; & $\vdots$ &\\
& &           & \; \\
\\
&       &   & \; $\hdots$ & 0 & 1 \\
& {\large 0}   &   & \; $\hdots$ & 0 & 0 \\
&       &   & \;  & $\vdots$ & $\vdots$
\end{tabular}
\right).\end{scriptsize}
\end{displaymath}
 The dual map $\psi^\star: \fu(mn)^\star
\rightarrow \fu(m)^\star \oplus \fu(n)^\star$ is simply given by
the transpose of the above matrix and sends a Hermitian matrix
$\rho^{AB}$ on $\complex^{m}\otimes \complex^n$ to its partial
traces:
    $$\psi^\star: \rho^{AB} \mapsto (\tr_B \rho^{AB}, \tr_A \rho^{AB}).$$
From the discussion above it follows that the composition
$\psi^\star \circ \Phi: \cO_\lambda \rightarrow \fu(m)^\star
\oplus \fu(n)^\star$ is a moment map. This sets the scene for the
application of Kirwan's theorem. It only remains to figure out
what the dual of the positive Weyl chamber of $\U(m)\times \U(n)$,
$\ft^\star_+(m)\oplus \ft^\star_+(n)$, is. The torus $\ft(m)$ of
$\fu(m)$ can be taken to be the diagonal matrices of $\fu(m)$. The
dual $\ft^\star(m)$ then consists of all real diagonal matrices on
$\complex^m$, and the positive Weyl chamber $\ft^\star_+(m)$ are
the elements in $\ft^\star(m)$ whose entries decrease down the
diagonal. The set $\ft^\star(m)$ therefore corresponds precisely
to the spectra of $\fu^\star(m)$. The map $\Gamma$, taking an
element of $\fu^\star(m)\oplus \fu^\star(n)$ to its pair of
spectra, is an $H$-invariant map as $H$ does not change the pair
of spectra at all. The image of the map $\Gamma \circ \psi^\star
\circ \Phi$ is then, according to Kirwan's theorem, a convex
polytope.
\end{proof}

\noindent The discussion of general theorems and properties of the
four problems has now come to an end. The next and second last
section of this chapter contains an application of the previous
work for the case of two qubits.

\subsection[The Two-Qubit Inequalities]{The Two-Qubit Inequalities\protect\footnote{This section contains collaborative work with Graeme Mitchison}}
\label{subsec-spec-two-qubit}

In this subsection, I derive all spectral inequalities for two
qubits, theorem~\ref{theorem-bravyi}, which serve to illustrate of
the work presented in this chapter. Recently, Sergei Bravyi found
these inequalities by a direct calculation~\cite{Bravyi04}, and
two more proofs of this result can be found in the work of
Klyachko. Both of Klyachko's proofs emerge from the wider context
of geometric invariant theory: the first one uses the Schubert
calculus, whereas the second proof is related to the one I give
below. This second proof employs
theorems~\ref{theorem-klyachko-our} and~\ref{theorem-converse}
(\cite[theorem 5.3.1.]{Klyachko04}) in order to reduce the problem
to the calculation of the Kronecker coefficient. Known results on
the Kronecker product of two-row shaped diagrams~\cite{RemWhi94,
Rosas01} are thereby transferred back and result in spectral
inequalities.

The proof I present below, employs theorem~\ref{ourtheorem} to
carry over the results by Klemm, Dvir and Clausen and
Maier~\cite{Klemm77, Dvir93, ClaMai93} on Kronecker coefficients
to the spectral realm.

\begin{theorem}[Klemm, Dvir, Clausen and Maier] \label{theorem-dvir}
For all $\mu, \nu$ and $\lambda$ with $\nu_1> |\lambda \cap \mu|$
it is true that $g_{\mu \nu \lambda}=0$. Conversely, there exists
$\nu$ with $\nu_1=|\lambda \cap \mu|$ s.th. $g_{\mu \nu
\lambda}\neq 0$.
\end{theorem}
For the sufficiency of the inequalities, a conceptually simplified
version of Bravyi's calculation is given here. More precisely,
instead of deriving a density operator for every point inscribed
by the inequalities, only the ones on the vertices are given.
Convexity of the solution (theorem~\ref{theorem-convex-polytope})
then extends this result to the whole polytope and completes the
proof of sufficiency.

The shifting and rescaling of the spectra of Hermitian operators
have analogous representation-theoretical transformations. This
has been discussed in subsection~\ref{subsec-LR} in the context of
the Littlewood-Richardson coefficients. Below,
theorem~\ref{theorem-dvir} is applied to a radically shifted
triple of Young diagrams: \emph{contragredient Young diagrams}. A
contragredient diagram $\lambda'$ of $\lambda$ is constructed as
follows:
\begin{itemize} \item[] Draw the rectangular Young diagram
with $d$ rows and $\lambda_1$ columns and consider the complement
of $\lambda$ in this rectangle. The diagram obtained after
rotation around 180 degrees is $\lambda'$. Formally, $\lambda'_i:=
\lambda_1-\lambda_{d-i+1}$.
\end{itemize}

\begin{lemma} \label{lemma-contragredient}
Let $\mu, \nu$ and $\lambda$ be diagrams of no more than $m, n$
and $mn$ rows and a total number of $k$ boxes with $g_{\mu \nu
\lambda} \neq 0$. The diagrams
\bea \mu''_i&=&n \lambda_1-\mu_{m-i+1}\\
     \nu''_i&=&m\lambda_1 - \nu_{n-i+1}\\
     \lambda''_i&=&\lambda_1-\lambda_{mn-i+1}
\eea are Young frames with $g_{\mu''\nu'' \lambda''} \neq 0$.
\end{lemma}

\begin{proof}
Notice that the diagrams $\mu''$ and $\nu''$ have positive row
length and are equivalent to the contragredient diagrams $\mu'$
and $\nu'$ since, as a consequence of theorem~\ref{theorem-dvir},
$\mu_1 \leq n\lambda_1$ and $\nu_1 \leq m \lambda_1$. $\lambda''$
equals the contragredient diagram $\lambda'$ of $\lambda$.

Given an irreducible representation $U$ of $\sunitaries (d)$ with
highest weight $\tau$, representing every $g \in \sunitaries(d)$
as
$$ g \rightarrow U(g),$$
the contragredient representation is given by the complex
conjugate
$$ g \rightarrow \bar{U}(g).$$
It is not difficult to see that $\bar{U}$ is also an irreducible
representation of the same dimension. It can be shown that the
highest weight of $\bar{U}$ is given by the contragredient diagram
(defined above) as follows: let $\ket{v}$ be the highest weight
vector of the representation $U$, then
$$ U(g) v= \prod_j u_j^{\tau_j} \ket{v},$$
for $g=\diag(u_1, \ldots, u_d) \in \sunitaries(d)$. Noting that
$\prod_j u_j=1$ we have
$$ \bar{U}(g) \ket{v}= \prod_j u_j^{n-k_j} \ket{v}$$
and therefore $\ket{v}$ is a weight vector with weight $\tau'$,
where $\tau'_j=n-\tau_j$, for $\bar{U}$. Application of the
raising operators $E_{ij}$ shows that it must be the weight vector
of highest weight for $\bar{U}$ and thus $\tau'$ is the highest
weight for $\bar{U}$ and therefore also the Young frame for
$\bar{U}$.

Taking the complex conjugate of
$$ U_\mu \otimes U_\nu \subset U_\lambda \downarrow^{SU(mn)}_{SU(m)\times
SU(n)}$$ therefore gives
    $$U_{\mu'} \otimes U_{\nu'}\subset U_{\lambda'}
        \downarrow^{SU(mn)}_{SU(m)\times SU(n)}.
    $$
Since $\mu''$ and $\nu''$ are equivalent to $\mu'$ and $\nu''$:
$$ U_{\mu''} \otimes U_{\nu''} \subset U_{\lambda'} \downarrow^{SU(mn)}_{SU(m)\times
SU(n)}.$$
\end{proof}
The following proposition contains the key group-theoretic
inequalities, which later, combined with theorem~\ref{ourtheorem},
lead to the two-qubit inequalities.

\begin{proposition} \label{lemma-ineq-kronecker} Let $\mu$ and $\nu$ be two-row diagrams. For
all four row diagrams $\lambda$ with $g_{\mu \nu \lambda}\neq 0$
the following hold:
\bea \label{eq-1-bravyi} \mu_2 &\geq& \lambda_3+\lambda_4\\
\label{eq-2-bravyi} \nu_2 &\geq& \lambda_3+\lambda_4\\
\label{eq-3-bravyi} \mu_2+\nu_2 &\geq&  \lambda_2+\lambda_3+2\lambda_4 \\
\label{eq-4-bravyi}|\mu_2-\nu_2|&\leq& \min \{\lambda_1-\lambda_3,
\lambda_2-\lambda_4\}.\eea
\end{proposition}

\begin{proof}
It suffices to restrict our attention to the case of three row
diagrams $\lambda$, i.e.~$\lambda_4=0$, since the mapping
    \beastar
        \mu_i &\rightarrow &\mu''_i=\mu_i+2 \lambda''_4\\
        \nu_i &\rightarrow &\nu''_i=\nu_i+2 \lambda''_4\\
        \lambda_i &\rightarrow &\lambda''_i =\lambda_i+ \lambda''_4
    \eeastar
does not change the value of the Kronecker coefficient.

We start by applying theorem~\ref{theorem-dvir} to $\mu$ and
$\nu$, which are hook-shaped two-row diagrams and $\lambda$, a
three row diagram. All diagrams have $n$ boxes. Unfortunately, it
is a tedious case-by-case study. Let us start with the first
inequality.

\vspace{0.2cm}

\noindent {\bf Inequality~(\ref{eq-1-bravyi}):}
theorem~\ref{theorem-dvir} implies that $\mu_1 \leq \min (
\lambda_1, \nu_1)+\min\{\lambda_2, \nu_2\}\leq
\lambda_1+\lambda_2$, which, invoking the equalities
$\mu_1+\mu_2=n$ and $\lambda_1+\lambda_2+\lambda_3=n$ leads to
$\mu_2 \geq \lambda_3$.

\vspace{0.2cm}

\noindent {\bf Inequality~(\ref{eq-2-bravyi}):} follows from
exchanging $\mu$ and $\nu$ in inequality~(\ref{eq-1-bravyi}).

\vspace{0.2cm}

\noindent {\bf Inequality~(\ref{eq-3-bravyi}):} three cases are to
be considered separately
\begin{enumerate}
\item[i)] $\lambda_1 \geq \mu_1 \Rightarrow \mu_2 \geq
\lambda_2+\lambda_3 \Rightarrow \mu_2 +\nu_2 \geq
\lambda_2+\lambda_3$
\item[ii)] $\lambda_1 \leq \mu_1$ and $\lambda_2 \leq \mu_2 \Rightarrow
\mu_2+\nu_2 \geq \lambda_2+\lambda_3$
\item[iii)] $\lambda_1 \leq \mu_1$
and $\lambda_2 \geq \mu_2$:
    \beastar &&\nu_1 \leq \lambda_1+\mu_2\\
        \Leftrightarrow &&n-\nu_2 \leq n-\lambda_2-\lambda_3+\mu_2\\
        \Leftrightarrow &&\mu_2+ \nu_2 \geq \lambda_2+\lambda_3
    \eeastar
\end{enumerate}

\vspace{0.2cm}

\noindent {\bf Inequality~(\ref{eq-4-bravyi}):} we start with
$\mu_2-\nu_2\leq \lambda_2$, which will be divided into three
cases:
\begin{itemize}
\item[i)] $\lambda_1 \geq \mu_1$:
    \beastar &&\nu_1 \leq \mu_1+\lambda_2\\
        \Leftrightarrow &&\nu_2 \geq \nu_2+\lambda_2
    \eeastar
\item[ii)] $\lambda_1 \leq \mu_1$ and $\mu_2 \leq \lambda_2$:
    \beastar && \nu_1 \leq \lambda_1+\mu_2\\
        \Leftrightarrow && \nu_2 \geq \lambda_2+\lambda_3+\mu_2\geq
        \lambda_2+\mu_2
    \eeastar
\item[iii)] $\lambda_1 \leq \mu_1$ and $\lambda_2 \leq \mu_2$:
    \beastar && \nu_1 \leq \lambda_1+\lambda_2 \leq \mu_1+\lambda_2\\
        \Leftrightarrow && n-\nu_2\leq n-\mu_2 +\lambda_2\\
        \Leftrightarrow && \mu_2-\nu_2\leq \lambda_2
    \eeastar
\end{itemize}
and by swapping $\mu$ and $\nu$:
    \be
        \label{eq-bravyi-4-firstpart} \nu_2-\mu_2 \leq \lambda_2.
    \ee
The inequality $ |\mu_2-\nu_2| \leq \lambda_1-\lambda_3$ can be
proven by applying inequality (\ref{eq-bravyi-4-firstpart}) to the
diagrams $\mu'', \nu''$ and $\lambda''$, as defined in
lemma~\ref{lemma-contragredient}. By
lemma~\ref{lemma-contragredient} $g_{\mu \nu \lambda}= g_{\mu''
\nu'' \lambda''}$ and since $\mu'', \nu''$ and $\lambda''$ are
equivalent to the contragredient diagrams $\mu^c, \nu^c$ and
$\lambda^c$, we have $|\mu_2^c-\nu_2^c|\leq \lambda^c_2$. This can
be rewritten as $|\mu_2-\nu_2|\leq \lambda_1-\lambda_3$.
\end{proof}

\noindent This completes the representation-theoretic preparation.
The inequalities in proposition~\ref{lemma-ineq-kronecker} are now
easily turned into the following two-qubit inequalities: by
theorem~\ref{ourtheorem} for every $\rho^{AB}$ there exists a
sequence $(\mu_i, \nu_i, \lambda_i)$ with $g_{\mu_i \nu_i
\lambda_i}\neq 0$ such that
\beastar \lim_{i\rightarrow \infty} \bar{\mu}_i = \spec \rho^A\\
     \lim_{i\rightarrow \infty} \bar{\nu}_i = \spec \rho^B\\
     \lim_{i\rightarrow \infty} \bar{\lambda}_i = \spec
     \rho^{AB}\eeastar
The inequalities in proposition~\ref{lemma-ineq-kronecker} thus
imply the same inequalities for the admissible spectra of
$\rho^{AB}$.

\begin{figure}[h] \label{figure-bravyi}
\begin{center}
\includegraphics[width=7cm]{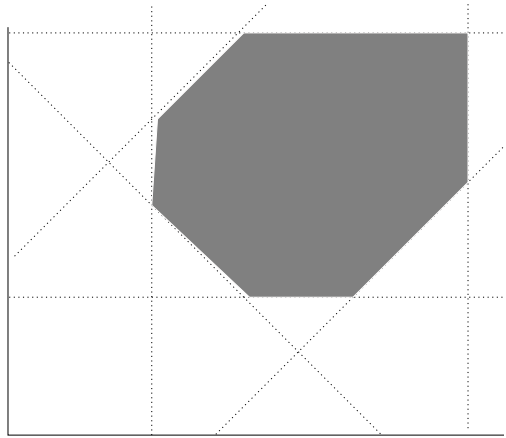}
\caption{The region defined by Bravyi's inequalities. Clockwise
starting from the top right corner: ABCDEFG. }
\end{center}
\end{figure}

\begin{theorem}[Bravyi\index{Bravyi's inequalities}] \label{theorem-bravyi} Let $\rho^{AB}$ be a density operator with
spectrum $r_i$ and local spectra $(a, 1-a)$ and $(b, 1-b)$, where
we take $a, b \leq \half$ without loss of generality. Then
\bea \label{eq-1-bravyi-spec} a &\geq& r_3+r_4\\
\label{eq-2-bravyi-spec} b &\geq& r_3+r_4\\
\label{eq-3-bravyi-spec} a+b &\geq& r_2+r_3+2r_4 \\
\label{eq-4-bravyi-spec}|a-b|&\leq& \min \{r_1-r_3, r_2-r_4\}.\eea
Conversely, if these inequalities hold, there is a density
operator of two qubits with spectrum $(r_1, r_2, r_3, r_4)$ and
local spectra $(a, 1-a)$ and $(b, 1-b)$ (see
figure~\ref{figure-bravyi}).
\end{theorem}

Let us now fix the spectrum of the joint density matrix to
$\lambda$. To prove the second claim of
theorem~\ref{theorem-bravyi}, the sufficiency of the inequalities,
Bravyi constructed a density matrix to every possible set of
spectra described by the inequalities. Since we already know that
$\qmp$ is a convex polytope
(theorem~\ref{theorem-convex-polytope}), it will suffice to
construct a density matrix for each point at which two
inequalities intersect. This will be done in the following.

The labeling of the edges is defined in the caption of
figure~\ref{figure-bravyi}. $\mu_2$ runs on the horizontal axis
and $\nu_2$ is on the vertical axis. By exchange of $\mu$ and
$\nu$ it suffices to find density matrices for the points $A, B,
C$ and $D$. We take the eigenbasis of a local basis to be the
computational basis $\{\ket{0}, \ket{1}\}$ such that the smaller
eigenvalue corresponds to the state $\ket{1}$.

\vspace{0.2cm}

 \noindent {\bf Point $A$:} $a=b=\half$:
$\rho^{AB}=\frac{\id}{d_A} \otimes \frac{\id}{d_B}$


\noindent {\bf Point $B$:}
    \be \label{eq-brayvi-point-A}
        a=\half \textrm{ and }
        a-b=\min (r_2, r_1-r_3)
    \ee
\bestar
\begin{split}
\ket{\psi_1}=&\alpha \ket{00}+\sqrt{1-\alpha^2}\ket{11}\\
\rho=\sum_{i=1}^3 r_i \proj{\psi_i} \qquad \quad   \ket{\psi_2}=&\beta \ket{01}+\sqrt{1-\beta^2}\ket{10} \\
\ket{\psi_3}=&\sqrt{1-\alpha^2}\ket{00}-\alpha \ket{11}
\end{split}
\eestar with
$\alpha, \beta \in [0, 1]$. Equality~(\ref{eq-brayvi-point-A}) is
then equivalent to
    \beastar \half=&a&=r_1 \alpha^2+r_2 \beta^2 +r_3
(1-\alpha^2)\\
\half-\min (r_2, r_1-r_3)=&b&=r_1 \alpha^2+(1-\beta^2) r_2 +r_3
(1-\alpha^2).
    \eeastar
There are two different cases to consider: $r_2 \leq r_1-r_3$:
$$
\beta^2=1 \qquad \alpha^2=\frac{\half-r_2-r_3}{r_1-r_3}$$
and $r_2\geq r_1-r_3$:
$$ \alpha^2=0 \qquad
\beta^2=\half+\frac{r_1-r_3}{r_2}$$


\noindent {\bf Point $C$:} $a-b= \min (r_2, r_1-r_3)$ and $b=r_3$.
Again we will first consider the case $ r_2 \leq r_1-r_3$:
$$ \rho= r_1 \proj{00}+r_2 \proj{10}+r_3
\proj{11}.$$ The second case is then $r_2 \geq r_1-r_3$:
$$ \rho=r_1 \proj{10}+r_2 \proj{00}+r_3 \proj{01}$$


\noindent {\bf Point $D$:} $b =r_3$ and $a+b=r_2+r_3$
$$ \rho=r_1 \proj{00}+r_2 \proj{10}+r_3 \proj{01}.$$

\noindent This concludes the construction of the density operators
that have spectra sitting at the vertices. By the convexity of the
spectral problem it also concludes the proof of
theorem~\ref{theorem-bravyi} and hence this section.

\section{Conclusion}
\label{sec-spectra-conclusion}

In this chapter I have proven a close connection between two
fundamental problems in physics and representation theory. The
physical problem asks for the compatibility of local quantum
states $\rho^A$ and $\rho^B$ with an overall quantum state
$\rho^{AB}$. The solution to this problem only depends on the
spectra of $\rho^A$ and $\rho^B$ as formulated in
problem~\ref{prob-comp}. This question is the simplest nontrivial
instance of the more general problem concerning the compatibility
of density matrices of more than two parties. Compatibility
questions arise from the study of entropy inequalities, such as
the strong subadditivity of von Neumann entropy~\cite{LieRus73PRL,
LieRus73JMP} or the $N$-Representability problem~\cite{ColYuk00},
whose solution is paramount for the efficient calculation of
energies of nearest-neighbour Hamiltonians.

The main results of this chapter are theorems~\ref{ourtheorem}
and~\ref{theorem-converse}, in which I have shown that
problem~\ref{prob-comp} is equivalent -- in an asymptotic sense --
to problem~\ref{prob-Kronecker}, an unsolved group-theoretic
problem. Problem~\ref{prob-Kronecker} raises the question of
whether or not an irreducible representation $V_\lambda$ of the
symmetric group is contained in the tensor product of two
irreducible representation $V_\mu$ and $V_\nu$ of the same group
-- i.e.~to decide whether or not the Kronecker coefficient $g_{\mu
\nu \lambda}$ in the Clebsch-Gordan decomposition
    \be
    \label{eq-decomp-concl}V_\mu \otimes V_\nu \cong \bigoplus_\lambda g_{\mu \nu \lambda} V_\lambda
    \ee
is nonzero.

The proofs of both theorems~\ref{ourtheorem}
and~\ref{theorem-converse} employ a connection between
representations of the symmetric group and spectra of density
operators (theorem~\ref{theorem-Keyl-Werner}). This was discovered
by Alicki, Rudnicki and Sadowski~\cite{AlRuSa87} and
independently, by Keyl and Werner in the context of quantum
information theory~\cite{KeyWer01PRA}. To keep the exposition
self-contained, I have given a short proof of this theorem based
on the majorisation property of Young symmetrisers. Apart from
theorem~\ref{theorem-Keyl-Werner}, the proof of
theorem~\ref{theorem-converse} needs a second, purely
group-theoretic ingredient: the verification that the set of
nonzero Kronecker coefficients forms a semigroup
(theorem~\ref{theorem-stability}). From a
representation-theoretical viewpoint this is most sophisticated
part of this chapter. It provides a positive resolution of a
recent conjecture by Klyachko~\cite[conjecture 7.1.4]{Klyachko04}
and generalises a recent announcement by
A.~N.~Kirillov~\cite[theorem 2.11]{Kirillov04}. This completes the
asymptotic equivalence of problem~\ref{prob-comp} and
problem~\ref{prob-Kronecker}.

In the first application, the equivalence is used to show how the
fact that the nonzero Kronecker coefficients form a finitely
generated semigroup (theorem~\ref{theorem-stability} and
corollary~\ref{corollary-finitely-generated}) implies that the
solution of the spectral problem forms a convex polytope
(theorem~\ref{theorem-convex-polytope}). An alternative proof of
this theorem is given in a Lie algebra setting using a theorem by
Kirwan. A second application of the equivalence shows how to
derive the inequalities of two qubits, previously established by
Bravyi~\cite{Bravyi04}, from a result on the Kronecker coefficient
by Klemm, Dvir and Clausen and Maier~\cite{Klemm77, Dvir93,
ClaMai93}. The sufficiency of the construction is established by a
direct construction of the vertices and by invoking the convexity
of the solution (theorem~\ref{theorem-convex-polytope}). This last
application concludes the chapter.

The chapter has one more facet to it, however: a novel and short
proof of the connection between Horn's problem and the
Littlewood-Richardson coefficients (problem~\ref{prob-Horn}
and~\ref{prob-LR}). To illustrate problems~\ref{prob-comp}
and~\ref{prob-Kronecker}, I have intertwined and contrasted the
presentation with an analogue discussion of
problems~\ref{prob-Horn} and~\ref{prob-LR}.

The ideas and results presented in this chapter show how methods
and thinking from quantum information theory can benefit both
representation theory and quantum information theory. A previously
unknown relation has been established (problem~\ref{prob-comp}
and~\ref{prob-Kronecker}), and it is shown how deep mathematical
results (equality of symplectic and GIT quotient) can be
circumvented in achieving known results on Horn's problem and the
Littlewood-Richardson problem. An immediate question that arises
from this work is how to generalise and understand the estimation
theorem, theorem~\ref{theorem-Keyl-Werner}, but also the proofs of
theorems~\ref{ourtheorem},~\ref{theorem-converse},
\ref{theorem-horn-lidskii-Klyachko}
and~\ref{theorem-Horn-converse} in the general framework of Lie
algebra theory. This way further connections of
representation-theoretic coefficients to problems in quantum
mechanics could be found and mutual benefits exploited.

The presented material gives a natural way of seeing typical
subspaces of a tensor product of density operators as irreducible
representations of the symmetric and unitary groups (Schur-Weyl
duality, theorem~\ref{theorem-Schur-duality}). Previously, this
point of view has been taken up by a number of works in quantum
information theory. It has been shown how projections onto the
symmetric subspace can be used to stabilise quantum
computation~\cite{BBDEJM97} and to achieve universal quantum data
compression~\cite{JHHH98}. Hayashi and Matsumoto have
significantly refined the analysis of data compression with the
help of Schur-Weyl duality~\cite{HayMat02, HayMat02b}.
Measurements on the total angular momentum have been used to
purify qubits, i.e.~to asymptotically transform mixed qubits into
pure qubits preserving the direction of the Bloch
vector,~\cite{CiEkMa99} and to estimate quantum
states~\cite{VLPT99}. Estimation schemes for $d$-dimensional
quantum systems and their spectra have been proposed
in~\cite{KeyWer01PRA, Keyl04}. Further applications of
representation theory to quantum information theory include
entanglement concentration~\cite{HayMat04} and the use of
irreducible representations as decoherence free
subspaces~\cite{ZanRas97, KnLaVi00, BaRuSp03, BGLPS04}. These
developments have recently received a complexity-theoretic
component as the Schur transform, the unitary transformation from
the standard basis into an orthonormal basis of the irreducible
representations appearing in the Schur-Weyl duality, has been
shown to be efficiently implementable in the number of subsystems
$k$~\cite{BaChHa04}. A comprehensive account of this algorithm and
its applications is contained in Aram Harrow's PhD
thesis~\cite{Harrow05}.

It is the contribution of this chapter to recognise that, in the
same spirit, relations of typical subspaces in multipartite
density matrices can be studied via group-theoretic
decompositions. This was explicitly carried out for the typical
subspaces of $(\rho^{AB})^{\otimes k}$, the reduced states
$(\rho^A)^{\otimes k}$ and $(\rho^{B})^{\otimes k}$ and the
corresponding group-theoretic decomposition.

\part{Insights from Cryptography}\label{part-crypto}
\chapter*{Prologue}
\vspace{-0.9cm}
    \begin{quote}
    ``The eavesdropper cannot elicit any information from the
    particles while in transit from the source to the legitimate
    users, simply because there is no information encoded there.
    The information ``comes into being'' after the legitimate
    users perform measurements and communicate in public
    afterwards. [...] [The eavesdropper's] intervention will be equivalent to
    introducing elements of \emph{physical reality} to the
    measurements of the spin components.''
    \end{quote}
\begin{flushright}\emph{Artur Ekert, in ``Quantum Cryptography Based on Bell's
Theorem''~\cite{Ekert91}.}
\end{flushright}
\noindent This quote illustrates the discovery of
entanglement-based quantum key distribution. The particles refer
to the parts of a Bell pair, a pair of maximally entangled states
of two spin-$\half$ particles, and will directly lead to secure
bits. Bell pairs are also the fundamental ingredient in
teleportation, superdense coding and, more generally, the
essential resource for quantum communication. The question arises:
how valuable are general bipartite quantum states when compared to
Bell states? The theory of entanglement measures has been
developed to answer this question.

This part of my PhD thesis addresses entanglement measures from a
cryptographic point of view and is divided into two chapters.
Chapter~\ref{chapter-entanglement} gives a review of entanglement
measures guided by the axiomatic approach, which focuses on the
general properties such as convexity, additivity and continuity.
This chapter contributes a number of tables and graphs that
summarise properties of and relations among entanglement measures,
as well as three specific examples.

Chapter~\ref{chapter-squashed} proposes a new measure for
entanglement called \emph{squashed entanglement}, which is
motivated by the intrinsic information, a quantity arising in
classical cryptography. The new measure possesses a large number
of properties discussed in the previous chapter, most notably
additivity. At the end of this chapter I will calculate squashed
entanglement for a class of quantum states and show how the tools
used in this calculation can lead to a new information-gain
disturbance tradeoff and the first cheat-sensitive quantum string
commitment scheme.




\chapter{The Zoo of Entanglement Measures}
\label{chapter-entanglement}
\section{Introduction}

This chapter is divided into four sections. After a historical
introduction, subsection~\ref{section-history}, I continue with a
few remarks on correlations in bipartite quantum states such as
total correlations, entanglement and secret key,
subsection~\ref{section-correlations-bipartite}. In
subsection~\ref{section-measuring-entanglement} I review the basic
approaches for measuring these correlations, with focus on the
axiomatic approach to entanglement measures.
Section~\ref{section-axiom-approach} contains extensive tables of
established measures that summarise their properties and mutual
relations. The focus is put on measures that are connected to the
resource-oriented approach highlighted in the introduction. The
work on the tables initiated the writing of
section~\ref{section-specific-results}, in which I will discuss
properties of three specific measures. The chapter is rounded off
with a conclusion, section~\ref{section-conjectures-etc}.

\subsection[A Historical Note on Entanglement]{A Historical Note on Entanglement\protect\footnote{Part of this subsection has appeared in~\cite{OiChr03}.}}
\label{section-history} In December 1900, Max Planck proposed that
the energy of a vibrational system cannot change continuously but
must jump by quanta of energy.  He explained on a later occasion
that ``it was only a formal assumption'', but his proposal was so
radical and fruitful that it influenced natural science throughout
the 20th century. Inspired by this idea, Albert Einstein was the
first to explain the photoelectric effect, Niels Bohr developed
what is nowadays known as {\em old quantum
  theory}, and Louis de Broglie discovered the wave nature of matter.  But it was
only through the joint effort of several theoretical physicists,
and particularly resting on the insights obtained by Wolfang
Pauli, Werner Heisenberg, Erwin Schr\"odinger and Paul Dirac, that
quantum theory took its present shape.  Once the mathematical
foundations of quantum mechanics were laid, it was developed into
a relativistic theory, which was thoroughly tested in experiments.
More than a hundred years after Planck's creation of the quantum
concept, quantum theory is now known to be the most precisely
tested theory in the history of natural science.

Quantum mechanics also entails several philosophical questions,
which are issues of ongoing discussion and which can be
encountered in research disciplines such as foundations of quantum
mechanics, quantum information theory and quantum computation.
This chapter looks into the phenomenon known as {\em
entanglement}\index{entanglement}, first described in 1935 by
Albert Einstein, Boris Podolsky and Nathan Rosen in a publication
in the Physical Review~\cite{EiPoRo35}. In this paper, the
so-called EPR-paper\index{EPR-paper} with the provoking title
``Can Quantum-Mechanical Description of Physical Reality Be
Considered Complete?'', the authors write ``If, without in any way
disturbing a system, we can predict with certainty [\ldots] the
values of a physical quantity, then there exists an element of
physical reality corresponding to this physical quantity.'' and
call such a theory `complete'. By constructing the famous
EPR-Paradox they show that, in general, the initial assumption was
not veritable. The conclusion is striking: the description of
reality by means of quantum mechanical wave functions will never
be complete. Einstein, Podolsky and Rosen expressed their
discontent as well as the belief that it is possible to find a
theory that satisfies their criteria. The paper gave rise to
numerous debates among theoretical physicists. Shortly after
publication, Schr\"odinger wrote a letter to Einstein in which he
expressed sincere appreciation that Einstein had initiated this
controversy. This letter started an intense correspondence between
the two, which culminated in two publications by Schr\"odinger
later that year. One was written in English and was published in
the \emph{Proceedings of the
  Cambridge Philosophical Society}~\cite{Schroedinger35a}, while the second
article, the famous tripartite `Cat'-paper, was written in German
and published in the journal
\emph{Naturwissenschaften}~\cite{Schroedinger35b}.  In these
publications he coined the term \emph{entanglement} or
\emph{Verschr\"ankung,} as he names it in German, in order to
describe a phenomenon which he regards to be not
``[\ldots]~\emph{one} but rather \emph{the} characteristic trait
of quantum mechanics, the one that enforces its entire departure
from classical lines of thought.''~\cite{Schroedinger35a} This
characteristic trait is the simple but intriguing fact that two
physical systems, which are described by their respective quantum
mechanical wavefunctions and undergo a temporary physical
interaction, can, in general, not be described by attributing to
each system a particular wavefunction. ``By the interaction the
two representatives (or $\psi$-functions) have become
entangled.''~\cite{Schroedinger35a}

This insight started a number of discussions concerning the
possible existence of `more complete' theories, which could
describe nature in a local and deterministic manner.  These local
theories would incorporate so called \emph{hidden variables} that
deterministically underlie the probabilistic nature of quantum
mechanical description.  However, all these discussions were
confined to a theoretical sphere until 1964, when John Bell
derived experimentally verifiable conditions that every local
hidden variable theory must satisfy.  It was exactly these
conditions, known as Bell's inequalities, which quantum mechanics
was expected to violate~\cite{Bell64}. Bell's inequalities are
commonly used in the form of CHSH-inequalities abbreviating the
surnames of John F. Clauser, Michael A. Horne, Abner Shimony, and
Richard A. Holt~\cite{CHSH69}. First experiments to test the
nonlocal nature of quantum mechanics were performed by Clauser and
Shimony~\cite{Clauser76, ClaShi78}, and Alain Aspect, Jean
Dalibard, Philippe Grangier and G{\'e}rard Roger~\cite{AsGrRo82,
AsDaRo82}, and later refined by many researchers
(see~\cite{Aspect99} for a short review). The measurement data
clearly violated Bell's inequalities and followed the predictions
given by quantum mechanics.  Thus, the first experimental evidence
for entanglement had been seen and it was only a matter of time
before its potential practical applications were proposed.

Some of the first people to realise its value as a resource were
David Deutsch, one of the fathers of quantum
computation~\cite{Deutsch85}, and Ekert, who proposed a
cryptographic scheme whose security was based on the violation of
Bell's inequalities~\cite{Ekert91}.  In recent years much progress
has been made, the most groundbreaking being Peter Shor's
factorisation algorithm~\cite{Shor97}, but also communication
protocols such as teleportation of quantum states~\cite{BBCJPW93}
and superdense coding~\cite{BenWie92} showed how entangled states
can be seen as a valuable resource. Unfortunately, the process of
decoherence~\cite{Zurek81}, a term describing the loss of
superposition due to the fact that the system interacts with the
environment, makes it difficult to implement these techniques. On
a theoretical level several techniques have been developed to
protect against decoherence, such as quantum error
correction~\cite{Shor95, Steane98} and decoherence free
subspaces~\cite{LiChWh98}. At the same time, progress has been
made to overcome the fundamental problem of controlling particles
at an atomic level.  While the advent of a quantum computer is not
expected in the near future, implementations of quantum
cryptographic schemes are already commercially
available~\cite{IdquantiqueWebsite, MagicQWebsite}.

Entangled quantum states are the basic ingredients for all of the
above presented tasks.  To make this statement precise, however,
it is necessary to focus on the theoretical aspects of this
problem and develop a mathematical definition as answer to the
questions: when is a quantum system in an entangled state, and
when does it exhibit this nonlocal phenomenon? The most simple
system allowing for such an effect consists of two subsystems
which can be spatially separated, as it is the case in
cryptography or teleportation. We denote the state of the system
by a vector $\kpsi$, which is an element of a Hilbert space $\cH$.
$\cH$ itself is the tensorial combination of its two parts $\cal
H_A$ and $\cal H_B$: $\cal H=\cal H_A \otimes \cal H_B$. $\kpsi$
is a \emph{product state} or \emph{separable} if there exist
vectors $|\phi^A \rangle \in \cal H_A$ and $|\phi^B \rangle \in
\cal H_B$, such that $|\psi \rangle$ can be written in the product
form $|\psi \rangle=|\phi^A \rangle \otimes |\phi^B\rangle$.
Otherwise $|\psi \rangle$ is said to be \emph{entangled}. An
example of a state which cannot be written in product form is the
famous \emph{Bell state} of two entangled spin-$\frac{1}{2}$
systems
$|\psi\rangle=|{\uparrow}\rangle\otimes|{\downarrow}\rangle-|{\downarrow}\rangle
\otimes|{\uparrow}\rangle$, where $|{\uparrow} \rangle$ and
$|{\downarrow}\rangle$ denote the wavefunctions corresponding to
the state spin-up and spin-down. If only part of a larger system
is accessible, the description by pure states must be replaced by
statistical mixtures of pure states, the so-called mixed states or
density operators (Preliminaries, page~\pageref{mixed-state}):
\begin{equation}
\rho=\sum_i p_i |\psi_i\rangle \langle \psi_i|,
\end{equation}
where $p_i,\, \sum_i p_i=1$ is the probability of finding the pure
state $|\psi_i\rangle$. Every such mixture is a positive operator
on $\cal H$ with unit trace, and every positive operator with
trace one can be written in the form of an ensemble $\{p_i,
|\psi_i\rangle\}$.  However, this ensemble is not unique: in
general, an infinite number of different ensembles will result in
the same density operator.

A state $\rho^{AB}$ on $\cH_A \otimes \cH_B$ is a \emph{product
state} if it is of the form $\rho^{AB}=\rho^A \otimes \rho^B$. If
the underlying system is clear from the context, I will drop the
superscript. The definition of separability for pure states
extends to a mixed state $\rho^{AB}$ in the following
way~\cite{Werner89}:
\begin{definition}\label{definition-entangled}
  Let $\rho^{AB}$ be a mixed quantum state on $\cal H=\cal
  H_A \otimes \cal H_B$. $\rho^{AB}$ is called {\em separable}\index{quantum state!separable}
  if there exists an
  ensemble $\{p_i, |\psi_i \rangle^{AB}\}$ with $|\psi_i\rangle^{AB}=|\phi_i^A\rangle
  \otimes|\phi_i^B\rangle$ and
\begin{equation} \label{sep} \rho^{AB}=\sum_i p_i \proj{\psi_i}^{AB}
=\sum_i p_i |\phi_i^A\rangle \langle \phi_i^A|\otimes
|\phi_i^B\rangle \langle\phi_i^B|
\end{equation}
otherwise $\rho^{AB}$ is said to be {\em entangled}\index{quantum
state!entangled}. The set of separable states, $\sep$, is a convex
set and by Carath\'eodory's theorem\index{Carath\'eodory's
theorem} any separable state can be written as a convex
combination of no more than $(\dim \cH)^2$ product
states~\cite{Horodecki97}.
\end{definition}

\noindent In the context of quantum information theory it is
natural to extend this definition to more than two parties by
saying that $\rho$ on $\cH_1 \otimes \cdots \otimes \cH_n$ is
separable if it is a convex combination of projectors onto pure
product states $\ket{\phi_1} \otimes \cdots \ket{\phi_n}$ and
entangled otherwise. Klyachko has reached a different
understanding of entanglement in the context of Geometric
Invariant Theory (GIT)~\cite{MuFoKi94}, where he defines an
entangled state as a semistable vector. His definition coincides
with the one given here in the case of a bipartite system, but
differs for systems made up of more than two
subsystems~\cite{Klyachko02}. This difference will be of no
concern to us since this work only deals with bipartite quantum
systems. Consequently, whenever I speak of entanglement I only
refer to bipartite entanglement.

There are two meaningful scenarios in which one can pose the
question of separability. i) The mathematical scenario: given a
description of a density matrix $\rho$ decide whether or not
$\rho$ is separable. ii) The physical scenario: given a quantum
system in state $\rho$ decide whether of not $\rho$ is separable.
Much effort has been invested in trying to answer these questions
and significant progress has been made. Milestones in the context
of the mathematical scenario include the Peres-Horodecci
separability criterion~\cite{Peres96, HoHoHo96} and the result by
Gurvits who showed the separability problem is computationally
intractable. More precisely, Gurvits showed that the weak
membership problem for separability is NP-hard~\cite{Gur02,
Gur03}. Algorithms for the separability problem have been proposed
and studied in~\cite{ITCE04, EHGC04, Ioannou05}. The physical
scenario can be reduced to the mathematical one with help of
estimation of quantum states, but also some clever direct ways to
decide the problem have been proposed (see e.g. ~\cite{HorEke02,
EMOHHK02, MHOKE03}).

From the first recognition of the entanglement phenomenon until
Bell's experiments, entanglement was of qualitative interest: does
$\kpsi$ violate local realism or does it not? With the advent of
quantum computation and quantum cryptography, however,
entanglement was turned into a resource that can be used to
perform real-life tasks such as teleportation or secret
communication. The question of deciding separability in the
mathematical setting therefore naturally extended to the
quantitative question: how much entanglement does $\rho$
contain?~\cite{BBPSSW96, BDSW96} Since different tasks require
different measures, the answer to the above question cannot be
unique. A striking demonstration of this fact was given by
Pawe{\l} Horodecki, who proved the existence of \emph{bound
entangled states}, i.e.~states from which no Bell states can be
extracted but which nevertheless require Bell states for their
construction~\cite{Horodecki97}.

\subsection{Correlations in Bipartite Quantum States}
\label{section-correlations-bipartite}

With the advent of quantum information theory, quantum systems
were looked at from an information-theoretical viewpoint.
Schumacher was the first to consider the quantum analogue of
classical data compression and posed the following question: given
an i.i.d.~quantum information source, i.e.~a source with signal
states drawn independently and with identical distribution from a
$\rho$-ensemble $\{p_i, \ket{\psi_i}\}$, at what rate is it
possible to reliably compress this source? The answer turned out
to be the von Neumann entropy of $\rho$, thus paralleling the
Shannon entropy of a random variable in the classical
case~\cite{Schumacher95, JozSch94}.

The scenario of quantum data compression is equivalent to the
transmission of classical information through a noiseless quantum
channel. In the same way that classical information theory studies
the conversion of noisy classical channels to noiseless classical
channels via coding theorems, quantum information theory deals
with questions of resource conversion in the quantum
realm~\cite{NieChu00Book}.

In this chapter, I will explore the quantum analogue to classical
correlation, focusing on the interconversion of noisy correlations
to noiseless ones as well as the converse task of simulating noisy
correlations with noiseless ones. The noiseless resource par
excellence is the maximally entangled state, a pure quantum state
of the form
    $$ \kpsi=\frac{1}{\sqrt{d}}\sum_{i=1}^d \ket{i}\ket{i}.$$
Frequently, local operations will be regarded as free of cost. The
state $\kpsi$ is therefore not unique, but just one representative
of the set of maximally entangled states, which is given by the
orbit of $\kpsi$ under local unitary operations $\{ U^A \otimes
U^B \kpsi \; | \ U^A \in \unitaries (\cH_A) \mbox{ and } U^B \in
\unitaries (\cH_B) \}$.

Maximally entangled states play an important role in quantum
information theory. An example is Nielsen's majorisation theorem,
which exhibits the maximally entangled state as the only state
which can be perfectly interconverted to any other pure state. In
the context of entanglement-based quantum key
distribution\index{quantum key distribution!entanglement-based},
maximally entangled states lead directly to secure bits, and the
teleportation of a qubit consumes exactly one maximally entangled
state of two qubits. Below, I introduce the primary scenarios that
guide us through the remainder of the chapter. Each scenario is
specified by a set of operations.

\begin{itemize}
\item {\bf Total Correlation (LOq)} \index{LOq}Alice and Bob are asked to approximately convert an
i.i.d.~sequence of one resource into an i.i.d.~sequence of another
resource with help of {\bf L}ocal {\bf O}perations and a sublinear
amount of noiseless bidirectional classical or {\bf q}uantum
communication. The noiseless resource are Bell states (maximally
entangled states of two qubits) and the noisy resource are (mixed)
quantum states.
\item {\bf Entanglement (LOCC)} \index{LOCC}Alice and Bob are asked to approximately convert an i.i.d.~sequence
of one resource into an i.i.d.~sequence of another resource with
help of {\bf L}ocal {\bf O}perations and an unlimited amount of
noiseless bidirectional {\bf C}lassical {\bf C}ommunication. The
noiseless resource is given by a Bell state and the noisy
resources are (mixed) quantum states. Some authors denote LOCC by
LQCC for {\bf L}ocal {\bf Q}uantum Operations and {\bf C}lassical
{\bf C}ommunication.
\item {\bf Secret Key (LOPC)}\index{LOPC} Alice and Bob are asked to approximately convert an i.i.d.~sequence
of one resource into an i.i.d.~sequence of another resource with
respect to a third part Eve. They are assisted by {\bf L}ocal {\bf
O}perations and an unlimited amount of noiseless bidirectional
{\bf P}ublic Classical {\bf C}ommunication. The public
communication is also accessed by Eve. The noiseless resource is a
secure state, i.e.~a state of the form $ \sum_{x=1}^d \frac{1}{d}
\proj{x}^A \otimes \proj{x}^B \otimes \rho^E$. Note that Eve's
state is independent of $x$. Below I will explain how this
scenario can be turned into an LOCC scenario with respect to a
different resource: the gamma states
(definition~\ref{def-gamma-secret}).
\end{itemize}

\noindent A number of other interesting scenarios have been
considered in the literature. In order to keep this review
concise, I have decided to restrict my attention to the above
three and refer the reader to a few references for resource
conversion under Positive-Partial-Transpose (PPT) preserving
operations (for a definition see
page~\pageref{PPT})~\cite{AEJPVD01, Rains01, AuPlEi02},
distillation of randomness~\cite{DevWin05} and destruction of
correlations~\cite{GrPoWi05}.

In all three scenarios it turns out that the rate of approximate
interconversion of an i.i.d.~sequence of pure quantum states
$\kpsi$ to Bell states is given by the \emph{entropy of
entanglement\index{entropy!of
entanglement}\label{entropy-of-entanglement}} $S(A)_\psi$, which
is denoted by $E(\kpsi)$. Curiously, the converse also holds and
the task of approximate generation of a sequence of $\kpsi$'s from
Bell states is given by same rate. This shows that $\kpsi$ and the
Bell state can be asymptotically and reversibly interconverted.
This question was first considered and answered in the
entanglement scenario~\cite{BBPS96}; for the LOq and LOPC
scenarios see~\cite{THLD02} and~\cite{DevWin05}.

The interconversion of mixed quantum states in all three scenarios
turns out to be more involved and leads to the distinction between
classical and quantum correlation. The discussion will be carried
out separately for the three scenarios, but before we start, some
general remarks on conversion protocols are necessary. $\rho$ is
said to be interconvertible to $\sigma$ at rate $R$ with the class
of operations $C$ if for all $\epsilon>0$ there is an $m_0 \equiv
m_0(\epsilon) \in \mathbb{N}$ such that for all $m\geq m_0$ there
is a sequence $n \equiv n(m)$ and operations $\Lambda_{n} \in C$
such that $\delta(\Lambda_{n}(\rho^{\otimes n}), \sigma^{\otimes
\floor{Rn}}) \leq \epsilon$. The best rate is given by
    \be
R_{\min}= \lim_{\epsilon \rightarrow 0} \inf_{\Lambda^n \in C} \{
\floor{Rn}| \delta(\Lambda_{n}(\rho^{\otimes n}), \sigma^{\otimes
\floor{Rn}}) \leq \epsilon  \},
    \ee
where $\sigma=P_{\kpsi}$ is a Bell state and $\Lambda^n \in C$
denotes the sequence $\{\Lambda_{n(m)}\}_{m=m_0(\epsilon)}^\infty$
with $\Lambda_n \in C$ (see table~\ref{table-operational-measures}
for examples).

\begin{table}
\begin{center}
\begin{footnotesize}
\begin{tabular}{@{} lll @{}}
{\bf Correlation Measures  }                                  & {\bf Acronym}       & {\bf Definition } \\
\\ \hline \hline \\
Entanglement Cost\index{entanglement!cost (LOq)} (LOq)~\cite{THLD02}                                                 & $E_{LOq}$      & $\lim_{\epsilon \rightarrow 0} \inf_{\Lambda \in \loq} $\\
    &&\quad$ \{\frac{m}{n}| \delta(\Lambda(\ppsiminus^{\otimes n}), (\rho^{AB})^{\otimes m})\leq \epsilon\} $\\
Entanglement Cost\index{entanglement!cost}~\cite{HaHoTe01}                                                      & $E_C$          & $\lim_{\epsilon \rightarrow 0} \inf_{\Lambda \in \locc}$\\
    &&\quad$ \{\frac{m}{n}| \delta(\Lambda(\ppsiminus^{\otimes n}), (\rho^{AB})^{\otimes m})\leq \epsilon\}$ \\
Distillable Entanglement\index{distillable entanglement}~\cite{BDSW96}        & $E_D$          & $\lim_{\epsilon \rightarrow 0} \inf_{\Lambda \in \locc} $\\
    \quad aka Entanglement of Distillation &&\quad$\{\frac{m}{n}| \delta(\ppsiminus^{\otimes n}, \Lambda((\rho^{AB})^{\otimes m}))\leq \epsilon\}$ \\
Distillable Key\index{distillable key}~\cite{DevWin05, HHHO05a}               & $K_D$          & $\lim_{\epsilon \rightarrow 0} \inf_{\Lambda \in \locc} $\\
    \quad aka Secret Key Rate&&\quad$\{\frac{m}{n}| \delta(\Lambda((\rho^{AB})^{\otimes n}), \gamma^m)\leq \epsilon\} $\\
 \\ \hline\hline \\
\end{tabular}
\end{footnotesize}
\end{center}
\caption{Definitions of operationally defined entanglement
measures\index{entanglement measures!operationally defined}. LOq
are Local Operations assisted by a sublinear amount of quantum
communication. LOCC (LQCC) are Local (Quantum) Operations and
Classical Communication. LOPC are Local Operations and Public
Communication.} \label{table-operational-measures}
\end{table}

\subsubsection{Total Correlation}
\label{subsubsection-total correlation} Let us start with an
excursion to the classical case of two random variables $X$ and
$Y$. Naturally, one would say that $X$ and $Y$ are uncorrelated if
they are independent and maximally correlated if they are
identical. An operational measure that interpolates between these
two points is the \emph{mutual information\index{mutual
information}} $I(X;Y)$. The mutual information is defined as
$H(X)-H(X|Y)=H(X)+H(Y)-H(XY)$, where $H(\cdot)$ denotes the
Shannon entropy. Given an i.i.d.~sequence of pairs of random
variables $XY$, where Alice has access to $X$ and Bob to $Y$,
$H(X|Y)$ equals the minimal rate of communication from Alice to
Bob in order for Bob to reconstruct $X$ from $Y$ and the
communication. One can therefore say that the mutual information
quantifies the total correlations present in the pair $XY$.

A similar result for the total correlation of a quantum state
$\rho^{AB}$ has recently been achieved in a scenario where
classical communication comes for free~\cite{HoOpWi05}. The total
correlation is now measured by the \emph{quantum mutual
information} $I(A;B)=S(A)-S(A|B)$. $S(A|B)$ is the conditional von
Neumann entropy and corresponds to the amount of quantum
information which is needed to transfer the state of Alice's
system to Bob while preserving the correlations with the
environment. Note that $S(A|B)$ could be negative, in which case
the transfer would result in a rate of $S(A|B)$ Bell states that
could later be used for other transmissions. One may take this as
justification that the quantum mutual information $I(A;B)$ is the
correct way of quantifying the total correlations. Support for
this view also comes from the fact that $I(A;B)$ vanishes if and
only if $\rho^{AB}$ is of the product form $\rho^A \otimes
\rho^B$. Recently, a destructive method has been presented where
it is shown that the amount of local randomness needed to destroy
the correlation equals $I(A;B)$~\cite{GrPoWi05}.

In this chapter, however, a different approach to quantify total
correlation is in the centre of attention. Namely, the conversion
of Bell states to quantum states $\rho$ under local operations and
a sublinear amount of quantum communication (LOq). This problem
had first been considered by Barbara Terhal, Micha\l\ Horodecki,
Debbie Leung and David DiVincenzo~\cite{THLD02}. In this paper,
the authors define
\begin{itemize}
\item \emph{entanglement cost under LOq} as the rate of conversion
of singlets to mixed states in the total correlation scenario,
which formally reads as
    $$
        E_{LOq}(\rho^{AB})=\lim_{\epsilon \rightarrow 0} \inf_{\Lambda \in \loq}\{\frac{m}{n}| \delta(\Lambda(\ppsiminus^{\otimes n}), (\rho^{AB})^{\otimes m})\leq
\epsilon\}.
    $$
\end{itemize}
The goal of information theory is the calculation of such a rate
in terms of a single-letter formula. Here, such a formula has been
found, unfortunately, still containing a regularisation:
\bestar
    E_{LOq}(\rho^{AB})= \lim_{n\rightarrow
\infty} \frac{1}{n} E_P\big((\rho^{AB})^{\otimes n}\big).
\eestar
$E_P$ is a correlation measure called the \emph{entanglement of
purification\index{entanglement!of purification}} and defined as
\bestar
    E_P(\rho^{AB})=\min_{\rho^{ABE} \in \ext(\rho^{AB})} S(AE),
\eestar
where the minimisation is performed over all extensions of
$\rho^{AB}$, i.e.~over all states $\rho^{ABE}$ with $\tr_E
\rho^{ABE}=\rho^{AB}$. \emph{A priori} this minimisation is very
difficult as it extends over an unbounded space. In the case of
entanglement of purification, the concavity of the conditional von
Neumann entropy and a theorem by Choi on extremal maps help to
reduce the problem to system extensions of bounded dimension:
$\dim \cH_E\leq \dim \cH_{AB}$. Still, however, a calculation
seems elusive and only numerical upper bounds for a class of
\emph{Werner states}~\cite{Werner89} have been
given~\cite{THLD02}. In subsection~\ref{sec-specific-purification}
I present the calculation of entanglement of purification for
states with support on the symmetric and antisymmetric subspace,
the first nontrivial calculation of $E_{LOq}$ and $E_P$.
Entanglement of purification also provides a formal connection to
squashed entanglement, which is introduced in
chapter~\ref{chapter-squashed}: squashed entanglement contains a
minimisation similar to the one in entanglement of purification
(see table~\ref{table-non-operational-measures}).

\subsubsection{Entanglement}

\begin{quote}
``When two systems, of which we know the states by their
respective representatives, enter into temporary physical
interaction due to known forces between them, and when after a
time of mutual influence the systems separate again, then they can
no longer be described in the same way as before, viz.~by endowing
each of them with a representative of its own. I would not call
that one but rather the characteristic trait of quantum mechanics,
the one that enforces its entire departure from classical lines of
thought. By the interaction the two representatives [the quantum
states] have become entangled.''
\end{quote}

\begin{quotewho}
        Erwin Schr\"odinger, in ``Die gegenw{\"a}rtige \linebreak Situation der
        Quantenmechanik''~\cite{Schroedinger35b} (translation in~\cite{WheelerZurek83}).
\end{quotewho}

\noindent As explained above, Bell states play an important role
in the basic quantum protocols for the tasks of teleportation, key
distribution and superdense coding. But what if the given resource
is not a tensor product of Bell states but a sequence of general
quantum states? How often can the desired task be executed per
quantum state? The arising rates have been studied for each task
individually; here, however, I only focus on the universal lower
bound -- the number of Bell states that can be extracted per
quantum state -- and the universal upper bound -- the rate of Bell
states needed to construct a sequence of states. As the above list
of protocols excludes the use of quantum communication, the
appropriate class of operations are local operations assisted by
classical communication (LOCC). The rates are known as:

\begin{itemize}
\item \emph{distillable entanglement}, the rate of conversion of
mixed states to singlets in the entanglement scenario.
\item \emph{entanglement cost}, the rate of conversion of singlets
to mixed states in the entanglement scenario.
\end{itemize}
The formal definitions are given in
table~\ref{table-operational-measures}.

\subsubsection{Secret Key}
\label{subsubsec-secret-key}

Ekert's discovery of entanglement-based quantum key
distribution\index{quantum key distribution!entanglement-based}
marks the start of the investigation of the connection between
security and bipartite quantum states~\cite{Ekert91}. It was soon
realised that the earlier quantum key distribution
protocol\index{quantum key distribution} by Bennett and
Brassard~\cite{BenBra84} can be cast in this way~\cite{BeBrMe92}
and has led to an interesting proof-technique based on
entanglement distillation~\cite{DEJMPS96, ShoPre00, Inamori00,
LoCha99, TaKoIm03}, which is the guiding principle when quantum
key distribution is efficiently extended to arbitrary lengths by
means of quantum repeaters~\cite{DBCZ99}.

Here, we are not concerned with the full real-life scenario in
which the eavesdropper provides Alice and Bob with a quantum state
of many particles from which they try to extract a secret key.
Rather, an information-theoretic scenario is considered in which
Alice and Bob receive an i.i.d.~sequence of copies of a fixed
$\rho$, whose mathematical description they know and from which
they attempt to extract secret bits. It is immediate that
distillable entanglement is a lower bound to the

\begin{itemize}
\item \emph{distillable key}, the rate of conversion of mixed
states to secure states in the secret key scenario.
\end{itemize}

\noindent A \emph{secure state} (of length $\log d$) is a state of
the form
$$\frac{1}{d} \sum_{x}^{d} \proj{x}^A \otimes \proj{x}^B \otimes \rho^E, $$
where $\rho^E$ is independent of $x$~\cite{DevWin05} (see
definition~\ref{def-gamma-secret}). Secure states are the secrecy
resource analogue of a maximally entangled state in $d\times d$
dimensions. Note that secure states are not pure and that a
purification of the state is not accessible to either Alice, Bob
or Eve. The key distillation scenario is therefore a tripartite
rather than a bipartite scenario. The class of operations used to
perform the distillation will be denoted by LOPC, which stands for
{\bf L}ocal {\bf O}perations assisted by {\bf P}ublic classical
{\bf C}ommunication, i.e.~the classical communication of $C$ from
Alice to Bob, also reaches Eve:
    \bestar
    \begin{split}
        \rho^{C_AABE}=\sum_c p_c &\proj{c}^{C_A} \otimes \rho^{ABE}\\
        &\longmapsto \rho^{C_A C_B C_E ABE}=\sum_c p_c \proj{ccc}^{C_AC_BC_E} \otimes
        \rho^{ABE}.
    \end{split}
    \eestar
\sloppy A recent result by Karol, Micha{\l} and Pawe{\l} Horodecki
and Jonathan Oppenheim shows that the key distillation rate $K_D$
can in fact be strictly larger than $E_D$~\cite{HHHO05a}. More
precisely, they were able to present a bound entangled quantum
state from which they extracted a secret bit. To do so, they
reformulated the tripartite secure state scenario with LOPC
transformations into an LOCC scenario. In this scenario they
introduced a new resource the $\gamma$-states or private states,
which play the analogue of Bell pairs in the secret key scenario.
In contrast to the entanglement distillation scenario, where
distillation can be considered with respect to a single Bell
state, the $\gamma$-states form a whole class of states and lead
to the rigorous definition of distillable key as given in
table~\ref{table-operational-measures}. In the following I will
review the definitions of the secrecy resources in the LOCC and
LOPC scenario and also discuss the equivalence between the
scenarios. The reader is referred to the detailed account of the
results in~\cite{HHHO05a, HHHO05c}, in which the following
definition and theorem are contained.

\fussy \begin{definition}\label{def-gamma-secret} Let
$\kpsi=\frac{1}{\sqrt{2^m}}\sum_i \ket{i}\ket{i}$ be a maximally
entangled state in dimension $2^m$. Then any state of the form
$$ \gamma_m^{ABA'B'}=U \proj{\psi}^{AB} \otimes \rho_{A'B'} U^\dagger$$
for arbitrary unitaries
$$ U=\sum_{ij}^{2^m} \proj{ij} \otimes U^{A'B'}_{ij}$$
is called a \emph{(private) gamma state}\index{gamma state} of
length $m$. Any state of the form
$$ \gamma_m^{ABE}=\sum_x \frac{1}{2^m} \proj{xx} \otimes \gamma^E$$
will be called a \emph{private ccq state}\index{private ccq state}
of length $m$.
\end{definition}

\begin{theorem}\label{theorem-gamma-states}
Private gamma states and private ccq states are equivalent,
i.e.~for $\gamma^{ABA'B'}$ a private gamma state, $\gamma^{ABE}$
is a private ccq state for any purification $\gamma^{ABA'B'E}$ of
$\gamma^{ABA'B'}$. Conversely, for any private ccq state
$\gamma^{ABE}$ and any purification $\gamma^{ABA'B'E}$ of it,
$\rho^{ABA'B'}$ is a private gamma state.
\end{theorem}

\noindent The following security definition captures both
uniformity and security of a key simultaneously (cf.
\cite{RenKoe05, DevWin05}).

\begin{definition}\label{def-RK-ext}
Let $\rho^{ABA'B'}$ be a quantum state. We say that
$\rho^{ABA'B'}$ is an \emph{$\epsilon$-private gamma state} of
length $m$ if $\delta(\rho^{ABA'B'}, \gamma^{ABA'B'}_m) \leq
\epsilon$ for some private gamma state $\gamma^{ABA'B'}_m$.
\par
 Let $\rho^{ABE}$ be a ccq state. We say that
$\rho^{ABE}$ is an \emph{$\epsilon$-private ccq state} of length
$m$, if $\delta(\rho^{ABE}, \gamma^{ABE}_m) \leq \epsilon$ for
some private ccq state $\gamma^{ABE}_m$.
\end{definition}

\noindent \sloppy Below I show that the equivalence between
private ccq states and private gamma state (theorem
\ref{theorem-gamma-states}) also holds in an approximate sense
(see also~\cite[theorem 7]{HHHO05c}).

\fussy
\begin{corollary} \label{cor-gamma-states}
If $\rho^{ABA'B'}$ is an $\epsilon$-private gamma state of length
$m$, then Alice and Bob hold an $\sqrt{2\epsilon}$-private ccq
state of length $m$. Conversely, if Alice and Bob hold an
$\epsilon$-private ccq state of length $m$, then they hold an
$\sqrt{2\epsilon}$-private gamma state of length $m$.
\end{corollary}

\begin{proof}
\sloppy The assumption implies that $\sqrt{F(\rho^{ABA'B'},
\gamma^{ABA'B'})} \geq 1-\epsilon$. According to Uhlmann's
theorem, eq.~(\ref{eq-uhlmann}), there are purifications
$\rho^{ABA'B'E}$ and $\gamma^{ABA'B'E}$ of $\rho^{ABA'B'}$ and
$\gamma_m^{ABA'B'}$, respectively, obeying $\sqrt{F(\rho^{ABABE},
\gamma^{ABA'B'E})}\geq 1-\epsilon$. Monotonicity under partial
trace over $A'B'$ and under measurements on $A$ and $B$ in the
computational basis then imply
$$\sqrt{F(\tilde{\rho}^{ABE},
\tilde{\gamma}^{ABE})} \geq 1-\epsilon.$$ Note that
$$ \tilde{\gamma}^{ABE}= \sum_{x}\frac{1}{2^m} \proj{xx} \otimes
\gamma^E$$ is a private ccq state by theorem
\ref{theorem-gamma-states}. Finally (see Preliminaries,
eq.~(\ref{eq-fidelity-55}))
$$ \delta(\tilde{\rho}^{ABE}, \tilde{\gamma}^{ABE}) \leq \sqrt{1-F(\tilde{\rho}^{ABE}, \tilde{\gamma}^{ABE})}\leq
\sqrt{2\epsilon},$$ which concludes the proof of the first part.
\par
\sloppy Conversely, if $\delta(\rho^{ABE}, \gamma^{ABE}) \leq
\epsilon$ there exist purifications with $$\sqrt{F(\rho^{ABA'B'E},
\gamma^{ABA'B'E})} \geq 1-\epsilon$$ and monotonicity under the
partial trace applied to system $E$ results in
$$\sqrt{F(\rho^{ABA'B'}, \gamma^{ABA'B'})} \geq 1-\epsilon$$ and
thus $\delta(\rho^{ABA'B'}, \gamma^{ABA'B'}) \leq
\sqrt{2\epsilon}.$
\end{proof}

\begin{definition}
\fussy Define the \emph{gamma distillable key} of a $\rho^{AB}$ as
the asymptotic ratio between $m$ and $n$, where $n$ is the number
of copies of $\rho^{AB}$ that can be converted into a gamma state
of length $m$, with asymptotically vanishing error. Formally,
\be K_D^\gamma(\rho^{AB})= \lim_{\epsilon \rightarrow 0} \sup_{\Lambda \in \locc} \{\frac{m}{n}| \delta(\Lambda((\rho^{AB})^{\otimes n}), \gamma^{ABA'B'}_m)\leq \epsilon\}. \ee
Likewise define the \emph{ccq distillable key} of a $\rho^{AB}$ as
the asymptotic ratio between $m$ and $n$, where $n$ is the number
of copies of $\rho^{AB}$ that can be converted into private ccq
states state of length $m$ by LOPC, with asymptotically vanishing
error. Formally,
\be K_D^{ccq}(\rho^{AB})= \lim_{\epsilon \rightarrow 0} \sup_{\Lambda \in \lopc} \{\frac{m}{n}| \delta(\Lambda((\rho^{AB})^{\otimes n}), \gamma^{ABE}_m )\leq \epsilon\}. \ee
\end{definition}

\noindent It follows from corollary~\ref{cor-gamma-states} that
the key distillation rates in both scenarios are identical.

\begin{corollary}
$$ K_D^\gamma(\rho^{AB})=K_D^{ccq}(\rho^{AB}).$$
I therefore drop the superscript and write $K_D(\rho^{AB})$.
\end{corollary}
This concludes the remarks on the secret key, which are mainly
used in chapter~\ref{chapter-squashed} where it is shown that
squashed entanglement is an upper bound to $K_D$.

\subsection{Measuring Entanglement}
\label{section-measuring-entanglement}

Apart from entanglement cost under LOq, all distillation rates
introduced in the previous section are \emph{entanglement
measures\index{entanglement measures}}, i.e.~they assign a
nonnegative number to each quantum state and vanish on separable
states. I use the term entanglement measure in this loose sense,
as there is no commonly agreed definition in the literature.
Entanglement cost under LOq does not qualify as an entanglement
measure, since it does not vanish on all separable states but only
on product states. Measures which behave this way will be called
\emph{correlation measures\index{correlation measure}}. Essential
to the quantities introduced in the previous section is
\emph{monotonicity}, i.e.~they are nonincreasing under their
respective class of operations. Entanglement of purification is a
\emph{LOq monotone} as it can only decrease under LOq, whereas
entanglement cost, distillable entanglement and also distillable
key are \emph{entanglement monotones\index{entanglement!monotone}}
or \emph{LOCC monotones\index{LOCC!monotone}} as they do not
increase under LOCC.

In addition to being monotones, the discussed measures have been
shown to obey a variety of other properties. This led to a
property-driven approach to entanglement measures, which focused
on the identification of good and bad properties as well as the
construction of numerous examples (see
table~\ref{table-non-operational-measures}). I refer to this
approach as the \emph{axiomatic approach\index{entanglement
measures!axiomatic approach}} to entanglement
measures.\footnote{Even though, strictly speaking, entanglement
cost under LOq and entanglement of purification are not
entanglement measures, they can be studied within the axiomatic
approach to entanglement measures, acknowledging that LOCC
monotonicity fails.} The judgement of good and bad followed the
guidance of the operational measures, i.e.~the measures that can
be defined in terms of a rate function and thus included the
investigation of additivity, LOCC monotonicity and continuity (for
a complete list see table~\ref{table-definitions}). This shows how
the axiomatic approach is rooted in but yet transcends the
\emph{operational approach\index{entanglement measures!operational
approach}}, which restricts its attention to operationally defined
entanglement measures. Taking a more pragmatic view, one observes
that most entanglement measures bound distillable entanglement
from above. Therefore, the study of distillable entanglement
itself can serve as justification to delve into the axiomatic
approach.

The next section contains an extensive review of the axiomatic
approach, with emphasis on the connection to operational measures.
With a few exceptions the review therefore excludes most distance
measures, as their connection to the operational approach is weak
or not present. A review of entanglement measures with more
emphasis on distance measures will be included in the forthcoming
book by Ingemar Bengtsson and Karol \.Zyczkowski on the geometry
of quantum states~\cite{BengZycz06Book}. For a function $d:
\states \times \states \rightarrow \mathbb{R}^+$, the
corresponding $\emph{distance measure\index{distance measure}}$
$E_d$ is defined as
    \bestar \label{meas-distance}
        E_d(\rho)=\inf_{\sigma \in \states'} d(\rho, \sigma),
    \eestar
where $\states'\subset \states$. Note that it is not required that
$d$ is a \emph{distance} in the mathematical sense. The distance
measures that have proved most useful in entanglement theory are
the ones based on the \emph{relative entropy $S(\rho||\sigma)$}
(see Preliminaries, page~\pageref{relative-entropy}). They are in
multiple ways connected to the operational and the axiomatic
approach and are therefore included in the review below. The most
extensively studied distance measure not based on relative entropy
is the \emph{robustness of entanglement\index{robustness of
entanglement}}~\cite{VidTar99}. $\states'$ is taken to be the set
of separable states and the distance function of this measure is
defined as the minimal $s$ such that $\frac{1}{1+s}(\rho+s
\rho_s)$ is separable.

The next section will develop the axiomatic approach further,
introduce the properties and discuss these in the context of the
most important measures.

\section{The Zoo of Entanglement Measures\index{zoo!of entanglement measures}}\label{section-axiom-approach}

\subsection{Introduction}
The aim of the axiomatic approach is to find, classify and study
all functions that capture our intuitive notion of what it means
to measure entanglement. The approach sets out axioms,
i.e.~properties, that an entanglement measure should or should not
satisfy. As discussed in the previous section, this intuitive
notion may be based on more practical grounds such as operational
definitions. The most striking applications of the axiomatic
approach are upper and lower bounds on operational measures such
as distillable entanglement, entanglement cost and most recently
distillable key.

First of all, however, the axiomatic approach has resulted in a
whole lot of different entanglement measures, each satisfying a
certain subset of the large number of properties listed in
table~\ref{table-properties}. In analogy to Scott Aaronson's
\emph{complexity zoo\index{zoo!complexity}}~\cite{ComplexityZoo}
and the \emph{particle
zoo\index{zoo!particle}}~\cite{ParticleZoo}, I was therefore
tempted to name this section the \emph{zoo of entanglement
measures}. In its very first edition the beasts are being tamed
with three parts

\begin{itemize}
\item List of definitions, table~\ref{table-definitions}
\item Property table, table~\ref{table-properties}
\item Relations graph, figure~\ref{figure-relations} and~\ref{figure-relations-qubits}.
\end{itemize}
Setting up this zoo serves in fact a double purpose: firstly, it
gives a structure to the many species of entanglement measures,
and secondly, it compresses the historical background for squashed
entanglement, the topic of chapter~\ref{chapter-squashed}.

The outline of this section is as follows. In
subsection~\ref{subsec-properties} the main functional properties
that have been considered in the literature of entanglement
measures are presented. The main part of this section,
subsection~\ref{subsec-property-table}, contains a table with the
data of eighteen properties of eleven entanglement measures as
well as a graph, which shows a tree of relations between the
different measures. The next subsection,
subsection~\ref{subsec-relations-among}, is devoted to explain
patterns in the table, i.e.~the less obvious relations among those
properties. Finally, the last part of this section,
subsection~\ref{subsec-between}, discusses the universal property
of entanglement cost and entanglement of distillation as well as
the uniqueness theorem for entanglement measures.

\subsection{Properties}
\label{subsec-properties}

Virtually every paper that introduced a new entanglement measure
has set out a number of properties that can be regarded as
natural. These were subsequently proved, disproved or conjectured
for the newly defined measure. Rather than judging whether or not
a property is natural or desirable, I have compiled a list of most
properties that have been discussed in the literature (table
\ref{table-definitions}). I now briefly discuss a few of the
entries in the table, complementing and explaining the definitions
in the table.

Property Norm requires a measure to be normalised on maximally
entangled states, thereby capturing the notion of entanglement in
basic quantum protocols: teleportation, quantum key distribution
and superdense coding, which take as resource maximally entangled
states. Property Van Sep demands that the entanglement measure
vanishes on separable quantum states, thereby
 essentially discriminating between a measure
of entanglement and a correlation measure. Many authors demand in
addition that there should exist an entangled state on which the
measure is strictly positive. The deeper motivation might be
similar to the one of invariant theory: to distinguish different
-- possibly topologically different -- objects. A more practical
consequence is that one excludes the trivial measure, i.e.~the
measures which are equal to zero on the whole of state space, from
the set of entanglement measures. All concrete functions under
consideration will satisfy Norm and are thus nonzero on maximally
entangled states; this subtlety in the definition of Van Sep will
therefore make no difference to us.

Property Norm together with monotonicity connects the axiomatic
approach with the operational approach. Monotonicity under a class
of operations has been informally introduced in
section~\ref{section-measuring-entanglement}. Formally, $E$ is
\emph{monotone} under a class of operations O (O Mon) if for any
operations in O, which sends $\rho$ to an ensemble $\{p_i,
\rho_i\}$ (which is in general not a $\rho$-ensemble)
$$ E(\rho) \geq \sum_i p_i E(\rho_i).$$
The smallest class of operations relevant here is the class of
{\bf Loc}al operations (Loc). A local operation is given by a
local operation on Alice's side and an independent local operation
on Bob's side. A local operation on, for instance, Alice's side is
a quantum instrument
$$ \rho \mapsto \{p_{i}, \rho_{i}\}$$
where
$$
\rho_{i}= A_i \otimes \id_B(\rho)/ p_{i}  \qquad p_{i} = \tr
A_i\otimes \id_B(\rho)
$$
and each $A_i$ is a CP map such that $\sum_i A_i$ is a CPTP map.
{\bf L}ocal {\bf O}perations and {\bf C}lassical {\bf
C}ommunication (LOCC) consist of finite sequences of operations in
Loc intertwined with classical communication, i.e.~transfer of the
index $i$ from Alice to Bob
    $$
        \sum_i p_i \proj{i}^{A'} \otimes \rho^{AB}_i \longmapsto  \sum_i p_i \proj{i}^{A'}\otimes \proj{i}^{B'} \otimes
        \rho^{AB}_i
    $$
and vice versa (see~\cite{DoHoRu02} for a parametrisation of an
LOCC operation with $n$ rounds of classical communication). LOCC
operations are contained in the (strictly larger~\cite{BDFMRSS99})
set SEP of {\bf SEP}arable operations. A separable operation
transforms $\rho$ into an ensemble $\{p_i, \rho_i\}$, where
$$
\rho_i=A_i \otimes B_i(\rho)/p_i   \qquad p_{i} = \tr A_i\otimes
B_i(\rho)
$$
for $A_i\otimes B_i$ a CP map such that $\sum_i A_i \otimes B_i$
is a CPTP map. Experience shows that explicit calculations
involving minimisations over the classes LOCC or SEP are rather
difficult. In many cases a way out is provided by the class of
\label{PPT}{\bf P}ositive {\bf P}artial {\bf T}ranspose preserving
(PPT)\index{PPT} operations, which encompasses (but does not
equal) SEP. The partial transpose of a quantum state $\rho$ is the
transpose of the second system, given in terms of the matrix
elements $\rho_{ij,kl}$ of $\rho$, where $ij$ are the indices for
system $A$ and $kl$ are the indices for system $B$ by
$$ \rho \rightarrow (\rho^\Gamma)_{ij,kl} =\rho_{ij,lk}.$$
A PPT operation is a quantum operation that transform states whose
partial transpose is a positive operator (PPT states) into other
PPT states. Minimisations with respect to PPT operations have led
to excellent upper bounds on distillable
entanglement~\cite{AEJPVD01, Rains01}. Furthermore, there is the
hope that the theory of entanglement measures can be significantly
simplified if LOCC operations are replaced by PPT
operations~\cite{EiAuPl03}. LOq is not contained in any of the
above; LOq monotonicity will, however, be equivalent to Loc
monotonicity for measures satisfying asymptotic continuity (As
Cont).

\begin{table}[tp]
\begin{footnotesize}
\begin{center}
\begin{tabular}{@{}l|ll@{}}
Acronym     & Property          & Definition\\
\hline \hline
Norm        & normalised on  & For all $\kpsi=\frac{1}{\sqrt{\dim \cA}}\sum_i \ket{i}^A\ket{i}^B$,   \\
            & max. ent. states   & \qquad $E(\proj{\psi})=\log \dim \cA$ with $\{ \ket{i}^A\}$ o.n.\\
\hline
Van Sep     & vanishing on      &For all $\rho \in \sep$, $E(\rho)=0$\\
            & separable states &\\
\hline
PPT Mon     & PPT monotone      & For all $PPT$, $\rho \rightarrow \{p_i, \rho_i\}$, $E(\rho) \geq \sum_i p_i E(\rho_i)$.\\
SEP Mon     & SEP monotone     & For all $SEP$, $\rho \rightarrow \{p_i, \rho_i\}$, $E(\rho) \geq \sum_i p_i E(\rho_i)$.\\
LOCC Mon    & LOCC monotone     & For all $LOCC$, $\rho \rightarrow \{p_i, \rho_i\}$, $E(\rho) \geq \sum_i p_i E(\rho_i)$.\\
Loc Mon     & local monotone    & For all (strictly) local instruments\footnote{i.e.~an instrument that acts either on $\cA$ or $\cB$}: $\rho \rightarrow \{p_i, \rho_i\}$\\
            &                   & \qquad  $E(\rho) \geq \sum_i p_i E(\rho_i)$.\\
\hline
LOq Mon    & LOq monotone     & For all $LOq$, $\rho \rightarrow \{p_i, \rho_i\}$, $E(\rho) \geq \sum_i p_i E(\rho_i)$.\\
\hline
As Cont     & asymptotic     & There is $c, c'\geq 0$ s.th. for all $\rho, \sigma$ with $\delta(\rho, \sigma) \leq \epsilon$,\\
            & continuous                  & \qquad $|E(\rho)-E(\sigma)|\leq c \epsilon \log d + c'$\\
As Cont Pure & asympt. cont.   & There is $c, c'\geq 0$ s.th. for all $\rho, \sigma=\proj{\psi}$ \\
            & near pure states      & \qquad with $\delta(\rho, \sigma) \leq \epsilon$, $|E(\rho)-E(\sigma)|\leq c \epsilon \log d + c'$\\
\hline
Conv         & convex       & For all $\rho, \sigma$ and $p \in [0,1]$,\\
            &                   & \qquad $pE(\rho)+(1-p)E(\sigma) \geq E(p \rho+(1-p)\sigma)$\\
Conv Pure   & convex on       & For all $\{p_i, \ket{\psi_i}\}$ with $p_i \geq 0$ and $\sum_i p_i =1$,\\
            & pure states   & \qquad $\sum_i p_i E(\proj{\psi_i}) \geq E(\rho)$\\
\hline
Strong Super & superadditive & For all $\rho^{AA'BB'}$, $E(\rho^{AA',BB'})\geq E(\rho^{AB})+E(\rho^{A'B'})$\\
  \; \; Add \\
\hline
Add         & additive     & For all $\rho, \sigma$, $E(\rho \otimes \sigma)= E(\rho)+E(\sigma)$\\
Ext (Add i.i.d.) & extensive   & For all $\rho$ and $N$, $NE(\rho)= E(\rho^{\otimes N})$\\
\hline
Sub Add     & subadditive    & For all $\rho, \sigma$, $E(\rho\otimes \sigma)\leq E(\rho)+E(\sigma)$\\
Sub Add i.i.d. & subadditive i.i.d.   & For all $\rho$ and $m, n$, $E(\rho^{\otimes (m+n)})\leq E(\rho^{\otimes m})+E(\rho^{\otimes n})$\\
Regu        & regularisable   & For all $\rho$, the limit $E^\infty(\rho)=\lim_{n \rightarrow \infty} \frac{E(\rho^{\otimes n})}{n}$ exists\\
\hline
Non Lock    & not lockable & There is $c\geq 0$ s.th. for all $\rho^{AA'B}$, \\
            &                   & \qquad $E(\rho^{AA'B}) \leq E(\rho^{AB})+ c \log \rank \rho^{A'}$\\
\hline \hline \\
\end{tabular}
\end{center}
\end{footnotesize}
\caption{Definitions of Properties\index{entanglement
measures!definitions of properties}. In each separate group, the
truth of one property implies the truth of the property below.
Additional straightforward connections between the properties are
listed in proposition~\ref{prop-trivial} followed by more advanced
connections.} \label{table-definitions}
\end{table}

On the topic of continuity, let $\rho$ and $\sigma$ be two states
that are close in trace distance, $\delta(\rho, \sigma) \leq
\epsilon$. $\rho$ will behave like $\sigma$ with probability
$1-\epsilon$, in the sense that the bias of guessing $\rho$ and
$\sigma$ correctly is smaller than $\delta(\rho, \sigma)/2$.
Fannes' inequality passes this difference in trace distance on to
von Neumann entropies (lemma~\ref{lemma-fannes}),
$$ |S(\rho)-S(\sigma)|\leq 2 \epsilon \log d + \mu(\epsilon). $$
Asymptotic continuity is therefore a strong continuity requirement
expected to be satisfied by an entanglement measure.

The next property in table~\ref{table-definitions} is convexity
(Conv), and is probably the most controversial. Motivated by the
physical intuition that loss of knowledge about a quantum state
should decrease the entanglement, it can be written as
$$ E(\rho) \leq \sum_i p_i E(\rho_i),$$
where $\{p_i, \rho_i\}$ is a $\rho$-ensemble. Convexity, together
with local monotonicity, implies LOCC monotonicity
(proposition~\ref{Proposition-Vidal}); it therefore is an
important proof tool in entanglement theory. The converse,
however, is true only in connection with additional continuity
requirements. This will be discussed in the context of logarithmic
negativity (proposition~\ref{prop-conv}). There is also evidence
for the nonconvexity of distillable entanglement~\cite{ShSmTe01},
which would follow from \emph{superactivation}: $E_D(\rho \otimes
\sigma)
> 0$ for a PPT-bound entangled states $\rho$ and a bound entangled
state $\sigma$ with non-positive partial transpose (NPT). The
existence of NPT-bound entangled states has not yet been settled,
but candidates for NPT-bound entangled states that would lead to
superactivation have been conjectured~\cite{DCLB00, DSSTT00}.

Additivity questions are much talked about in quantum information
theory, last but not least because of Shor's proof of equivalence
of four major additivity questions: the additivity of the
classical capacity of a quantum channel, the additivity of the
minimum output entropy, the strong superadditivity (Strong Super
Add) of entanglement of formation and the additivity (Add) of
entanglement of formation~\cite{Shor03}. This is has been recently
extended by two more conjectures~\cite{Matsumoto05}.

The question of additivity of entanglement measures, such as
entanglement of formation is concerned with the behaviour of the
measure when several systems are tensored together. By definition
operationally defined measures are extensive (Ext), i.e.~they are
additive on tensor products as is indicated here for entanglement
cost:
$$ E_C(\rho^{\otimes n})=n E_C(\rho) \qquad \mbox{ for all } n \in \naturals \mbox{ and } \rho.$$
But already the truth of the additivity in general (Add)
$$ E_C(\rho\otimes \sigma)\stackrel{?}{=} E_C(\rho)+E_C(\sigma) \qquad \mbox{ for all } \rho \mbox{ and } \sigma$$
is an open question. In chapter~\ref{chapter-squashed} additivity
questions will be taken up again, as the importance of squashed
entanglement stems from its exceptional additivity properties.
Squashed entanglement satisfies strong superadditivity (Strong
Super Add) as well as additivity (Add), two properties that are
not known to hold simultaneously for any other measure.

Most non-operationally defined entanglement measures are given by
minimisations over certain sets of states
(cf.~table~\ref{table-non-operational-measures}) and are thus
subadditive. Let $E$ be such a measure, then it can easily be
turned into the extensive measure
$$
    E^\infty(\rho)=\lim_{n\rightarrow \infty} E(\rho^{\otimes n})/n,
$$
the \emph{regularisation\index{regularisation}} of $E$. The role
of regularisations in entanglement measures is two-fold. Firstly,
the operational measures $E_C$ and $E_{LOq}$ can be expressed as
the regularisation of $E_F$ and $E_P$, respectively. Secondly, if
$E$ is a subadditive measure, which provides an upper bound to
distillable entanglement, $E^\infty$ will improve this bound.
Unfortunately, regularisations are difficult to handle and only
few nontrivial calculations are known~\cite{AEJPVD01}. For
subadditive measures the existence of the regularisation is
guaranteed.

The youngest addition to the property table arose through the
observation that certain entanglement measures can be
\emph{locked}, i.e.~that there exist quantum states which, under
loss of a single qubit, can change their value by an arbitrary
amount~\cite{HHHO05b} (see corollary~\ref{acc-lock} for a related
effect occurring for the accessible information). This property
has entered the tables in the form of a converse, the
non-lockability of an entanglement measure (Non Lock). It can be
seen as a type of continuity with respect to tensor products.

\subsection{The Measures, their Properties and their Relations}
\label{subsec-property-table}

In addition to the operationally defined measures from
table~\ref{table-operational-measures}, a number of \emph{ad hoc}
definitions for entanglement measures will be considered (see
table~\ref{table-non-operational-measures}). I briefly go through
the list to make clear the connection among the different
measures, and to highlight the individual measure's significance.
The first on the list is entanglement of purification $E_P$,
$$
E_P(\rho^{AB})=\min_{\kpsi^{AA'BB'}: \tr_{A'B'}
\proj{\psi}^{AA'BB'}=\rho^{AB}} S(AA'),
$$
the minimum entropy of entanglement of the purifications of
$\rho^{AB}$. This definition is easily seen to be equivalent to
the definition in table~\ref{table-non-operational-measures}.
Since the conversion of Bell states into such a purification can
be done with a sublinear amount of classical
communication,\footnote{The entire protocol can be performed in a
number of steps, which is polynomial in the number of constructed
states and uses the Schur transform~\cite{HayMat04, BaChHa04}.}
and since the rate of Bell states is given by the entropy of
entanglement, it becomes clear that $E_P$ is an upper bound to
$E_{LOq}$. Moreover, its regularisation $E^\infty_P$ equals
$E_{LOq}$. No counter example is known to the conjecture that the
regularisation can be removed and $E_P$ equals $E_{LOq}$ (see also
subsection~\ref{subsubsection-total correlation}).

\begin{table}
\begin{center}
\begin{footnotesize}
\begin{tabular}{@{}lll @{}}
{\bf Correlation Measures  }                                  & {\bf Acronym}       & {\bf Definition } \\
\\ \hline \hline \\
Entanglement of Purification\index{entanglement!of purification}~\cite{THLD02}              & $E_P$        & $\min_{\rho^{ABE} \in \ext} S(AE)$\\
\\
Entanglement of Formation\index{entanglement!of formation}~\cite{BDSW96}                 & $E_F$        & $\min_{\{p_i, \rho_i\}\in \ens} \sum_i p_i S(A)_i$\\
\\
Relative Entropy of Ent. C\index{relative entropy!of entanglement}~\cite{VPRK97}                & $E_R^{C}$    & $\min_{\sigma \in \states_C} S(\rho||\sigma)$ \\
\\
Regularised Relative Entropy of Ent. C                  & $E_R^{C \infty}$ & $\lim_{n \rightarrow \infty} E_R^C(\rho^{\otimes n})/n$\\
\\
Reverse Relative Entropy C\index{reverse relative entropy of entanglement}~\cite{EiAuPl03}              & $E_{RR}^{C}$ & $\min_{\sigma \in \states_C} S(\sigma||\rho)$ \\
\\
Logarithmic Negativity\index{logarithmic negativity}~\cite{VidWer02}                  & $E_N$        & $\log |\rho^\Gamma|$\\
\\
Rains' Bound\index{Rains' bound}~\cite{Rains01}                              & $E_{Rains}$  & $\min_{\sigma \in \states} \left( S(\rho|| \sigma)+\log |\sigma^\Gamma|\right)$\\
\\
Squashed Entanglement\index{squashed entanglement}~\cite{ChrWin04}                   & $E_{sq}$     & $\inf_{\rho^{ABE}\in \ext} \half I(A;B|E)$\\
\\
 \hline\hline\\
\end{tabular}
\end{footnotesize}
\end{center}
\caption{Definitions of non-operationally defined entanglement
measures\index{entanglement measures!non-operationally defined}.
$\ext$ is the set of extensions of $\rho$, $\ens$ is the set of
ensembles of $\rho$ (Preliminaries, page~\ref{ensemble}). C is the
index of a set of states $\states_C$. The superscript will be
dropped if $\states_C$ is the set of separable states $\sep$.
$\rho^\Gamma$ denotes the partial transpose of $\rho$.}
\label{table-non-operational-measures}
\end{table}

In the same way that entanglement of purification arises from a
protocol for state construction from Bell states with LOq,
entanglement of formation arises from a protocol for state
construction from Bell states under LOCC. Let $\rho^{\otimes n}$
be the approximate state to be constructed. Then any ensemble
$\{p_i, \ket{\psi_i}\}$ of $\rho$ will lead to an LOCC protocol in
the following way. Firstly, for each $i$, construct a fraction of
$p_i N+o(N)$ states $\ket{\psi_i}$ from $E(\ket{\psi_i})p_i
N+o(N)$ Bell states. Secondly, permute the systems, disregard the
label $i$ and trace out over $o(N)$ of the systems. For large $N$,
these two steps will construct a state arbitrarily close to
$\rho^{\otimes N}$. Note that it is the last step, which requires
communication proportional to $N$. The rate of consumed Bell
states equals $\sum_i p_i E(\ket{\psi_i})$. The best protocol has
a rate given by $E_F(\rho)$, and it has been shown
in~\cite{HaHoTe01} that the regularised entanglement of formation
equals entanglement cost
$$ E^\infty_F(\rho)=E_C(\rho).$$
As discussed in subsection~\ref{subsec-properties} no counter
example to the conjecture $E_F(\rho)=E_C(\rho)$ is known to date.
A corollary to this conjecture is that $E_C(\rho)>0$ for all
entangled states, a fact which has recently been verified by a
direct argument~\cite{YHHS05}.

I do not present a comprehensive account of the instances in which
entanglement of formation has been calculated, but before
continuing, let me mention one milestone: the two-qubit formula by
Scott Hill and William K. Wootters~\cite{HilWoo97,
Wootters98}\label{wootters-formula}. For a state $\rho$ on
$\complex^2 \otimes \complex^2$,
\be E_F(\rho)=h \left( \frac{1+\sqrt{1-C^2}}{2}\right) \ee
where $h(\cdot)$ is the binary entropy function and $C(\rho)$ is
the \emph{concurrence\index{concurrence}}\footnote{The concurrence
$C(\rho)$ should not be confused with the index $C$ labeling a set
of quantum states $\states_C$.} of $\rho$. The concurrence is
defined in terms of the eigenvalues $\lambda_i$ of the `spin
flipped' matrix $\sqrt{\rho \sqrt{\tilde{\rho}} \sqrt{\rho}}$ as
$$C(\rho)=\max\{0, \lambda_1-\lambda_2-\lambda_3-\lambda_4\}.$$
\sloppy A natural interpretation of this formula has been obtained
in an invariant-theoretic context by Frank Verstraete, Jeroen
Dehaene and Bart De Moor~\cite{VeDeMo02}.

\begin{figure}
\begin{center}
\[
\xymatrix{
                &       & E_P \ar[d]^\neq \ar[dr]^{\stackrel{?}{=}}  \\
                &       & E_F \ar[dd]^\neq \ar[dr]^{\stackrel{?}{=}} & E_{LOq} \ar[d]^\neq \\
                &       &                         & E_C  \ar[ddl]^\neq   \ar[d]^\neq     \\
                &       & E_R \ar[d]^\neq \ar[dl]^\neq       &          E_{sq} \ar[dd]^\neq   &               \\
E_N \ar@{.}[uuuurr]^{\not\lessgtr} \ar[dr]^\neq & E_R^{PPT}\ar[dr] \ar[d]^{\stackrel{?}{=}}& E_R^\infty \ar[d]^\neq \ar[dr]& \\
                & E_{Rains}\ar[drr] \ar@{.}[r]^?& E_R^{PPT\infty}\ar[dr]         & K_D \ar[d]^\neq  & \\
                &       &                         & E_D
}
\]
\end{center}
\caption{Relations graph of entanglement
measures\index{entanglement measures!relations graph}. $E_A
\rightarrow E_B$ indicates that $E_A(\rho) \geq E_B(\rho)$ for all
$\rho$. Conjectures and strict inequalities are marked out as
well.} \label{figure-relations}
\end{figure}
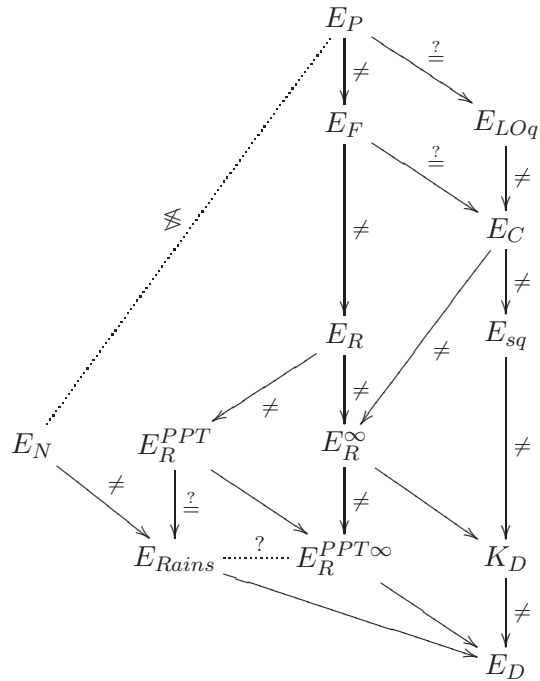

A whole collection of measures is based on the relative entropy.
\fussy Since these are distance measures, they are taken relative
to a set of quantum states $S_C$. Depending on the choice of this
set, they provide good upper bounds to key as well as entanglement
distillation rates (see figure~\ref{figure-relations}).
In~\cite{AuPlEi02} a subset of $\sep$ and $\ppt$ is considered.
The authors define the set of states with margins equal to the
margins of $\rho$, $\states_{M}(\rho)=\{\sigma \, | \,
\sigma^A=\rho^A \mbox{ and } \sigma^B=\rho^B\}$, and focus their
attention on the intersections of this set with $\sep$ or $\ppt$.
Note that the set $\states_M(\rho)$ is characterised precisely by
the spectral relations that have been the topic of
chapter~\ref{chapter-relation} and problem~\ref{prob-comp} in
particular. The resulting variants of $E^C_R$ remain interesting
in their own right. The audacious idea of reversing the entries of
the relative entropy results in the additive measure
$$E_{RR}^{C}(\rho) =\min_{\sigma \in \states_C \cap \states_{M}(\rho)} S(\sigma||\rho),$$
for $C \in \{PPT, SEP\}$. Unfortunately, this measure is not
continuous and diverges on pure states. In
table~\ref{table-properties}, I have restricted the attention to
the measures $E_R\equiv E^{SEP}_R$ and $E_R^\infty \equiv
E_R^{SEP\infty}$ as well as to $E_{RR}\equiv E_{RR}^{SEP}$; the
other cases are similar.

The next measure in the list, the logarithmic negativity, was
introduced as a `computable measure of
entanglement'~\cite{VidWer02}, and remains the only such measure
which is normalised on pure states and defined for states of all
dimensions\footnote{The logarithmic negativity also has a close
cousin, the negativity
$\cN(\rho)=\frac{|\rho^\Gamma|-1}{2}$~\cite{VidWer02}. It is an
LOCC monotone~\cite{Eisert01, VidWer02, Plenio05}, though not
normalised to $\log d$ on maximally entangled states in $d\times
d$ dimensions.}. The logarithmic negativity does not connect very
well to the approximate resource conversion scenario, which we
have taken here. This is indicated by the failing of convexity and
asymptotic continuity, and expresses itself clearly in the fact
that it does not coincide with the entropy of entanglement on pure
states. Rather than to approximate resource conversion, the
logarithmic negativity connects to an \emph{exact} resource
conversion scenario: $E_N$ is directly connected to the cost of
exactly prepare states with PPT operations~\cite{AuPlEi02,
Ishizaka04}.

Combining both relative entropies and the logarithmic negativity,
Rains' bound
$$
E_{Rains}(\rho)=\min_{\sigma \in \states} \left( S(\rho||
\sigma)+\log |\sigma^\Gamma|\right),
$$
is probably the best known upper bound for distillable
entanglement. Curiously, on Werner states it coincides with
$E^{PPT \infty}_R$, but no firm connection has been derived
between the two measures -- although one might consider the
two-qubit inequality by Ishizaka as biased
evidence~\cite{Ishizaka04}:
$$ E_R^\infty(\rho) \leq E_{Rains}(\rho) \mbox{ for } \rho \in \states(\complex^2 \otimes \complex^2).$$
This inequality is incorporated in
figure~\ref{figure-relations-qubits}. Note also that the
minimisation is performed over the whole state space with the
logarithmic negativity as penalty; hence it cannot be larger than
both $E^{PPT}_R$ and $E_N$.

The last measure in the list is squashed entanglement
$$E_{sq}(\rho)=\inf_{\rho^{ABE}\in \ext} \half I(A;B|E).$$
As mentioned previously, squashed entanglement has good additivity
properties. The proof of additivity along with squashed
entanglement's other properties will be given in
chapter~\ref{chapter-squashed}.


\begin{sidewaystable}
\begin{tiny}
\begin{tabular}{@{}l||l|l|l|l|l|l|l|l|l|l|l@{}}
 Measure    &  $E_{LOq}$ & $E_P$   & $E_C$   & $E_F$     & $E_D$    & $E_R$           & $E_R^\infty$     & $E_{RR}$     & $E_N$       & $E_{sq}$     & $K_D$\\
Main Ref.    &  \cite{THLD02} & \cite{THLD02}  & \cite{BDSW96} & \cite{BDSW96} & \cite{BDSW96}  & \cite{VPRK97}  &        & \cite{EiAuPl03} & \cite{VolWer01}& \cite{ChrWin04}  & \cite{HHHO05a}\\
            &            &         &   \cite{HaHoTe01} &              &  \cite{Rains99} &        &                 &                  &              &     &\cite{DevWin05}            \\
\hline \hline
Norm        &  y         & y       & y       & y         & y         & y               & y              & n            & y           & y             & y   \\
\hline
Van Sep     &  n         & n       & y        & y        & y         & y               &  y             & y              & y           & y           & y   \\
\hline
PPT Mon     &  n         & n       & n        & n        & ?         & n                & n             & ?              & y \cite{Plenio05}  & n    & n           \\
SEP Mon     &  n         & n       & ?        & ?        & ?         & ?                & ?             & ?              & y \cite{Plenio05}       & ?  & ?          \\
LOCC Mon    &  n         & n       & y        & y        & y         & y                & y              & y              & y \cite{Plenio05} & y  & y          \\
Loc Mon     &  y         & y       & y        & y        & y         & y               &  y             & y              & y                  & y    & y \\
\hline
As Cont     &  ?         & y       & ?        & y        & ?         & y                & y Prop.~\ref{prop-rel-ent-cont} & n              & n Cor.~\ref{cor-log-neg-as-cont}  & y \cite{AliFan04} & ?            \\
As Cont Pure&  y         & y       & y        & y        & y Prop.~\ref{prop-as-con-pure} & y & y Prop.~\ref{prop-as-con-pure} & n   & n Cor.~\ref{cor-log-neg-pure-cont}  & y  & y Prop.~\ref{prop-as-con-pure}          \\
\hline
Conv        &  n         & n       & y \cite{DoHoRu02} & y        & ?   &  y               &  y \cite{DoHoRu02} & y              & n \cite{VidWer02}  & y     &  ?      \\
Conv Pure   &  n         & n       & y        & y        & y \cite{DoHoRu02} & y       & y \cite{DoHoRu02}  & y              & ?           & y       &  y Prop.~\ref{prop-conv-pure}    \\
\hline
Strong Super &  ?         & ?       & ?        & ?, $\Leftrightarrow$ Add  & y         & n  \cite{VolWer01}  & ?                 & ?             & ?           & y     & y       \\
\quad  Add  &           &        &             & \cite{Shor03}       &                 &                    &                    &               &             &       &         \\
\hline
Add         &  ?         & ?       & ?        & ?        & ?         &  n  \cite{VolWer01}&  ?                & y              & y           & y         & ?  \\
Ext (Add i.i.d.)& y         & ?       & y        & ?        & y          &  n  \cite{VolWer01} & y               & y             & y           & y          & y  \\
\hline
Sub Add     &  y         & y       & y        & y        & ?         & y                &  y \cite{DoHoRu02}   & y              & y           & y        & ?    \\
Sub Add i.i.d. &  y         & y       & y        & y        & y         & y                &  y \cite{DoHoRu02}   & y              & y           & y        & y    \\
Regu        &  y         & y       & y        & y        & y         & y                &  y \cite{DoHoRu02}   & y              & y           & y        & y    \\
\hline
Non Lock    &  n \cite{ChrWin04}&n \cite{ChrWin04}& n \cite{HHHO05a}  & n \cite{HHHO05a} & ? & y \cite{HHHO05a} & y \cite{HHHO05a}  & ? & n \cite{HHHO05a} & n \cite{ChrWin05} & ?\\
\end{tabular}
\end{tiny}
\vspace{0.3cm} \caption{Properties of Entanglement
Measures\index{entanglement measures!properties of}. If no
citation is given, the property either follows directly from the
definition or was derived by the authors of the main reference.}
\label{table-properties}
\end{sidewaystable}


This concludes the presentation of the entanglement measures. A
few comments on the structure of the table of properties of
entanglement measures, table~\ref{table-properties}, and the graph
of relations, figure~\ref{figure-relations}, are still needed. The
table lists the status of the properties in
table~\ref{table-definitions} for the measures listed in
tables~\ref{table-operational-measures}
and~\ref{table-non-operational-measures}: `y' stands for `yes' the
measure satisfies the property, `n' stands for `no, it does not'
and `?' indicates that the status is unknown. References are given
next to the properties, and in the case where there is no
reference the property either follows from the definition or is
proven in the main reference.

The relations among the entanglement measures are displayed in
figure~\ref{figure-relations}. An arrow from $E_1$ to $E_2$,
$\xymatrix{E_1 \ar[r] &E_2}$, stands for
\bestar  E_1(\rho)\geq E_2(\rho) \qquad \mbox{ for all } \rho.\eestar
If in addition
\bestar E_1(\rho)\neq E_2(\rho) \qquad \mbox{ for some } \rho,\eestar
the table will show $\xymatrix{E_1 \ar[r]^{\neq}& E_2}$, and no
relation between two measures, i.e.~$E_1(\rho)>E_2(\rho)$ for some
$\rho$ as well as $E_2(\sigma)>E_1(\sigma)$ for some $\sigma$, is
written as $\xymatrix{E_1 \ar@{.}[r]^{\not\lessgtr}& E_2}$. By a
careful look one will also notice that measures based on relative
entropy as well as the distillation rates (lower left of the
diagram) are separated from the top right part. This separation
indicates that the value of the measures on the lower left can
differ by a large amount (on particular quantum states) when
compared to its respective value in the top right corner. This
effect is related to the locking of entanglement measures.

\subsection{Relations among the  Properties}
\label{subsec-relations-among} In this subsection, I discuss
general connections among the properties. The items listed in
table~\ref{table-definitions} have been grouped so that in each
group the truth of a property implies the truth of the property
below. A number of other obvious relations are stated in
proposition~\ref{prop-trivial}. For simplicity, the notation $A
\implies B$ should be interpreted as `the truth of $A$ implies the
truth of $B$'. Likewise, $A \notimplies B$ means `the truth of $A$
alone does not imply the truth of $B$'.

\begin{proposition} \label{prop-trivial}
For any entanglement measure $E$,
\begin{itemize}
\item LOq $\implies$ Loc
\item Add $\implies$ Sub Add
\item Ext~$\implies$ Sub Add i.i.d.~\item Sub Add i.i.d.~$\implies$ Regu
\item Strong Super Add and Sub Add $\implies$ Add
\item Strong Super Add and Sub Add i.i.d. $\implies$ Ext
\item LOCC Mon and Norm $\implies$ Van Sep.
\end{itemize}
\end{proposition}

\noindent Vidal has given an argument in \cite{Vidal00} that a
function $E$ satisfies LOCC Mon if and only if it satisfies Loc
Mon and Conv. Recently, however, it has been shown that the
logarithmic negativity $E_N$ is a counterexample to this claim of
equivalence: $E_N$ satisfies LOCC Mon~\cite{Plenio05} but fails to
obey property Conv~\cite{VidWer02}. This example seems to have
become possible due to $E_N$'s lacking of As Cont. Prior to the
results on logarithmic negativity, evidence for the non-convexity
of distillable entanglement have been put forward~\cite{ShSmTe01}.

Below I prove that Loc Mon and Conv together imply LOCC Mon.
Conversely, I show how LOCC Mon, together with As Cont and Ext,
leads to Conv.

\begin{proposition}[Vidal~\cite{Vidal00}] \label{Proposition-Vidal}
Conv and Loc Mon $\implies$ LOCC Mon.
\end{proposition}

\begin{proof}
Since $E$ is Loc Mon and every LOCC protocol can be written as an
alternating sequence of local operations and classical
communication, it remains to show that $E$ is non-increasing under
classical communication. Classical communication can be modeled as
\bestar
    \begin{split}
    \rho^{C_A Q_A Q_B}:=\sum_c p_c &\proj{c}^{C_A} \otimes \rho^{Q_A Q_B}_c \\
    &  \longmapsto  \rho^{C_A C_B
    Q_A Q_B}:=\sum_c p_c \proj{c}^{C_A} \otimes \proj{c}^{C_B} \otimes
    \rho^{Q_A Q_B}_c,
    \end{split}
\eestar
in analogy to the copying of a random variable $C$,
$$
C \rightarrow CC.
$$
It will now be shown that the conditions of the proposition imply
that $E(\rho^{C_A Q_A Q_B}) =E(\rho^{C_A C_B Q_A Q_B})$. One
direction is easily seen by Loc Mon
$$ E(\rho^{C_A Q_A Q_B}) \leq E(\rho^{C_A C_B Q_A Q_B}).$$
The relevant direction, however, is the opposite one. Loc Mon
implies
$$  E(\rho^{C_A Q_A Q_B}) \geq \sum_c p_c E( \rho^{Q_A Q_B}_c).$$
Both parties now append ancillas in the pure state $\ket{c}$ to
their system. Since appending and removing of a pure ancilla is a
local operation it holds that
$$ E(\rho^{Q_A Q_B}_c)=E(\rho^{Q_A Q_B}_c \otimes \proj{c}^{C_A} \otimes \proj{c}^{C_B}).$$
Finally, convexity gives
$$ \sum_c p_c E( \rho^{Q_A Q_B}_c \otimes \proj{c}^{C_A} \otimes \proj{c}^{C_B}) \geq E( \sum_c p_c \rho^{Q_A Q_B}_c \otimes \proj{c}^{C_A} \otimes \proj{c}^{C_B}),$$
which concludes the proof of
$$ E(\rho^{C_A Q_A Q_B}) \geq E(\rho^{C_A C_B Q_A Q_B}).$$
\end{proof}
Since LOCC Mon implies Loc Mon, it suffices to find the necessary
continuity requirements to also imply Conv. The proposition below
shows that asymptotic continuity, together with extensitivity and
subadditivity, is sufficient. In contrast, additivity requirements
alone cannot suffice. The question remains open to determine
whether or not additivity requirements are strictly necessary.

\begin{proposition} \label{prop-conv} For any entanglement measure
$E$,
\begin{itemize}
\item Sub Add, Ext, As Cont and LOCC Mon $\implies$ Conv
\item Sub Add,
Ext and LOCC Mon $\notimplies$ Conv.
\end{itemize}
\end{proposition}

\begin{proof}
Let $\rho$ and $\sigma$ be two density matrices on $\complex^d$.
It will be shown that the assumptions imply convexity, i.e.~for
all $p \in [0, 1]$,
    $$
        pE(\rho)+(1-p)E(\sigma) \geq E(\gamma) \mbox{ where } \gamma=p
        \rho+(1-p) \sigma.
    $$
Start by expanding $n$ copies of $\gamma$ binomially:
\beastar \gamma^{\otimes n}
    &=& \sum_k p^k (1-p)^{n-k} \binom{n}{k} \rho^{n}_p\\
    &=& \underbrace{\sum_{k: k/n \in \cB^{\epsilon}(p)} p^k (1-p)^{n-k}\binom{n}{k}\rho^{n}_p}_{=:\Gamma^n}  + \sum_{k: k/n \not\in \cB^\epsilon(p)} p^k (1-p)^{n-k} \binom{n}{k}
    \rho^n_p,
\eeastar
where $\rho^n_p =\frac{1}{n!} \sum_{\pi \in S_n} \pi
(\rho^{\otimes (n-k)}\otimes \sigma^k)$ and $\pi ( \cdot )$
denotes the permutation of the $n$ factors with $\pi$. Note that
for all $\epsilon
>0$ and large enough $n \equiv n(\epsilon)$
    \beastar
\delta_n:=\tr \sum_{k: k/n \not\in \cB^\epsilon(p)} p^k
        (1-p)^{n-k} \binom{n}{k} \rho^n_p &\leq& \sum_{k: k/n \not\in
        \cB^\epsilon(p)} e^{-n \ \min_{k: k/n \not\in \cB^\epsilon(p)}
        D(k/n||p)} \\
    &\leq& (n+1) e^{-n \frac{2}{\ln 2} \epsilon^2}\\
    &\leq& \epsilon,
    \eeastar
where the second line is Pinsker's inequality
(lemma~\ref{lemma-pinsker}). Hence
$\gamma^n:=\frac{1}{1-\delta_n}\Gamma^n$ obeys
$$ || \gamma^n-\gamma^{\otimes n} ||_1 \leq 3\epsilon$$
for $\epsilon \leq \half$ and asymptotic continuity implies
    \bestar
        |E(\gamma^{\otimes n})- E(\gamma^n)| \leq \delta(\epsilon) n \log
        d
    \eestar
for some $\delta(\epsilon)\rightarrow 0$ for $\epsilon \rightarrow
0$. Note that it is possible to construct $\gamma^n$ with
classical communication, and at most $\lceil(p+\epsilon)n\rceil$
copies of $\sigma$ and $\lceil(1-p+\epsilon)n \rceil$ copies of
$\rho$. Property Ext, LOCC Mon and Sub Add therefore show
\beastar n E(\gamma) &\stackrel{\textrm{Ext}}{=}& E(\gamma^{\otimes n}) \\
                 &\stackrel{\textrm{As Cont}}{\leq} &  E(\gamma^n)+ \delta(\epsilon) n \log d \\
                 &\stackrel{\textrm{LOCC Mon}}{\leq} &  E(\rho^{\otimes \lceil(p+\epsilon)n\rceil} \otimes \sigma^{\otimes \lceil(1-p+\epsilon)n \rceil}) + \delta(\epsilon) n \log d\\
                 &\stackrel{\textrm{Sub Add}}{\leq} &   \lceil(p+\epsilon)n\rceil E(\rho)\\
                 &&\qquad \quad +\lceil(1-p+\epsilon)n \rceil E(\sigma)+\delta(\epsilon) n \log
                 d.
\eeastar
Since $\epsilon$ was arbitrary, this implies $E(\gamma) \leq p
E(\rho)+(1-p) E(\sigma)$. Conversely, Sub Add, Ext and LOCC Mon do
not suffice to imply Conv. This can be seen in the example of
logarithmic negativity, which satisfies all of the assumptions but
is not convex (see subsection~\ref{subsec-log-neg}).
\end{proof}

\noindent If a measure does not satisfy Ext,
proposition~\ref{prop-conv} is not applicable. As stated in the
proposition below, one can at least assert that the regularisation
of $E$ is convex. The proof of this fact follows from inspection
of the proof of proposition~\ref{prop-conv}.

\begin{proposition}\label{prop-conv-infty}
If $E$ is Sub Add, As Cont and LOCC Mon, then $E^\infty$ is Conv,
whereas only Sub Add and LOCC Mon are not sufficient to conclude
that $E^\infty$ is Conv.
\end{proposition}
This concludes the general remarks about the relations among the
properties. The next section deals with the exceptional role of
entanglement cost and distillable entanglement.

\subsection{Between Distillable Entanglement and Entanglement Cost} \label{subsec-between} This section starts with the uniqueness of the entropy of
entanglement. For mixed states there is no such simple behaviour,
but the special role of distillation and cost rates remains in the
form of a betweenness theorem.

The entropy of entanglement $E(\kpsi)=S(A)_\psi$ equals the
minimal rate of Bell states required to construct a sequence of
$\kpsi$. $E(\kpsi)$ also equals the rate of Bell states that can
be extracted from a sequence of states $\kpsi$. But what is the
role of the entropy of entanglement when pure state entanglement
is regarded in the axiomatic approach? The answer is that
$E(\kpsi)$ is the unique measure as long as only Norm, LOCC Mon,
As Cont Pure and Ext are imposed~\cite[theorem 23]{DoHoRu02}.

\begin{theorem}[Uniqueness theorem for entanglement measures\index{entanglement measures!uniqueness theorem}]
\label{theorem-uniqueness} If $E$ is defined on pure states the
following two statements are equivalent
\begin{itemize}
\item[i)] Norm, LOCC Mon, As Cont Pure and Ext
\item[ii)] $E(\kpsi)=S(A)_\psi$,
\end{itemize}
where all properties are only demanded on pure states and an LOCC
operation takes pure states to pure states. Furthermore,
\begin{itemize}
\item[iii)] Norm, LOCC Mon and Ext
\end{itemize}
is not equivalent to i) and ii).
\end{theorem}
This uniqueness theorem does not hold for entanglement measures of
mixed quantum states, a fact which I explained earlier in this
text by referring to the difference between distillable
entanglement and entanglement cost. However, if Norm, LOCC Mon,
Ext and As Cont hold for all mixed states, then
    \be \label{eq-uniqueness}
        E_D(\rho) \leq E(\rho) \leq E_C(\rho) \qquad \mbox{ for all }
        \rho.
    \ee
Unfortunately, Ext is quite a strong requirement, which is often
either false or uncertain. It therefore turns out that squashed
entanglement is the only non-regularised, non-operationally
defined measure known to satisfy the assumptions and thus
eq.~(\ref{eq-uniqueness}). In the following, more details relating
to the sequence of inequalities~(\ref{eq-uniqueness}) will be
discussed, starting with the lower bound.
Propositions~\ref{prop-dist-lower}, \ref{prop-cost-upper} and
\ref{prop-cost-infty} can be found in the papers~\cite{HoHoHo00}
and \cite{DoHoRu02}.

\begin{proposition}
\label{prop-dist-lower} If $E$ satisfies Norm, LOCC Mon and As
Cont Pure, then
$$ E_D \leq E.$$
If in addition Regu is valid, then
$$ E_D \leq E^\infty.$$
\end{proposition}

\noindent All known ways to make entanglement cost an upper bound
need in addition extensitivity and at least to some degree of
continuity.

\begin{proposition}
\label{prop-cost-upper} Let
\begin{itemize}
\item[i)] Norm, LOCC Mon, Ext and As Cont
\item[ii)] Norm, LOCC Mon, Ext, As Cont Pure and Conv Pure
\end{itemize}
If $E$ satisfies i) or ii) then
$$ E \leq E_C$$
is true. Conversely, Norm, LOCC Mon and Ext are not sufficient to
draw this conclusion.
\end{proposition}

\noindent If $E$ does not satisfy property Ext, at least a
statement about the relation of $E^\infty$ and $E_C$ is possible.

\begin{proposition}
\label{prop-cost-infty} Let
\begin{itemize}
\item[i)] Norm, LOCC Mon and As Cont and Regu
\item[ii)] Norm, LOCC Mon, As Cont Pure, Sub Add and Conv Pure.
\end{itemize}
If $E$ satisfies i) or ii)
$$ E^\infty \leq E_C$$
holds. Conversely, Norm, LOCC Mon and Regu are not sufficient to
draw this conclusion.
\end{proposition}

\noindent The negative statements in
propositions~\ref{prop-cost-upper} and~\ref{prop-cost-infty} is a
consequence of the failing of As Cont for the logarithmic
negativity (see subsection~\ref{subsec-log-neg}). Opposite to the
previous propositions, one can also infer continuity properties
starting from a betweenness requirement; Fannes' inequality and
the hashing inequality make it possible.

\begin{proposition}\label{prop-as-con-pure}
Every measure $E$ with $E_D(\rho) \leq E(\rho) \leq S(A)_\rho$ is
As Cont Pure (and likewise when interchanging $A$ and $B$).
\end{proposition}

\begin{proof}
Let $\epsilon>0$ and $||\rho-\proj{\psi}||_1 \leq \epsilon$ for
some $\kpsi$. By assumption we have $-S(B|A)_\rho \leq E_D(\rho)
\leq E(\rho) \leq S(A)_\rho$, where the first inequality is the
hashing inequality proven in~\cite{DevWin05}. Since
$E_D(\psi)=S(A)_\psi$ one can now use Fannes' inequality
(lemma~\ref{lemma-fannes}) to conclude
$$ |E_D(\psi)-E_D(\rho)|\leq \delta \log d_A + \delta \log d_A d_B$$
for some $\delta\equiv\delta(\epsilon)\rightarrow 0$ as $\epsilon
\rightarrow 0$.
\end{proof}
Note that this proposition is not a consequence of
theorem~\ref{theorem-uniqueness}, which only shows that $E$ must
be asymptotically continuous \emph{on} pure states and not
necessarily in the mixed neighbourhood. A similar statement, now
for the property Conv Pure, has been derived in~\cite[Lemma
25]{DoHoRu02}.
\begin{proposition} \label{prop-conv-pure}
    Any quantity satisfying $E \leq E_C$ and $E(\psi)=S(A)_\psi$
    satisfies Conv Pure.
\end{proposition}
\begin{proof}
    $$E(\sum_i p_i \proj{\phi_i}) \leq E_C(\sum_i p_i \proj{\phi_i}) \leq \sum_i p_i E_C \proj{\phi_i})=\sum_i p_i E(\proj{\phi_i})$$
\end{proof}
Taken together, on the one hand
propositions~\ref{prop-as-con-pure} and~\ref{prop-conv-pure}
portray very accurately the connection of asymptotic continuity,
and on the other hand they also portray the uniqueness theorem and
the extremal positions of distillable entanglement and
entanglement cost. This concludes the subsection on the role of
entanglement cost and distillable entanglement, as well as their
restriction to pure states (the entropy of entanglement).

\section{Three Specific Correlation Measures}
\label{section-specific-results}

Here, I discuss recent progress on three selected correlation
measures. The first subsection presents a calculation of
entanglement of purification for a class of quantum states. It is
the first nontrivial calculation of its kind and also proves the
additivity of $E_P$ on the considered states. The second and third
subsections deal with continuity properties of the logarithmic
negativity and the regularised relative entropy of entanglement.
Logarithmic negativity can be used to illustrate both the
importance of asymptotic continuity in the uniqueness theorem for
entanglement measures and the betweenness property of entanglement
measures. The third subsection gives a proof of the asymptotic
continuity of the regularised entropy of entanglement with respect
to PPT as well as separable states.

\subsection[Entanglement of Purification]{Entanglement of Purification\protect\footnote{The result presented in this subsection has
appeared in~\cite{ChrWin05, ChrWin05Proc}.}}
\label{sec-specific-purification} In the following, I calculate
entanglement of purification for symmetric and antisymmetric
states. As the simplest instance of Schur-Weyl duality
(theorem~\ref{theorem-Schur-duality}), the space $\cH_{AB}=
(\complex^d)\otimes (\complex^d)$ of system $AB$ falls into two
parts, the symmetric and the antisymmetric space,
$$\mathbb{C}^d \otimes \mathbb{C}^d \cong \cH_{\rm sym} \oplus \cH_{\rm anti}.$$

\begin{proposition}
  \label{purification-prop}
  For all states $\rho^{AB}$ with support entirely within
 the symmetric or the antisymmetric subspace,
  $$E_{LOq}(\rho^{AB})=E_P^{\infty}(\rho^{AB})=E_P(\rho^{AB})=S(\rho^A).$$
  In fact, for another such state $\rho^{\prime AB}$,
  $$E_P(\rho^{AB}\otimes\rho^{\prime AB})
           = E_P(\rho^{AB}) + E_P(\rho^{\prime AB}).$$
\end{proposition}
\begin{proof}
  To every quantum state that is entirely supported on the symmetric
  subspace we can find a purification of the form
  $\ket{\Psi} = \sum_i \sqrt{p_i} \ket{\zeta_i}^{AB} \ket{\psi_i}^C$,
  with $F\ket{\zeta_i}=\ket{\zeta_i}$, where $F$ is the
  flip operator $F$ swapping the two systems.
  A similar form exists for states on the
  antisymmetric subspace,
  $\ket{\Psi} = \sum_i \sqrt{p_i} \ket{\alpha_i} \ket{\psi_i}$
  with $F\ket{\alpha_i}=-\ket{\alpha_i}$. Any other
  state extension of $\rho^{AB}$ can be obtained by the application of a CPTP map
  $\Lambda: C \longrightarrow E$, i.e.
  $$\rho^{ABE}=(\id\otimes\Lambda)\Psi^{ABC}.$$
  From the symmetry of $\rho^{AE}$ and $\rho^{BE}$, it immediately
  follows that $S(A|E)=S(B|E)$ and by weak monotonicity of the von
  Neumann entropy: $2 S(A|E)=S(A|E)+S(B|E)\geq 0$. Hence for
  every extension $\rho^{ABE}$, $S(AE)\geq S(A)$ holds with equality
  for the trivial extension.

  Another way of arriving at this conclusion is via the no-cloning
  principle. Assume that $\rho^{AE}$ is one-way distillable from Eve to
  Alice, i.e. distillable via local operations and one-way classical communication sent from Eve to Alice.
  Then, by symmetry, $\rho^{BE}$ is also one-way distillable from Eve to Bob, whereby Eve
  uses the same instrument and qubits
  for both directions. Hence, Eve would
  share the same maximally entangled state with both Alice and Bob,
  which is impossible by the monogamy of entanglement. By the
  hashing inequality \cite{DevWin05} vanishing one-way distillability
  implies $S(A|E)\geq 0$ and $S(B|E)\geq 0$ and the conclusion
  on $E_P$ follows.

  The same reasoning applies to a tensor product of a state
  $\rho$ supported on the \mbox{(anti-)}\-symmetric subspace with a state $\sigma$ supported on
  the \mbox{(anti-)}\-symmetric subspace. Additivity follows and therefore
    $$
    E_{LOq}(\rho)=E^\infty_P(\rho)=E_P(\rho)
    $$
  for such states.
\end{proof}

\noindent The above proof using monogamy and the hashing
inequality has
  the advantage of giving a slightly more general result: assume that
  for a purification $\kpsi^{ABC}$ of $\rho^{AB}$,
  $\rho^{AC}$ is not one-way distillable (from $C$ to $A$).
  Then for every channel $\Lambda: C \longrightarrow E$,
  $\rho^{AE} = (\id\otimes\Lambda)\rho^{AC}$ is still
  one-way nondistillable, hence $S(\rho^{AE}) \geq S(\rho^A)$,
by the hashing inequality.

The local monotonicity of entanglement of purification will lead
to a neat consequence of proposition~\ref{purification-prop},
namely to the fact that entanglement of purification can be
locked.

\begin{corollary}
  Let
  $$\rho^{A'\!AB}
        = p \proj{0}^{A'}\otimes\sigma^{AB} + (1-p) \proj{1}^{A'}\otimes\alpha^{AB},$$
  with states $\sigma$ and $\alpha$ supported on the symmetric
  and antisymmetric subspace, respectively.
  Then,
  $$E_{LOq}(\rho^{A'\!AB})=E_P^{\infty}(\rho^{A'\!AB}) = E_P(\rho^{A'\!AB})
                   \geq p S(\sigma^A) + (1-p) S(\alpha^A).$$
  In particular, we have
  $$E_{LOq}(\omega^{A'\!AB})=E_P(\omega^{A'\!AB}) = \log d \text{\ \ and\ \ \ }
    E_{LOq}(\omega^{AB})=E_P(\omega^{AB}) = 0$$
  for
  \begin{equation*}\begin{split}
    \omega^{A'\!AB} &= \frac{d+1}{2d}\proj{0}^{A'}\otimes\frac{2}{d(d+1)}P_{\rm sym}^{AB} \\
                    &\phantom{=}
                      + \frac{d-1}{2d}\proj{1}^{A'}\otimes\frac{2}{d(d-1)}P_{\rm anti}^{AB},
  \end{split}\end{equation*}
  with the projectors $P_{\rm sym}$ and $P_{\rm anti}$ onto the symmetric
  and antisymmetric subspace, respectively. This shows that both entanglement of
  purification as well as entanglement cost under
LOq do not satisfy Non Lock.
\end{corollary}
\begin{proof}
  Local monotonicity, a fact which can be easily verified, together with proposition \ref{purification-prop}
  implies that $E_P(\rho^{A'\!AB})\geq p S(\sigma^{AB}) + (1-p) S(\alpha^{AB})$,
  and the same for $E_P^\infty$.

  The dimensions of the symmetric and antisymmetric subspace are
  given by $\frac{d(d+1)}{2}$ and $\frac{d(d-1)}{2}$, respectively.
  The state $\omega^{A'\!AB}$ is constructed such that $\omega^{AB}$
  is maximally mixed on $AB$, with evidently zero
  entanglement of purification. On the other hand, by the above,
  \begin{equation*}\begin{split}
    E_P^\infty(\omega^{A'\!AB})
                   &\geq \frac{d+1}{2d}E_P\left( \frac{2}{d(d+1)}P_{\rm sym} \right)    \\
                   &\phantom{=}
                         + \frac{d-1}{2d}E_P\left( \frac{2}{d(d-1)}P_{\rm anti} \right) \\
                   &=    \log d.
  \end{split}\end{equation*}
  This bound is attained since
  $E_P(\omega^{A'\!AB}) \leq S(\omega^B) = \log d$.
\end{proof}

\noindent This concludes the first nontrivial analytical
calculation of entanglement of purification and entanglement cost
under LOq. It thereby also provides the first proof of nontrivial
additivity result for entanglement of purification and confirms
the numerical calculation of the totally symmetric state on
two-qubits~\cite{THLD02}.

\subsection{Logarithmic Negativity} \label{subsec-log-neg} The
logarithmic negativity~\cite{VidWer02}
    $$E_N(\rho)= \log | \rho^\Gamma|$$
is a quantity satisfying a large number of the properties in
table~\ref{table-properties}. Most interestingly, however, is the
failing of asymptotic continuity (As Cont), even near pure states
(As Cont Pure) and on pure states. This behaviour arises from the
non convexity of $E_N$ and will be used to illustrate
subsection~\ref{subsec-relations-among} and~\ref{subsec-between}.

Notice that by definition $E_N$ satisfies property Add and is thus
also Ext and Sub Add. By the work of Plenio we know further that
$E_N$ is LOCC Mon \cite{Plenio05}. So, if $E_N$ satisfied As Cont,
then one could apply proposition~\ref{prop-conv} and conclude that
$E_N$ is Conv. This, however, is not true as was already pointed
out in~\cite{VidWer02} and one is led to the conclusion that $E_N$
cannot be asymptotically continuous.

    \begin{corollary} \label{cor-log-neg-as-cont}
        $E_N$ is not As Cont.
    \end{corollary}
In~\cite{VidWer02}, $E_N$ has been calculated for pure states and
does \emph{not} coincide with the entropy of entanglement on pure
states.

    \begin{proposition} \label{prop-log-neg-unique}
        $E_N(\proj{\psi}) \geq S(A)_\rho$ with equality if and only if
        $\kpsi$ is a maximally entangled state.
    \end{proposition}
Therefore, as a corollary to proposition~\ref{prop-cost-upper} and
theorem~\ref{theorem-uniqueness} one finds the following stronger
statement.

    \begin{corollary} \label{cor-log-neg-pure-cont}
        $E_N$ is not asymptotically continuous
        near pure states (As Cont Pure) nor \emph{on}
        pure states.
    \end{corollary}
In the view of proposition~\ref{prop-dist-lower} one may wonder
how Vidal and Werner could prove that the logarithmic negativity
is an upper bound to distillable entanglement. The answer is in
fact hidden in the precise formulation of
proposition~\ref{prop-log-neg-unique}: it is sufficient to demand
`good' behaviour of a measure close to maximally entangled states
to ensure that it is an upper bound for distillable entanglement.
In contrast, this is not sufficient to imply a lower bound to
entanglement cost or an upper bound on distillable key.

As discussed, logarithmic negativity has good monotonicity and
additivity properties, whereas it fails to satisfy even the
weakest continuity bounds. This had already been used to show that
it is lockable~\cite{HHHO05b}. It follows from this analysis that
the logarithmic negativity demonstrates the necessity of the
continuity assumption in propositions~\ref{prop-conv},
\ref{prop-conv-infty}, \ref{prop-cost-upper},
\ref{prop-cost-infty} and theorem~\ref{theorem-uniqueness}.

\subsection{Regularised Relative Entropy of Entanglement}
\label{subsec-regularised}

In this section, I will show that the regularised relative entropy
is asymptotically continuous with respect to either the set of
separable states or the set of PPT states. It is the first proof
of asymptotic continuity for the regularisation of a measure,
which is not known to be extensive.

\begin{proposition} \label{prop-rel-ent-cont}
The relative entropy of entanglement $E^{C\infty}_R$ with respect
to a convex set $C$ that includes the maximally mixed state,
satisfies property As Cont. I.e.~there is a function
$\delta(\epsilon)$ with $\delta(\epsilon) \rightarrow 0$ for
$\epsilon \rightarrow 0$ such that for all $||\rho-\sigma||_1 \leq
\epsilon$
$$ |E^\infty_R(\rho)-E^\infty_R(\sigma)|\leq \delta(\epsilon) \log
d,$$ where $d$ is the dimension of the system supporting $\rho$
and $\sigma$. In particular this proves that $E^{PPT \infty}_R$ as
well as $E^\infty_R$ are asymptotically continuous.
\end{proposition}

\begin{proof}
Let $||\rho- \sigma||_1 =\epsilon>0$, where $\rho$ and $\sigma$
are $d$-dimensional states. According to Alicki and Fannes
\cite{AliFan04}, there are states $\gamma$, $\tilde{\rho}$ and
$\tilde{\sigma}$ with $\gamma=(1-\epsilon)\rho+\epsilon
\tilde{\rho}=(1-\epsilon) \sigma+\epsilon \tilde{\sigma}$. If we
succeed to prove asymptotic continuity on mixtures, i.e.~\be
\label{eq-rel-cont-mixt} |E^\infty_R(\rho)-E^\infty_R(\gamma)|
\leq \delta(\epsilon) \log d, \ee then continuity for $\rho$ and
$\sigma$ follows by use of the triangle inequality:
$$|E^\infty_R(\rho)-E^\infty_R(\sigma)| \leq |E^\infty_R(\rho)-E^\infty_R(\gamma)|+|E^\infty_R(\gamma)-E^\infty_R(\sigma)| \leq 2\delta(\epsilon) \log d.$$
The main step in the proof of the
estimate~(\ref{eq-rel-cont-mixt}) is the following inequality for
an ensemble $\{ p_i, \tau_i\}$,
    \be \label{ineq-relent-ineq}
        \sum_i p_i E_R(\tau_i) - E_R(\sum_i p_i \tau_i) \leq S(\sum_i
        p_i \tau_i) -\sum_i p_i S(\tau_i) \leq H(X)
    \ee
where the random variable $X$ has distribution $p_i$.
Inequality~(\ref{ineq-relent-ineq}) has first been proven for the
relative entropy with respect to the set of separable
states~\cite{LPSW99} (see also \cite{EFPPW00}). Very recently this
result has been extended to hold for any convex set that includes
the maximally mixed state~\cite{SynHor05}. Here, it implies the
following estimate
$$E_R(\gamma^{\otimes N}) \geq \sum_k \epsilon^k
(1-\epsilon)^{N-k} \binom{N}{k} E_R(\rho^{\otimes (N-k)} \otimes
\tilde{\rho}^{\otimes k})- N h(\epsilon), $$ where $h(\epsilon)$
is the Shannon entropy of the distribution $(\epsilon,
1-\epsilon)$. I will now replace all $\tilde{\rho}$'s on the RHS
by $\rho$'s. This is done in two steps: i) remove the states of
the form $\tilde{\rho}$ on the RHS, since the partial trace
operations is an LOCC operation the RHS can only decrease, ii)
append the states $\rho$ and apply the inequality
$$ E_R(\rho^{\otimes N}) \leq E_R(\rho^{\otimes (N-k)})+kE_R(\rho),$$
which holds by subadditivity of $E_R$. This gives
\beastar E_R(\gamma^{\otimes N})
        &\geq& \sum_k \epsilon^k (1-\epsilon)^{N-k} \binom{N}{k}
            E_R(\rho^{\otimes (N-k)} \otimes \tilde{\rho}^{\otimes k})- N h(\epsilon) \\
        &\stackrel{i)}{\geq}& \sum_k \epsilon^k (1-\epsilon)^{N-k} \binom{N}{k}
            E_R(\rho^{\otimes (N-k)} )- N h(\epsilon) \\
        &\stackrel{ii)}{\geq}& \sum_k \epsilon^k (1-\epsilon)^{N-k} \binom{N}{k}
            (E_R(\rho^{\otimes N})-k E_R(\rho) )- N h(\epsilon) \\
        &=& E_R(\rho^{\otimes N}) - \sum_k k \epsilon^k (1-\epsilon)^{N-k} \binom{N}{k} E_R(\rho)- N h(\epsilon) \\
        &=& E_R(\rho^{\otimes N}) - N\epsilon E_R(\rho)- N h(\epsilon) \\
        &\geq& E_R(\rho^{\otimes N}) - N (\epsilon \log d+ h(\epsilon) ) \\
        &\geq& E_R(\rho^{\otimes N}) - N (\epsilon + h(\epsilon))\log d.
\eeastar
The last equality sign is the evaluation of the mean value of the
binomial distribution. Since the above calculation holds for all
$N$, this shows
$$E_R^\infty(\gamma) \geq E_R^\infty(\rho)-\delta(\epsilon) \log d$$
for $\delta(\epsilon) :=\epsilon + h(\epsilon)$. Conversely, the
convexity of $E_R^\infty$ \cite{DoHoRu02} implies
$$E_R^\infty(\gamma) \leq (1-\epsilon) E_R^\infty(\rho)+\epsilon
E_R^\infty (\tilde{\rho}) \leq E_R^\infty(\rho) +\epsilon \log
d.$$ This concludes the proof of the
estimate~(\ref{eq-rel-cont-mixt}) and the proposition.
\end{proof}

\noindent A vital ingredient in the proof was
inequality~(\ref{ineq-relent-ineq}), which bounds the strength of
the convexity of the relative entropy. Prior to this work, the
same inequality has been used in~\cite{HHHO05b} to prove property
Non Lock for the relative entropy. As both entanglement of
purification and formation are lockable, a simple translation of
inequality~(\ref{ineq-relent-ineq}) to these measures is not
possible. Other ways to verify property As Cont for entanglement
cost under LOCC and LOq will have to be found -- if As Cont holds.

\section{Conclusion}
\label{section-conjectures-etc}

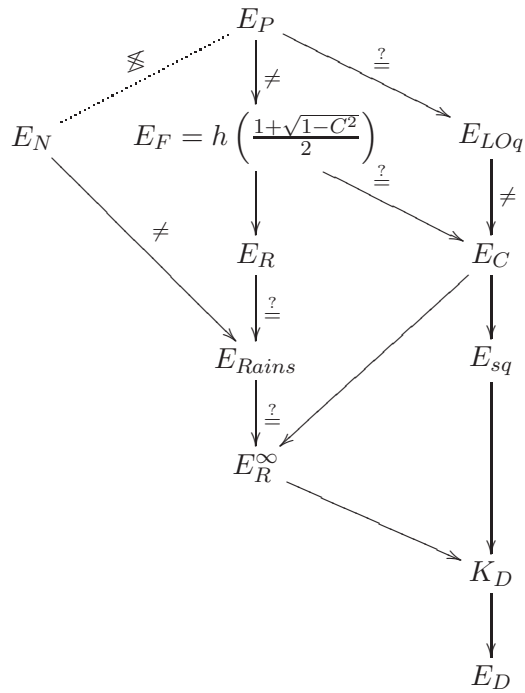
\begin{figure}
\begin{center}
\[
\xymatrix{
                &                       & E_P \ar[d]^\neq \ar[dr]^{\stackrel{?}{=}}  \\
                & E_N \ar@{.}[ur]^{\not\lessgtr} \ar[ddr]^\neq       & E_F=h \left( \frac{1+\sqrt{1-C^2}}{2}\right) \ar[d] \ar[dr]^{\stackrel{?}{=}}   & E_{LOq} \ar[d]^\neq \\
                &                       & E_R \ar[d]^{\stackrel{?}{=}}      & E_C  \ar[ddl]  \ar[d]     \\
                &                       & E_{Rains} \ar[d]^{\stackrel{?}{=}}          & E_{sq} \ar[dd] &               \\
                &                       & E_R^\infty   \ar[dr]                        & \\
                &                       &                                             & K_D \ar[d] & \\
                &                       &                                             & E_D
}
\]
\end{center}
\caption{Relations between entanglement measures when only
considered on states of two qubits. An arrow $E_A \rightarrow E_B$
indicates that $E_A(\rho) \geq E_B(\rho)$ for all $\rho$ on
$\complex^2\otimes \complex^2$. Through the restrictions to states
of two qubits the following simplifications arise when compared to
figure~\ref{figure-relations}. Since $\sep=\ppt$ holds for qubits,
$E_R=E_R^{PPT}$ as well as $E_R^\infty=E_R^{\infty PPT}$.
Furthermore one has the formula for $E_F$ by Hill and Wootters
(see page~\pageref{wootters-formula}) and the relation
$E_R^\infty(\rho) \leq E_{Rains}(\rho)$ by
Ishizaka~\cite{Ishizaka04}. Many examples that show the inequality
of two measures are constructions for higher dimensional systems.
Here, these are not applicable, which results in less ``$\neq$''
signs when compared to figure~\ref{figure-relations}.
\index{entanglement measures!qubit relations graph}}
\label{figure-relations-qubits}
\end{figure}

In this chapter I have given a review of the axiomatic approach to
entanglement measures. The bulk of the work is contained in
several tables and a graph, which are meant to tame the beasts in
this zoo of measures. Tables~\ref{table-operational-measures}
and~\ref{table-non-operational-measures} contain definitions of
measures, and table~\ref{table-definitions} introduces the
properties of entanglement measures. The main table,
table~\ref{table-properties}, contrasts measures with properties
and is to my knowledge the most comprehensive summary of its kind.

Many of the measures are connected to another in one way or the
other; a graph displaying the hierarchy in the zoo is contained in
figure~\ref{figure-relations}. As a bonus, I have included a graph
with the relations of measures two-qubits
(figure~\ref{figure-relations-qubits}).

Compiling table~\ref{table-properties} has unavoidably led to
contemplation about some of its question marks. As a result, I
have been able to prove the asymptotic continuity of the
regularised relative entropy of entanglement with respect to both
PPT and separable states (subsection~\ref{subsec-regularised}).
Alerted by a recent paper by Plenio~\cite{Plenio05}, I have also
decided to add subsection~\ref{subsec-log-neg} discussing the
continuity and convexity properties of logarithmic negativity,
which also illustrates the theorems in
subsection~\ref{subsec-relations-among}. Last but not least, this
chapter featured the first calculation of entanglement of
purification for a class of symmetric and antisymmetric states.

\chapter{Squashed Entanglement}
\label{chapter-squashed}

\section{Introduction}

The previous chapter provided a review of the theory of
entanglement and a number of existing entanglement measures. In
this chapter I will propose a new measure of entanglement called
\emph{squashed entanglement}. The focus will be put on the
cryptographic motivation of squashed entanglement, its properties
and the consequences for quantum information theory.

The chapter is structured as follows. In the introduction, I will
highlight the open questions that arose during the review of
entanglement measures in chapter~\ref{chapter-entanglement} and
that spurred the need for further research. Subsequently, I will
introduce a scenario from classical cryptography that inspired the
proposal of squashed entanglement. In
section~\ref{section-proposal}, I will define squashed
entanglement and give proof for its properties. In
section~\ref{section-evaluating-comm-gain} squashed entanglement
will be evaluated on a class of quantum states and the tools used
in this calculation will find application in two other quantum
cryptographic contexts. In
section~\ref{section-squashed-conclusion} I will round off the
discussion and provide an outlook into future research.

\subsection{Entanglement Measures}
\label{subsec-squashed-entangl-measures} In
chapter~\ref{chapter-entanglement} I have motivated the study of
entanglement measures and reviewed a large number of examples
within the axiomatic, or property-driven, approach. Remarkably,
most proposed measures satisfy only (or are only known to satisfy)
a small number of properties. Additivity for instance is only
known to hold for the logarithmic negativity ($E_N$) and the
reverse relative entropy of entanglement ($E_{RR}$), whereas
asymptotic continuity may only be true for entanglement of
formation ($E_F$), the relative entropy of entanglement ($E_R$)
and its regularised version ($E^\infty_R$, see
proposition~\ref{prop-rel-ent-cont}). In particular, asymptotic
continuity fails for $E_N$ and $E_{RR}$. This discussion shows
that from the outset it is not clear whether asymptotic continuity
and additivity can go hand in hand. It also shows that the result
on entanglement cost as an extremal measure,
proposition~\ref{prop-cost-upper}, is currently only applicable to
distillation rates (and of course entanglement cost itself).
Furthermore, we do not know of any other strongly superadditive
entanglement measure apart from distillable entanglement and
distillable key.

One result of this chapter is an answer to the above questions by
proving that squashed entanglement is strongly superadditive,
additive and asymptotically continuous. This not only provides a
new insight into the axiomatic approach, but also gives a new tool
to quantum information theory in form of an upper bound to
distillable entanglement and a lower bound to entanglement cost.

\subsection{Secret Key Agreement\index{secret key agreement}}
\label{subsec-secret-key} Cryptography\index{cryptography},
originally only the art of secret writing, is nowadays a subject
encompassing all aspects of communication which contain elements
of secrecy and mistrust. As electronic communication pervades our
daily life and increasingly replaces mail, visits to banks,
libraries and casinos, the need for secure communication is no
longer restricted to secret services, but is of direct concern to
the individuum in society.

A cryptographic system consists of a number of players who wish to
execute a communication protocol in order to solve a cryptographic
task. The oldest and certainly most well-known cryptographic task
is \emph{secure communication\index{secure communication}}. Here,
two honest parties, usually known as Alice and Bob, wish to
communicate in secrecy via a communication line to which an
eavesdropper, Eve, has access. This scenario has been analysed
under computational and physical limitations imposed on the
players.

The most widely used systems are public key cryptosystems, where
the security is based on computational assumptions. Here, the
sender Alice encrypts a message with Bob's public key and sends
the cryptogram (or cipher) to him. Bob receives the cipher and
uses his own private key to decrypt the cipher and retrieve the
message. A well-known public key
cryptosystem\index{cryptography!public key} is the RSA
cryptosystem~\cite{RSA78}, which is based on the assumption that
the factoring of large integers is classically intractable. A
large number of other public key cryptosystems have been designed.
Oded Regev has presented one recently which, if broken, would
result in an efficient quantum algorithm to solve a certain
lattice problem~\cite{Regev05}.

\begin{figure}
\begin{center}
\includegraphics{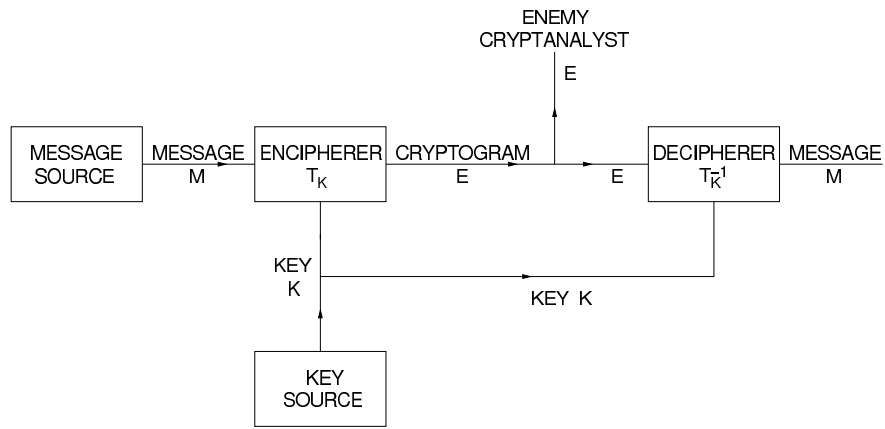} 
\end{center}
\caption{Schematic of a general secrecy system~\cite{Shannon49}.}
\label{figure-shannon}
\end{figure}

If a cryptosystem can be shown to be secure without any
computational assumptions, it is said to be
information-theoretically\index{security!information-theoretical}
or unconditionally secure\index{security!unconditional}. In 1949,
Shannon investigated such a scenario (see
figure~\ref{figure-shannon}).
The sender and receiver, Alice and Bob, are supplied with an
independently and uniformly distributed key $K$. Alice uses $K$ to
encipher the message $M$ and obtains a cipher $E$. $E$ is sent via
a classical channel\footnote{Here and in the following it is
assumed that all communication is authenticated.} to Bob who
deciphers the message using the same $K$. Shannon said that the
cryptosystem is \emph{perfectly secure\index{security!perfect}} if
the message equivocation $H(M|E)$ equals the entropy of the
message, $H(M|E)=H(M).$ Of course, Bob should also be able to
reconstruct the message from $E$ and $K$, $H(M|EK)=0.$ A secure
system must therefore obey $H(K) \geq H(M)$, in other words, the
key must be longer than the message itself (Shannon's
theorem\index{Shannon's theorem}), while equality can be achieved
with Vernam's one-time pad\index{Vernam's one time
pad}~\cite{Vernam26}. The validity of Shannon's negative result on
the key length has been extended by Ueli Maurer to allow for a
two-way communications channel between Alice and
Bob~\cite{Maurer93}.

\begin{figure}\begin{center}
\xymatrix{
        &      & & Satellite \ar[ddddll]^{\alpha} \ar[ddddrr]^{\varepsilon} \ar[ddrr]^{\beta}  & \\
        &      & &                                              & & \\
        &      & &                                              & & Bob \\
        &      & &  \ar@{.>}[drr]                              & & \\
        & Alice \ar@{<~>}[rrrruu]  & &                            & & Eve
} \end{center} \caption{The Satellite\index{satellite scenario}
broadcasts bits to Alice, Bob and Eve. The individual channels
flip the bits independently with probability $\alpha$, $\beta$ and
$\varepsilon$, respectively. The wavy double arrow indicates
two-way communication between Alice and Bob which is intercepted
by Eve (dotted arrow). Secret key agreement is possible whenever
$\alpha,
\beta \neq \half$ and $\epsilon >0$~\cite{Maurer93}. }
\label{figure-satellite}
\end{figure}
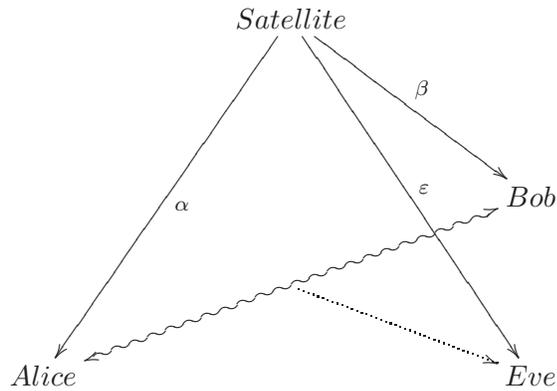

The essential question that remains from Shannon's work is how to
distribute a long key without making any computational
assumptions. One approach to this question is to impose reasonable
physical constraints. Here, I will focus on work sparked off by
Aaron D.~Wyner in 1975. Wyner proposed to give Eve a degraded
version of the signal sent from Alice to Bob; he called this
degrading a \emph{wiretap channel}~\cite{Wyner75}. Wyner's
scenario has been developed further by Imre Csisz\'ar and Janos
G.~K\"orner~\cite{CsiKoe78} and has been generalised by Ueli
Maurer, and Rudolf Ahlswede and Imre Csis{\'a}r to a scenario
known as \emph{secret key agreement} from common randomness by
public discussion~\cite{Maurer93, AhlCsi93}. Alice, Bob and Eve
have access to an i.i.d.~sequence of correlated triples of random
variables: $X$ is accessible for Alice, $Y$ for Bob and $Z$ for
Eve. The ranges of all three random variables is assumed to be
finite. Alice and Bob have at their disposal an unlimited amount
of public classical two-way communication and wish to convert the
noisy correlation contained in $X$ and $Y$ into a secret key $K$
that is virtually unknown to Eve. The maximal yield of secret key
bits per realisation of $XYZ$ is called the \emph{secret key
rate}\index{secret key rate} or \emph{distillable
key}\index{distillable key} $K_D(X;Y|Z)$. For a precise definition
of the secret key rate, which is frequently denoted by
$S(X;Y||Z)$, see~\cite{MauWol00b}. Figure~\ref{figure-satellite}
illustrates the scenario with the `satellite scenario'.

In analogy to bound entangled states in quantum theory, Nicolas
Gisin and Stefan Wolf conjectured that there are triples $XYZ$
from which no secret key can be distilled, but which nevertheless
require secret key bits for their formation; such distributions
are said to contain \emph{bound information\index{bound
information}}~\cite{GisWol00}. A solution to this conjecture
remains to be found. A formal definition of the \emph{key cost}
rate $K_C(X;Y|Z)$ of secret bits needed in order to establish a
sequence of $XYZ$ by public discussion is given
in~\cite{RenWol03}\footnote{In~\cite{RenWol03} the key cost is
called \emph{information of formation} and denoted by
$I_{form}(X;Y|Z)$.}. Recently, Andreas Winter discovered the
formula
$$
K_C(X;Y|Z)=\min_{\begin{subarray}{c} XY \rightarrow Z \rightarrow
\bar{Z} \\ X \rightarrow W\bar{Z} \rightarrow Y \end{subarray}}
I(XY;W|\bar{Z}),
$$
where $A \rightarrow B \rightarrow C$ indicates a Markov
chain~\cite{Winter05}. This solves the formation problem. In
contrast, the distillation problem remains largely unsolved. An
upper bound on the secret key rate is given by the mutual
information $I(X;Y)$ and can be seen as a generalisation of
Shannon's theorem. The mutual information, conditioned on $Z$ or
any random variable $\bar{Z}$ that can be obtained from $Z$,
remains an upper bound to the secret key rate; this means that
$K_D(X;Y|Z) \leq I(X;Y|\bar{Z})$ for all $Z \rightarrow
\bar{Z}$~\cite{AhlCsi93, MauWol99}. Alike the relations graph of
entanglement measures (figure~\ref{figure-relations}) one can
order the measures of secret correlations:
    $$
        K_D(X;Y|Z) \leq I(X;Y\downarrow\downarrow Z) \leq
        I(X;Y\downarrow Z) \leq K_C(X;Y|Z).
    $$
As discussed
$$I(X;Y\downarrow Z)=\inf_{XY \rightarrow Z \rightarrow \bar{Z}} I(X;Y| \bar{Z}),$$
the \emph{intrinsic information}\index{intrinsic information} of
$X$ and $Y$ with respect to $Z$, is an upper bound to the
distillable key and, as has been shown in~\cite{RenWol03}, a lower
bound to the key cost. The \emph{reduced intrinsic
information}\index{intrinsic information!reduced}
$$I(X;Y\downarrow\downarrow Z)=\inf_{XYZ \rightarrow U} I(X;Y\downarrow ZU)$$
improves the previous bound on $K_D(X;Y|Z)$. This fact follows
from a continuity property of the secret key rate which is not
possessed by the intrinsic information. The absence of this type
continuity related to the locking effect, which has been discussed
earlier in this thesis in the context of entanglement measures
(see table~\ref{table-definitions} and
subsection~\ref{sec-specific-purification})~\cite{RenWol03}.

The conjecture of bound information has highlighted a parallel
between secret key agreement and entanglement theory, which has
proven very beneficial in recent years. In fact, it is natural to
extend both the secret key agreement from random variables and the
key distillation from bipartite quantum states
(subsection~\ref{section-correlations-bipartite},
page~\pageref{subsubsec-secret-key}) to a unified secret key
agreement\slash key distillation scenario from tripartite quantum
states $\rho^{ABE}$. Such a scenario has been considered
in~\cite{DevWin05, ChrRen04, CHHHLO05}.

In the next section, I define a quantum analogue to intrinsic
information, which is called squashed entanglement. But before I
do so, let me mention that the invention of quantum key
distribution\index{quantum key
distribution}\index{cryptography!quantum}~\cite{BeBrEk92} has
offered an alternative way out of Shannon's pessimistic theorem:
the classical communication line between Alice and Bob is simply
replaced with a quantum communication line. No additional
assumptions need to be made, since the natural restrictions
imposed on the eavesdropper by the validity of quantum mechanics
suffice to imply the security of the proposed protocols (see
e.g.~\cite{Mayers96, LoCha99,
 ShoPre00, ChReEk04})

\section[Proposal for a New Measure]{Proposal for a New Measure\protect\footnote{With exception of proposition~\ref{prop-squashed-key}, the results in this section have been obtained in collaboration with Andreas Winter and have appeared in~\cite{ChrWin04}.}}

\label{section-proposal} This section forms the main part of this
chapter. Up to now I have reviewed and illustrated the role of
entanglement measures in quantum information theory and argued
that they bear a resemblance to a scenario in classical
cryptography: the secret key agreement. Here I propose a new
entanglement measure called squashed entanglement, which is
motivated by the intrinsic information, a quantity that arises in
secret key agreement.

The section is divided into three subsections. In
subsection~\ref{subsec-squashed-definition} I define squashed
entanglement and clarify the origin of this definition.
Subsequently, in section~\ref{section-squashed-properties}, I
prove that squashed entanglement possesses a number of the
properties that have been introduced in the context of
entanglement measures. The topic of
subsection~\ref{section-squashed-between}, the last part of this
section, is the relation of squashed entanglement to other
measures, such as distillable key, distillable entanglement and
entanglement cost.

\subsection{Definition and Motivation}
\label{subsec-squashed-definition}

\begin{figure}
\begin{center}
\includegraphics[width=0.7\textwidth]{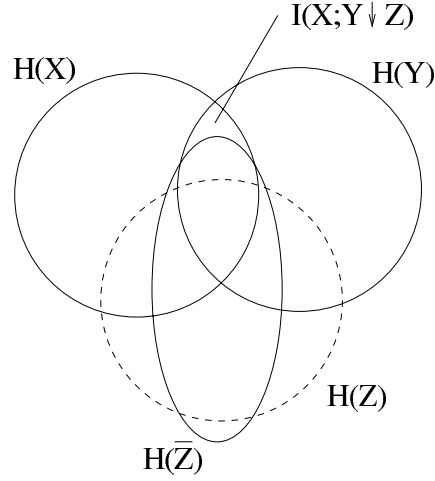}
\end{center}\vspace{-1cm}
\caption{Visualisation of the Intrinisic
Information~\cite{Wolf99PhD}.} \label{figure-venn}
\end{figure}

Intrinsic information emerges in the context of secret key
agreement and measures the correlations between random
variables~\cite{MauWol99}: The \emph{intrinsic (conditional
mutual) information} between two discrete random variables $X$ and
$Y$, given a third discrete random variable $Z$, is defined as
\begin{equation*}\begin{split}
  I(X;Y\downarrow Z) = \inf_{XY \rightarrow Z \rightarrow \bar{Z}}
  I(X;Y|\bar{Z}),
\end{split}\end{equation*}
where the infimum extends over all $\bar{Z}$, such that $XY
\rightarrow Z \rightarrow \bar{Z}$ is a Markov chain (see
figure~\ref{figure-venn}). In other words, a minimisation is
performed over all discrete channels mapping $Z$ to $\bar{Z}$ that
are specified by a conditional probability distribution
$P_{\bar{Z}|Z}$. In~\cite{ChReWo03} it is shown that the range of
$\bar{Z}$ can be taken to be equal to the range of $Z$; hence if
$Z$'s range is finite the minimum will be
achieved\footnote{Similarly, the minimum in Winter's formula for
$K_C(X;Y|Z)$ is achieved~\cite{Winter05}.}.

A first idea to utilise intrinsic information for measuring
quantum correlations was mentioned in~\cite{GisWol00}. This
inspired the proposal of a quantum analog to intrinsic
information~\cite{Christandl02}. Here, part $E$ of a purification
$\ket{\psi}^{ABE}$ of a quantum state $\rho^{AB}$ is given to Eve.
The conditional mutual information is then maximised over
measurements by Alice and Bob, followed by a minimisation over
Eve's measurements. This proposal possesses certain beneficial
properties demanded of an entanglement measure, and it opened the
discussion that has resulted in the current work.

Instead of using the classical conditional mutual information and
connecting it via measurement to quantum states, I propose to use
quantum information directly. The \emph{quantum conditional mutual
information\index{mutual information!quantum conditional}} of a
tripartite quantum state $\rho^{ABE}$ is given by
$$I(A;B|E):=S(AE)+S(BE)-S(ABE)-S(E),$$
and has first been considered in~\cite{CerAda97}. $S(\cdot)$
denotes the von Neumann entropy and by virtue of the \emph{strong
subadditivity of von Neumann entropy}, $S(AE)+S(BE) \geq
S(ABE)+S(E)$, the quantum conditional mutual information is
nonnegative. Strong subadditivity of von Neumann entropy is a
result of high importance to theoretical physics and was first
proven by Lieb and Ruskai in 1973~\cite{LieRus73PRL, LieRus73JMP}
(see also Preliminaries, page~\pageref{strong-subadditivity}).
This leads us to the following definition:

\begin{definition} \label{definition-squashed-ext} The
\emph{squashed entanglement\index{squashed entanglement}} of a
quantum state $\rho^{AB}$ on $\cH_A \otimes \cH_B$ is given by
    $$
        \sq (\rho^{AB}):=\inf_{\rho^{ABE} \in \ext(\rho^{AB})}
        \half I(A;B|E),
    $$
where the infimum is taken over the set $\ext(\rho^{AB})$ of all
extensions of $\rho^{AB}$, i.e.~over all quantum states
$\rho^{ABE}$ on $\cH_A \otimes \cH_B \otimes \cH_E$ with
$\rho^{AB}=\tr_E\rho^{ABE}$. The dimension of system $E$ is
\emph{a priori} unbounded.
\end{definition}
Since every extension $\rho^{ABE}$ can be generated by some CPTP
map $\Lambda$ applied on a purifying system $C$ of $\rho^{AB}$,
i.e.
\be \rho^{ABE}=\id \otimes \Lambda (\proj{\psi^{ABC}})\ee
where $\rho^{AB}=\tr_E \proj{\psi^{ABC}}$ and conversely every
CPTP map gives rise to a state $\rho^{ABE}$, one obtains the
following equivalent formulation of
definition~\ref{definition-squashed-ext}.
\begin{definition}
\label{definition-squashed-CP} The \emph{squashed entanglement} of
a quantum state $\rho^{AB}$ is given by
    $$
        \sq (\rho^{AB}):=\inf_{\Lambda} \frac{1}{2}
        I(A;B|E),
    $$
where the infimum is taken over all CPTP maps $\Lambda: C
\rightarrow E$ acting on the purifying part $C$ of a purification
$\ket{\psi^{ABC}}$ of $\rho^{AB}$. Since all purifications are
identical up to a unitary operation acting on $C$, this definition
is independent of the choice of the purification. The dimension of
$E$ is \emph{a priori} unbounded.
\end{definition}

\noindent Related work on the relation between entanglement
measures and the quantum mutual information has been conducted by
Robert R.~Tucci~\cite{Tucci99, Tucci00, Tucci02}. In the context
of key distillation from tripartite mixed quantum states, a
definition encompassing both the intrinsic information and
squashed entanglement (in the sense of
definition~\ref{definition-squashed-CP} and without the prefactor)
has been used~\cite{ChrRen04, CHHHLO05}. The normalisation factor
$1/2$ is chosen so that squashed entanglement assumes the value
$\log d$ on maximally entangled states in dimension $d\times d$
(see property Norm, table~\ref{table-definitions}). This points to
the next section, where I start by evaluating squashed
entanglement on pure states.

\subsection{Properties\index{squashed entanglement!properties}} \label{section-squashed-properties}

I go through the properties roughly in the order in which they
appear in table~\ref{table-definitions}.
\begin{proposition}
\label{prop-sq-pure} Let $\ket{\psi}^{AB}$ be a pure quantum
states. Then
$$
    \sq(\proj{\psi}^{AB})=S(A)_{\kpsi},
$$
i.e.~squashed entanglement equals the entropy of entanglement for
pure states. In particular $\sq$ satisfies Norm (see
table~\ref{table-definitions}).
\end{proposition}
\begin{proof}
  Let $\rho^{AB}=\proj{\psi}^{AB}$ be a pure state.
  All extensions of $\rho^{AB}$ are of the form $\rho^{ABE}=\rho^{AB}\otimes\rho^E$;
  therefore
  $$\frac{1}{2}I(A;B|E) = S(\rho^A) = E(\ket{\psi}^{AB}),$$
  which implies $\sq(\proj{\psi}^{AB}) = E(\kpsi^{AB})$.
\end{proof}

\begin{proposition}
  \label{value-zero}
Let $\rho^{AB}$ be a separable quantum state. Then
    $$
        \sq(\rho^{AB})=0,
    $$
i.e.~$\sq$ satisfies Van Sep (see table~\ref{table-definitions}).
\end{proposition}
\begin{proof}
Every separable $\rho^{AB}$ can be written as a convex combination
of separable pure states
\[  \rho^{AB}=\sum_i p_i \proj{\psi_i}^A \otimes
    \proj{\phi_i}^B.
\]
The quantum conditional mutual information of the extension
\[ \rho^{ABE}:=\sum_i p_i \proj{\psi_i}^A \otimes
    \proj{\phi_i}^B \otimes \proj{i}^E, \]
with orthonormal states $\{\ket{i}^E\}$, is zero. Squashed
entanglement thus vanishes on the set of separable states.
\end{proof}

\noindent The opposite, namely that every entangled quantum state
has strictly positive squashed entanglement, has yet defied any
proof.

\begin{conjecture} \label{conjecture-squashed-pos}
$\sq(\rho)>0$ for all entangled states.
\end{conjecture}

\noindent The following proposition may be seen as support for
this conjecture:

\begin{proposition} \label{prop-qmi-pos}
For all entangled states $\rho$ and all extensions $\rho^{ABE}$
with $\dim E < \infty$
$$ I(A;B|E)_\rho >0.$$
\end{proposition}

\begin{proof}
Recently, the structure of states that satisfy equality in the
strong subadditivity of von Neumann entropy has been
investigated~\cite{HJPW04}: it was shown that if $I(A;B|E)=0$ and
$\dim E < \infty$, then, with a suitable basis transformation
$E\rightarrow EE'E''$, $\rho^{ABE}$ can be rewritten in the form
$$\rho^{ABE} = \sum_i p_i \rho_i^{AE'} \otimes \rho_i^{E''B} \otimes \proj{i}^E.$$
Clearly $\rho^{AB} = \sum_i p_i (\tr_{E'}\rho_i^{AE'})
\otimes (\tr_{E''}\rho_i^{E''B})$ is separable.
\end{proof}

\noindent The minimisation in squashed entanglement ranges over
extensions of $\rho^{AB}$ with \emph{a priori} unbounded size.
$\sq(\rho)=0$ may therefore be possible, even if any finite
extension has strictly positive quantum conditional mutual
information. Therefore, without a bound on the dimension of the
extending system, proposition~\ref{prop-qmi-pos} does not suffice
to conclude that $\sq(\rho^{AB})$ implies separability of
$\rho^{AB}$. A different approach to this question could be
provided by a possible approximate version of the main result
of~\cite{HJPW04}: if there is an extension $\rho^{ABE}$ with small
quantum conditional mutual information, then $\rho^{AB}$ is close
to a separable state.

The interest in answering this conjecture lies in its implications
for entanglement cost, which are discussed later. The next
property on the list is monotonicity.

\begin{proposition} \label{prop-squashed-loc-mon}
Squashed entanglement is nonincreasing under local operations (Loc
Mon).
\end{proposition}
\begin{proof}
Without loss of generality, assume that the instrument
$\{\Lambda_k\}$ acts locally on $A$, i.e.
  \[ \rho^{ABE} \rightarrow \sigma^{AA'BE}:=\sum_k (\Lambda_k \otimes\idmap_{BE})(\rho^{ABE})
                                                                          \otimes \proj{k}^{A'}, \]
  with $\{\ket{k}^{A'}\}_k$ being an orthonormal basis on $A'$.
  It will be convenient to define $p_k:= \tr \Lambda_k \otimes
  \idmap_{BE} (\rho^{ABE})$ and $\sigma_k:=\Lambda_k \otimes
  \idmap_{BE} (\rho^{ABE})/p_k$. In order to unitarily
implement the quantum operation one can perform the following
steps: (i) Attach two ancilla systems $A'$ and $A''$ in states
$\ket{0}^{A'}$ and $\ket{0}^{A''}$ to the system $ABE$. (ii)
Perform a unitary transformation $U$ on $AA'A''$ followed by (iii)
a partial trace operation over system $A''$. For any extension of
$\rho^{AB}$ this leads to
  \begin{align*}
    I(A;B|E)_\rho &\stackrel{\text{(i)}}{=}      I(AA'A'';B|E)_\rho                                         \\
             &\stackrel{\text{(ii)}}{=}     I(AA'A'';B|E)_\sigma \\
             &\stackrel{\text{(iii)}}{\geq} I(AA';BE)_\sigma            \\
             &\stackrel{\text{(iv)}}{=}     I(A';B|E)_\sigma
                                              +I(A;B|EA')_\sigma         \\
             &\stackrel{\text{(v)}}{\geq}   \sum_k p_k I(A;B|E)_{\sigma_k}  \\
             &\stackrel{\text{(vi)}}{\geq}  \sum_k 2 p_k \sq(\sigma_k).
  \end{align*}
  The steps are justified as follows:
  attaching auxiliary pure systems does not change the entropy of a system,
  step (i). The unitary evolution affects only the systems $AA'A''$ and
  therefore does not affect the quantum conditional mutual information in step (ii). To
  show that discarding quantum systems cannot increase the quantum conditional mutual
  information, step (iii), expand
  \[ I(AA';B|E)_\sigma \leq
              I(AA'A'';B|E)_\sigma \]
   into
  \begin{equation*}\begin{split}
    &S(AA'E)+S(BE)-S(AA'BE)-S(E)                  \\
    &\phantom{==}
     \leq S(AA'A''E)+S(BE)-S(AA'A''BE)-S(E),
  \end{split}\end{equation*}
  which is equivalent to
  \[ S(AA'E)-S(AA'BE) \leq S(AA'A''E)-S(AA'A''BE), \]
  by strong subadditivity of von Neumann entropy.
Step (iv) is known as the chain rule\index{chain rule} and can be
seen by expanding both sides of the equation. In order to verify
step (v), note that $I(A';B|E)$ is nonnegative and that
$I(A;B|EA')$ can be written as the expectation value $\sum_k p_k
I(A;B|E)_{\sigma_k}$, since system $A'$ is classical. The
inequality in step (vi) holds since $\sigma_k^{ABE}$ is a valid
extension of $\sigma_k$. This concludes the proof of the
proposition as the original extension of $\rho^{AB}$ is arbitrary.
\end{proof}

\noindent Monotonicity under local operations and classical
information is implied by convexity combined with local
monotonicity.

\begin{proposition}
Squashed entanglement is convex (Conv), i.e.~for all quantum
states $\rho^{AB}$, $\sigma^{AB}$ and $p \in [0,1]$
$$ \sq(\gamma^{AB}) \leq p \sq(\rho^{AB})
+(1-p)\sq(\sigma^{AB}) $$ for $\gamma^{AB} :=
p\rho^{AB}+(1-p)\sigma^{AB}.$
\end{proposition}

\begin{proof}
 Consider any extensions $\rho^{ABE}$
  and $\sigma^{ABE}$ of the states
  $\rho^{AB}$ and $\sigma^{AB}$, respectively. Without loss of generality,
the extensions are defined on identical systems $E$.
  Combined, $\rho^{ABE}$ and $\sigma^{ABE}$ form an extension
  $$\gamma^{ABEE'} := p\rho^{ABE}\otimes\proj{0}^{E'}
                      + (1-p)\sigma^{ABE}\otimes\proj{1}^{E'}$$
  of $\gamma^{AB}$. The convexity of
  squashed entanglement then follows from the observation
\bestar
    p I(A;B|E)_{\rho} + (1-p) I(A;B|E)_{\sigma} = I(A;B|EE')_{\gamma} \geq 2\sq(\gamma^{AB}).
\eestar
\end{proof}

\begin{corollary}
  \label{monotone}
Squashed entanglement is nonincreasing under LOCC operations (LOCC
Mon).
 \end{corollary}

\begin{proof}
Proposition~\ref{Proposition-Vidal} says that a quantity which is
Conv and Loc Mon must necessarily be LOCC Mon, a result which
first appeared in~\cite{Vidal00}.
\end{proof}

\noindent This concludes the proof of the monotonicity properties.
The next set of properties concerns the additivity of squashed
entanglement, starting with superadditivity. Apart from
distillable entanglement and distillable key, which are both
superadditive by definition, no other entanglement measure is
known to satisfy this property.

\begin{proposition}
  \label{prop-super-additive}
Squashed entanglement is strongly superadditive (Strong Super
Add), i.e.
   \[ \sq(\rho^{AA'BB'}) \geq \sq(\rho^{AB}) + \sq(\rho^{A'B'}) \]
  is true for every density operator $\rho^{AA'BB'}$,
  $\rho^{AB}=\tr_{A'B'} \rho^{AA'BB'}$,
\end{proposition}
\begin{proof}
  Let
  $\rho^{AA'BB'E}$ be an extension of $\rho^{AA'BB'}$, i.e.~$\rho^{AA'BB'}= \tr_E \rho^{AA'BB'E}$.
  Then
  \begin{align*}
     I(AA';BB'|E)&=    I(A;BB'|E)+I(A';BB'|EA)  \\
                 &=    I(A;B|E)+I(A;B'|EB)      \\
                 &\phantom{=}
                      +I(A';B'|EA)+I(A';B|EAB') \\
                 &\geq I(A;B|E)+I(A';B'|EA)     \\
                 &\geq 2\sq(\rho^{AB})+2\sq(\rho^{A'B'}).
  \end{align*}
  The first inequality is due to strong subadditivity of the von Neumann
  entropy. Note that $E$ is an extension for system $AB$ and that $EA$ extends system
  $A'B'$. Hence, the last inequality holds since squashed entanglement
  is defined as a minimisation over all extensions of the respective states.
The claim follows because the calculation was independent of the
choice of the extension.
\end{proof}

\begin{proposition}
Squashed entanglement is subadditive (Sub Add), i.e.
  \[ \sq(\rho^{AB} \otimes \rho^{A'B'}) \leq \sq(\rho^{AB}) + \sq(\rho^{A'B'}), \]
  for all $\rho^{AB}$ and $\rho^{A'B'}$.
\end{proposition}
\begin{proof}
  Let $\rho^{ABE}$ be an extension of
  $\rho^{AB}$ and let $\rho^{A'B'E'}$ be an extension for $\rho^{A'B'}$. It is evident that
  $\rho^{ABE} \otimes \rho^{A'B'E'}$ is a valid extension for
  $\rho^{AA'BB'} \equiv \rho^{AB} \otimes \rho^{A'B'}$, hence
  \begin{align*}
    2\sq(\rho^{AA'BB'}) &\leq I(AA';BB'|EE')                                 \\
                       &=    I(A;B|EE')+\underbrace{I(A;B'|EE'B)}_{=0}      \\
                       &\phantom{=}
                            +I(A';B'|EE'A)+\underbrace{I(A';B|EE'AB')}_{=0} \\
                       &=    I(A;B|E)+I(A';B'|E').
  \end{align*}
  This inequality holds for arbitrary extensions of $\rho^{AB}$ and
  $\rho^{A'B'}$ and brings the argument to a close.
\end{proof}

\noindent Additivity on tensor products follows now directly from
superadditivity and subadditivity.

\begin{corollary}
Squashed entanglement is additive (Add), i.e.
    $$
        \sq(\rho^{AB} \otimes \sigma^{A'B'}) =\sq(\rho^{AB})
        +\sq(\sigma^{A'B'}),
    $$
extensive (Ext), i.e. for all $n$
 $$\sq(\rho^{AB})= \frac{\sq(\rho^{\otimes n})}{n}$$
 and therefore it coincides with its regularisation
 $$ \sq^\infty(\rho^{AB})=\lim_{n \rightarrow \infty} \frac{\sq(\rho^{AB \otimes n})}{n}=\sq(\rho^{AB}).$$
\end{corollary}

\noindent This shows that, apart from the logarithmic negativity
($E_N$) and the reverse relative entropy of entanglement
($E_{RR}$), squashed entanglement is the only entanglement measure
known to be additive. Since $E_{RR}$ and $E_N$ are not
asymptotically continuous -- not even on pure states -- the
question arises whether or not additivity and continuity may
contradict each other. This is not so as
propositions~\ref{prop-squashed-cont-pure} and
corollary~\ref{cor-squashed-continuous} show.

\begin{proposition} \label{prop-squashed-cont-pure}
Squashed entanglement is asymptotically continuous near pure
states (As Cont Pure).
\end{proposition}

\begin{proof}
Let $\kpsi$ be a purification of $\rho^{AB}$ and
$\rho^{ABE}=\idmap_{AB}\otimes \Lambda(\proj{\psi})$. Taking a
quick look at the Venn diagram in figure~\ref{figure-venn} shows
that $I(A;B)_\rho \geq 2\sq(\rho^{AB}) \geq
I(A;B)_\rho-I(A;E)_\rho$. The RHS of this inequality is lower
bounded by $I(A;B)_{\kpsi}
-I(A;E)_{\kpsi}=I(A;B)_\rho-2S(AB)_\rho$, a fact that follows from
the monotonicity of the mutual information. For $\rho$
$\epsilon$-close in trace distance to a pure state, by Fannes'
inequality: $S(AB)_\rho \leq \delta \log d$, for some
$\delta\equiv \delta(\epsilon)$. This which concludes the proof of
asymptotic continuity near pure states.
\end{proof}

\noindent In order to prove full asymptotic continuity, a general
Fannes-type inequality for the conditional von Neumann entropy is
needed (lemma~\ref{lemma-fannes}). This inequality was conjectured
in~\cite{ChrWin04} and proven for the special case where system
$AB$ is in a $qc$-state. A full proof of the inequality has been
obtained by Robert Alicki and Mark Fannes~\cite{AliFan04}. I will
now give their argument starting with a lemma for mixtures of
quantum states, which was discovered independently from Alicki and
Fannes.

\begin{lemma} \label{lemma-hishult}
Let $\rho^{AB}=(1-\epsilon) \sigma^{AB} + \epsilon \sigma'^{AB}$.
Then
    $$
        |S(A|B)_{\rho}- S(A|B)_{\sigma}| \leq 2\epsilon \log d
        +h(\epsilon),
    $$
where $d$ is the dimension of system $A$ and $h(\cdot )$ the
binary entropy, holds.
\end{lemma}

Note that this inequality is not a consequence of Fannes'
inequality (lemma~\ref{lemma-fannes}). Bounding conditional
entropies with Fannes' inequality would result in a bound
dependent on the dimension of $AB$ rather than on $A$ alone.

\begin{proof}
Since
$$ \rho^{AB}=(1-\epsilon) \sigma^{AB} + \epsilon \sigma'^{AB}  $$ one can estimate
    \beastar
        S(A|B)_\sigma-S(A|B)_\rho &= & S(\rho^{AB}||\frac{\id_A}{d} \otimes \rho^B)-S(\sigma^{AB}||\frac{\id_A}{d} \otimes \sigma^B)\\
            &\leq & (1-\epsilon)S(\sigma^{AB}||\frac{\id_A}{d} \otimes \sigma^B)+\epsilon S(\sigma'^{AB}||\frac{\id_A}{d} \otimes \sigma'^B)\\
            && \qquad \qquad -S(\sigma^{AB}||\frac{\id_A}{d} \otimes \sigma^B)\\
            &=& \epsilon \big(S(\sigma'^{AB}||\frac{\id_A}{d} \otimes \sigma'^B)-S(\sigma^{AB}||\frac{\id_A}{d} \otimes \sigma^B)\big)\\
            &=& \epsilon \big(S(A|B)_\sigma-S(A|B)_{\sigma'}\big)\\
            &\leq & \epsilon \big( S(A)_\sigma+S(A)_{\sigma'}\big)\\
            &\leq & 2 \epsilon \log d
    \eeastar
The first line follows from the identity $S(\rho||\frac{\id_A}{d}
\otimes \rho^B)=-S(A|B)_\rho+\log d$. Joint convexity of the
relative entropy (lemma~\ref{joint-convexity-rel-ent}) implies the
first inequality. The remaining two inequalities follow by
inserting the estimates $-S(A)\leq S(A|B)\leq S(A)$ and $S(A)\leq
\log d$. Conversely,
$$ S(A)_\rho \geq (1-\epsilon) S(A)_\sigma+\epsilon
S(A)_{\sigma'}$$ and
$$S(AB)_\rho \leq (1-\epsilon) S(AB)_\sigma+\epsilon
S(AB)_{\sigma'}+h(\epsilon)$$ can be combined to give
$$S(A|B)_\rho -S(A|B)_\sigma \leq \epsilon (S(A|B)_{\sigma'}-  S(A|B)_\sigma) +h(\epsilon)\leq 2 \epsilon \log d +h(\epsilon).$$
\end{proof}

\noindent With a nice trick, Alicki and Fannes extended this lemma
to two arbitrary quantum states $\rho$ and $\sigma$.

\begin{lemma}[Conditional Fannes' inequality~\cite{AliFan04}\index{Fannes' inequality!conditional}]
\label{lemma-cond-fannes} Let $\rho^{AB}$ and $\sigma^{AB}$ be
quantum states with $||\rho^{AB}-\sigma^{AB}||_1 \leq \epsilon$
and $d$ be the dimension of system $A$ only. Then
    \be
        |S(A|B)_{\rho} -S(A|B)_\sigma|\leq 4 \epsilon \log d+
        2h(\epsilon)
    \ee
where $h(\cdot)$ is the binary entropy function.
\end{lemma}

\begin{proof}
\sloppy Let $||\rho-\sigma||_1=\epsilon >0$ and define
\beastar \gamma&=&(1-\epsilon) \rho+|\rho-\sigma|\\
\tilde{\rho}&=&|\rho-\sigma|\frac{1}{\epsilon}\\
\tilde{\sigma}&=&\frac{1-\epsilon}{\epsilon}(\rho-\sigma)+\frac{1}{\epsilon}
|\rho-\sigma|\eeastar in order to write $\gamma$ in the form
$$ \gamma=(1-\epsilon) \rho+\epsilon
\tilde{\rho}=(1-\epsilon)+\epsilon \tilde{\sigma},$$ analogous to
the theorem by Thales of Milete. The claim follows directly from
lemma~\ref{lemma-hishult}:
    \bestar
    \begin{split}
        |S(A|B)_{\rho} -S(A|B)_\sigma| &\leq |S(A|B)_{\rho} -S(A|B)_\gamma |+ |S(A|B)_\gamma
        -S(A|B)_\sigma| \\
        &\leq 4\epsilon \log d +2h(\epsilon).
    \end{split}
    \eestar
\end{proof}

\noindent \fussy A straightforward calculation shows that this
lemma implies the asymptotic continuity of squashed entanglement.

\begin{corollary} \label{cor-squashed-continuous}
$\sq$ satisfies As Cont, more precisely: for all $\rho, \sigma$
with $\delta(\rho, \sigma) \leq \epsilon$,
$|\sq(\rho)-\sq(\sigma)| \leq 16\sqrt{2 \epsilon} \log d +4 h(2
\sqrt{2\epsilon})$
\end{corollary}

\begin{proof}
It suffices to show that for all extensions $\rho^{ABE}$ of
$\rho^{AB}$ there is an extension $\sigma^{ABE}$ of $\sigma^{AB}$
with $|I(A;B|E)_{\rho}-I(A;B|E)_\sigma| \leq \epsilon' \log d$ for
some $\epsilon'(\epsilon) \rightarrow 0$ as $\epsilon\rightarrow
0$. Since $\delta(\rho, \sigma) \leq \epsilon$, for every
purification $\ket{\psi}^{ABC}$ of $\rho^{AB}$, there is a
purification $\ket{\phi}^{ABC}$ of $\sigma^{AB}$ such that
$|\bra{\phi}^{ABC}\kpsi^{ABC}|^2=F(\rho, \sigma)$. Apply
monotonicity of the fidelity for a channel $\Lambda: C \rightarrow
E$:  $F(\rho^{ABE}, \sigma^{ABE}) \geq |\braket{\phi}
{\psi}^{ABC}|^2 $ and combine it with the
inequalities~(\ref{fidelity-trace-inequ}):
\beastar \delta(\rho^{ABE}, \sigma^{ABE}) &\leq& \sqrt{1-F(\rho^{ABE},
\sigma^{ABE})} \\
&\leq&  \sqrt{1-F(\rho^{AB}, \sigma^{AB})} \\&\leq&
\sqrt{1-(1-\delta(\rho, \sigma))^2} \\&\leq& \sqrt{2\delta(\rho,
\sigma)}.\eeastar Write $I(A;B|E)=S(A|E)-S(A|BE)$ and estimate
\beastar |I(A;B|E)_{\rho}-I(A;B|E)_\sigma|
&\leq& |S(A|E)_\rho-S(A|E)_\sigma| \\
&&\quad + |S(A|BE)_\rho-S(A|BE)_\sigma| \\
&\leq& 16\sqrt{2 \epsilon} \log d +4 h(2
\sqrt{2\epsilon}).\eeastar
\end{proof}

\noindent This concludes the proof of the monotonicity, additivity
and continuity properties of squashed entanglement. Please see
table~\ref{table-properties} for a summary. In the next subsection
I discuss the relation between squashed entanglement and other
measures of entanglement.

\subsection{Relations to other Entanglement Measures\index{squashed entanglement!relations to other measures}}
\label{section-squashed-between} Instead of invoking the abstract
results from proposition~\ref{prop-dist-lower} and
\ref{prop-cost-upper}, a direct calculation is carried out in
order to show the betweenness relation
$$ E_D \leq \sq \leq E_C,$$
or in fact the longer chain of inequalities
    \be \label{eq-chain}
        E_D \leq K_D \leq \sq \leq E_C \leq E_F.
    \ee
Let us start by showing that squashed entanglement is a lower
bound to entanglement of formation.

\begin{proposition}
  \label{EoF}
  $\sq$ is upper bounded by entanglement of formation:
  $$\sq(\rho^{AB}) \leq E_F(\rho^{AB}).$$
\end{proposition}
\begin{proof}
  Let $\{p_k, \ket{\psi_k}\}$ be an ensemble for $\rho^{AB}$:
  $$\sum_k p_k \proj{\psi_k}^{AB} = \rho^{AB}.$$
  The purity of the states implies
  \[ \sum_k p_k S(A)_{\Psi_k} = \frac{1}{2} \sum_k p_k I(A;B)_{\Psi_k}. \]
  Consider the following extension $\rho^{ABE}$ of $\rho^{AB}$:
  \[ \rho^{ABE}:= \sum_k p_k \proj{\psi_k}^{AB} \otimes \proj{k}^E \]
for which
  \[\frac{1}{2} I(A;B|E)=\frac{1}{2} \sum_k p_k I(A;B)_{\Psi_k}=\sum_k p_k S(A)_{\Psi_k}. \]
  Thus, it is clear that entanglement of formation can be regarded as an
  infimum over a certain class of extensions of $\rho^{AB}$. Squashed
  entanglement is an infimum over \emph{all} extensions of $\rho^{AB}$,
  evaluated on the same quantity $\frac{1}{2}I(A;B|E)$ and therefore smaller or
  equal to entanglement of formation.
\end{proof}

\noindent This result extends to entanglement cost by virtue of
the additivity of squashed entanglement.

\begin{corollary}
  \label{E-cost}
  $\sq$ is upper bounded by \emph{entanglement cost}:
  $$\sq(\rho^{AB}) \leq E_C(\rho^{AB}).$$
\end{corollary}
\begin{proof}
\sloppy Entanglement cost is equal to the regularised entanglement
of formation \cite{HaHoTe01},
  \[ E_C(\rho^{AB})=\lim_{n\rightarrow\infty} \frac{1}{n}E_F\left( (\rho^{AB})^{\otimes n}\right). \]
  This, together with proposition~\ref{EoF}, and the additivity of
  the squashed entanglement (proposition~\ref{prop-super-additive}) implies
  \begin{align*}
    E_C(\rho^{AB}) &=    \lim_{n\rightarrow\infty}
                                 \frac{1}{n}E_F\left( (\rho^{AB})^{\otimes n}\right) \\
                   &\geq \lim_{n\rightarrow\infty}
                                 \frac{1}{n}\sq\left( (\rho^{AB})^{\otimes n}\right)  \\
                   &=    \sq(\rho^{AB}).
  \end{align*}
\end{proof}

\noindent \fussy It is worth noting that in general $\sq$ is
strictly smaller than $E_F$ and $E_C$: consider the totally
antisymmetric state $\sigma^{AB}$ of a two-qutrit system
$$\sigma^{AB} = \frac{1}{3}\bigl( \proj{I}+\proj{II}+\proj{III} \bigr),$$
with
\begin{align*}
  \ket{I}   &= \frac{1}{\sqrt{2}}\left( \ket{2}^A\ket{3}^B-\ket{3}^A\ket{2}^B \right), \\
  \ket{II}  &= \frac{1}{\sqrt{2}}\left( \ket{3}^A\ket{1}^B-\ket{1}^A\ket{3}^B \right), \\
  \ket{III} &= \frac{1}{\sqrt{2}}\left( \ket{1}^A\ket{2}^B-\ket{2}^A\ket{1}^B \right).
\end{align*}
On the one hand, it is known from~\cite{Yura03} that
$E_F(\sigma^{AB})=E_C(\sigma^{AB})=1$, though, on the other hand,
one may consider the trivial extension in squashed entanglement
and find
$$\sq(\sigma^{AB}) \leq \frac{1}{2} I(A;B) = \frac{1}{2}\log 3 \approx 0.792.$$
The best known upper bounds on $E_D$ for this state, the Rains
bound ($E_{Rains}$) and the regularised relative entropy of
entanglement ($E_R^{PPT \infty}$), give the only slightly smaller
value $\log \frac{5}{3} \approx 0.737$. It remains open if there
exist states for which squashed entanglement is smaller than
$E_{Rains}$ or $E_R^{PPT \infty}$.

The strict positivity of squashed entanglement for entangled
states would, via corollary~\ref{E-cost}, imply strict positivity
of entanglement cost for all entangled states. This has first
recently been proven in~\cite{YHHS05}.

Below I will show that squashed entanglement is an upper bound to
the distillable key. The relative entropy of entanglement with
respect to separable states is the only other known bound for this
quantity.

\begin{proposition} \label{prop-squashed-key}
\be E_{sq}(\rho^{AB})\geq K_D(\rho^{AB})\ee
\end{proposition}
\begin{proof}
By the definition of $K_D(\rho^{AB})$, for every $\epsilon>0$
there is a number $n$ such that there exists an LOCC protocol
given by a CPTP map $\Lambda_n$ with
    \bestar
        \delta(\Lambda_n(\rho^{\otimes n}), \gamma_m) \leq
        \epsilon.
    \eestar
The gamma states $\gamma_m$ have been defined in
definition~\ref{def-gamma-secret}. Since squashed entanglement is
a monotone under LOCC and asymptotically continuous
(corollaries~\ref{monotone} and~\ref{cor-squashed-continuous})
    \bestar
        \sq(\rho^{\otimes n}) \geq \sq(\Lambda(\rho^{\otimes n}) \geq \sq(\gamma_m)-16 \sqrt{2 \epsilon} n \log d-4 h(2\sqrt{2\epsilon}).
    \eestar
Fix $m$ and consider the state $\gamma \equiv \gamma_m$ in its
natural representation $\gamma=U \rho^{AA'BB'}U^\dagger$ with
$U=\sum_i \proj{ii}^{AB} \otimes U_i^{A'B'}$ and
$\rho^{AA'BB'}=\proj{\psi}^{AB} \otimes \rho^{A'B'}$, where
$\ket{\psi}=\frac{1}{\sqrt{m}} \sum_{i=1}^m \ket{i}\ket{i}$, and
$\rho^{A'B'}$ is an arbitrary state on $A'B'$. In order to show
that $\sq (\gamma) \geq m$, consider an extension $\rho^{AA'BB'E}$
of $\rho^{AA'BB'}$ and the induced extension $\gamma^{AA'BB'E}=U
\otimes \id_E \rho^{AA'BB'E} U^\dagger \otimes \id_E$. Clearly,
    \bestar
        S(AA'BB'E)_\gamma=S(AA'BB'E)_\rho= S(A'B'E)_\rho=S(A'B'E)_{\gamma_i},
    \eestar
with $\gamma_i^{A'B'E}:= U_i \otimes \id_E \rho^{A'B'E}
U_i^\dagger \otimes \id_E$. Furthermore
    \bestar
        S(E)_{\gamma_i}=S(E)_\rho \quad \textrm{and} \quad S(AA'E)_\gamma=S(A)_\gamma+\sum_i p_i S(A'E)_{\gamma_i}
    \eestar
and similarly for $S(BB'E)_\gamma$. Altogether this gives
    \bestar
        I(AA';BB'|E)_\gamma \geq S(A)_\gamma+S(B)_\gamma+\sum_i p_i I(A';B'|E)_{\gamma_i} \geq 2m,
    \eestar
where the non-negativity of the quantum mutual information was
used in the last inequality. This shows that $\sq(\gamma_m)\geq m$
and therefore $\sq(\rho) \geq \frac{m}{n}-16 \sqrt{2 \epsilon}
\log d-\frac{4}{n} h(2\sqrt{2\epsilon})$, with the RHS converging
to $K_D(\rho^{AB})$.
\end{proof}

\noindent Since maximally entangled states are a special class of
$\gamma$ states, namely those with $\rho^{A'B'}$ trivial, squashed
entanglement is also an upper bound on distillable entanglement.

\begin{corollary}
\label{distillable-bound}
\be E_{sq}(\rho^{AB})\geq E_D(\rho^{AB})\ee
\end{corollary}

\noindent The direct proof of this fact only needs continuity near
pure states, proposition~\ref{prop-as-con-pure}, and is given
in~\cite{ChrWin04}. Hence all distillable states have strictly
positive squashed entanglement.

It has recently been shown that there exist bound entangled states
which have positive key rate~\cite{HHHO05a}, which implies:

\begin{corollary}
There exist bound entangled states $\rho^{AB}$ with
$E_{sq}(\rho^{AB})>0$. In particular, squashed entanglement is not
a PPT monotone.
\end{corollary}

\noindent The results in this section are summarised in the
relations graph of in chapter~\ref{chapter-entanglement}
(figure~\ref{figure-relations}). We see that squashed entanglement
is arranged in a chain of operationally defined measures
(ineqs.~(\ref{eq-chain})), whereas the relation to other
entanglement measures remains unknown. It is a challenge raised in
this chapter to discover these relations, in particular the
relation to the relative entropy of entanglement.

\section{Evaluating, Committing and Gaining Information}
\label{section-evaluating-comm-gain}

In the following, three specific topics related to squashed
entanglement are discussed. In subsection~\ref{subsec-flower},
squashed entanglement is evaluated on a class of quantum states,
known as \emph{flower states}, and follows that squashed
entanglement can be locked. The tool used to perform this
calculation is a new type of entropic uncertainty relation for
quantum channels. In subsection~\ref{subsec-info-gain-dist} I show
how to use this uncertainty relation to obtain an information-gain
disturbance tradeoff. In subsection~\ref{subsec-cheat-sensitive}
the uncertainty relation is used to prove for the first time the
cheat sensitivity of a quantum string commitment scheme.

\subsection[The Squashed Entanglement of Flower States]{The Squashed Entanglement of Flower States\protect\footnote{The results presented in this subsection have appeared
in~\cite{ChrWin05, ChrWin05Proc}.}} \label{subsec-flower} Consider
a uniform ensemble $\cE_0=\{ \frac{1}{d}, \ket{i} \}_{i=1}^{d}$ of
basis states of a Hilbert space $\cH$ and the rotated ensemble
$\cE_1=\{ \frac{1}{d}, U\ket{i} \}_{i=1}^{d}$ with a unitary $U$.
Application of the CPTP map $\Lambda$ (with output in a
potentially different Hilbert space) results in the two ensembles
\begin{align*}
  \Lambda(\cE_0) &= \left\{ \frac{1}{d}, \Lambda(\proj{i}) \right\}         \\
  \Lambda(\cE_1) &= \left\{ \frac{1}{d}, \Lambda(U\proj{i}U^\dagger) \right\}
\end{align*}
with Holevo information for $\cE_0$ given by
$$\chi(\Lambda(\cE_0)) = S\left( \frac{1}{d}\sum_i \Lambda(\proj{i}) \right)
                         - \frac{1}{d}\sum_i S\bigl( \Lambda(\proj{i}) \bigr)$$
and similarly for $\cE_1$. Consider also the quantum mutual
information of $\Lambda$ relative to the maximally mixed state
$\tau = \frac{1}{d}\id$, which is the average state of either
$\cE_0$ or $\cE_1$:
$$I(\tau;\Lambda) = S\bigl(\tau\bigr)
                   + S\bigl(\Lambda(\tau)\bigr)
                   - S\bigl((\id\otimes\Lambda) (\proj{\psi_d})\bigr),$$
where $\ket{\psi_d}$ is a maximally entangled state in dimension
$d$ purifying $\tau$.

\begin{lemma}[Channel Uncertainty Relation\index{channel uncertainty relation}]
  \label{lemma-uncertainty}
  Let $U$ be the Fourier transform of dimension $d$, i.e. of
  the Abelian group $\integers_d$ of integers modulo $d$. More
  generally,
  $U$ can be a Fourier transform of any finite Abelian group labeling
  the ensemble $\cE_0$, e.g. for $d=2^\ell$, and the group
  $\integers_2^\ell$, $U=H^{\otimes\ell}$
  with the Hadamard transform $H$ of a qubit.
  Then for all CPTP maps $\Lambda$,
  \be
    \label{channel-ineq}
    \chi\bigl(\Lambda(\cE_0)\bigr) + \chi\bigl(\Lambda(\cE_1)\bigr)
                                                   \leq I(\tau; \Lambda).
  \ee
\end{lemma}
\begin{proof}
Define $\rho^{SC}=(\id\otimes\Lambda)\proj{\psi_d}$, and let $M_0$
be the projection onto the basis $\{\ket{i}\}$ and $M_1$ the
projection onto the conjugate basis $\{U\ket{i}\}$,
$$M_0(\sigma) = \sum_{i=1}^d \proj{i} \sigma \proj{i},$$
$$M_1(\sigma) = \sum_{i=1}^d U\proj{i}U^\dagger \sigma U\proj{i}U^\dagger.$$
Let $X$ be the cyclic shift operator of the basis $\{\ket{i}\}$,
and $Z=UXU^\dagger$ the cyclic shift of the conjugate basis
$\{U\ket{i}\}$. The significance of taking $U$ as the Fourier
transform lies in the fact that $\{\ket{i}\}$ is the eigenbasis of
$Z$ and $\{U\ket{i}\}$ is the eigenbasis of $X$. Hence,
\beastar
  M_0(\sigma) &=& \frac{1}{d}\sum_{l=1}^d Z^l \sigma Z^{-l}, \\
  M_1(\sigma) &=& \frac{1}{d}\sum_{k=1}^d X^k \sigma X^{-k}.
\eeastar
A central role will be played by the correlated state
    $$\Omega^{S_0S_1SC} := \frac{1}{d^2}\sum_{k,l=1}^d
        \proj{k}^{S_0}\otimes\proj{l}^{S_1}\otimes\rho_{kl}^{SC},
    $$
where
$$\rho_{kl}^{SC} := (X^kZ^l\otimes\id) \rho^{SC}
  (Z^{-l}X^{-k}\otimes\id).$$
With these definitions it is straightforward to check that
\beastar
   I(S_0S;C)&=&\chi\bigl(\Lambda(\cE_0)\bigr)\\
   I(S_1S;C)&=&\chi\bigl(\Lambda(\cE_1)\bigr)\\
   I(S_0S_1S;C)&=&I(\tau; \Lambda)
\eeastar
  and the assertion is the consequence of a short calculation:
  \begin{equation*}\begin{split}
    I(S_0S_1S;C) &=    I(S_0S;C) + I(S_1S;C|S_0) \\
            &=    I(S_0S;C) + I(S_1S;S_0C)  \\
            &\geq I(S_0S;C) + I(S_1S;C),
  \end{split}\end{equation*}
  where I have used only standard identities and strong subadditivity,
  and in the second line the independence of $S_0$ and $S_1$,
  expressing itself as
  $$I(S_1S;C|S_0)=\underbrace{I(S_1S;S_0)}_{=0} + I(S_1S;C|S_0) =I(S_1S;S_0C). $$
In the general case of an Abelian group, one has to replace
  the operators $X$ and $Z$ by the regular representation of the
  group and its conjugate via the Fourier transform.
\end{proof}

\noindent This proof as well as a related argument involving dense
coding capacities can be found in~\cite{ChrWin05, ChrWin05Proc}.

This lemma results in the following corollary, which is an
instance of \emph{locking of classical information in quantum
states}. The accessible information of the ensembles $\cE_0$ and
$\cE_1$ are each equal to $\log d$. If, however, the identity of
the ensemble is lost, i.e. if one is presented with states from
the ensemble $\cE=\half \cE_0+\half \cE_1$, then $\I=\half \log
d$. Losing a single bit of information can therefore result in an
arbitrary decrease of the accessible information. This result has
been proven in~\cite{DHLST04} using the entropic uncertainty
relation of~\cite{MaaUff88}. Here, it emerges as a consequence of
the channel uncertainty relation (lemma~\ref{lemma-uncertainty}).

\begin{corollary}
  \label{acc-lock}
  For the Fourier transform $U$ and the ensemble
  $\cE=\frac{1}{2}\cE_0+\frac{1}{2}\cE_1$, defined as in
  lemma~\ref{lemma-uncertainty}, the following equality holds:
  $$I_{acc}(\cE) = \frac{1}{2}\log d.$$
\end{corollary}
\begin{proof}
  Let $X$ denote a random variable uniformly distributed over the
  labels $ij$ ($i=1,\ldots,d$, $j=0,1$) of the ensemble $\cE$.
  The left hand side of inequality (\ref{channel-ineq})
  equals $2 I(X;Y)$ in the special case where the CPTP map $\Lambda$
  is a measurement with outcome $Y$. Measuring the system $B$ of
  $\ket{\psi}^{AB}=\frac{1}{\sqrt{2d}}\sum_{i,j} \ket{ij}^A U_j\ket{i}^B$
  gives $2 I(X;Y) \leq I(\tau;\Lambda) \leq \log d$.
  Clearly, a measurement performed
  in one of the two bases will achieve this bound.
\end{proof}

\noindent The tools are now prepared to tackle the calculation of
squashed entanglement for a class of states considered in
\cite{HHHO04}.

\begin{proposition}
  \label{squash-lock}
  For \emph{flower states\index{flower states}}~\cite{HHHO04} $\rho^{AA'BB'}$ defined via their purification
  \be
    \label{flower-states}
    \ket{\Psi}^{AA'BB'C}=\frac{1}{\sqrt{2d}}\sum_{\substack{i=1\ldots d \\ j=0,1}}
                           \ket{i}^A\ket{j}^{A'}\ket{i}^B \ket{j}^{B'}
                           U_j\ket{i}^C,\ee
  where $U_0=\id$ and $U_1$ is a Fourier transform, it is true
  that
  $$E_{sq}(\rho^{AA'BB'})=1+\frac{1}{2}\log d\text{\ \ and\ \ \ }
    E_{sq}(\rho^{ABB'})=0.$$
This shows that squashed entanglement can be locked, i.e. does not
possess property Non Lock (see table~\ref{table-definitions})
\end{proposition}
\begin{proof}
According to definition~\ref{definition-squashed-CP}, squashed
entanglement can be regarded as a minimisation over CPTP channels
$\Lambda: C \longrightarrow E$ acting on the purifying system $C$
for $\rho^{AA'BB'}$:
    $$
        \rho^{AA'BB'E} = (\id_{AA'BB'}\otimes\Lambda)\kpsi^{AA'BB'C}.
    $$
The reduced state of $\kpsi$ on $C$ is maximally mixed:
$\tr_{AA'BB'}\proj{\psi}=\tau=\frac{1}{d}\id$, hence
  \begin{align}
    \label{eq-S-E}
    S(\rho^E)         &= S(\Lambda(\tau)),                       \\
    \label{eq-S-ABE}
    S(\rho^{AA'BB'E}) &= S\bigl( (\id\otimes\Lambda)\Phi_d \bigr).
  \end{align}
Since $\rho$ is maximally correlated the reduced states of $\rho$
on $ZZ'E$, for $ZZ' \in \bigl\{ AA',\ BB'\bigr\}$ read
$$\rho^{ZZ'\!E}=\frac{1}{2d}\sum_{i,j} \proj{i}^Z\otimes\proj{j}^{Z'}
                                       \otimes\Lambda(U_j\proj{i}U_j^\dagger)^E.$$
The remaining two entropy terms of the conditional mutual
information are of the form:
\begin{align}
  S\bigl(\rho^{AA'E}\bigr)
                 &= S\bigl(\rho^{BB'E}\bigr)                                    \nonumber\\
                 &= \log d + 1
                   + \frac{1}{2d}\sum_{i,j}
                                 S\bigl( \Lambda(U_j\proj{i}U_j^\dagger) \bigr) \nonumber\\
                 &= 1 + S(\tau) + S(\Lambda(\tau))                              \nonumber\\
                 &\phantom{=}                                              \label{eq-S-AE}
                   -\frac{1}{2}\chi\bigl(\Lambda(\cE_0)\bigr)
                   -\frac{1}{2}\chi\bigl(\Lambda(\cE_1)\bigr).
\end{align}
Combining equations~(\ref{eq-S-E}), (\ref{eq-S-ABE}) and
(\ref{eq-S-AE}) it follows
\begin{equation*}\begin{split}
  I(AA';BB'|E) &= S(AA'E) + S(BB'E)  \\
               &\quad- S(AA'BB'E) - S(E) \\
               &= 2 + \log d + S(\tau) + S(\Lambda(\tau)) \\
               & \quad-\chi\bigl(\Lambda(\cE_0)\bigr)-\chi\bigl(\Lambda(\cE_1)\bigr) -S((\id\otimes\Lambda)\Phi_d)            \\
               &= 2+ \log d + I(\tau;\Lambda)             \\
               & \quad  - \chi\bigl(\Lambda(\cE_0)\bigr) - \chi\bigl(\Lambda(\cE_1)\bigr)         \\
               &\geq 2 + \log d,
\end{split}\end{equation*}
where the last inequality is an application of lemma
\ref{lemma-uncertainty}. The bound is achieved for trivial $E$,
since $I(A;B)=2+\log d$. This concludes the calculation of
squashed entanglement for $\rho^{AA'BB'}$. $\rho^{ABB'}$ is
evidently separable, and thus has zero squashed entanglement.
\end{proof}

\noindent It is an open question whether or not the minimisation
in squashed entanglement can be taken over POVMs only. If so, the
simpler argument $I(AA';B'|E) \geq I(AA';BB')-I(A;E)
             =    2\log d + 2 - \log d$,
only using corollary~\ref{acc-lock}, proves proposition
\ref{squash-lock}.

In~\cite{HHHO04} $E_C\bigl(\rho^{AA'BB'}\bigr)$ was not explicitly
calculated, but it was observed that it is larger than
$\frac{1}{2}\log d$. Notice that the argument in~\cite{HHHO04}
actually proves
$$E_C\bigl(\rho^{AA'BB'}\bigr) = E_F\bigl(\rho^{AA'BB'}\bigr)
                               = 1 + \frac{1}{2}\log d$$
as a consequence of corollary~\ref{acc-lock} and the relation
$$E_F\bigl(\rho^{AA'BB'}\bigr) = S(\rho^A) - \max_M \chi.$$
The maximisation ranges over all measurements $M$ on $BB'$ and
$\chi$ is the Holevo quantity of the induced ensemble on $AA'$. In
fact, this result is a direct consequence of
proposition~\ref{squash-lock} by observing that $E_C(\rho)\geq
E_{sq}(\rho)$ (proposition~\ref{E-cost}). Equality is achieved --
even for $E_F$ -- for $\Lambda$ being a complete measurement in
one of the mutually conjugate bases.

The gap between entanglement of formation and squashed
entanglement as well as between squashed entanglement and
distillable entanglement can be made simultaneously large. This
was pointed out by Patrick Hayden, Karol, Micha{\l} and Pawe{\l}
Horodecki, Debbie Leung and Jonathan Oppenheim and mentioned with
their kind permission in~\cite{ChrWin05, ChrWin05Proc}. As I show
below even the gap between squashed entanglement and distillable
key can be made arbitrarily large.

\begin{proposition}
  \label{squash-lock-gap}
  Let $\rho^{AA'BB'}$ be defined by the purification
  $$\ket{\psi}=\frac{1}{\sqrt{2dm}}
               \sum_{\substack{i=1\ldots d \\ j=0,1,k=1\ldots m}}
                   \ket{i}^A\ket{jk}^{A'} \ket{i}^B\ket{jk}^{B'} V_k U_j\ket{i}^C,$$
  where $U_0=\id$ and $U_1$ is a Fourier transform.
  For all $\epsilon>0$ and large enough $d$, there exists
  a set of $m=(\log d)^3$ unitaries $V_k$ such that
  \begin{align*}
    E_C   \bigl(\rho^{AA'BB'}\bigr) &\geq (1-\epsilon)\log d + 3\log\log d - 3, \\
    E_{sq}\bigl(\rho^{AA'BB'}\bigr) &=     \frac{1}{2}\log d + 3\log\log d + 1, \\
    K_D   \bigl(\rho^{AA'BB'}\bigr) &\leq 6\log\log d + 2.
  \end{align*}
  Hence, $ K_D \ll E_{sq} \ll E_C$ is possible.
\end{proposition}
\begin{proof}
  Define ensembles $\cE=\{\frac{1}{2 m d}, V_k U_j \ket{i}\}_{ijk}$ and
  $\tilde{\cE}=\{\frac{1}{m d}, V_k \ket{i}\}_{ik}$.
  As observed before, for the states under consideration,
  \begin{equation*}\begin{split}
    E_F\bigl(\rho^{AA'BB'}\bigr) &= S(\rho^A) - \max_M \chi            \\
                                 &= \log d + \log m + 1 - I_{acc}(\cE),
  \end{split}\end{equation*}
  and since $I_{acc}(\cE)$ is additive~\cite{H73}
  (see also~\cite{DLT02}),
  $$E_C\bigl(\rho^{AA'BB'}\bigr) = \log d + \log m + 1 - I_{acc}(\cE).$$
  It was shown in~\cite{HLSW04} that for all $\epsilon>0$
  and large enough $d$, there exists a set of $m=(\log d)^3$
  unitaries $V_k$ such that
  $I_{acc}(\tilde{\cE}) \leq \epsilon\log d + 3$ (This behaviour is also known as the \emph{strong locking effect}). Clearly the mixing of two such ensembles cannot increase the
  accessible information by more than $1$. This can be seen operationally, since even if
  the bit identifying the ensemble was known, a measurement
  would still face an ensemble isomorphic to $\cE$:
  $I_{acc}(\cE) \leq I_{acc}(\tilde{\cE}) + 1$.
  Therefore,
  $$E_C\bigl(\rho^{AA'BB'}\bigr) \geq (1-\epsilon) \log d + 3\log\log d - 3.$$
  Essentially the same calculation as in the proof of
  proposition~\ref{squash-lock} shows that
  $$E_{sq}\bigl(\rho^{AA'BB'}\bigr)=\frac{1}{2}\log d + \log m + 1,$$
where one has to use lemma~\ref{lemma-uncertainty} for each of the
  pairs of ensembles
  $$\{\frac{1}{d}, V_k\ket{i}\}_i \mbox{ and } \{\frac{1}{d}, V_k U_1\ket{i}\}_i, \mbox{ for } k=1, \ldots, m.$$
  Finally, the process of discarding the $1+\log m$ qubits of register $A'$ (which
  leaves a separable state $\rho^{ABB'}$) cannot
  decrease the relative entropy of entanglement by more than
  $2(1+\log m)$~\cite{HHHO04}, and since the latter is a bound on
  distillable key~\cite{HHHO05a}: $K_D(\rho) \leq 2(3\log\log d + 1)$.
\end{proof}

\noindent Interestingly, the flower states
(eq.~(\ref{flower-states})) can be understood as quantum analogues
of the distributions analysed in~\cite{RenWol03}. The latter have
been constructed for a similar purpose, namely to show that the
gap between intrinsic information and the secret key rate can be
arbitrarily large.

\subsection[An Information-Gain versus Disturbance Tradeoff]{An Information-Gain versus Disturbance
Tradeoff~\protect\footnote{Part of the results presented in this
section have been presented at QIP 2005 in Boston (MA) in the
context of~\cite{BCHLW05}.}} \label{subsec-tradeoff} \sloppy In
this subsection I make use of the channel uncertainty relation
(lemma~\ref{lemma-uncertainty}) in order to derive a novel
information-gain versus disturbance
tradeoff\index{information-gain vs.~disturbance}. The task is the
following: Given a set of non-orthogonal quantum states, derive
the tradeoff between information-gain from a measurement on the
set of states and the disturbance caused by this measurement. To
make the task precise, an information and a disturbance measure
must be chosen. Usual candidates include the accessible
information and the trace distance. The latter measures the
disturbance, i.e. the distance, between pre- and post-measurement
states. Tradeoffs of this type have been considered by Christopher
A.~Fuchs and Asher Peres~\cite{FucPer96, Fuchs95, Fuchs98}, and
are motivated by the study of `prepare and measure quantum key
distribution\index{quantum key distribution!prepare and
measure}'~\cite{BenBra84, Bennett92}. Here, Alice is the sender of
a set of non-orthogonal signal states which are intercepted by Eve
and then forwarded to Bob. The tradeoff in this scenario is
between Eve's information-gain from a measurement on the
intercepted states and the disturbance observed by Alice and Bob
when they check the integrity of the communication line.

\fussy The described scenario, however, has a conceptual problem:
there is no way to ensure that Eve has actually performed a
measurement. She could simply branch off some of the quantum
information that she receives and forward only the remaining part.
The following analysis avoids this problem by getting rid of
measurements by Eve altogether. Her information-gain is quantified
in terms of quantum information, i.e. either in terms of quantum
mutual information or Holevo $\chi$ information. By Holevo's
theorem the presented result will also lead to a tradeoff for the
accessible information.

The following lemma is a technical consequence of Fannes'
inequality and is needed in the proof of the tradeoff.

\begin{lemma} \label{lemma-Holevo-relation}
Let $\cE=\{p_i, \rho_i=\proj{\psi_i}\}$ be an ensemble of pure
states and $\tilde{\cE}=\{p_i, \sigma_i\}$ be an ensemble of mixed
states, both on $\complex^d$. If $\sum_i \bra{\psi_i} \sigma_i
\ket{\psi_i} \geq 1-\epsilon$, then
$$ |\chi(\tilde{\cE}) - \chi(\cE)| \leq 4\sqrt{\epsilon} \log d  +2 \mu(2\sqrt{\epsilon}),$$
where $\mu(x):=\min \{-x\log x, \frac{1}{e}\}$.
\end{lemma}

\begin{proof}
The justification of the estimate
$$ \epsilon \geq \sum_i p_i (1-\tr \rho_i \sigma_i) \geq \sum_i
p_i \delta_i^2 \geq \big(\sum_i p_i \delta_i \big)^2,$$ where
$\delta_i:= \delta(\rho_i, \sigma_i)$ is as follows: the second
inequality is identical to inequality~(\ref{fidelity-trace-inequ})
(Preliminaries, page~\pageref{fidelity-trace-inequ}), whereas the
third follows from the convexity of the square function. Strong
convexity of the trace distance implies $\delta(\rho, \sigma) \leq
\sqrt{\epsilon}$. Fannes' inequality (lemma~\ref{lemma-fannes})
will be applied to the overall state
$$ |S(\rho)-S(\sigma)| \leq 2\sqrt{\epsilon} \log d +  \min \{\eta(2 \sqrt{\epsilon}), \frac{1}{e}\}$$
where $\eta(x):=-x\log x$, and to the individual ones
\beastar \sum_i p_i |S(\sigma_i)- S(\rho_i)|
&\leq& \big(\sum_i p_i \delta_i \big) 2\log d  +\sum_i p_i \min
\{\eta(2 \delta_i), \frac{1}{e}\}\\
&\leq & \sqrt{\epsilon} 2\log d + \min \{\eta(2 \sqrt{\epsilon}),
\frac{1}{e}\}
\eeastar
where the last inequality is true by the concavity of $\eta(x)$.
Inserting these estimates in the Holevo $\chi$ quantities
$\chi(\cE)=S(\rho)$ and $\chi(\tilde{\cE})=S(\sigma)-\sum_i p_i
S(\sigma_i)$ concludes the proof.
\end{proof}

\noindent Let $\cE_0=\{ \frac{1}{d}, \ket{i} \}_{i=1}^{d}$ be an
ensemble of orthogonal states in $\complex^d$ and $\cE_1=\{
\frac{1}{d}, U\ket{i} \}_{i=1}^{d}$ the ensemble rotated with the
Fourier transform $U$ of dimension $d$. More generally, $U$ can be
the Fourier transform of any finite Abelian group labeling the
states in $\cE_0$. Further, let $\cE=\half \cE_0+\half
\cE_1=\{\frac{1}{2d}, U^r \ket{i}\}_{i=1\ldots d, r=0, 1}$ be the
combined ensemble ($U^0=\id$ and $U^1=U$).

\begin{figure}
\begin{center}
\unitlength 0.8mm
\begin{picture}(100.00,60.00)(10,30)

\linethickness{0.15mm}
\put(20.00,70.00){\line(1,0){10.00}}
\put(20.00,70.00){\line(0,1){10.00}}
\put(30.00,70.00){\line(0,1){10.00}}
\put(20.00,80.00){\line(1,0){10.00}}

\put(23,73){\makebox{$A$}}

\linethickness{0.15mm}
\put(20.00,50.00){\line(1,0){10.00}}
\put(20.00,50.00){\line(0,1){10.00}}
\put(30.00,50.00){\line(0,1){10.00}}
\put(20.00,60.00){\line(1,0){10.00}}

\put(23,53){\makebox{$S$}}

\linethickness{0.15mm}
\multiput(25.00,60.00)(0,1.82){6}{\line(0,1){0.91}}

\linethickness{0.15mm}
\put(50.00,40.00){\line(1,0){20.00}}
\put(50.00,40.00){\line(0,1){20.00}}
\put(70.00,40.00){\line(0,1){20.00}}
\put(50.00,60.00){\line(1,0){20.00}}

\put(58,48){\makebox{${\bf U}$}}

\linethickness{0.15mm}
\put(70.00,55.00){\line(1,0){20.00}}

\put(93,53){\makebox{$\tilde{S}$}}

\linethickness{0.15mm}
\put(70.00,45.00){\line(1,0){20.00}}

\linethickness{0.15mm}
\put(30.00,55.00){\line(1,0){20.00}}

\linethickness{0.15mm}
\put(90.00,35.00){\line(0,1){10.00}}

\linethickness{0.15mm}
\put(90.00,50.00){\line(1,0){10.00}}
\put(90.00,50.00){\line(0,1){10.00}}
\put(100.00,50.00){\line(0,1){10.00}}
\put(90.00,60.00){\line(1,0){10.00}}

\linethickness{0.15mm}
\put(85.00,25.00){\line(1,0){10.00}}
\put(85.00,25.00){\line(0,1){10.00}}
\put(95.00,25.00){\line(0,1){10.00}}
\put(85.00,35.00){\line(1,0){10.00}}
\put(88,28){\makebox{$E$}}
\end{picture}
\end{center}
\caption{The states of the ensemble $\cE$ ($\tilde{\cE}$) are in
system $S$ ($\tilde{S}$) and system $A$ purifies system $S$. Eve's
quantum operation is modeled by a unitary $U$; she keeps system
$E$ and forwards system $\tilde{S}$.} \label{figure-tradeoff}
\end{figure}
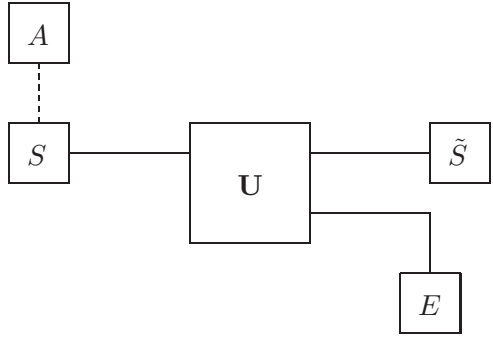

\label{subsec-info-gain-dist}
\begin{theorem} \label{theorem-quantum-mutual-gain-disturbance}
Let $d$ be the dimension. Assume that Alice sends quantum states
drawn from the ensemble $\cE$ to Eve and that she keeps a
purifying reference system $A$ at her place. Eve performs a
quantum operation on the received states, keeps a system $E$ and
forwards the ensemble $\tilde{\cE}=\{\frac{1}{2d}, \sigma_i\}$ to
Bob (see figure~\ref{figure-tradeoff}). If a disturbance of the
ensemble of at most $\epsilon$ is detected, i.e. if $\sum_i
\frac{1}{2d} \bra{\psi_i} \sigma_i \ket{\psi_i} \geq 1- \epsilon$,
then the quantum mutual information gain of an eavesdropper Eve
with respect to Alice's reference system obeys the bound:
$$ I(A;E) \leq 8 \sqrt{\epsilon} \log d +4 \mu(2\sqrt{\epsilon}),$$
where $\mu(x):=\min \{-x\log x, \frac{1}{e}\}$.
\end{theorem}

\begin{proof}
Let $\ket{\psi}^{AS}$ be the state of the system before Alice
sends $S$ to Eve. Eve's interaction is modeled by a unitary $U$
which splits system $S$, the carrier of the ensemble $\cE$, into
$\tilde{S}$ and $E$:
$$
U: \ket{\psi}^{AS}\ket{0}^E  \mapsto \ket{\tilde{\psi}}^{A\tilde{S}E}:=U^{SE} \otimes
\id_A \ket{\psi}^{AS}\ket{0}^E.
$$
This induces a CPTP map from $S$ to $\tilde{S}$, the systems
carrying $\cE$ and $\tilde{\cE}$, respectively. From
lemma~\ref{lemma-Holevo-relation} and~\ref{lemma-uncertainty}
follow the estimates
$$ \chi(\tilde{\cE}) \geq (1-4 \sqrt{\epsilon})\log d
-2\mu(2\sqrt{\epsilon})$$ and
$$I(A;S) \geq 2 \chi(\tilde{\cE}). $$
Since $ASE$ is a pure state, this leads to
\beastar
    I(A;E)  &=& 2S(A)-I(A;S)\\
            &\leq& 2 \log d  - 2(1-4 \sqrt{\epsilon})\log d
                +4\mu(2\sqrt{\epsilon}) \\
            &=&4\mu(2\sqrt{\epsilon})+ 8 \sqrt{\epsilon} \log d,
\eeastar
which is the claim that was set out to prove.
\end{proof}

\begin{corollary}
\label{cor-Holevo-gain-disturbance} For a setup that is identical
to the one in
theorem~\ref{theorem-quantum-mutual-gain-disturbance}, the Holevo
information gain of an eavesdropper Eve is bounded by
$$ \chi(\tilde{\cE}_E) \leq 4 \sqrt{\epsilon} \log d
+2\mu(2\sqrt{\epsilon})$$ and so is the accessible information
gain:
$$\I(\tilde{\cE}_E) \leq 4 \sqrt{\epsilon} \log d
+2\mu(2\sqrt{\epsilon}).$$ $\tilde{\cE}_E$ denotes the ensemble on
Eve's system $E$.
\end{corollary}

\begin{proof}
Let $\Gamma$ be the channel from $S$ to $E$. Then
$\cE_E=\Gamma(\cE)$ and by lemma~\ref{lemma-uncertainty}
$$ 2\chi(\cE_E) \leq I(A;E).$$
Inserting this estimate in
theorem~\ref{theorem-quantum-mutual-gain-disturbance} concludes
the proof of the first bound. The second bound follows from the
first one by Holevo's bound, theorem~\ref{theorem-holevo}.
\end{proof}

\noindent Recently, P. Oscar Boykin and Vwani P.~Roychowdhury
discovered a slightly better tradeoff for the accessible
information~\cite{BoyRoy04}. They proved
$$\I(\tilde{\cE}_E) \leq 4 \sqrt{ \epsilon} \log d$$
via a direct analysis of Eve's measurements. In contrast, the
calculation that led to
corollary~\ref{cor-Holevo-gain-disturbance} has emerged from a
novel uncertainty relation which is not restricted to measurements
but deals with general quantum channels. In the view of the
results on locking of classical information in quantum
states~\cite{DHLST04}, which show that the accessible information
can be significantly smaller than the Holevo information, one can
therefore regard
theorem~\ref{theorem-quantum-mutual-gain-disturbance} and the
first part of corollary~\ref{cor-Holevo-gain-disturbance} as a
significant strengthening of the behaviour found by Boykin and
Roychowdhury.

\subsection[Cheat Sensitive Quantum String Commitment]{Cheat Sensitive Quantum String Commitment\protect\footnote{Part of the results presented in this section
have been obtained in collaboration with Harry Buhrman, Hoi-Kwong
Lo, Patrick Hayden and Stephanie Wehner.}}
\label{subsec-cheat-sensitive}

An important building block in modern cryptography is \emph{bit
commitment\index{bit commitment}}. Here, two mutually mistrustful
parties wish to execute the following two-phase procedure: in the
commit phase, Alice chooses a bit $x$, jots it down on a piece of
paper, puts it in a safe, locks
 the safe and hands the locked safe over to Bob. Bob, in possession of the
safe, is certain that Alice cannot change the value of the bit,
whereas Alice is pleased to see that Bob cannot read the value as
she owns the only copy of the key to the safe. At a later point in
time, the reveal phase, Alice gives Bob the key. He can then open
the safe and retrieve the value which is written on the paper.

A direct application of bit commitment is a sealed-bit auction.
Each bidder commits to the amount of money he is willing to pay
for the item at stake. Once the auction is closed, the auctioneer
opens the commitments and determines the winner. There are a
number of more sophisticated applications of bit commitment such
as zero-knowledge proofs~\cite{Goldreich01Book} and quantum
oblivious transfer~\cite{Yao95, Crepeau94}. A more straightforward
application is coin tossing~\cite{Blum83}. Here, Alice commits to
a randomly chosen bit value $x$, then Bob announces a random bit
$y$. Subsequently, Alice reveals her value and the outcome of the
coin toss is defined to be $x\oplus y$. The coin toss is fair as
long as one party is honest and the commitment is secure.

Unfortunately, unconditionally secure classical bit
commitment\index{bit commitment!unconditionally secure} is
impossible. In the quantum realm, where Alice and Bob use quantum
computers and are connected via a quantum channel, no secure
scheme can be designed either\index{quantum bit
commitment!impossibility}\index{bit
commitment!quantum}~\cite{Mayers96BC, LoChau96BCprl,
LoChau96BCphyscom, Mayers97BC, BCMS97BC}. Alternative routes to
bit commitment have been suggested: one example is Adrian Kent's
protocol for bit commitment in a relativistic\index{bit
commitment!relativistic} setting~\cite{Kent06}. Bit commitment is
also possible in a setting in which correlations are stronger than
quantum correlations (but nevertheless in accordance with the
no-signalling principle), and where they are provided in the form
of trusted \emph{non-local boxes\index{non-local
box}}~\cite{BCUWW05}. As a third example I would like to mention
Louis Salvail's construction. He works in a quantum mechanical
scenario and has designed a scheme based on the assumption that
the number of particles that can be measured coherently is
limited~\cite{Salvail98}.

Suggestions to weaken the security demand of quantum bit
commitment have also been made. Adrian Kent and Lucien Hardy, as
well as Dorit Aharonov, Amnon Ta-Shma, Umesh Vazirani and Andrew
Yao have introduced \emph{cheat sensitive quantum bit
commitment\index{quantum bit commitment!cheat sensitive}} schemes
where cheating is allowed but will be detected with nonvanishing
probability~\cite{HarKen04, ATVY00}. Recently, I have pursued a
different direction in collaboration with Harry Buhrman, Patrick
Hayden, Hoi-Kwong Lo and Stephanie Wehner~\cite{BCHLW05}. We
investigated commitments to $n$ bits simultaneously, but allow a
small amount of cheating by Alice and Bob. Our results are
two-fold: if the Holevo information is used to quantify the
cheating, we show that no meaningful protocol is possible, thereby
extending the Mayers-Lo-Chau no-go theorem. If instead the
accessible information is used, we prove that locking of classical
information in quantum states can lead to a class of protocols
that significantly restrict the cheating of both parties. Quantum
commitments to strings have also been considered by
Kent~\cite{Kent03}. His scenario, however, differs significantly
from ours as he assumes that Alice does not commit to a
superposition of strings.

\sloppy In this subsection I introduce the framework for quantum
string commitment as introduced in~\cite{BCHLW05} and prove that
the protocol LOCKCOM($\log d, \{\id, U\}$) from this paper, where
$U$ is the quantum Fourier transform or -- for $d=2^n$ -- the
Hadamard transform $H^{\otimes n}$, is cheat sensitive against
Bob. As in the proof of the information-gain versus disturbance
tradeoff in the previous subsection, this proof is based on the
channel uncertainty relation (lemma~\ref{lemma-uncertainty}) that
has been discovered in connection with squashed entanglement.

\fussy The work presented in this section uses the multi-party
quantum communication model by Yao~\cite{Yao95} and simplified by
Lo and Chau~\cite{LoChau96BCprl}. Let Alice and Bob each have a
quantum computer. In a two-party quantum communication protocol
the two computers (initially in pure states) interact a finite
number of rounds via a quantum channel: Let $A$ denote Alice's
system, $B$ Bob's system and $C$ the channel, i.e. a system that
Alice and Bob have in turn access to. The total system is
initialised in state $\ket{0}^A\ket{0}^B\ket{0}^C$. When it is
Alice's (Bob's) turn, she (he) performs a unitary $U^{AC}$
($U^{BC}$), chosen from a set of possible unitary transformations
known to both parties beforehand. As the initial state is pure and
the operations are unitary, the state of the total system is pure
at any time. Since every measurement can be modeled by a unitary
operation followed by tracing out part of the system, without loss
of generality, the total protocol can be regarded as a sequence of
unitary transformations with a partial trace operation spared
until the end.

The classical outcome of a measurement is saved in a designated
part of the system, which can be read off at the end of the
protocol.

A \emph{quantum string commitment (QSC) protocol}\index{quantum
string commitment} is a quantum communication protocol between two
parties, Alice (the committer) and Bob (the receiver), which
consists of three phases:
\begin{itemize}
\item (Commit Phase) If both parties are honest Alice chooses
a string $x \in \01^n$. Alice and Bob communicate and in the end
Bob holds evidence state $\rho = \sum_x p_x \rho_x$. ($p_x$ is the
probability of $x$ for Bob, $\rho_x$ the state corresponding to a
commitment of $x$)
\item (Reveal Phase)
If both parties are honest Alice sends information to Bob which is
sufficient to reveal $x$.
\item (Confirmation Phase) If both parties are honest they `accept'.
\end{itemize}
Let $K$ denote the quantity which measures the amount of quantum
information (in bits) of the ensemble $\cE=\{p_x, \rho_x\}$. Since
we use the standard model for two-party quantum protocols, there
is such an ensemble $\cE$ for any QSC protocol. Later, $K$ will be
the accessible information $\I(\cE)=\max_Y I(X;Y)$. The maximum is
taken over all random variables which are outcomes of a
measurement on $\cE$.

In the following, the definitions have been altered slightly in
order to accommodate for cheat sensitivity against Bob.
Disregarding the remarks on cheat sensitivity, the results
in~\cite{BCHLW05} remain valid for this definition.

A quantum string commitment protocol is an
\emph{$(n,a,b)$-$K$-cheat sensitive quantum string commitment
protocol against Alice (Bob)}\index{quantum string
commitment!cheat sensitive} if
\begin{itemize}
\item (Concealing) Bob's information at the end of the commit phase measured
in terms of $K$ is no larger than $b$: $K(\cE) \leq b$.
\item (Binding)
$\sum_{x \in \01^n} \tilde{p}_x \leq 2^a$ where $\tilde{p}_x$ is
the probability that Alice is able to successfully reveal $x \in
\01^n$ at the reveal stage.
\item (Cheat sensitivity against Alice (Bob)) If Alice (Bob) does not follow
the protocol, there is a nonzero probability that she (he) will be
detected by Bob (Alice).
\end{itemize}

\noindent In~\cite{BCHLW05} a class of protocols, based on locking
of classical information in quantum states~\cite{DHLST04}, has
been introduced. The protocols are defined in terms of a set of
unitaries ${\cal U}$ acting on a $d$-dimensional space. The
following presentation incorporates cheat sensitivity against Bob.

\begin{protocol}{CS-Bob-LOCKCOM($n,
{\cal U}$)}{}\index{LOCKCOM}\label{lockcom}
\item Commit phase: Alice randomly chooses the string
$x \in \01^n$ and a unitary $U_r$ from a set of unitaries ${\cal
U}$ known to both Alice and Bob. She sends the state $U_r
\ket{x}$.
\item Reveal phase: Alice sends $r$ to Bob, he applies
$(U_r)^\dagger$ to the state that he received from Alice and
measures in the computational basis. His outcome is denoted by
$y$.
\item Confirmation phase: Bob sends $y$ to Alice. If Alice is honest, and if
$x=y$ she declares `accept' otherwise `abort'.
\end{protocol}

\noindent In~\cite{BCHLW05} it has been proven that
CS-Bob-LOCKCOM($\log d, \{\id, U\}$), where $U$ is the quantum
Fourier transform or -- for $d=2^n$ -- the Hadamard transform
$H^{\otimes n}$ is a $(\log d, 1, \frac{\log d}{2})$-$\I$-quantum
string commitment protocol. Here, I prove that a dishonest Bob is
detected whenever he has obtained a non-zero amount of
\emph{classical} information about $x$ \emph{before} the reveal
stage. More precisely, I give a tradeoff for cheat detection
versus accessible information gain against a dishonest Bob, with
the property that every nonzero classical information gain leads
to a nonzero detection probability of Bob. This means that the
scheme is cheat sensitive against Bob.

\begin{figure}
\begin{center}
\includegraphics[height=0.55\textheight]{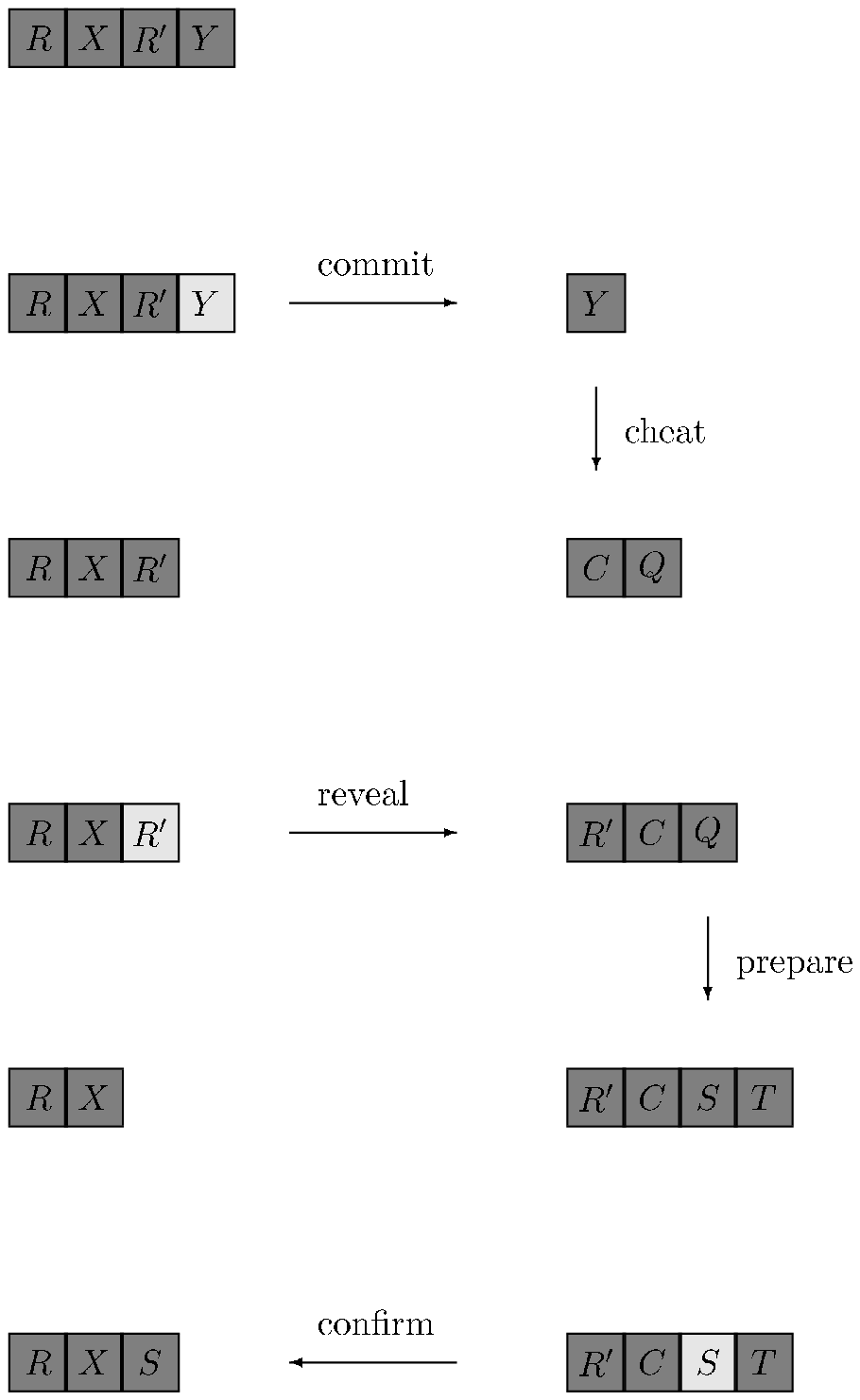}
\vspace{-0.5cm}
\end{center}
\caption{Execution of LOCKCOM with honest Alice on the left and
cheating Bob on the right. Time flows downwards.}
\label{figure-cheat}
\end{figure}

The following is a description of the sequence of events if Alice
is honest and if Bob applies a general cheating strategy (see also
figure~\ref{figure-cheat}).

\begin{itemize}
\item The commit phase of the protocol $LOCKCOM(\log d, \{\id, U\})$ is equivalent to the following procedure:
Alice prepares the state
$$\ket{\psi}:=\frac{1}{\sqrt{2d}} \sum_{x,r} \ket{x}^X\ket{r}^R\ket{r}^{R'}U^r
\ket{x}^Y$$ on the system $XRYR'$ and sends system $Y$ (over a
noiseless quantum channel) to Bob. It is understood that $U^0=\id$
and $U^1=U$. Note that $R'$ contains an identical copy of $R$ and
corresponds to the \emph{reveal information}.
\item Bob's most general cheating operation is given by a unitary matrix $V_{cheat}$ that splits the system $Y$
into a classical part $C$ and a quantum part $Q$. The classical
part contains by definition the information gathered during
cheating. I will assume that $Q$ contains a copy of $C$. This does
not restrict the claim for generality of the attack, since $C$ is
assumed to be classical.
$$V_{cheat}: Y \rightarrow CQ$$
The map $V_{cheat}$ followed by the partial trace over $Q$ is
denoted by $\Lambda^C$ and likewise $V_{cheat}$ followed by the
partial trace over $C$ is denoted by $\Lambda^Q$.
\item Alice sends the reveal information $R'$ to Bob.
\item Bob applies a preparation unitary $V_{prepare}$ to his
system. Since $Q$ contains a copy of $C$, the most general
operation can be taken to act only on $R'Q$:
$$V_{prepare}: R'Q \rightarrow R'ST.$$
Bob then sends $S$ to Alice and keeps $T$.
\item Alice measures $S$ in the computational basis and compares
the outcome to her value in $X$. If the values do not agree, we
say that \emph{Alice has detected Bob cheating}. The probability
for this happening is given by
$$ \frac{1}{d}\sum_{x=1}^d \left(1-\tr \proj{x} \rho^S_x\right),$$
where $\rho^S_x:=\tr_{XRR'T} \proj{x} \proj{\psi}^{XRR'ST}$, where
$\ket{\psi}^{XRR'ST}$ is the pure state of the total system after
Bob's application of $V_{prepare}$.
\end{itemize}

\noindent Note that Alice measures in the computational basis
since for honest Bob $V_{prepare}=\sum_{r' \in \01} \proj{r'}
\otimes (U^r)^\dagger$, in which case his outcome agrees with the
committed value of an honest Alice.

Before continuing let me define ensembles in dependence of the
classical information contained in $XR$, i.e. for $Z \in \{C, Q
\}$, define

    \bestar
        \cE^Z_{r}:=\{p_{x}, \rho_{xr}^Z\} \textrm{ with }\rho_{xr}^Z:=\frac{1}{p_x p_r}\tr_{XRR'CQ\backslash Z} \proj{xr}
        \proj{\psi}^{XRR'CQ}
    \eestar
and for $Z \in \{S, T \}$ let
    \bestar
        \cE^Z_{r}:=\{p_{x}, \rho_{xr}^Z\} \textrm{ with } \rho_{xr}^Z:=\tr_{XRR'CST\backslash Z} \proj{xr}
        \proj{\psi}^{XRR'CST}.
    \eestar
Sometimes we are only interested in the ensemble averaged over the
values of $r$: for $ Z \in \{C, Q, S, T\}$
    \be
        \cE^Z:=\{p_{x},  \rho_x^Z\} \textrm{ where } \rho_x^Z= \half \left(\rho_{x0}^Z+
        \rho_{x1}^Z \right).
    \ee

\begin{theorem} \label{theorem-Bob-cheat-sensitive}
If Bob is detected cheating with probability less than $\epsilon$,
then his \emph{classical information gain} obeys
$$\chi(\cE^C) \leq
4\sqrt{\epsilon}\log d +2\mu(2\sqrt{\epsilon})$$.
\end{theorem}

\begin{proof}
Let $\cE_0$ and $\cE_1$ be defined as in lemma
\ref{lemma-uncertainty}. In the commit phase of the protocol,
Alice chooses one of the ensembles (each with probability
$\half$), and one of the states in the ensemble (each with
probability $\frac{1}{d}$). The justifications for the following
estimate are given in a list below.
\bea
 \label{eq-rewriting1}\chi(\cE^C_{0})+\chi(\cE^C_{1}) &=&\chi(\Lambda^C(\cE_{0}))+\chi(\Lambda^C(\cE_{1}))\\
 \label{eq-rewriting2}&\leq& I(XRR';C) \\
 \label{eq-rewriting3}&=&2S(XRR')-I(XRR';Q)\\
 \label{eq-rewriting4}&\leq& 2S(XRR')-\chi(\Lambda^Q(\cE_{0}))-\chi(\Lambda^Q(\cE_{1}))\\
 \label{eq-rewriting5}&=& 2S(XR)-\chi(\cE^Q_{0})-\chi(\cE^Q_{1})\\
 \label{eq-rewriting6}&\leq& 2S(XR)-\chi(\Lambda^S_0(\cE^Q_{0}))-\chi(\Lambda^S_1(\cE^Q_{1}))\\
 \label{eq-rewriting7}&=& 2S(XR)-\chi(\cE^S_{0})-\chi(\cE^S_{1})\\
 \label{eq-rewriting8}&\leq&2S(XR)-2\chi(\cE^S).
\eea
The justifications:
\begin{itemize}
\item Equality~(\ref{eq-rewriting1}): By definition of the string commitment scheme and the map $\Lambda^C$: $\cE^C_r=\{ p_x, \rho_{xr}^C\}=\{ p_x,
\Lambda^C( U^r\proj{x}(U^\dagger)^r)\}=: \Lambda^C(\cE_r)$.
\item Inequality~(\ref{eq-rewriting2}): Application of lemma~\ref{lemma-uncertainty} for the map $\Lambda^C$. Note that system
$XRR'$ is a reference system for the completely mixed state on
system $Y$ on which the channel $\Lambda^C$ is applied. Hence
$I(\tau; \Lambda^C)=I(XRR';C)$.
\item Equality~(\ref{eq-rewriting3}): Simple rewriting of the entropy
terms making use of the definition of quantum mutual information
and the purity of $XRR'CQ$.
\item Inequality~(\ref{eq-rewriting4}): Application of lemma
\ref{lemma-uncertainty} for the map $\Lambda^Q$. Note that system
$XRR'$ is a reference system for the completely mixed state on
system $Y$ on which the channel $\Lambda^Q$ is applied. Hence
$I(\tau; \Lambda^Q)=I(XRR';Q)$.
\item Equality~(\ref{eq-rewriting5}): $R'$ is a copy of $R$: $S(XRR')=S(XR)$. By definition of the string commitment scheme and the map $\Lambda^Q$: $\cE^Q_r=\{ p_x,
\rho_{xr}^Q\}=\{ p_x, \Lambda^Q( U^r\proj{x}(U^\dagger)^r)\}.$
\item Inequality~(\ref{eq-rewriting6}) and equality~(\ref{eq-rewriting7}): follow from the data processing inequality
$\chi(\Lambda^S(\cE^Q_r)) \leq \chi(\cE^Q_r)$ and from the
definition $\Lambda^S(\cE^Q_r)=\cE^S_r$.
\item Inequality~(\ref{eq-rewriting8}): Finally $\cE^S:=\{p_x,
\rho^S_x:=\half \left(\rho^S_{x0}+\rho^S_{x1} \right)\}$, which by
the concavity of von Neumann entropy implies $\chi(\cE^S) \leq
\half \left(\chi(\cE^S_0)+\chi(\cE^S_1)\right)$.
\end{itemize}
If Bob is detected cheating with probability less than $\epsilon$,
then by lemma~\ref{lemma-Holevo-relation} the Holevo quantity
$\chi(\cE_S)$ of the ensemble given in $S$ that Bob sends to Alice
obeys
    \be \label{eq-0}
        \chi(\cE_S) \geq (1-4 \sqrt{\epsilon})\log
        d-2\mu(2\sqrt{\epsilon}).
    \ee
Inserting inequality~(\ref{eq-0}) into
inequality~(\ref{eq-rewriting8}) and noting that $S(XR)=S(Y)=\log
d$ proves the claim.
\end{proof}

\noindent This proves cheat sensitivity against Bob of the
simplest protocol in the family LOCKCOM. It has been shown
in~\cite{HLSW04} that a higher number of bases can lead to a
stronger locking effect (see proof of
proposition~\ref{squash-lock-gap}). I expect strong locking to
lead to novel information-gain versus disturbance tradeoffs and --
concerning the quantum string commitment protocols based on this
effect~\cite{BCHLW05} -- to allow for improved cheat sensitivity.

\section{Conclusion}
\label{section-squashed-conclusion} In
chapter~\ref{chapter-entanglement} I have reviewed the literature
on entanglement measures with the aim to consolidate the knowledge
in this field. This has been done with help of tables and graphs,
and led to the identification of specific open problems (see
table~\ref{table-properties}) -- some of which were resolved later
on (section~\ref{section-specific-results}). Furthermore, I have
carried out an investigation of more general connections among the
properties with focus on continuity and convexity properties
(subsection~\ref{subsec-relations-among}). The introduction of
squashed entanglement in chapter~\ref{chapter-squashed} also
contributes to this discussion: it shows that superadditivity,
additivity and asymptotic continuity are compatible properties of
entanglement monotones. This puts the main feature of squashed
entanglement, its additivity properties, in the centre of the
attention. Here, the simplicity of the proofs is remarkable --
proofs that are mostly based on strong subadditivity of von
Neumann entropy combined with the unboundedness of the
minimisation (see e.g.~proposition~\ref{prop-super-additive}). An
open question that emerges from this work is whether or not this
minimisation can -- without loss of generality -- be restricted to
finite-size extensions. Support for this conjecture comes from the
intrinsic information (the classical counterpart to squashed
entanglement), because the minimisation in the intrinsic
information can -- as a consequence to Carath{\'e}odory's
theorem\index{Carath\'eodory's theorem} -- be restricted to
channels whose output dimension equals the input
dimension~\cite{ChReWo03}. It is therefore natural to conjecture
that taking $\dim \cH_E \leq (\dim \cH_{AB})^2$ is always
sufficient in the minimisation of squashed entanglement. If this
is true, then squashed entanglement is strictly positive on all
entangled states. This has implications for the study of
entanglement cost and the separability of quantum states. The
monogamy of entanglement in a three party scenario has been
examined by Masato Koashi and Andreas Winter who used the fact
that squashed entanglement obeys
$$ \sq(\rho^{AB})+\sq(\rho^{AC})\leq \sq(\rho^{A,BC})$$
in order to conclude that
$$E_D(\rho^{AB})+E_D(\rho^{AC})\leq E_C(\rho^{A,BC})$$
holds for any state $\rho^{ABC}$~\cite{KoaWin04}.

Squashed entanglement -- and in particular the observation that it
is bounded by $\half I(A;B)$ -- has also emerged as a useful tool
in proving separations among channel capacities~\cite{BDSS04} and
generic properties of bipartite quantum states~\cite{HaLeWi04}.
Proposition~\ref{squash-lock-gap} gives an example for the
behaviour found by Patrick Hayden, Debbie Leung and Andreas Winter
that randomly chosen states typically have a large gap between
distillable entanglement and entanglement cost. In terms of upper
and lower bounds, the chain of inequalities
$$ E_D \leq K_D \leq \sq \leq E_C \leq E_F$$
places squashed entanglement in an even better position
(figure~\ref{figure-relations}), which implies that not only $E_D
<< E_C$ but even $K_D << E_C$ happens generically.

\vspace{0.5cm}
\noindent Motivation for the presented study was the
relation between entanglement distillation and secret key agreement
that emerged from Gisin and Wolf's conjecture of bound
information~\cite{GisWol00}. More generally, this work aims at
improving the understanding of the relation between bipartite
quantum states and secrecy, which was sparked off by the invention
of entanglement-based key distribution~\cite{Ekert91}. I see the
last part of chapter~\ref{chapter-squashed} as return of value from
the study of entanglement measures to cryptography. Here, a
calculation of squashed entanglement on a set of quantum states
motivated the search for an entropic uncertainty relation for
quantum channels (lemma~\ref{lemma-uncertainty}). In
subsection~\ref{subsec-cheat-sensitive} this relation was shown to
imply a novel information-gain versus disturbance tradeoff which
removes an obstacle in the application of such tradeoffs to prepare
and measure quantum key distribution\index{quantum key
distribution!prepare and measure} schemes: it removes the
unrealistic assumption that Eve actually performs a measurement and
considers general quantum instruments instead. A second application
of the channel uncertainty relation resulted in a proof of the cheat
sensitivity of a class of string commitment protocols proposed
in~\cite{BCHLW05} (subsection~\ref{subsec-tradeoff}), and a
conjecture predicting this behaviour for much wider class of
protocols.

\backmatter

\chapter{Concluding Remarks} \label{chapter-conclusion}

In this PhD thesis I have presented a study of the structure of
bipartite quantum states. The results have been obtained in the
context of quantum information theory, a theory that unifies
quantum mechanics with classical information theory. The research
presented in both parts of this thesis is motivated by recent
developments in classical information theory, and brought to
quantum mechanical grounds with tools from representation,
estimation and entanglement theory.

Part~\ref{part-group} of this thesis is inspired by the way
Shannon entropy inequalities reveal the structure of a set of
correlated random variables; a concrete starting point being Chan
and Yeung's recent discovery that entropy inequalities stand in a
one-to-one relation to inequalities among the sizes of finite
groups and subgroups (see
subsection~\ref{subsection-classical-analogue}).
Part~\ref{part-crypto} has its origin in unconditionally-secure
classical cryptography and the quantification of the amount of
secret correlations in a triple of random variables with help of
the intrinsic information, a function of Shannon entropies (see
subsection~\ref{subsec-secret-key} and
~\ref{subsec-squashed-definition}).

Thus, both parts originate in the classical concept of Shannon
entropy and quickly develop into independent entities, each
following the mandate of their respective topics: group theory and
cryptography. The unifying element is the von Neumann entropy, the
quantum analogue to Shannon entropy. Part~\ref{part-group} shows
how subadditivity of von Neumann entropy arises from relations
among representations of the symmetric group
(corollary~\ref{corollary-subadditive}). This extends to a new
approach in the study of entropy inequalities with the immediate
challenge to discover a representation-theoretic proof of strong
subadditivity (of von Neumann entropy),
$$ S(AE)+S(BE) \geq S(ABE)+S(E).$$
In part~\ref{part-crypto} the strong subadditivity of von Neumann
entropy is omnipresent: the new entanglement measure is defined as
(one half times) the difference of the inequality from equality
and almost all proofs of its properties are directly based on
strong subadditivity. A challenge that arises from this work is to
analyse the set of states that are close to achieving equality.
Such a result would generalise the known exact equality
conditions~\cite{HJPW04}, resolve
conjecture~\ref{conjecture-squashed-pos} and result in new
quantitative insights into entangled bi- as well as multipartite
quantum states.

This conclusion ends with separate discussions of the two parts
that make up this thesis.

\paragraph*{Insights from Group Theory.} In part~\ref{part-group} the
spectra of quantum states, the quantum analogue to probability
distributions, are analysed by means of representation theory of
finite groups and Lie groups. I have focused on the following
question: Given a triple $(r^A, r^B, r^{AB})$ does there exist a
state $\rho^{AB}$ with
\bestar \label{eq-concl-triple}
    (r^A, r^B, r^{AB})=(\spec \rho^A, \spec \rho^B, \spec
    \rho^{AB})\; ?
\eestar
If the answer is affirmative, $(r^A, r^B, r^{AB})$ is called an
admissible spectral triple. The main result presented here is the
discovery of the asymptotic equivalence between the problem of
deciding whether or not a given spectral triple is admissible
(problem~\ref{prob-comp}) and the problem of determining when the
Kronecker coefficient of the symmetric group is nonvanishing
(problem~\ref{prob-Kronecker}). The Kronecker coefficient $g_{\mu
\nu \lambda}$ is defined by the expansion
    $$
        V_\mu \otimes V_\nu \cong \bigoplus_\lambda g_{\mu \nu \lambda}
        V_\lambda,
    $$
where $V_\mu$, $V_\nu$ and $V_\lambda$ are irreducible
representations of the symmetric group $S_k$. By `asymptotic
equivalence' I mean that, given an admissible spectral triple
$(r^A, r^B, r^{AB})$, there is a sequence $g_{\mu_i \nu_i
\lambda_i} \neq 0$ of $S_{k_i}$ such that $(\frac{\mu_i}{k_i},
\frac{\nu_i}{k_i}, \frac{\lambda_i}{k_i})$ converges to $(r^A,
r^B, r^{AB})$ and, conversely, if $g_{\mu \nu \lambda} \neq 0$,
then $(\frac{\mu}{k}, \frac{\nu}{k}, \frac{\lambda}{k})$ is an
admissible spectral triple.

The discovery of this equivalence is astonishing as it ties
together two apparently remote -- and moreover prominent -- open
problems. Problem~\ref{prob-comp} is closely related to the
one-particle instance of the $N$-representability problem, a
problem which is of fundamental importance to quantum chemistry
and the theory of condensed matter (see
section~\ref{sec-spectra-intro} and the book by Coleman and
Yukalov~\cite{ColYuk00}). Recently, problem~\ref{prob-comp} has
attracted much attention in its own right. In the context of
quantum information theory, spectral inequalities that determine
the set of admissible spectral triple in special dimensions have
been discovered: The two-qubit instance of problem~\ref{prob-comp}
has been solved by Bravyi~\cite{Bravyi04} (see
subsection~\ref{subsec-spec-two-qubit} for a proof using the newly
discovered asymptotic equivalence), the compatibility of $n$
qubits with an $n$-qubit pure state by Atsushi Higuchi, Tony
Sudbery, and Jason Szulc~\cite{HiSuSz03} and
Bravyi~\cite{Bravyi04} and the compatibility of three qutrits with
a $3 \times 3 \times 3$-dimensional pure state by
Higuchi~\cite{Higuchi03}. In contrast, the asymptotic equivalence
demonstrated here is valid for any $m \times n$-dimensional system
and also extends to the question of compatibility of a finite set
of $m_i$-dimensional states $\rho_i$ with an $m_1 \times m_2
\times \cdots $-dimensional pure state. Needless to say I have not
given a description of the set of admissible spectral triple. The
mapping of problem~\ref{prob-comp} to
problem~\ref{prob-Kronecker}, however, opens a new way to its
solution and can be regarded as a significant progress in
understanding compatibility requirements for local and global
spectra.

\sloppy Shortly after announcing the core of the presented work,
Klyachko made public a complete set of inequalities that describe
the polytope of solutions~\cite{Klyachko04} and, using related
methods, Daftuar and Hayden showed how to calculate the
inequalities for the compatibility of only one margin with the
overall state~\cite{DafHay04}. Klyachko achieves his encompassing
result by noting that problem~\ref{prob-comp} amounts to the
``decomposition of [the] projection of a coadjoint orbit of group
$\SU(\cH_A \otimes \cH_B)$ into coadjoint orbits of subgroup
$\SU(\cH_A)\times \SU(\cH_B)$''. This places
problem~\ref{prob-comp} in the framework of geometric invariant
theory (GIT) and the work of Arkady Berenstein and Reyer Sjamaar
in particular~\cite{BerSja00}. As an illustration of this method I
have applied a theorem by Kirwan in order to show that the
admissible spectral triple form a convex polytope, a surprisingly
nontrivial result (subsection~\ref{subsec-spec-convexity}).
Furthermore, Klyachko observes that Heckman's
work~\cite{Heckman82} (see also~\cite{BerSja00}) can be used to
derive a form of asymptotic equivalence of problem~\ref{prob-comp}
and problem~\ref{prob-Kronecker} similar to the one presented
here. Allen Knutson has kindly explained to me how this proof
emerges from the equivalence of symplectic and GIT quotient, the
``big gun'' as he writes in his paper on Horn's
problem~\cite{Knutson00}.

\fussy In contrast to the advanced mathematical tools applied by
Klyachko, the work presented here relies almost exclusively on
standard textbook material from the representation theory of
finite and unitary groups. More surprising than the result itself
is therefore the conceptual simplicity and the elementary nature
of the way it is proved: Asymptotic equivalence is reached through
an application of a spectral estimation theorem to the tensor
power of a bipartite state and its margins
(subsection~\ref{subsec-spec-comp}).

Curiously, the presented method bypasses the use of geometry. How
is this possible? Knutson has suggested that the answer lies in
the use of the symmetric group. This points to the Schur-Weyl
duality, which moves us from the unitary group, the natural object
when considering the problem in the geometric context, to the
symmetric group where the proof is reduced to an analysis of the
growth of dimensions of irreducible representations.

Since the equality of symplectic and GIT quotient (Kirwan and
Ness) can be applied in many contexts, it is natural to pose the
following questions: Can the core ideas of the proof presented in
this thesis be applied elsewhere, and can they lead to a new
understanding of Kirwan's and Ness' theorem? A partial answer to
these questions is given in section~\ref{subsec-spectra-horn},
where I have proved the asymptotic equivalence of Horn's problem
(problem~\ref{prob-Horn}),\footnote{In Horn's problem the question
is asked whether or not to a triple $(r^A, r^B, r^{AB})$ one can
find Hermitian operators $A$ and $B$ such that $(r^A, r^B,
r^{AB})=(\spec A, \spec B, \spec A+B)$} and the problem of
deciding whether or not the Littlewood-Richardson coefficient is
nonzero (problem~\ref{prob-LR}). I have thereby given a novel
proof for this well-known asymptotic equivalence~\cite{Lidskii82,
Heckman82, Klyachko98, Knutson00}\footnote{In fact, if integral
spectra are considered, equivalence holds as a consequence of the
proof of the saturation conjecture for $\GL(d)$ by Knutson and
Tao~\cite{KnuTao99} (see end of
subsection~\ref{subsec-spectra-horn},
page~\pageref{saturation-conjecture})} and shown how versatile the
developed method is.

With the asymptotic equivalence of problem pair~\ref{prob-comp}
and~\ref{prob-Kronecker}, and problem pair~\ref{prob-Horn}
and~\ref{prob-LR} there exist two examples which have allowed a
bypassing of the equality of GIT and symplectic quotient. It is
the challenge of this work to explore and understand these results
better and to add new examples in order to develop this work into
a general technique that can take our understanding of spectral
and group-theoretic problems on a new level.

\paragraph*{Insights from Cryptography.}

Part~\ref{part-crypto} of this thesis has its starting point in
secret key agreement from random variables by public discussion.
Here, Alice and Bob wish to extract a common bit-string starting
from correlated randomness such that Eve, a wiretapper with access
to related correlated randomness, is ignorant of this bit-string.
In order to achieve agreement on a secure bit-string -- or key --
Alice and Bob have an unlimited amount of public communication at
their disposal. The work of Gisin and Wolf points out the analogy
of this scenario to entanglement distillation, where three players
share a pure tripartite quantum state from which Alice and Bob
wish to extract states of the form $\frac{1}{\sqrt{2}}
\ket{00+11}$. This formulation of entanglement distillation is
commonly adopted when discussing entanglement-based quantum key
distribution and contrasts with the predominant view that
entanglement distillation is a bipartite mixed state scenario.

In this work I have immersed myself in the adversarial tripartite
scenario. Inspired by the intrinsic information, a correlation
measure in secret key agreement, Eve is given the role to squash
Alice and Bob's quantum mutual information. The result is the
definition of a new measure for entanglement, \emph{squashed
entanglement} (subsection~\ref{subsec-squashed-definition}).
Squashed entanglement is at the centre of part~\ref{part-crypto}
of this thesis and dictates its build-up.

I have started this part with a review of the axiomatic -- or
property-driven -- approach to entanglement measures
(chapter~\ref{chapter-entanglement}). A list of properties
including monotonicity, additivity and continuity has been
compiled, their mutual relation reviewed and their importance from
the point of view of operationally-defined measures, such as
distillable entanglement and entanglement cost, has been
discussed. The bulk of the review is contained in a table that
summarises the knowledge of whether or not a specific entanglement
measure satisfies a given property. A large number of measures
fitting this approach is thereby characterised, and their mutual
relations are exhibited in two graphs. To my knowledge this review
is the most comprehensive of its kind. The tables and graphs point
out the positions where our data is incomplete and are intended to
serve as a resource for further study. Regarding this work, the
assembly of table~\ref{table-properties} has stimulated my
interest in continuity requirements that has led to the proof of
asymptotic continuity of the regularised relative entropy of
entanglement.

\sloppy The review has also set the scene for the proposal of the
new measure. In chapter~\ref{chapter-squashed}, squashed
entanglement
    $$
        \sq(\rho^{AB})=\inf_{\tr_E
        \rho^{ABE}=\rho^{AB}} \half I(A;B|E)_\rho
    $$
has been introduced and all properties from the previous
discussion, apart from monotonicity under separable
operations,\footnote{Monotonicity under separable operations
remains also undecided for any other LOCC monotone that is not
known to be a PPT monotone.} have either been proved to be
satisfied for squashed entanglement or they have been shown to
fail. Squashed entanglement is the only known entanglement
monotone, which is asymptotically continuous, convex, strongly
superadditive, subadditive and therefore additive. As a
consequence, for the first time, superadditivity, additivity and
asymptotic continuity have been shown to be compatible properties
of an entanglement monotone. Moreover, the proofs for the
exceptional additivity properties have been obtained with little
effort as they are only based on strong subadditivity of von
Neumann entropy combined with the unboundedness of the
minimisation (see conjecture~\ref{conjecture-squashed-pos} and
subsection~\ref{section-squashed-conclusion} for the possibility
to restrict the minimisation to finite-dimensional $E$). This
stands in sharp contrast to the unsolved additivity (or even
extensitivity)\footnote{The additivity conjecture reads
$E(\rho\otimes \sigma)\stackrel{?}{=} E(\rho)+E(\sigma)$, whereas
extensitivity only requires $E(\rho^{\otimes n})=nE(\rho)$ for all
$n \in \naturals$.} conjectures for correlation measures such as
entanglement of purification, entanglement of formation and the
related channel capacity conjectures~\cite{Shor03, Matsumoto05}.
Remarkably, additivity is also not known to be satisfied for any
-- naturally extensive -- operationally-defined measure and might
even fail for distillable entanglement~\cite{ShSmTe01}.

\fussy Insights have been discussed in this thesis which are not
only \emph{from} cryptography but also \emph{for} cryptography
(section~\ref{section-evaluating-comm-gain}). A calculation of
squashed entanglement for a set of quantum states has motivated
the search for an entropic uncertainty relation for quantum
channels. This inequality has led to a novel information--gain
versus disturbance tradeoff which has removed an obstacle in the
application of such tradeoffs to quantum key distribution: the
assumption that Eve performs a measurement is replaced by the most
general operation, a quantum instrument. A second application of
the channel uncertainty relation has resulted in a proof of cheat
sensitivity of a class of string commitment protocols proposed
in~\cite{BCHLW05}.

Squashed entanglement's relation to other entanglement measures is
summarised in the following sequence of inequalities
(section~\ref{section-squashed-between}):
$$ E_D \leq K_D \leq \sq \leq E_C \leq E_F.$$
Squashed entanglement takes the central position mainly due to its
additivity ($\sq \leq E_C$) and, but not only, its asymptotic
continuity ($K_D \leq \sq$). The only other measure sharing this
place is the displeasing regularisation of the relative entropy of
entanglement with respect to separable states. The middle position
unfolds in several applications for squashed entanglement, such as
separating channel capacities~\cite{BDSS04}, the study of generic
properties of bipartite quantum states~\cite{HaLeWi04} and
investigating the monogamy of entanglement~\cite{KoaWin04}.

The second inequality, $K_D \leq \sq$, is worth a closer look.
Recall that Gisin and Wolf have suggested a close analogy between
key distillation from random variables and entanglement
distillation from bipartite quantum states, which started the
search for \emph{bound information}, the classical analogue to
bound entanglement~\cite{GisWol00}. This conceptual analogy has by
now become folklore and in this light, squashed entanglement --
the analogue to intrinsic information -- would be expected to
bound distillable entanglement just as the intrinsic information
bounds the secret key rate. It does, but in fact in a much
stronger way: it bounds the distillable key. Should we therefore
amend the picture suggested by Gisin and Wolf in favour of an
analogy between key distillation from random variables and key
distillation from bipartite quantum states? The failing to prove
the existence of bound information or the discovery that bound
entangled states can lead to a secure bit-string could be seen as
further hints to an affirmative answer~\cite{HHHO05a}. Admittedly,
these indications are far from being conclusive, but they will
influence our understanding of key distillation, which is
currently transforming into a true tripartite mixed state scenario
in which an answer might be concealed~\cite{DevWin05, ChrRen04,
CHHHLO05}.

%

\newcommand{\etalchar}[1]{$^{#1}$}

\printindex
\newpage
\thispagestyle{empty} \vspace*{3.8cm}

\begin{figure}[h]
\begin{center}

\includegraphics[height=12cm]{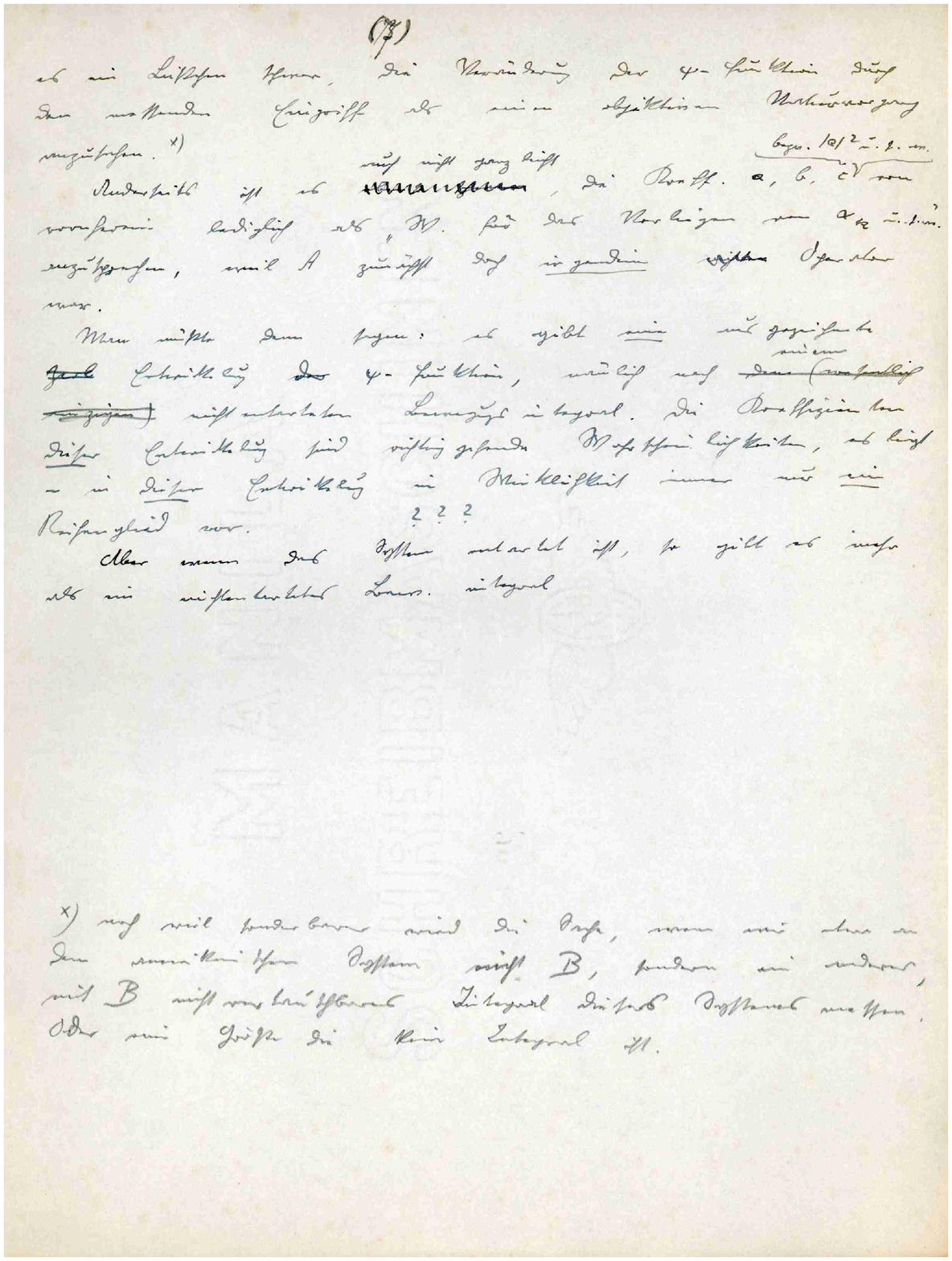}
\end{center}
\end{figure}

\end{document}